\ifx\texorpdfstring\undefined\def\texorpdfstring#1#2{#1}\fi
\ifx\unichar\undefined\def\unichar#1{#1}\fi
\documentclass[aps,prd,10pt,twocolumn,showpacs,showkeys,preprintnumbers,floatfix,nofootinbib,superscriptaddress,unicode]{revtex4-1}
\usepackage{silence}
\usepackage{amsfonts} 
\usepackage{amssymb} 
\usepackage{amsmath} 
\usepackage{blkarray}
\usepackage{graphicx} 
\usepackage{subfigure} 
\usepackage{array} 
\usepackage{dcolumn} 
\usepackage{bm} 
\usepackage{latexsym} 
\usepackage{longtable} 
\usepackage{hyperref} 
\usepackage{bm}
\usepackage{slashed}
\usepackage{bbold}
\usepackage{color}
\usepackage{multirow}
\usepackage{rotating}
\usepackage{comment}
\usepackage{adjustbox}

\graphicspath{{./figs_2pt/}{./figs_AFF/}{./figs_AFF-deltaM/}{./figs_AFF_CCFV/}{./figs_AFF_Q2/}{./figs_AFF_comp/}{./figs_Aesc/}{./figs_Aesc/dup/}{./figs_Fpi_CCFV/}{./figs_PCAC/}{./figs_PPD/}{./figs_VFF/}{./figs_VFF-deltaM/}{./figs_VFF_CCFV/}{./figs_VFF_Q2/}{./figs_VFF_all/}{./figs_VFF_strategy/}{./figs_charges/}{./figs_charges/3pt/}{./figs_charges/CCFV/}{./figs_charges/CCFVz/}{./figs_charges/UandD/}}

\newcommand{\gpNN}{\mathop{g_{\pi { NN}}}\nolimits}
\newcommand{\MeV}{\mathop{\rm MeV}\nolimits}

\newcommand{\rAsq}{\langle  r_A^2 \rangle}

\newcommand{\expv}[1]{\langle#1\rangle}

\newcommand{\rEsq}{\mathop{ \langle r_E^2 \rangle }\nolimits}
\newcommand{\rMsq}{\mathop{ \langle r_M^2 \rangle }\nolimits}

\providecommand{\abs}[1]{\lvert#1\rvert}
\providecommand{\matrixe}[3]{\langle#1\lvert#2\rvert#3\rangle}

\definecolor{green}{rgb}{0.1, 0.8, 0.1}

\newcolumntype{.}[1]{D{.}{.}{#1}}

\newcommand\suptwo{\unichar{"B2}}
\usepackage{textalpha}
\newcommand\tildeabove{\unichar{"303}}

\begin{document}

\title{Precision Nucleon Charges and Form Factors Using 2+1-flavor Lattice QCD}
%
%


\author{Sungwoo Park}
\email{sungwoo@jlab.org}
\affiliation{Los Alamos National Laboratory, Theoretical Division T-2, Los Alamos, NM 87545}
\affiliation{Center for Nonlinear Studies, Los Alamos National Laboratory, Los Alamos, New Mexico 87545, USA}

\author{Rajan Gupta}
\email{rajan@lanl.gov}
\affiliation{Los Alamos National Laboratory, Theoretical Division T-2, Los Alamos, NM 87545}

\author{Boram Yoon}
\email{boram@lanl.gov}
\affiliation{Los Alamos National Laboratory, Computer Computational and Statistical Sciences, CCS-7, Los Alamos, NM 87545}

\author{Santanu Mondal}
\email{santanu.sinp@gmail.com}
\affiliation{Los Alamos National Laboratory, Theoretical Division T-2, Los Alamos, NM 87545}

\author{Tanmoy Bhattacharya}
\email{tanmoy@lanl.gov}
\affiliation{Los Alamos National Laboratory, Theoretical Division T-2, Los Alamos, NM 87545}

\author{Yong-Chull Jang}
\email{ypj@bnl.gov}
\affiliation{Physics Department, Columbia University, New York, NY 10027, USA}

\author{B\'alint~Jo\'o}
\email{joob@ornl.gov}
\affiliation{Oak Ridge Leadership Computing Facility, Oak Ridge National Laboratory, Oak Ridge, TN 37831, USA}


\author{Frank~Winter}
\email{fwinter@jlab.org}
\affiliation{Jefferson Lab, 12000 Jefferson Avenue, Newport News, Virginia 23606, USA}

\collaboration{Nucleon Matrix Elements (NME) Collaboration}
\noaffiliation
\preprint{LA-UR-21-20526}
\date{\today}
\begin{abstract}
We present a high statistics study of the isovector nucleon charges
and form factors using seven ensembles of 2+1-flavor Wilson-clover
fermions. The axial vector and pseudoscalar form factors obtained on
each of these ensembles satisfy the partially conserved axial current
relation between them once the lowest  energy $N \pi$ excited state is
included in the spectral decomposition of the correlation functions used for 
extracting the ground state matrix elements. Similarly, we find evidence
that the $N\pi \pi $ excited state contributes to the correlation
functions with the insertion of the vector current, consistent with
the vector meson dominance model. The resulting form factors are
consistent with the Kelly parameterization of the experimental
electric and magnetic data. Our final estimates for the isovector
charges are $g_{A}^{u-d} = 1.32(6)(5)_{\rm sys}$, $g_{S}^{u-d} =
1.06(9)(6)_{\rm sys}$, and $g_{T}^{u-d} = 0.97(3)(2)_{\rm sys}$,
where the first error is the overall analysis uncertainty and the
second is an additional combined systematic uncertainty. The form
factors yield: (i) the axial charge radius squared, ${\langle
r_A^2 \rangle}^{u-d}=0.428(53)(30)_{\rm sys} \ {\rm fm}^2$, (ii) the induced
pseudoscalar charge, $g_P^\ast=7.9(7)(9)_{\rm sys}$, (iii) the pion-nucleon
coupling $g_{\pi {\rm NN}} = 12.4(1.2)$, (iv) the electric charge
radius squared, ${\langle r_E^2 \rangle}^{u-d} = 0.85(12)(19)_{\rm
sys} \ {\rm fm}^2$, (v) the magnetic charge radius squared, ${\langle
r_M^2 \rangle}^{u-d} = 0.71(19)(23)_{\rm sys} \ {\rm fm}^2$, and (vi)
the magnetic moment $\mu^{u-d} = 4.15(22)(10)_{\rm sys}$. All our
results are consistent with phenomenological/experimental values but
with larger errors. Last, we present a  Pad\'e
parameterization of the axial, electric and magnetic form factors over
the range $0.04 < Q^2 < 1$ GeV${}^2$ for phenomenological studies.
\end{abstract}
\maketitle
%
%
%
%
\section{Introduction}
\label{sec:into}

The success of high precision experiments such as DUNE at
Fermilab~\cite{Abi:2018dnh,Acciarri:2015uup} and the T2T-HyperK in
Japan~\cite{Abe:2018uyc,hyperk} is predicated on precise determination
of the flux of the neutrino beam, incident neutrino energy and their
cross sections off nuclear targets.  A major source of uncertainty in
the analysis of neutrino-nucleus interactions is the axial vector form
factors of the nucleon and appropriate nuclear corrections.  Steady
improvements in lattice quantum chromodynamics (QCD) calculations are
expected to provide first principle results with control over all
systematics~\cite{Kronfeld:2019nfb}. In this paper, we present high
statistics results for the matrix elements of the isovector axial and
vector current between ground state nucleons. From these we extract
the axial, electric, and magnetic form factors and charges that are
inputs in the analysis of the charged current lepton-nucleus
scattering utilizing electron, muon and neutrino beams.  A heuristic 
parameterization of the form factors for phenomelogical analyses 
is summarized in Eqs.~\eqref{eq:GAPade},~\eqref{eq:GAz2} 
and~\eqref{eq:GEMPade}.\looseness-1

In previous publications, we have presented results for the isovector
charges, $g_A^{u-d}$, $g_S^{u-d}$ and
$g_T^{u-d}$~\cite{Gupta:2018qil}; axial, $G_A(Q^2)$, induced
pseudoscalar, $\widetilde{G}_P(Q^2)$ and pseudoscalar, $G_P(Q^2)$,
form factors~\cite{Rajan:2017lxk,Jang:2019vkm}; and the electric and
magnetic form factors, $G_{E}(Q^2)$ and
$G_{M}(Q^2)$~\cite{Jang:2019jkn}.  Those calculations were done using
the clover-on-HISQ formulation, i.e., the Wilson-clover fermion action
was used to construct correlation functions on background gauge
configurations generated with 2+1+1 flavors of the highly improved
staggered quark (HISQ) action by the MILC
Collaboration~\cite{Bazavov:2012xda}. They exposed a number of issues
that require attention: The central value for the isovector axial
charge $g_A^{u-d}=1.218(25)(30)$~\cite{Gupta:2018qil}, a key parameter
that encapsulates the strength of weak interactions of nucleons, is
about 5\% below the accurately measured value $\lambda = g_A/g_V =
1.27641(45)_{\rm stat}(33)_{\rm
sys}$~\cite{Markisch:2018ndu,Brown:2017mhw,Mendenhall:2012tz,Mund:2012fq}.
Second, the axial and pesudoscalar form factors, $G_A$,
$\widetilde{G}_P$ and $G_P$, did not satisfy the relation imposed on
them by the partially conserved axial current (PCAC)
relation~\cite{Rajan:2017lxk}, whereas the original three-point
correlation functions did. Third, the electric and magnetic form
factors, $G_{E}$ and $G_{M}$, showed significant deviations from the
Kelly parameterization,  which accurately describes the experimental
data~\cite{Jang:2019jkn}. Last, while the uncertainty in the scalar
and tensor charges, $g_S^{u-d}=1.022(80)(60)$ and
$g_T^{u-d}=0.989(30)(10)$, was reduced to $O(10\%)$ as required to put
constraints on novel scalar and tensor interactions at the
$O(10^{-3})$ level~\cite{Bhattacharya:2011qm} that can arise at the
TeV scale, future experiments targeting $O(10^{-4})$ sensitivity
require the reduction of errors to a few percent level.\looseness-1

In this paper, we revisit these issues with high-statistics
calculations on seven ensembles with similar lattice parameters but
generated using 2+1 flavor Wilson-Clover fermions by the
JLab/W\&M/LANL/MIT Collaborations~\cite{JLAB:2016}.  Three important
improvements are made over those presented in our previous
papers~\cite{Gupta:2018qil,Rajan:2017lxk,Jang:2019vkm,Jang:2019jkn}.
First, these calculations have been done using a unitary,
clover-on-clover, lattice formulation, whereas possible systematics in
the clover-on-HISQ mixed action calculations due to the nonunitarity
formulation were not explored. Second, the results are based on much
higher statistics, $O(2\hbox{--}6 \times 10^5)$ measurements on
$O(2\hbox{--}5 \times 10^3)$ configurations. The resulting smaller
errors in the raw data provide more reliable control over the
systematics. Last, we compare several analysis strategies to control
excited-state contamination (ESC) and quantify the sensitivity of the
results to different theoretically motivated values of the mass gaps,
and investigate the possible excited states that may be contributing.

Results for the nucleon charges from a subset of the ensembles
analyzed here have been presented in
Refs.~\cite{Yoon:2016dij,Yoon:2016jzj}.  In parts of the paper, we will
drop, for brevity, the superscripts $(u-d)$ to denote isovector
quantities since all the analyses presented here are restricted to
this case. We will, however, include this superscript in the final
results and at appropriate places to avoid confusion for the general
reader.  For the overall methodology used to calculate the two- and
three-point correlation functions, we refer the reader to our previous
work~\cite{Gupta:2018qil, Rajan:2017lxk,Jang:2019jkn}.

This paper is organized as follows.  After a review of the
phenomenology and known results in Sec.~\ref{sec:Pheno} and the
lattice setup and error analysis strategy in Sec.~\ref{sec:setup}, we briefly summarize
the main systematics that need to be resolved in
Sec.~\ref{sec:systematics}.  The analysis of excited states in the
two-point functions is discussed in Sec.~\ref{sec:spectrum}, and in
three-point functions in Sec.~\ref{sec:ESC}. The relations for the
extraction of form factors from ground state matrix elements are given
in Sec.~\ref{sec:RFF} and the results for the isovector charges
$g_{A, S, T}^{u-d}$ in Sec.~\ref{sec:charges}. The analysis of the
$A_4$ correlator, $\langle \Omega | {\mathcal N}(\tau) A_4(t)
{\mathcal N}(0)|\Omega\rangle$, and the consequent description of the
strategies used for controlling ESC in the axial channel are discussed
in Sec.~\ref{sec:A4}. The extraction of the axial form factors is
then presented in Sec.~\ref{sec:AFF} followed by the
parameterization of the $Q^2$ dependence of $G_A(Q^2)$ and the
extraction of $g_A$ and $\langle r_A^2 \rangle$ in
Sec.~\ref{sec:AFFFinalFits}, and of the induced pseudoscalar form
factor ${\widetilde G}_P(Q^2)$ and the couplings
$g_P^\ast$ and $g_{\pi N N}$ in Sec.~\ref{sec:GP}.
Sec.~\ref{sec:VFF} is devoted to the electromagnetic form factors.
Final estimates at the physical point defined by $a = 0$, $M_\pi =
135$~MeV and $M_\pi L = \infty$ are obtained using simultaneous
chiral-continuum-finite-volume (CCFV) fits in Sec.~\ref{sec:CCFV}. An
alternate heuristic parameterization of the form factors is given in
Sec.~\ref{sec:FFfits}, and the comparison with previous work and
phenomenology in Sec.~\ref{sec:CompLQCD}. Our conclusions are
presented in Sec.~\ref{sec:conclusions}.  Further details of the data,
analyses, and figures are presented in eight appendixes.

\section{Phenomenology}
\label{sec:Pheno}

One of the  main uncertainties in the phenomenological analyses of neutrino-nucleon
scattering is the knowledge of the axial form factors. Direct
experiments using liquid hydrogen (proton) targets are not being
carried out due to safety concerns. Thus, phenomenologists are looking
to lattice QCD to provide first principle estimates. A good validation
of the lattice methodology for the calculation of form factors is to demonstrate
agreement between the simultaneously calculated isovector electric and
magnetic form factors with the Kelly (or other good) parameterization of
the accurate experimental data (see Sec.~\ref{sec:VFF}). Furthermore,
calculating the full set of axial and electromagnetic form factors is
the first step in the analysis of the charged current neutrino-nucleon 
cross section {with all required input taken} from lattice QCD. Our results in
Eqs.~\eqref{eq:GAPade}~\eqref{eq:GAz2} and~\eqref{eq:GEMPade} represent significant
progress toward this goal.

The matrix element of the isovector axial vector current $A_\mu =
\overline{u} \gamma_\mu \gamma_5 d$ between ground state nucleons,
which describes neutron $\beta$-decay and the weak charged current of
the interaction of the neutrino with the nucleon, has the following
relativistically covariant decomposition in terms of two form factors:
\begin{multline}
\label{eq:AFFdef}
\left\langle N({\bm p}_f,s_f) | A_\mu (\bm {q}) | N({\bm p}_i,s_i)\right\rangle  = \\
{\overline u}_N({\bm p}_f,s_f)\left( G_A(q^2) \gamma_\mu
+ q_\mu \frac{\widetilde{G}_P(q^2)}{2 M_N}\right) \gamma_5 u_N({\bm p}_i,s_i),
\end{multline}
where $G_A(q^2)$ is the axial vector form factor,
$\widetilde{G}_P(q^2)$ the induced pseudoscalar form factor,
$|N({\bm p}_f,s_f) \rangle$ the nucleon state with momentum ${\bm p}_f$ and
spin $s_f$, and the momentum transfer is $\bm{q}={\bm p}_f-{\bm
p}_i$. Throughout this paper, all data for the form factors are 
presented in terms of $Q^2 \equiv {\bf p}^2 - (E-m)^2 = -q^2$, i.e.,
the spacelike four-momentum squared. We use the DeGrand-Rossi
basis for the gamma matrices~\cite{DeGrand:1990dk}, and assume isospin
symmetry, $m_u=m_d$. Thus, we neglect the induced tensor form factor
$\widetilde{G}_T$ that vanishes in the isospin
limit~\cite{Bhattacharya:2011qm}. The axial charge $g_A
\equiv G_A(q^2=0)$ is obtained from both the forward matrix element
and by extrapolating $G_A(Q^2)$ to $Q^2 = 0$ as discussed in
Secs.~\ref{sec:charges}, and~\ref{sec:CCFVcharges}, respectively.

The pseudoscalar form factor, $G_P$, is defined by 
\begin{equation}
\left\langle N({\bm p}_f) | P (\bm{q}) | N({\bm p}_i)\right\rangle  = 
{\overline u}_N({\bm p}_f) G_P(q^2) \gamma_5 u_N({\bm p}_i) \,, 
\label{eq:PSdef}
\end{equation}
where $P = \overline{u} \gamma_5 d$ is the pseudoscalar density. 

The discrete lattice momenta are given by $2 \pi {\bf n} / L
a$ with the components of the vector ${\bf n} \equiv (n_1, n_2, n_3)$
taking on integer values, $|n_i| \in \{0,L\}$.  The normalization of the nucleon spinors $u_N(\bm  p,s)$ in Euclidean space is
\begin{equation}
\sum_{s} u_N(\bm  p,s) \bar{u}_N({\bm p},s) =
   \frac{E({\bm p})\gamma_4-i\bm \gamma\cdot {\bm p} + M}{2 E({\bm p})} \,.
\label{eq:spinor}
\end{equation}

The three form factors, $G_A(Q^2)$, $\widetilde{G}_P(Q^2)$ and
${G}_P(Q^2)$, are not independent because of the PCAC operator
identity, $\partial_\mu A_\mu - 2 {\widehat m} P=0$. By contracting
Eq.~\eqref{eq:AFFdef} with $q^\mu$ and using Eq.~\eqref{eq:PSdef},
this identity gives the following relation between them:
\begin{equation}
2 {\widehat m} G_P(Q^2) = 2 M_N G_A(Q^2) - \frac{Q^2}{2M_N} {\widetilde G}_P(Q^2) \,,
\label{eq:PCAC}
\end{equation}
where ${\widehat m} \equiv Z_m Z_P (m_u +m_d)/(2 Z_A)$ is the average
bare PCAC mass of the $u$ and $d$ quarks, 
$Z_m$, $Z_P$ and $Z_A$ are the renormalization constants
for the quark mass, the pseudoscalar and the axial currents, respectively.
Table~\ref{tab:gpiNN} gives the results for ${\widehat m} $ calculated
using the PCAC relation within the pseudoscalar two-point correlation
function, i.e., by requiring that, up to lattice artifacts, the
relation $\Gamma(\tau) = \langle \Omega|(\partial_\mu A_\mu - 2 {\widehat
  m} P)_\tau P_0 | \Omega \rangle=0$ holds for all Euclidean times
$\tau \neq 0$.  It can also be measured using the three point functions by 
inserting the operator $\partial_\mu A_\mu = 2 {\widehat m} P$ between 
any state including the nucleon. Estimates of ${\widehat m}$ from 
two- and three-point correlation functions with the same
bare lattice operators should agree up to discretization artifacts.

The pseudoscalar two-point function also gives the pion decay constant $F_\pi$ 
through the matrix element
$\langle \Omega|A_4^{\rm point}| \pi\rangle \allowbreak= {\sqrt{2}}
M_\pi F_\pi$, which is obtained from a simultaneous fit to data in the
plateau region of 
$\langle \Omega|A_4^{\rm point}(\tau)\allowbreak P^{\rm smeared}(0) |\Omega \rangle$ 
and $\langle \Omega|P^{\rm smeared}(\tau)\allowbreak P^{\rm smeared}(0) |\Omega \rangle$. 
These values for $F_\pi$ are given in
Table~\ref{tab:gpiNN}, and their CCFV extrapolation is shown in the
bottom row of Fig.~\ref{fig:gpiNN}. The result is consistent with the
experimental value. The largest contributor to the error, $1\sigma
\approx 4\%$, is the CCFV extrapolation. Since the calculations
of $F_\pi$ on the lattice are among the most
reliable~\cite{Aoki:2019cca}, it is reasonable to expect a 4\%
uncertainty in results from CCFV fits to seven points for all other quantities
analyzed in this work.

Last, Table~\ref{tab:gpiNN} also gives the product $M_N g_A/F_\pi$,
which is equal to the pion-nucleon coupling $g_{\pi N N}$ by the
Goldberger-Treiman relation, for three estimates of $g_A$ given in
Table~\ref{tab:gAdP2z2}, i.e., from $\{4,3^*\}$,
$\{4^{N\pi},2^\text{sim},P_2\}$ and $\{4^{N\pi},2^\text{sim},z^2\}$
strategies used to control ESC that are defined in
Sec.~\ref{sec:CCFVcharges} (also see Appendix~\ref{sec:glossary} 
for their definitions). The nucleon mass, $M_N$, is given in
Table~\ref{tab:Ensembles}. \looseness-1

\begin{table*}   
\begin{ruledtabular}
\begin{tabular}{l|l|llll|lll}
& & & $F_\pi^\text{bare}$ & $F_\pi|_{R1}$ & $F_\pi|_{R2}$ & \multicolumn{3}{c}{$g_{\pi NN} = M_N g_A/F_\pi$} \\ 
ID & $a{\widehat m}_\text{PCAC}$ & $aF_\pi^\text{bare}$ &[MeV]&[MeV]&[MeV]& $\{4,3^*\}$ & $\{4^{N\pi},2^\text{sim},P_2\}$ & $\{4^{N\pi},2^\text{sim},z^2\}$ \\ 
\hline
$a127m285$  & 0.009304(34) & 0.07115(15)  & 110.5(1.8)   & 97.5(2.1)    & 95.5(2.0)    & 12.46(12)    & 12.42(28)    & 12.32(19)    \\
$a094m270$  & 0.005726(29) & 0.05182(12)  & 108.8(1.2)   & 96.0(1.7)    & 95.1(1.4)    & 12.92(48)    & 12.49(45)    & 12.46(30)    \\
$a094m270L$ & 0.005724(05) & 0.05204(05)  & 109.2(1.2)   & 96.8(1.9)    & 97.2(1.4)    & 12.45(09)    & 12.63(16)    & 12.55(13)    \\
$a091m170$  & 0.002104(09) & 0.04743(06)  & 102.8(1.1)   & 90.7(1.7)    & 90.2(1.4)    & 12.45(19)    & 12.55(37)    & 12.63(33)    \\
$a091m170L$ & 0.002123(10) & 0.04754(05)  & 103.1(1.1)   & 90.2(1.7)    & 89.8(1.3)    & 12.55(16)    & 13.19(33)    & 13.17(31)    \\
$a073m270$  & 0.004328(04) & 0.04016(04)  & 108.9(1.2)   & 97.9(1.6)    & 97.8(1.4)    & 12.70(14)    & 12.63(18)    & 12.58(14)    \\
$a071m170$  & 0.001522(04) & 0.03661(04)  & 102.2(1.2)   & 91.6(1.3)    & 91.8(1.3)    & 12.60(32)    & 13.08(39)    & 13.10(36)    \\
\hline
CCFV       &              &              &              & 93.0(3.8)    & 95.9(3.5)    & 12.65(38)    & 13.60(65)    & 13.58(49)    \\
\end{tabular}
\end{ruledtabular}
\caption{Results for the PCAC quark mass ${\widehat m}$ defined in the
  text and the pion decay constant $F_\pi$ with the two
  renormalization methods defined in Sec.~\ref{sec:gV}. The $\sim 1\%$ uncertainty in $F_\pi$ comes
  mainly from that in the scale $a$ given in
  Table~\ref{tab:Ensembles}. The combination $M_N g_A/F_\pi $, which is
  independent of $Z_A$ and dimensionless, is equal to $ g_{\pi NN} $ by the
  Goldberger-Treiman relation. It is evaluated using three ways of 
  calculating $g_A$ discussed in Secs.~\ref{sec:charges} and \ref{sec:CCFVcharges}: $\{4,3^*\}$ in which $g_A$ is taken from
  the forward matrix element, $\{4^{N\pi},2^\text{sim},{P_2}\}$  and $\{4^{N\pi},2^\text{sim},{z^2}\}$
  that uses $P_2$ Pad\'e and $z^2$ fits to $G_A(Q^2)$ given in
  Table~\protect\ref{tab:gAdP2z2}. The last row gives the continuum
  result from CCFV fits to these data as discussed in Sec.~\ref{sec:CCFVcharges}.  }
\label{tab:gpiNN}
\end{table*}

A large part of the analysis presented in this work is influenced by
the recent understanding and resolution~\cite{Jang:2019vkm} of why the 
axial form factors calculated in the ``standard'' way do not satisfy the PCAC relation given in
Eq.~\eqref{eq:PCAC}, a problem that {afflicts} previous lattice
calculations~\cite{Rajan:2017lxk}. We show that a much lower energy
excited state, with a mass gap much smaller than obtained from
$n$-state fits to the two-point nucleon correlator and used in the
standard analysis of three-point functions, contributes in the axial
channel. Including these states in the fits, with masses consistent
with the noninteracting $N(\bm{p}=0) \pi(\bm{p})$ and $N(-\bm{p}) \pi(\bm{p})$ states on the
lattice, gives form factors that show much better agreement with the
PCAC relation, Eq.~\eqref{eq:PCAC}, and satisfy other consistency
checks discussed in Sec.~\ref{sec:AFFconsistency}. While the need for
including such low-energy multihadron states has, so far, been
demonstrated only in the axial and pseudoscalar channels, it behooves
us to determine whether such multihadron states also contribute in
other channels. In this paper, we build on the discussion in
Ref.~\cite{Jang:2019vkm}, and investigate the dependence of various
matrix elements on the spectrum of excited states obtained from
different fits.

The decomposition in Minkowski space of the matrix element of the
electromagnetic current $V_\mu^{\rm em} = \frac{2}{3} {\overline
u} \gamma_\mu u - \frac{1}{3} {\overline d} \gamma_\mu d $ within the
nucleon ground state into the Dirac, $F_{1}$, and Pauli, $F_{2}$, form
factors is:
\begin{multline}
\label{eq:VFF}
\left\langle N({\bm p}_f,s_f) | V_\mu^{\rm em} ({\bm q}) | N({\bm p}_i,s_i)\right\rangle  = \\
{\overline u}_N({\bm p}_f,s_f)\left( F_1(q^2) \gamma_\mu
+ i \sigma_{\mu \nu} q_\nu
\frac{F_2(q^2)}{2 M_N}\right)u_N({\bm p}_i,s_i),
\end{multline}
where $\sigma_{\mu \nu} = i/(\gamma_\mu \gamma_\nu
- \gamma_\nu \gamma_\mu)/2$ and the induced scalar form factor is
neglected since we work in the isospin limit. Throughout this paper,
we will present results in terms of the isovector Sachs electric,
$G_{E}$, and magnetic, $G_{M}$, form factors that are related to the
Dirac and Pauli form factors in Euclidean space as
\begin{align}\label{eq:sachs}
G_E(Q^2) &= F_1(Q^2) - \frac{Q^2}{4M_N^2}F_2(Q^2) \,, \\
G_M(Q^2) &= F_1(Q^2) + F_2(Q^2).
\end{align}
These are very well measured experimentally, and from them 
one gets the vector charge 
\begin{equation}
g_V = G_E|_{Q^2=0} = F_1|_{Q^2=0}
\end{equation}
which satisfies the conserved vector current relation $g_V Z_V =1$,
where $Z_V$ is the renormalization constant for the local vector
current used on the lattice. The isovector form factor $G_M$ gives 
the difference between the magnetic moments of the proton
and the neutron:
\begin{equation}
\mu^{p} - \mu^{n} = G_M|_{Q^2=0} = (F_1 + F_2)|_{Q^2=0} = 1 + \kappa_p - \kappa_n \,.
\label{eq:Pmmdef}
\end{equation}
The anomalous magnetic moments of
the proton and the neutron, $\kappa_p$ and $\kappa_n$, in units of the Bohr magneton, 
are known very precisely~\cite{PhysRevD.98.030001}:
\begin{eqnarray}
\kappa_p  &=& \phantom{-}1.79284735(1) \,   \qquad\ ({\rm proton})    \,,  \nonumber \\
\kappa_n  &=& -1.91304273(45)   \qquad ({\rm neutron}) \,.
\label{eq:mu_expt}
\end{eqnarray}

In phenomenological studies, it is customary to parameterize the form
factors to obtain their value and slope at $Q^2=0$. These give the
charges, $g_A$, $g_V$ and $\mu$, and the charge radii squared,
$\langle r_{A,E,M}^2\rangle$, defined as
\begin{equation}
\langle r^2\rangle = -6\frac{d}{dQ^2}\left.\left(\frac{G(Q^2)}{G(0)}\right)\right|_{Q^2=0} \,.
\label{eq:rdef}
\end{equation}
For the electromagnetic form factors, the Kelly
parameterization provides a good fit to the experimental data~\cite{Kelly:2004hm} and gives
\begin{align}
r_E^{p-n}|_{\rm exp} &= 0.926(4) \,, \nonumber \\
r_M^{p-n}|_{\rm exp} &= 0.872(7) \,. 
\label{eq:isovectorradiiKelly}
\end{align}
In this study, we analyze various systematics and provide results for
both axial and electromagnetic form factors over a range of $Q^2$,
especially the region $\lesssim 1$~GeV${}^2$ where nonperturbative
effects are large.  These data are analyzed using the dipole, Pad\'e
and model-independent $z$-expansion parameterizations.  Control over various systematics
in the extraction of the form factors is illustrated by comparing the
lattice data for $G_{E,M}$ with the Kelly parameterization in
Secs.~\ref{sec:VFF} and~\ref{sec:FFfits}. For the purpose of 
comparison, and given the much larger errors in the lattice data, 
one can equally well use other parameterizations, for example, the recent 
rational fraction discussed in Ref.~\cite{Xiong:2019umf}, without 
{a change in} our conclusions.

\section{Lattice and Error Analysis Methodology}
\label{sec:setup}

The parameters of the seven ensembles with 2+1-flavors of $O(a)$
improved Wilson-clover fermions generated by the
JLab/W\&M/LANL/MIT Collaboration are given in
Table~\ref{tab:Ensembles} in Appendix~\ref{sec:compQ2}.  The
parameters used to calculate the quark propagators are given in
Table~\ref{tab:cloverparams}.  We have made $O(2\hbox{--}6\times
10^5)$ measurements of each observable on these ensembles using the
truncated solver with bias correction~\cite{Bali:2009hu,Blum:2012uh}
and the coherent sequential
propagator~\cite{Bratt:2010jn,Yoon:2016dij} methods. Even with these
statistics, because of the $e^{-(M_N - 3M_\pi/2)\tau}$ decay of the
signal-to-noise ratio, the three-point correlation functions are
well-measured only up to source-sink separation $\tau \sim 1.5$~fm.
At these separations, excited state contamination is significant and
we fit the data using the spectral decomposition of the correlation
functions to isolate the ground state value as discussed in
Sec.~\ref{sec:ESC}. In the calculation of form factors, the signal
also degrades with momentum transfer $Q^2$, and the errors at the
larger momentum transfers are sizable in some cases.

The central values and errors are calculated using a
single-elimination jackknife method.  We make $O(100)$ measurements on
each configuration with randomly selected but widely separated source
points to maximize decorrelations. From these, bias corrected averages
are constructed for each configuration, which are then binned over
5--11 configurations to further reduce correlations. These $O(500)$
binned values are then analyzed using the jackknife procedure.  All
fits using minimization of $\chi^2$ 
are made using the full covariance matrix calculated using the
binned values. This procedure is followed for all observables, values
of momentum insertion, and ensembles.  Note that even when using a
Bayesian procedure including priors to stabilize the fits, the errors
are calculated using the jackknife method and are thus the usual
frequentist standard errors\footnote{When priors are used, the
augmented $\chi^2$ is defined as the standard correlated $\chi^2$ plus
the square of the deviation of the parameter from the prior mean
normalized by the prior width. This quantity is minimized in the
fits. In the following, we quote this augmented $\chi^2$ divided by the
degrees of freedom, and call
it $\chi^2$/dof for brevity. In the jackknife process, we keep the
prior and its width fixed. This is a consistent strategy as the errors
quoted are frequentist errors and do not represent a Bayesian
credibility interval. The $p$-value listed in figures showing fits is
given for convenience only as it is calculated from the also listed $\chi^2$
value using the standard $\chi^2$ distribution.}.\looseness-1

We use two criteria to determine whether the fits, for example, those
used to remove ESC or the CCFV fits, are overparameterized: (i) the
Akaike Information Criteria (AIC)~\cite{1100705} which requires that
the total $\chi^2$ decreases by two units for every extra free
parameter in the fit ansatz, and (ii) whether the errors in the
additional parameters introduced to include, for example, the third state have more
than 100\% uncertainty. The AIC weights are calculated to assess whether the fits 
are overparameterized. The actual choice of the averaging
performed to get final results is discussed in the individual
sections.

Overall, the errors in data from three ensembles need to be reduced to
improve precision: on $a094m270$ due to the small volume and on
$a091m170L$ and $a071m170$ due to the lighter pion mass. Of these, the
latter two ensembles are important for the chiral extrapolation, and
we plan to double their statistics in the future.

In our previous work using the clover-on-HISQ formulation, we observed
that some observables that should vanish by the parity symmetry show a
nonzero signal at the 2.5--3$\sigma$ level. Even though such
deviations are most likely statistical fluctuations, we improved the
realization of parity symmetry in our clover-on-clover work by
applying a random parity transformation on each gauge configuration as
follows: For a randomly chosen direction $\mu \in 1\hbox{--}4$, each
gauge configuration is parity transformed by implementing
\begin{align}
  U_\nu(x) &\longrightarrow U_\nu(P_\mu(x) - \hat{\nu})^\dag \textrm{ for } \nu \neq \mu, \\
  U_\mu(x) &\longrightarrow U_\mu(P_\mu(x))
\label{eq:parity}
\end{align}
where $P_\mu(x)$, the parity transformation acting on the vector $x$
labeling the sites, flips the sign of all components, except for
$x_\mu$~\cite{Yoon:2016dij,Yoon:2016jzj}.

\section{Systematics in the extraction of nucleon matrix elements}
\label{sec:systematics}

\begin{figure*}[tbp] 
\subfigure
{
    \includegraphics[width=0.41\linewidth]{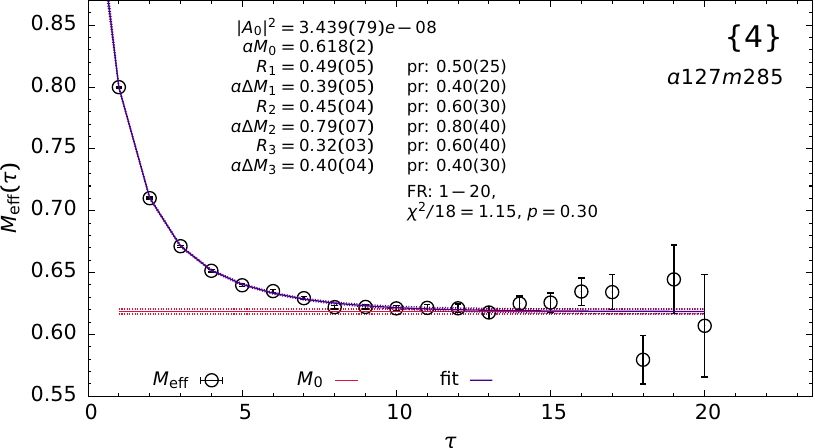} \ \hspace{0.4cm} 
    \includegraphics[width=0.41\linewidth]{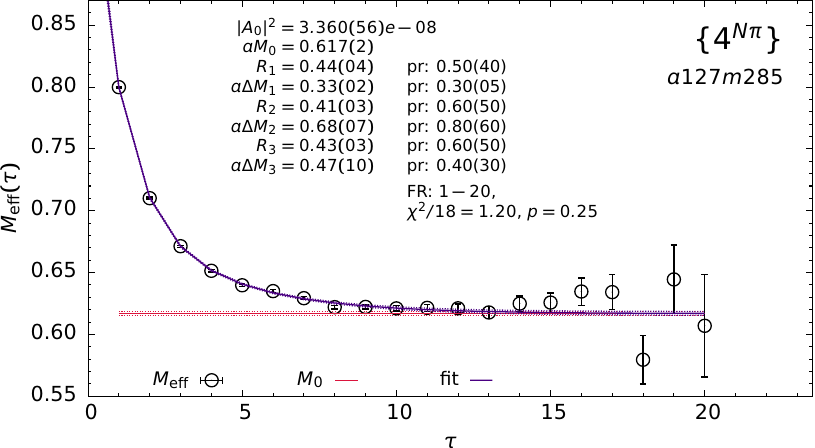} 
}
{
    \includegraphics[width=0.41\linewidth]{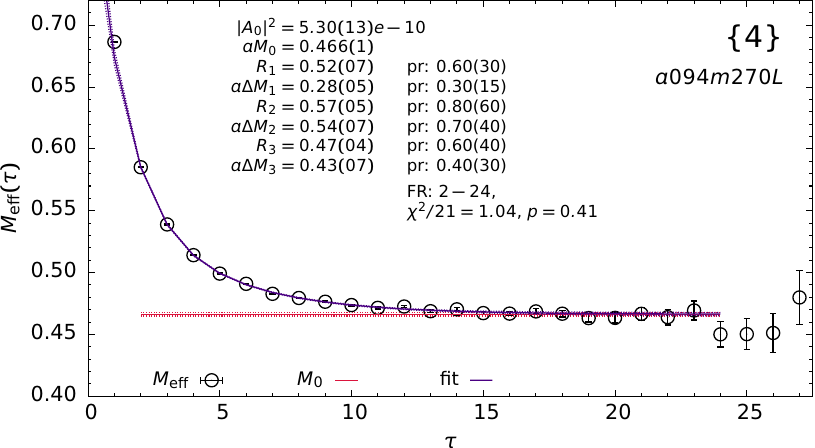} \ \hspace{0.4cm} 
    \includegraphics[width=0.41\linewidth]{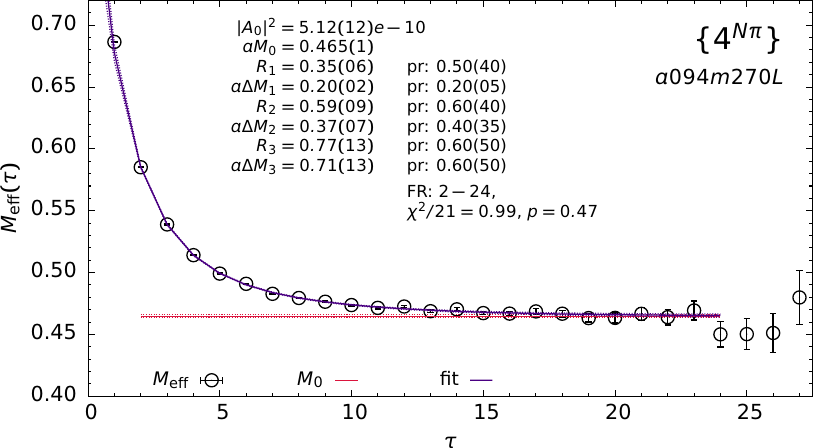} 
}
{
    \includegraphics[width=0.41\linewidth]{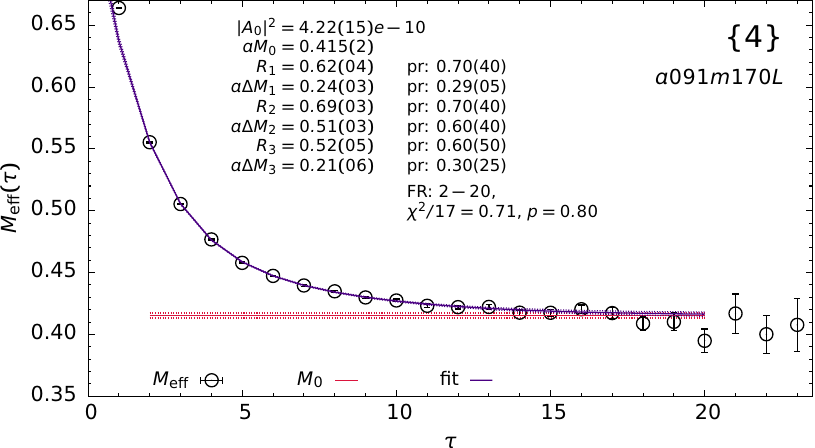} \ \hspace{0.4cm} 
    \includegraphics[width=0.41\linewidth]{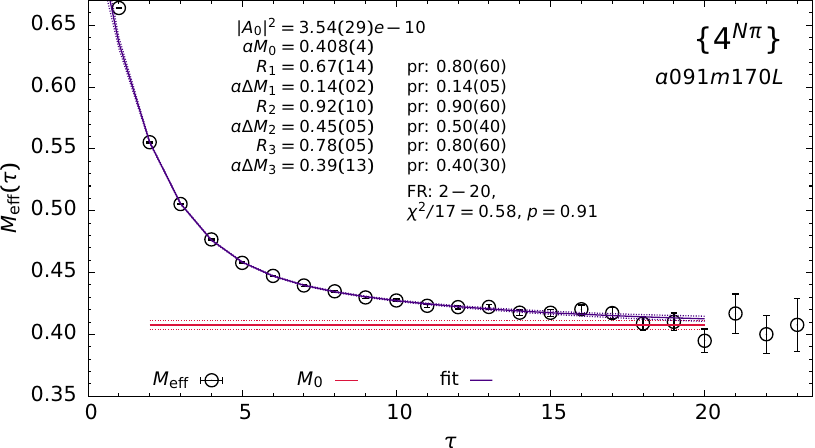} 
}
{
    \includegraphics[width=0.41\linewidth]{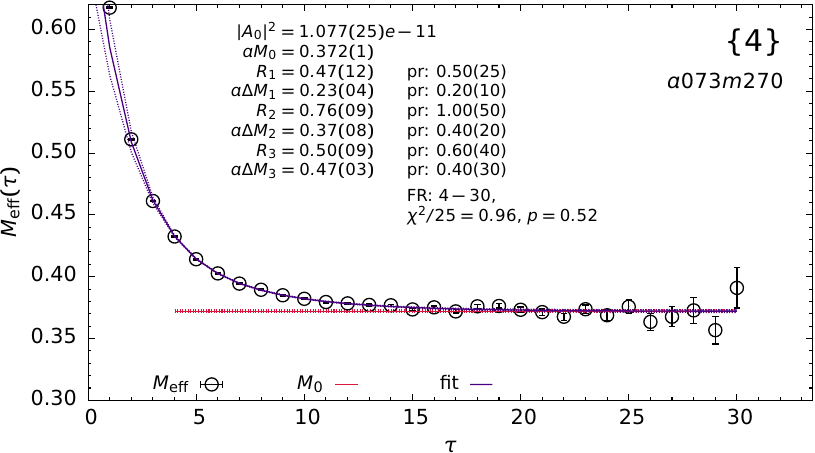} \ \hspace{0.4cm} 
    \includegraphics[width=0.41\linewidth]{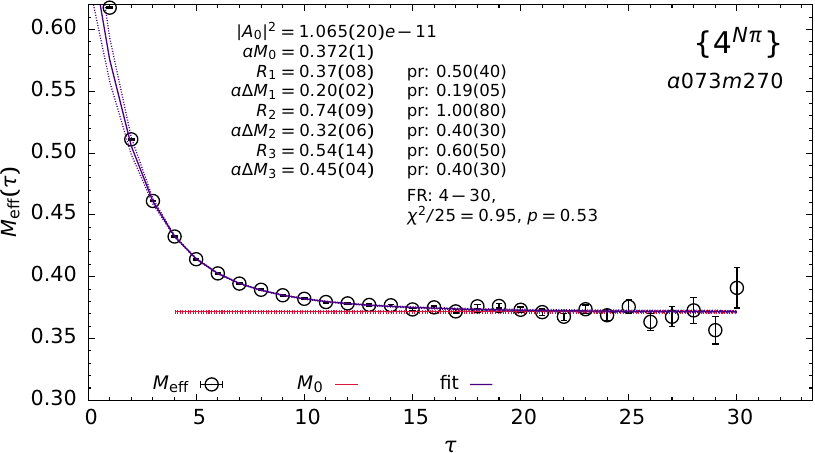} 
}
{
    \includegraphics[width=0.41\linewidth]{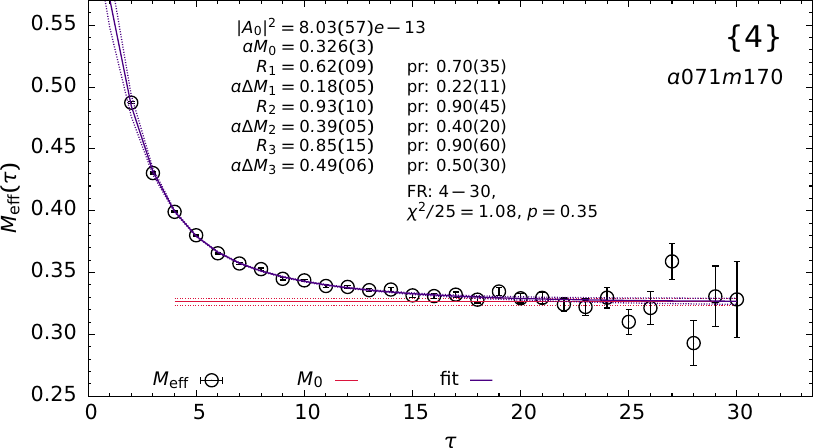} \ \hspace{0.4cm} 
    \includegraphics[width=0.41\linewidth]{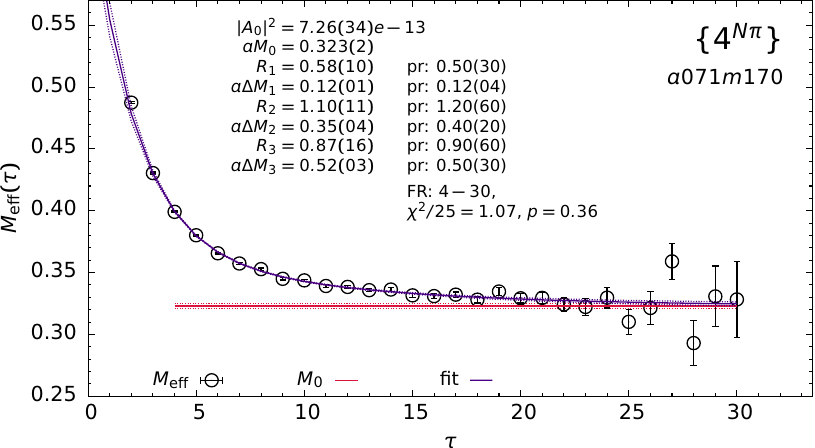} 
}
\vspace*{-1pt}
\caption{The effective mass $M_{\rm eff}$ plotted
  versus the source-sink separation $\tau/a$ for 5 ensembles.  The
  left panels show the standard four-state fits, $\{4\}$, while  
  the right panels show $\{4^{N\pi}\}$, in which the 
  noninteracting energy of $N\pi$ state is input as the
  central value of the prior for $\Delta M_1$. The legends 
  give the ground state amplitude ${\mathcal{A}}_0$
  and mass $a M_0$, the excited-state amplitude ratios $R_i =
  |{\mathcal{A}}_i|^2/|{\mathcal{A}}_0|^2$ and mass gaps $a \Delta
  M_i = a(M_i - M_{i-1})$, the prior value and width (pr) used, the
  fit range FR, the $\chi^2/$dof and the ensemble ID.  The
  signal-to-noise grows rapidly after \(\tau=1.8\hbox{--}2.2\)~fm
  depending on the statistics and the ensemble.  Note that for the
  $170$~MeV ensembles, even the ground state mass and amplitude differ
  by about 2--3$\sigma$ between the two fit strategies, and the
  relative contribution, $R_1 e^{-{\Delta M_1 \tau}}$, of the low mass
  $N(\bm 1)\pi(-\bm 1)$ state in the $\{4^{N\pi}\}$ fit is still about $3\%$ at $\tau \approx
  1.8$~fm.  \label{fig:2ptCOMP}}
\end{figure*}

There are four challenges to high precision calculations of nucleon
charges and form factors (or their primitives, the ground state matrix
elements) at a given value of $\{a, M_\pi, M_\pi L\}$. The first and
key challenge is the exponentially decreasing signal-to-noise in all
nucleon correlation functions---the signal falls off as
$e^{-(M_N-1.5M_\pi)\tau}$ with increase in the source-sink separation
$\tau$. As shown in Fig.~\ref{fig:2ptCOMP}, with $O(2\hbox{--}6 \times
10^5)$ measurements, a good signal in the two-point functions extends
to $\sim 2$~fm. Similarly, in the three-point functions, it extends to
$\sim 1.5$~fm as illustrated in
Figs.~\ref{fig:gAcomp},~\ref{fig:gScomp} and~\ref{fig:gTcomp}.  At
$\sim 1.5$~fm, ESC is still significant in all three-point functions
as shown in
Appendixes~\ref{sec:compgAST},~\ref{sec:AFFESC},~\ref{sec:compAFF},
and~\ref{sec:TcompVFF}.  As a result, for given fixed statistics, one
has to balance statistical uncertainty against a systematic bias due
to the values of $\tau$ picked to control ESC.

The second challenge is determining all the excited states that
contribute significantly to a given three-point function and isolating
their contribution by making fits to a truncated spectral
decomposition---a sum of exponentials as shown in Eqs.~\eqref{eq:2pt}
and~\eqref{eq:3pt}. While the contribution of a given excited state is
exponentially suppressed by its mass gap, we are, however, confronted
by a tower of low-lying multihadron excited states starting with
$N(\bm p=0) \pi(-\bm p)$, $N(\bm p) \pi(-\bm p)$, $N({\bm 0})\pi({\bm
  0})\pi({\bm 0})$. On $M_\pi = 135$~MeV ensembles, the tower, as a function of
$\bm p$, starts at $\approx 1200$~MeV, and gets arbitrarily dense as
$\bm{p} \to 0$. Thus, the suppression of excited-state contributions
due to the mass gap is smaller than in mesons and decreases as $M_\pi
\to 0$ and $\bm p \to 0$. In short, possible contributions of the many
multihadron states that lie below the first two radial excitations, 
N(1440) and N(1710), need to be evaluated.

It is typical to reduce the contributions of excited states by
smearing the delta-function source used to generate the quark
propagators. We use the gauge-invariant Wuppertal
method~\cite{Gusken:1989ad} with parameters given in
Table~\ref{tab:cloverparams} in Appendix~\ref{sec:compQ2}.  However,
in this approach, one does not have detailed control over the size of
the coupling to a given excitation since there is only one tunable
parameter, the smearing size given by $\sigma$ in
Table~\ref{tab:cloverparams}.  Second, for a given three-point
function, couplings to certain states can get enhanced. A case in
point is the contribution of the $N(\bm p=0) \pi(-\bm p)$  and $N(\bm p) \pi(-\bm p)$ states in the
axial channel as discussed in Sec.~\ref{sec:AFF}.

The third issue is calculating the renormalization factor, including
operator mixing, to connect to a continuum scheme such as
$\overline{\rm MS}$. This systematic, for the calculations presented
in this work, is considered to be under control to within about 2\% as
discussed in Ref.~\cite{Aoki:2019cca} and in Sec.~\ref{sec:gV}.

Once data with control over the statistical and the above systematic
uncertainties are obtained at multiple values of $\{a, M_\pi, M_\pi
L\}$, simultaneous chiral-continuum-finite-volume (CCFV) fits, which
include corrections with respect to $M_\pi$, $a$ and $M_\pi L$, are
used to extract the physical result in the limits $M_\pi \to 135$~MeV,
$a \to 0$, and $M_\pi L \to \infty$. Having only seven ensembles introduces the fourth challenge: 
only leading order corrections in each variable can be included without
overparameterization, hence residual corrections may be
underestimated. The analyses performed, using appropriate CCFV fit
ansatz, are described in Sec.~\ref{sec:CCFV}.

Of these four issues, the most serious is excited state contributions,
which is exacerbated by the exponentially falling signal-to-noise
ratio with $\tau$.  To summarize, while the overall methodology for
all the lattice calculations presented here is well-established, a
clear strategy for controlling excited state contamination that can be
applied to all nucleon matrix elements remains elusive as discussed
below. We, therefore, analyze the data using multiple strategies, each
of which should converge and give the correct result with perfect
data. At appropriate places, we give reasons for picking the strategy
used to quote the final results and estimates of possible remaining
systematic uncertainties.

%
\section{The nucleon spectrum from fits to the two-point function}
\label{sec:spectrum}

To determine the nucleon spectrum, we keep four states in the spectral
decomposition of the two-point functions $C^{2\text{pt}}$ with momentum $\bm p$:
\begin{equation}
  C^{2\text{pt}}(\tau;\bm{p}) = \sum_{i=0}^3 \abs{\mathcal{A}_i(\bm p)}^2 e^{-E_i(\bm p) \tau}  \,.
\label{eq:2pt}
\end{equation}
Here $E_i$ are the energies and ${\mathcal{A}}_i$ are the corresponding amplitudes
for the creation/annihilation of a given state $|i\rangle$ by
the interpolating operator $\cal N$ chosen to be
\begin{align}
\mathcal{N}(x) = \epsilon^{abc} \left[ {q_1^a}^T(x) C \gamma_5  \frac{(1 \pm \gamma_4)}{2} q_2^b(x) \right] q_1^c(x) \,, 
\label{eq:nucl_op}
\end{align}
with color indices $\{a, b, c\}$, charge conjugation matrix
$C=\gamma_4 \gamma_2$ in the DeGrand-Rossi
basis~\cite{DeGrand:1990dk}, and $q_1$ and $q_2$ denoting the two
different flavors of light Dirac quarks. The $E_i$ and the
${\mathcal{A}}_i$ are extracted from a fit to a large range,
$[\tau_{\rm min} , \tau_{\rm max}]$. The starting time, $\tau_{\rm
min}/a$ is taken to be small, between 1 and 4, and $\tau_{\rm max}$ is
$ \sim 2$~fm with the current statistics as shown in
Fig.~\ref{fig:2ptCOMP}. For brevity, throughout this paper, it should
be assumed that the values of $t$ and $\tau$ are in lattice units.

There are two nagging issues with this ``standard'' analysis. First
the mass gaps, $\Delta E_1 \equiv (E_1 - E_0)$, shown in
Table~\ref{tab:deltaM} are slightly larger than even of N(1440). This could be
explained away by assuming that the lower-energy states, such as
$N\pi$ or even N(1440), do not couple significantly.  Second, the axial vector and
pseudoscalar form factors obtained using this spectrum to remove the
ESC do not satisfy the PCAC relation, Eq.~\eqref{eq:PCAC}, to a much larger
extent than observed in the original three-point correlation
functions in which the size of deviation is consistent with that expected
due to discretization errors~\cite{Rajan:2017lxk}.

The likely reason for both issues is that standard fits to the
two-point function do not expose the lighter multihadron, $N\pi$,
$N\pi\pi,  \ldots$, states that are needed in the analysis of three-point functions~\cite{Jang:2019vkm}. In
Fig.~\ref{fig:2ptCOMP}, we show results of four-state fits to
$C^{2\text{pt}}(\tau;\bm{p}=0)$ along with the data for the effective
energy defined as
\begin{equation}
E_{\rm eff}(\tau) =  \log \frac{C^{\rm 2pt}(\tau)}{C^{\rm 2pt}(\tau+1)} \,.
\label{eq:effmass}
\end{equation}
It converges to the ground state energy for $\tau \to \infty$ and for
$\bm{p}=0$ reduces to $M_{\rm eff}(\tau)$.  The criteria used for judging
the quality of the fits is $\chi^2$/dof.  The panels on the left show
fits with the standard strategy labeled $\{4\}$, in which empirical
Bayesian priors with wide widths are used only to stabilize the fits.  The initial
central values for the priors for $\Delta M_1 \equiv M_{1}-M_{0}$,
$\Delta M_2 \equiv M_{2}-M_{1} $, and for the corresponding amplitude
ratios, $R_i \equiv |{\mathcal{A}}_i|^2/|{\mathcal{A}}_0|^2$, are
taken from an unconstrained three-state fit. Prior widths are set at
$\sim 50\%$ of the value. The fit is repeated and resulting values are
used as central values for the priors in a four-state fit. This
process is iterated one more time to adjust the priors for the three
excited states. The final fit parameters for the $\bm p = 0$ case, the
prior value and width, the fit range (FR) and the augmented
$\chi^2$/dof of the fit are given in the labels.

The second strategy, labeled $\{4^{\rm N\pi}\}$, uses a prior for the
mass gap, $\Delta E_1$, with value given by the lowest relevant state,
$N({\bm 1})\pi(-\bm 1)$ or $N({\bm 0})\pi({\bm 0})\pi({\bm 0})$, with a narrow width. (The priors
and their widths for the five larger volume ensembles are given in the
labels in Fig.~\ref{fig:2ptCOMP}.) No narrow prior is put on the
amplitude $R_1$. The rest of the procedure is the same as for $\{4\}$.

We stress an important clarification regarding the notation $\Delta
E_1$ and it ``representing'' the first excited state that is implicit
throughout this paper.  The value of $\Delta E_1$ given by a
four-state fit is a number that minimizes $\chi^2$/dof and, most
likely, represents an ``effective'' combination of a set of the lowest
contributing states. Fits to different correlation functions can,
therefore, give different ``effective'' $\Delta E_1$ (in fact $\Delta
E_i$) depending on the couplings of and spacings between the contributing
states.

There are two reasons for
stopping at four-state fits. First, in the three-state fits to the
three-point functions we use $E_0$, $E_1$ and $E_2$. The ignored
$E_3$, which is most contaminated by all the higher neglected states,
acts as a buffer. Second, including more than four states
overparameterizes the fits.  A summary of the ground-state mass and
the mass gap of the first excited state obtained from different fits
is given in Table~\ref{tab:deltaM}. Note that in most cases, the
$a\Delta M_1^{\{4\}}$ is a little larger but close to that expected for the 
N(1440). The one exception is the low value on the $a094m270$ ensemble that 
should be the same, modulo finite volume corrections, as from  $a094m270L$. \looseness-1

\begin{table*}[htbp]   
\begin{ruledtabular}
  \begin{tabular}{l|ll|lll|lll}
  ID          & $aM_N^{\{4\}}$ & $aM_N^{\{4^{N\pi}\}}$ 
& $a\Delta M_1^{\{2\}}$ & $a\Delta M_1^{\{4\}}$ & $a\Delta M_1^{\{4^{N\pi}\}}$ 
& $a\Delta {\widetilde M}_1^{\{2^\text{free}\}}|_{g_A}$ & $a\Delta {\widetilde M}_1^{\{2^\text{free}\}}|_{g_S}$ & $a\Delta {\widetilde M}_1^{\{2^\text{free}\}}|_{g_T}$ \\
\hline
$a127m285$ & 0.618(2) & 0.617(2) & 0.43(5) & 0.39(5) & 0.33(2) & 0.15(7) & 0.71(11) & 0.60(10)  \\
$a094m270$ & 0.468(5) & 0.470(2) & 0.31(6) & 0.22(8) & 0.25(1) & 0.09(13) & 0.51(6) & 0.54(3)  \\
$a094m270L$ & 0.466(1) & 0.465(1) & 0.35(2) & 0.28(5) & 0.20(2) & 0.13(3) & 0.52(2) & 0.50(1)  \\
$a091m170$ & 0.416(2) & 0.413(3) & 0.34(2) & 0.29(5) & 0.16(1) & 0.08(13) & 0.39(8) & 0.46(6)  \\
$a091m170L$ & 0.415(2) & 0.408(4) & 0.31(3) & 0.24(3) & 0.14(2) & 0.14(9) & 0.54(9) & 0.44(4)  \\
$a073m270$ & 0.372(1) & 0.372(1) & 0.32(2) & 0.23(4) & 0.20(2) & 0.06(3) & 0.40(2) & 0.40(2)  \\
$a071m170$ & 0.326(3) & 0.323(2) & 0.25(3) & 0.18(5) & 0.12(1) & 0.08(4) & 0.41(7) & 0.38(2)  \\
  \end{tabular}
\end{ruledtabular}
\caption{Results for the nucleon mass in lattice units, $aM_N^{\{4\}}$
  and $aM_N^{\{4^{N \pi}\}}$, obtained from the two four-state fits to
  the two-point functions.  The next six columns give the values of
  the mass gap, $a\Delta M_1 \equiv a(M_1-M_0)$, of the first excited
  state obtained from different fits studied in this work. The
  notation \{2\} (\{4\}) denotes a two-state (four-state) fit to the
  two-point functions, ($\{4^{N\pi}\}$) is a four-state fit to the
  two-point functions with a prior for $a\Delta M_1$ with a narrow
  width corresponding to the noninteracting $N(\bm q)\pi(-\bm q)$ (or
  the $N(\bm 0)\pi(\bm 0)\pi(\bm 0)$) state (see also 
  Appendix~\ref{sec:glossary}). In the three $\{2^\text{free}\}$
  cases, the mass gaps $a\Delta {\widetilde M}_1$ are determined from
  fits to the three-point functions used to extract the three charges
  $g_{A,S,T}$ as explained in Sec.~\protect\ref{sec:charges}. }
\label{tab:deltaM}
\end{table*}

We find, illustrated by the zero-momentum case in
Fig.~\ref{fig:2ptCOMP}, that (i) the final value of $\Delta M_1^{\rm
N\pi}$ tracks the prior in $\{4^{\rm N\pi}\}$ and (ii) the two fits,
$\{4\}$ and $\{4^{\rm N\pi}\}$, are not distinguished on the basis of
the augmented $\chi^2/$dof, which are similar.  In fact, for each $\bm p$ there is a
flat direction in $E_1$, i.e., a whole region of
parameter values between $\{4\}$ and $\{4^{\rm N\pi}\}$ gives similar
augmented $\chi^2/$dof. Since the $E_1^{\rm N\pi}$ corresponds to
roughly the value for the lowest theoretically allowed state and is
much smaller than the radial excitation N(1440) or $E_1^{\rm 2pt}$, we
will assume it is a good estimate of the lower end of possible
values. Similarly, the data derived $E_1^{\rm 2pt}$ is taken to be an
estimate of the upper end when probing the sensitivity of results for
the ground state matrix elements to $E_1$. Later we will discuss other
estimates of $E_1$ obtained from fits to the three-point functions.

The values of $Q^2 = {\bm p}^2 - (E-M_N)^2$ for the two strategies are
given in Table~\ref{tab:Q2}, and are essentially the same.
Nevertheless, all the analyses and plots presented use the values
of $Q^2$ appropriate to the fits, $\{4\}$ or $\{4^{\rm N\pi}\}$.

An important point to note from Fig.~\ref{fig:2ptCOMP} is that the
$M_{\rm eff}$ data from the $M_\pi \approx 170$~MeV ensembles do not
show a plateau over the range $1 \lesssim \tau \lesssim 2$~fm, in
contrast to what is commonly assumed. Concomitantly, we find a
systematic difference in $M_0$ and ${\mathcal{A}}_0$ between the two
strategies, $\{4\}$ and $\{4^{\rm N\pi}\}$, with $\{4^{\rm N\pi}\}$
giving a 1--2$\sigma$ smaller value for both $M_0$ and
$|{\mathcal{A}}_0|^2$, and the relative difference growing as $M_\pi$
is reduced. Note that the correlated decrease in $M_0$ and
$|{\mathcal{A}}_0|^2$ under
$\{4\} \to \{4^{\rm N\pi}\}$ is consistent with both fits preserving
the asymptotic, $\tau \to \infty$, value of $C^{2\text{pt}}(\tau)$.
Such a variation implies that one has to re-examine the strategy for
even extracting $M_0$ in calculations where percent precision
is needed, such as in the calculation of the pion-nucleon sigma term,
$\sigma_{\pi N}$, using the Hellmann-Feynman
theorem~\cite{Aoki:2019cca,Gupta:2021ahb} and in the extraction of matrix elements
discussed here.  Consequently, we consider a number of strategies for
the analysis of charges and axial and vector form factors in
Sec.s~\ref{sec:ESC},~\ref{sec:charges},~\ref{sec:AFF}
and~\ref{sec:VFF}.

\section{Controlling excited-state contamination in three-point functions}
\label{sec:ESC}

The spectral decomposition of the three-point
functions, $C_\mathcal{O}^{3\text{pt}}$, truncated at 3 states is:
\begin{equation}
  C_\mathcal{O}^{3\text{pt}}(\tau;t) =
   \sum_{i,j=0}^2 {\mathcal{A}_i^{\bm p}} {\mathcal{A}_j}\matrixe{i^{\bm p}}{\mathcal{O}}{j} e^{-E_i t - E_j(\tau-t)}\,,
   \label{eq:3pt}
\end{equation}
%
where $O$ is the operator, $\mathcal{A}_i$ are the amplitudes
with which the states $|i \rangle$ are created by the interpolating operator
${\mathcal N}$ with energies $E_i$ as defined in
Eq.~\eqref{eq:2pt}. The source point has been translated to $t=0$, the
operator is inserted at time $t$, and the nucleon is annihilated at
the sink time slice $\tau$. In Eq.~\eqref{eq:3pt},
$\mathcal{A}_i^{\bm p} $ and $|i^{\bm p} \rangle$ denote that these
states could have nonzero momentum $\bm{p}$, whereas the momentum at the sink
is fixed to zero in all three-point functions. Thus, for 
momentum transfer $\bm q = \bm p$, the initial nucleon's momentum
is $-\bm p$.

In principle, the spectrum of the transfer matrix that contributes to
the three-point functions, Eq.~\eqref{eq:3pt}, should be obtainable from
the two-point function, Eq.~\eqref{eq:2pt}, however, the relative
contributions can vary significantly as mentioned above, particularly
in different 3-point functions. As a result, their contribution may be
manifest in some correlators but not in all. This is demonstrated for
the axial channel in Sec.~\ref{sec:A4} and for the vector current in
Sec.~\ref{sec:VFF}.

It is important for the reader to note that individual excited state
amplitudes ${\mathcal{A}_i^{\bm p}}$ and $ {\mathcal{A}_j}$ 
with $i,j > 0$, and their
values determined from fits to two-point functions,
$C^{2\text{pt}}(\tau)$, are never used in fits to
$C_\mathcal{O}^{3\text{pt}}(\tau;t)$. The reason is that in fits to
$C_\mathcal{O}^{3\text{pt}}(\tau;t)$, only the combinations
${\mathcal{A}_i^{\bm p}} {\mathcal{A}_j}\matrixe{i^{\bm
p}}{\mathcal{O}}{j}$ enter. Furthermore, while these combinations are 
unknown parameters in fits to $C_\mathcal{O}^{3\text{pt}}(\tau;t)$ 
to remove ESC, they are not used any further in the analysis. 

Data for the three-point functions have been accumulated for the 4--6
values of $\tau$ specified in Table~\ref{tab:Ensembles}, and for each
$\tau$ for all values $0 < t < \tau$. In much of the subsequent analyses, we make
$3^\ast$-state fits. These are three-state fits with the term
proportional to $\matrixe{2^{\bm p}}{\mathcal{O}}{2}$ set to zero as
it is not resolved with the current data and including it
overparameterizes the fit.

The spectral decomposition, given in Eqs.~\eqref{eq:2pt}
and~\eqref{eq:3pt}, forms the basis of all analyses of excited-state
contamination in two- or three-point functions.  In order to extract the
ground state matrix element
$\matrixe{0^{\bm p}}{\mathcal{O}_\Gamma}{0}$ for a given $\bm{p}$
using the three-state ansatz given in Eq.~\eqref{eq:3pt}, one has to, {\it a priori}, resolve 16
parameters from fits to $C_\mathcal{O}^{3\text{pt}}$ calculated as a
function of $t$ and $\tau$.  These are $\mathcal{A}_0,
\mathcal{A}_0^{\bm p}$, the three each $M_i$ and $E_i$, and the eight
products of the type $\abs{\mathcal{A}_0^{\bm p}}
\abs{\mathcal{A}_1}\matrixe{0^{\bm p}}{\mathcal{O}_\Gamma}{1}$
involving excited state transition matrix elements.  The ideal
situation occurs  when $\mathcal{A}_0, \mathcal{A}_0^{\bm p}$ and the three
$M_i$ and $E_i$ can be obtained from, say, fits to the two-point
functions for then the fit ansatz reduces to a sum of terms with a
linear dependence on the unknowns. This, however, requires the
states that provide significant contributions to two- and three-point
functions at the simulated values of $t$ and $\tau$ are the same---naively a
reasonable expectation since the same interpolating operator $\cal N$ is
used in both.

In Ref.~\cite{Jang:2019vkm}, we showed that, operationally, this
expectation fails for the form factors in the axial vector and pseudoscalar channels. In
fact, taking the three $M_i$ and $E_i$ from $\{4\}$-fits to
$C^{2\text{pt}}(\tau;\bm{p})$ to extract the axial vector form factors
from $C_{A_\mu}^{3\text{pt}}$ and $C_{P}^{3\text{pt}}$ gave results
that do not satisfy the PCAC relation between them.  Since the original
correlation functions, $C_{A_\mu}^{3\text{pt}}$ and
$C_{P}^{3\text{pt}}$, do satisfy PCAC up to discretization errors, the problem was shown to be
introduced while extracting the ground state matrix elements from the
correlation functions. We showed that the lower-energy excited states
$N(\bm q) \pi(-\bm q)$ and $N(\bm 0) \pi(-\bm q)$ contribute to the two sides of the 
operator insertion in the three-point functions even though they are not manifest in
straightforward fits to the two-point function.  The lesson was, one
cannot just take the spectrum obtained from the two-point function
with current statistics and apply it to all the three-point
functions. One has to explore and validate, both numerically and
theoretically, the relevant values of $M_i$ and $E_i$ to use in the
extraction of the various ground state matrix elements.

Theoretically, $N(\bm q) \pi(-\bm q)$ and $N(\bm 0) \pi(-\bm q)$ states
have much smaller energy, $E_1$, compared to that obtained from standard fits
to the two-point function.  (The 
noninteracting energies of multiparticle states in a finite box are taken to be
the sum of lattice single particle energies assuming a
relativistic dispersion relation.)  The clue to their relevance came
from fits to the three-point function with the insertion of the time
component of the axial current, $\langle \Omega | {\mathcal N}(\tau)
A_4(t) {\mathcal N}(0)|\Omega\rangle$~\cite{Jang:2019vkm}.  Fits to it
using Eq.~\eqref{eq:3pt} with the $E_i$ from standard fits to
$C^{2\text{pt}}(\tau;\bm{p})$ gave large $\chi^2/$dof. Consequently
these data were ignored in previous works (see
Ref.~\cite{Rajan:2017lxk}) because $G_A$ and ${\widetilde G}_P$ can be
determined from the $A_i$ correlators as defined in
Eqs.~\eqref{eq:r2ff-GPGA1}--\eqref{eq:r2ff-GPGA3}, i.e., the $A_4$ data
were superfluous because the system of equations,
Eqs.~\eqref{eq:r2ff-GPGA1}--\eqref{eq:r2ff-GPGA4}, is overdetermined.
The reason for the poor signal was that the ESC in this channel is
very large, in fact it dominates the signal. Exploiting this last fact led
us to determine the relevant mass gap[s], which are  much smaller than the
standard $\Delta E_1$, i.e., from $\{4\}$. \looseness-1

To analyze $\langle \Omega | {\mathcal N}(\tau) A_4(t) {\mathcal
  N}(0)|\Omega\rangle$ we, instead, used the two-state version of
Eq.~\eqref{eq:3pt} with the excited state energy $E_1$ left as a free
parameter~\cite{Jang:2019vkm}. The resulting value, labeled $E_1^{\rm
  A4}$, was close to the noninteracting $N\pi$ state, and much
smaller than what the fits to the two-point function gave (labeled
$E_1^{\rm 2pt}$).  The three form factors $G_A$, ${G}_P$ and
${\widetilde G}_P$, extracted using $E_1^{\rm A4}$, satisfied PCAC to
within expected lattice systematics. This resolution has, however, created
a conundrum for the analysis of all nucleon matrix elements---what are
the relevant excited-state energies, $E_i$, that contribute to a given
matrix element, how to determine them, and how to deal with the towers
of multiparticle states such as $N\pi$, $N\pi\pi$, $ \cdots$ that have
the same quantum numbers as the nucleon and become increasingly dense
as the lattice size $L \to \infty$.  Addressing these questions is
particularly hard for channels that do not have an independent check
such as PCAC.

The tools available include extracting the $E_i$ from fits
to the three-point functions themselves, getting guidance from heavy baryon
chiral perturbation theory, evaluating the full tower of excited
states that could contribute, and satisfying relations such as PCAC.
In this paper, we attempt to develop a framework to determine the
relevant $E_i$ for each matrix element considered and, if possible,
associate them with [multi]hadron states for a deeper understanding of
the excited states that contribute. For the axial channel, this is
done in Appendix~\ref{sec:AFFESC}, and for the vector channel in 
Sec.~\ref{sec:VFF}

Throughout the paper, we will use $M_i$ and $E_i$ for first 
excited-state energies determined from four-state fits to 
the two-point functions, and
${\widetilde M}_1$ and $ {\widetilde E}_1$ for the 
values obtained from two-state fits to the three-point functions.

\section{Extracting form factors from ground state matrix elements}
\label{sec:RFF}

All matrix elements are obtained from fits to the three-point
correlators with the insertion of the various components of the axial,
pseudoscalar, scalar, tensor and vector currents.  To display these three-point correlator data we construct the ratio, ${\cal R}_{\mathcal O}$,
of the three-point to the two-point correlation functions,
\begin{align}
{\cal R}_\mathcal{O}&(t, \tau, \bm{p}, \bm{0}) =   \frac{C^{3\text{pt}}_\mathcal{O}(t,\tau;\bm{p},\bm{0})}{C^{2\text{pt}}(\tau,\bm{p})} \, \times \, \nonumber \\
&
  \left[ \frac{C^{2\text{pt}}(t,\bm{p}) C^{2\text{pt}}(\tau,\bm{p}) C^{2\text{pt}}(\tau-t,\bm{0})}{C^{2\text{pt}}(t,\bm{0}) C^{2\text{pt}}(\tau,\bm{0}) C^{2\text{pt}}(\tau-t,\bm{p})}
  \right]^{1/2} \,,
\label{eq:ratio}
\end{align}
where $C^{2\text{pt}}$ and $C_\mathcal{O}^{3\text{pt}}$ are defined in
Eqs.~\eqref{eq:2pt} and~\eqref{eq:3pt}.  This ratio gives the desired
ground state matrix element in the limits $t \to \infty$ and $(\tau-t)
\to \infty$.  For all the two-point correlation functions in
Eq.~\eqref{eq:ratio}, we use the results of the appropriate four-state fit
instead of the measured values. When calculating the three-point
correlation functions, we use the spin projection  ${\cal P} =
(1 + \gamma_4)(1 + i\gamma_5 \gamma_3)/2$.  As a result, the ``3''
direction is special  while ``1'' and ``2'' are equivalent under the
rotational cubic symmetry. For the axial vector current, $\overline q
\gamma_5\gamma_\mu q$, the imaginary part of the $A_i$ and real part of $A_4$ 
have a signal in the following four ratios and give the desired form factors in the
limit $t$ and $(\tau-t) \to\infty$:
\begin{align}
  {\cal R}_{51} \rightarrow&\; \frac{1}{\sqrt{(2 E_p (E_p+M))}} \left[ -\frac{q_1 q_3}{2M} {\widetilde G}_P \right] \,,
  \label{eq:r2ff-GPGA1} \\
  {\cal R}_{52} \rightarrow&\; \frac{1}{\sqrt{(2 E_p (E_p+M))}} \left[ -\frac{q_2 q_3}{2M} {\widetilde G}_P \right] \,,
  \label{eq:r2ff-GPGA2} \\
  {\cal R}_{53} \rightarrow&\; \frac{1}{\sqrt{(2 E_p (E_p+M))}} \left[ -\frac{q_3^2}{2M} {\widetilde G}_P + (M+E) G_A \right]  \,,
  \label{eq:r2ff-GPGA3} \\
  {\cal R}_{54} \rightarrow&\; \frac{ q_3}{\sqrt{(2 E_p (E_p+M))}} \left[ \frac{M-E}{2M} {\widetilde G}_P + G_A \right] \,.
  \label{eq:r2ff-GPGA4}
\end{align}
The ${\widetilde G}_P$ can be determined from ${\cal R}_{51}$ with momenta 
$\bm q = (i, 0, j) \times (2\pi /La)$ and from ${\cal R}_{52}$ 
with $\bm  q = (0, i, j) \times (2\pi /La)$. In
practice, cases equivalent under the cubic symmetry 
are averaged before we make the ESC fits. The
$G_A$ can be determined uniquely from ${\cal R}_{53}$ with $q_3 = 0$. 
In the other momentum channels, the coupled set of equations,
Eqs.~\eqref{eq:r2ff-GPGA1}--\eqref{eq:r2ff-GPGA3}, are solved for $G_A$
and ${\widetilde G}_P$ using the full covariance matrix.  The $A_4$
correlator gives a second, and {so far considered} redundant because of
the much larger errors, linear combination of $G_A$ and ${\widetilde
  G}_P$. As discussed below, it will play an important role in
determining the first excited state parameters, and thus in the
overall analysis.

The pseudoscalar form factor $G_P(Q^2)$ is given by the real part of
${\cal R}_{5}$, i.e., with $\mathcal{O} = \overline q {\gamma_5} q$ in
Eq.~\eqref{eq:ratio}:
\begin{align}
  {\cal R}_{5} \rightarrow&\; \frac{1}{\sqrt{(2 E_p (E_p+M))}} \left[ q_3 {G}_P \right] \,. 
  \label{eq:r2ff-GP} 
\end{align}

For the electric and magnetic form factors, 
the following quantities, with $\mathcal{O}= (2 \overline u \gamma_\mu
u - \overline d \gamma_\mu d)/3$, have a signal: 
\begin{align}
  \sqrt{2E_p(M_N+E_p)} \, \Re ({\cal R}_{i}) =&\; - \epsilon_{ij3} q_j {G}_M  \,,
  \label{eq:GM1} \\
  \sqrt{2E_p(M_N+E_p)} \, \Im ({\cal R}_{i}) =&\;   q_i {G}_E  \,,
  \label{eq:GE1} \\
  \sqrt{2E_p(M_N+E_p)} \, \Re ({\cal R}_{4}) =&\;   (M_N+E_p) {G}_E  \,. 
  \label{eq:GE4} 
\end{align}
Exploiting the cubic symmetry under spatial rotations, we construct
two averages over equivalent three-point correlators before doing fits
to get the ground-state matrix elements: over $ \Re ({ C}_{1}^{\rm 3pt})$ 
and $\Re ({ C}_{2}^{\rm 3pt})$ for $G_M(Q^2)$ and over $ \Im ({ C}_{1}^{\rm 3pt})$, $ \Im ({
  C}_{2}^{\rm 3pt})$ and $ \Im ({ C}_{3}^{\rm 3pt})$ for $G_E(Q^2)$. We label these form
factors as $ G_M^{V_i}$ and $ G_E^{V_i}$. Together with $ G_E^{V_4}$
extracted from Eq.~\eqref{eq:GE4}, they constitute the three form
factors analyzed. Each is obtained from a distinct correlation
function, and it is important to note that the discretization artifacts
and the excited-state contaminations in these can be very different. 

We remind the reader that these ratios are used only to plot the
data. Our results are obtained by making $n$-state fits to the
correlation functions themselves. In making these fits we attempt to
balance statistical and systematic uncertainties. Data at smaller
$\tau$ have smaller statistical errors but larger ESC because a larger
number of states contribute.  Similarly, data close to the source and
the sink have larger ESC. Therefore, for each $\tau$ we neglect data
on $t_{\rm skip}$ time slices at either end, and we make fits to data
with the largest $\tau$ values that have statistically precise  
data. By skipping the same number of points, $t_{\rm skip}$, at all
$\tau$ fit, we increase the weight of the larger $\tau$ data to partially
compensate for the larger statistical weight given to the lower error
points at smaller $\tau$. \looseness-1

\begin{figure*}[htbp] 
\subfigure
{
    \includegraphics[width=0.24\linewidth]{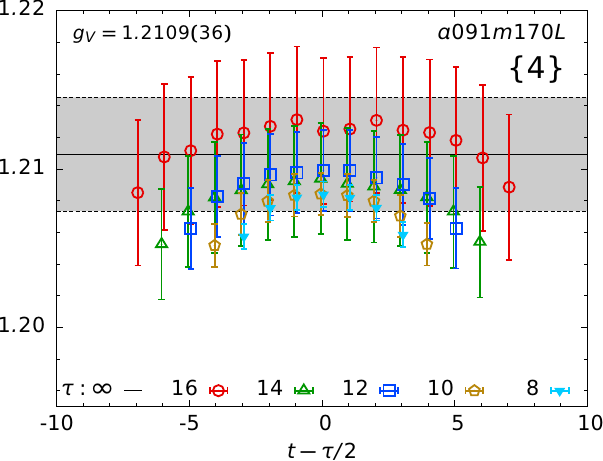}  
    \includegraphics[width=0.24\linewidth]{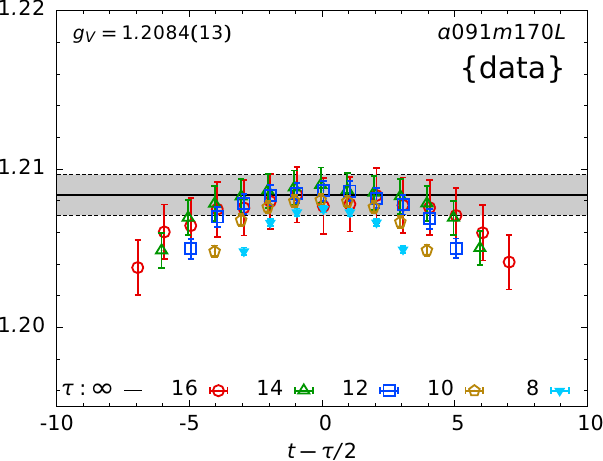}  
    \includegraphics[width=0.24\linewidth]{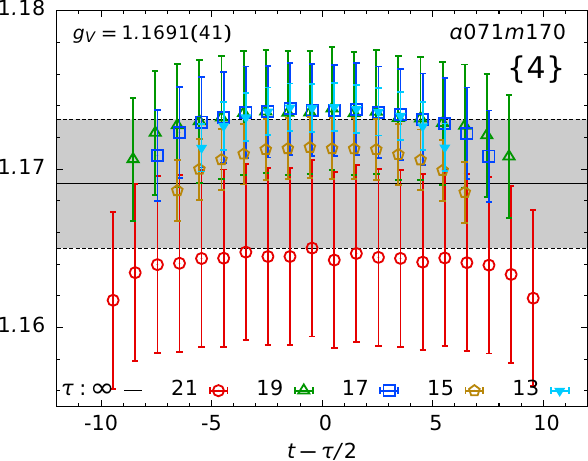}  
    \includegraphics[width=0.24\linewidth]{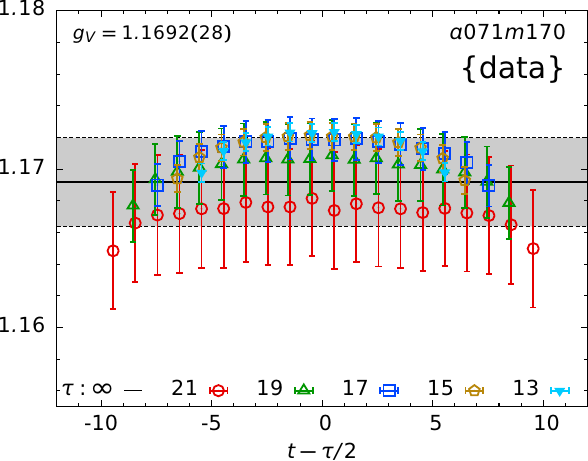}  
}
\caption{Data for the ratio $R_{V_4}$, which give $g_V$ in the limit
  $\tau \to \infty$, are  plotted versus $t-\tau/2$ for the $a091m170L$
  (left two) and $a071m170$ (right two) ensembles. Panels 1 and 3 
  show results using the $\{4\}$-state fit for $C^{\rm 2pt}$ in
  $R_{V_4}$, while in panels 2 and 4 the data for $C^{\rm 2pt}$ are
  used. The final estimate shown by the gray band is the average of
  the five (six) central points for $\tau=14,16$ ($\tau=19,21$) for
  the $a091m170L$ ($a071m170$) ensemble.
\label{fig:gVd7}}
\end{figure*}


\section{Extracting nucleon charges}
\label{sec:charges}

This section covers the calculations of the isovector nucleon charges,
$g_\Gamma^{u-d}$, from the forward matrix elements:
\begin{equation}
 \langle N(p, s) \vert \mathcal{O}_\Gamma \vert N(p, s) \rangle
 = g_\Gamma \bar{u}_s(p) \Gamma u_s(p) \,.
\label{eq:defcharges}
\end{equation}
For $ \bm{q} =0$ in the three-point functions, Eq.~\eqref{eq:ratio}
simplifies to ${\cal R}_{\mathcal{O}_\Gamma}(t, \tau, \bm{0}, \bm{0})
=
C^{3\text{pt}}_{\mathcal{O}_\Gamma}(t,\tau;\bm{0},\bm{0})/C^{2\text{pt}}(\tau,\bm{0})
$.  With the spin projection in the ``3'' direction, the Dirac matrix structure of  
operators used to calculate the scalar, vector, axial and tensor charges are 
$\Gamma = 1, \gamma_4, \gamma_3\gamma_5$ and $\sigma_{12}$,
respectively.  Since the nucleon states and all four operators, which
commute with $\gamma_4$, have positive parity, therefore all possible
excited states with positive parity are theoretically allowed in all
four channels: axial, scalar, tensor and vector. Based on conserved
symmetries alone, the ones with the smallest mass gap are
$N(0) \pi(0) \pi(0) $ or $N({\bm 1}) \pi (-{\bm 1})$. As mentioned
above, their noninteracting energies are roughly the same on each of
the seven ensembles. The unknown is their coupling in the
various channels.  Furthermore, the analysis of the two-point function
in Sec.~\ref{sec:spectrum} showed that there is a large range of
$M_1$ values with similar $\chi^2$/dof in four-state
fits. This range includes the $N \pi$ and $N \pi \pi$ states. We will,
therefore, investigate the impact on ground state matrix elements of
choosing values of $M_1$ over this interval, the lower end of which
is taken to be the approximately degenerate energy of these two
states ignoring interactions.

The question is how to determine, nonperturbatively, which of the
possible states contribute significantly?  In chiral perturbation theory, $N({\bm
  1}) \pi (-{\bm 1})$ arises at one-loop~\cite{Bar:2016uoj} and $N(0)
\pi(0) \pi(0) $ at two loops~\cite{Bar:2018wco} in the axial
channel. Similarly, the vector current couples to the $\rho$ meson
(vector meson dominance), or equivalently the two-pion state it decays
into for sufficiently small pion mass (see also the discussion in
Ref.~\cite{Hoferichter:2016duk}).  As will be shown later, the
contribution of these multihadron states increases with decreasing
$M_\pi$ (and $\bm q$ in the case of form factors) in both the axial
and the vector channels. More generally, it is, {\it a priori}, not
straightforward to narrow down the states that give significant
contributions to a particular correlation function. Again, the
criterion we will use is the $\chi^2/$dof of the fits, input from
$\chi$PT, and the sensitivity of the observables to the value of the
mass gaps used in the fit to judge the best strategy.

To include the effect of either of the two kinds of states, $N(0)
\pi(0) \pi(0) $ or $N({\bm 1}) \pi (-{\bm 1})$, we use the spectrum
from the $\{4^{N\pi}\}$ fit noting that the fit to the three-point
function only cares about $\Delta M_1$ and not the identity of the
state[s].  So, in the current analysis, the contributions from all three
possibilities, $N(0) \pi(0) \pi(0) $ or $N({\bm 1}) \pi (-{\bm 1})$ or
both, are included under the same label $\{4^{N\pi}\}$.  

We examine two more strategies, which we call $\{4,2^{\rm free}\}$ and
$\{4^{N\pi},2^{\rm free}\}$, in which $E_1$ is left a free parameter
to be fixed by a two-state fit to the three-point functions. Note that
these two strategies differ only in the ground state parameters
${\mathcal{A}}_0$ and $M_0$ (or $E_0$), which are slightly different
between the $\{4^{N\pi}\}$ and $\{4\}$ fits as shown in
Fig.~\ref{fig:2ptCOMP}.

Furthermore, in Appendix~\ref{sec:anatomy}, we examine the ESC in each
charge from operator insertion on the $u$ and $d$ quarks
separately. These data provide additional understanding of the
statistical precision of the data, and how the errors and ESC in the
isovector ($u-d$) and in the connected part of the isoscalar ($u+d$)
combinations, arise.

\begin{table*}[htbp]  
\centering
\begin{ruledtabular}
\begin{tabular}{l|ccc|ccc|ccc}
Ensemble ID & $g_V$ & $Z_V$ & $Z_V g_V$ & $Z_A$  & $Z_S$  & $Z_T$  & $Z_A/Z_V$  & $Z_S/Z_V$ & $Z_T/Z_V$  \\ \hline
$a127m285$  & 1.260(04) & 0.806(23) & 1.016(30) & 0.882(13) & 0.829(15) & 0.892(16) & 1.089(14) & 1.017(40) & 1.106(11)\\ 
$a094m270$  & 1.213(05) & 0.828(17) & 1.005(21) & 0.883(12) & 0.789(11) & 0.928(17) & 1.065(09) & 0.946(25) & 1.121(08)\\ 
$a094m270L$ & 1.203(02) & 0.829(19) & 0.997(23) & 0.886(14) & 0.796(14) & 0.929(19) & 1.070(10) & 0.958(29) & 1.122(09)\\ 
$a091m170$  & 1.210(03) & 0.832(20) & 1.006(24) & 0.882(13) & 0.790(15) & 0.931(20) & 1.061(11) & 0.947(27) & 1.122(08)\\ 
$a091m170L$ & 1.211(04) & 0.827(18) & 1.001(22) & 0.875(14) & 0.783(11) & 0.926(15) & 1.056(09) & 0.943(24) & 1.120(08)\\ 
$a073m270$  & 1.171(02) & 0.857(15) & 1.003(17) & 0.899(11) & 0.779(10) & 0.961(18) & 1.052(09) & 0.911(30) & 1.124(07)\\ 
$a071m170$  & 1.169(04) & 0.853(13) & 0.998(16) & 0.896(07) & 0.767(13) & 0.965(15) & 1.051(09) & 0.897(28) & 1.132(07)\\ 
\end{tabular}
\end{ruledtabular}
\caption{Results for the bare vector charge $g_V$ and the renormalization
  constants $Z_{A,S,T,V}$ calculated nonperturbatively on the lattice
  using the RI-sMOM scheme. The value of the product $Z_V g_V $ is
  consistent with unity and the errors in it are dominated by those in
  $Z_V$. Note that the errors in the ratios $Z_A/Z_V$ and $Z_T/Z_V$
  are smaller than those in $Z_A$ and $Z_T$, respectively, while those in $Z_S/Z_V$
  are larger than in $Z_S$.  }
\label{tab:gVZV}
\end{table*}

A comparison between fits with these four strategies is shown in
Figs.~\ref{fig:gAcomp},~\ref{fig:gScomp} and~\ref{fig:gTcomp} for the
three charges $g_{A,S,T}$.  The data show the
following common features:
\begin{itemize}
\item
The symmetry of $R_{\mathcal{O}}^{\rm 3pt}$ (and
$C_{\mathcal{O}}^{\rm 3pt}$) about the midpoint of the interval,
$t=\tau/2$, improves with statistics as expected. The observed
deviations, mostly in the largest $\tau$ data for $g_S$, are
statistical fluctuations (see also the discussion in
Appendix~\ref{sec:anatomy}).
\item
The value of $R_{\mathcal{O}}^{\rm 3pt}$ at each $t$
(especially at the midpoint, $t=\tau/2$) converges monotonically
toward the $\tau = \infty$ value. Having a clear monotonic behavior, i.e., 
not obscured by the errors, is important for choosing the
values of $\tau$ to keep in the $n$-state fits to remove ESC, and 
it improves the stability of the fits with respect to variations in $\tau$ and $t_{\rm skip}$. .
\end{itemize}
Having data with these features, hallmarks of high statistics calculations, improves the
reliability of three-state fits that we make to the largest three
(four) values of $\tau$ listed in Table~\ref{tab:Ensembles} to obtain
results in the limit $\tau \to \infty$ for $g_A$ and $g_T$ ($g_S$).
To evaluate the convergence of estimates for $g_{A,S,T}$ on each ensemble,
we compared results from the two- and $3^\ast$-state fits.  Using this
framework, and the methodology for statistical analysis given in
Sec.~\ref{sec:setup}, the four charges, $g_{A,S,T,V}$, are analyzed next.

\subsection{\texorpdfstring{$g_V$}{gV} and Operator Renormalization}
\label{sec:gV}

The data for the vector charge obtained from the correlator $\langle N(\tau,{\bm 0})
V_4(t, {\bm 0}) N(0,{\bm 0}) \rangle$ show a small (about 1\%)
variation over the range of $\tau$ values investigated as illustrated
in Fig.~\ref{fig:gVd7} for the $a091m170L$ and $a071m170$
ensembles. We show two versions of the ratio ${\cal R}_V(t,\tau,0,0)$:
$C^{3\text{pt}}_V(t,\tau;\bm{0},\bm{0})/C^{2\text{pt}}(\tau,\bm{0})|_{\rm
fit}$ and
$C^{3\text{pt}}_V(t,\tau;\bm{0},\bm{0})/C^{2\text{pt}}(\tau,\bm{0})$,
where in the first case we use the result of the $\{4\}$ fit,
$C^{2\text{pt}}(\tau,\bm{0})|_{\rm fit}$, while in the second case we use the
two-point function itself.  In both cases, the data are essentially
flat about $\tau/2$, so for the final value of $g_V$, we take the
average of 5--6 central points at the largest two values of $\tau$
using the first version. The errors in these estimates cover the
spread in the values at $\tau/2$ for the various $\tau$.

A check on these estimates of $g_V$ is that the product $Z_V g_V = 1$
within $O(a)$ discretization errors, where $Z_V$ is the renormalization constant for 
the local vector current used in this study. Values of $Z_V g_V $ are shown in
Table~\ref{tab:gVZV} and deviate from unity by $\lesssim 1\%$, i.e., by
an amount smaller than the errors in the product that come mainly from
$Z_V$.

The calculation of the renormalization constants
$Z_{A,S,T,V}$ for the local axial, scalar, tensor and vector quark bilinear
operators on the lattice is done using the regularization 
independent symmetric momentum (RI-sMOM) 
scheme~\cite{Martinelli:1994ty,Sturm:2009kb}. Results are then converted 
to the $\overline{MS}$ scheme at scale 2~GeV using two-loop matching
and three-loop running as described in Ref.~\cite{Yoon:2016jzj}.  The
calculation is done on all seven ensembles.  Using these estimates,
together with the conserved vector charge relation $Z_V g_V = 1$, we
present renormalized quantities calculated in two ways. In the first
method, labeled ${\rm Z}_1$, the renormalized results for operator $O$ are
given by $Z_O O$. In the second method, labeled ${\rm Z}_2$, we construct
the two ratios: $Z_O/Z_V$ and $g_O/g_V$ for the charges. For constructing $Z_O/Z_V$,
we start with the ratio of the two amputated three-point functions in the
RI-sMOM scheme, and for $g_O/g_V$, the ratio of the matrix element after 
making the excited-state fits for each. In both cases, these ratios are taken 
within the jackknife process. For $Z_2$, the expectation is a cancellation of
correlated fluctuations in each of the two ratios leading to smaller
overall errors. The data summarized in Table~\ref{tab:gVZV} show that
the errors in $Z_A/Z_V$ and $Z_T/Z_V$ are smaller than in $Z_{A,T}$
but not in $Z_S/Z_V$ versus $Z_S$. Furthermore, data in
Tables~\ref{tab:gAdP2z2} and~\ref{tab:gSTs4e7} show smaller errors from
${\rm Z}_2$ for $g_{A,T}$ and from ${\rm Z}_1$ for $g_S$. Results from 
the two methods, after the CCFV fits
carried out in Sec.~\ref{sec:CCFVcharges}, 
differ by $\sim 0.03$. When quoting the central 
values, we will choose to renormalize 
$g_{A,T}$ using ${\rm Z}_2$ and $g_S$ using ${\rm Z}_1$. The 
difference between the two estimates will be used to assign
an appropriate systematic uncertainty in the three charges.

\subsection{\texorpdfstring{$g_A$}{gA}}

\begin{figure}[tbp] 
\subfigure
{
    \includegraphics[width=0.92\linewidth]{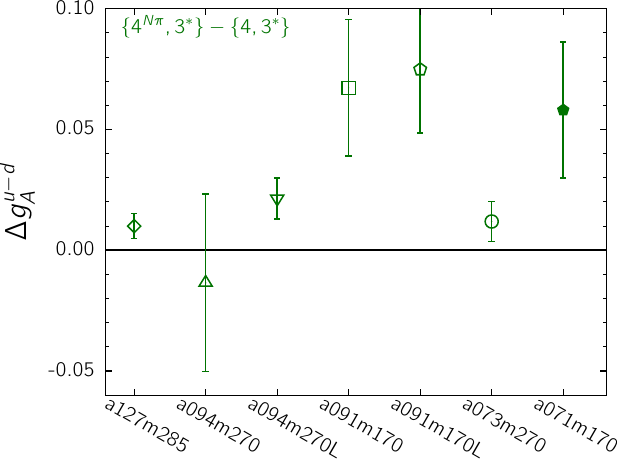}  
}
{
    \includegraphics[width=0.92\linewidth]{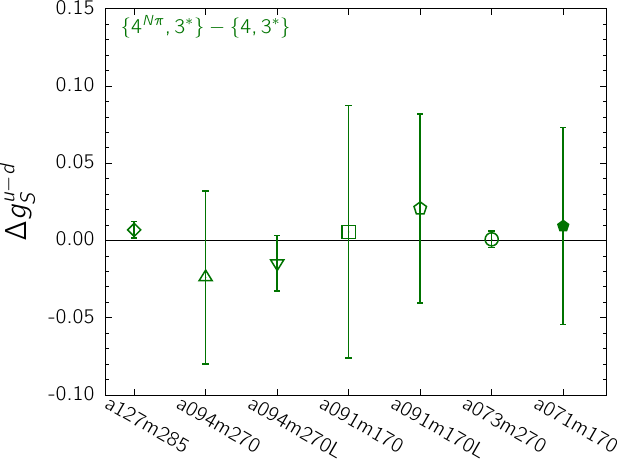}  
}
{
    \includegraphics[width=0.92\linewidth]{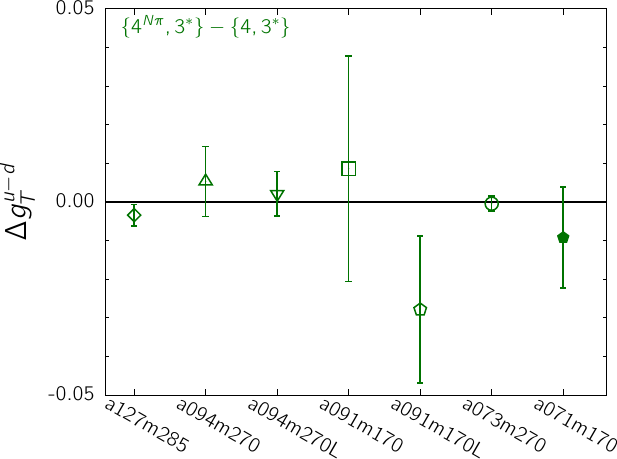}  
}
\caption{The difference in the
  renormalized (${\rm Z}_1$ method) axial (top), scalar (middle) and the tensor (bottom)
  charges between the two strategies, $\{4^{N\pi},3^\ast\} - \{4,3^\ast\} $. The data are shown 
  for all the seven ensembles.
\label{fig:diffcharges}}
\end{figure}

The findings from the four fit strategies, $\{4,3^\ast\}$,
$\{4^{N\pi},3^\ast\}$ (and their two-state versions $\{4,2\}$,
$\{4^{N\pi},2\}$ to check for overparameterization), 
$\{4^{},2^{\rm free}\}$ and $\{4^{N\pi},2^{\rm free}\}$ are the following: 
\begin{itemize}
\item 
The results from the $\{4^{},2^{\rm free}\}$ or $\{4^{N\pi},2^{\rm
free}\}$ fits are shown in Fig.~\ref{fig:gAcomp} by the broad gray bands 
and given in the labels. The output values of $\Delta
{\widetilde M}_1$ on all but the $a091m170L$ ensemble have large
errors and are much smaller than even those for the $N\pi$ state as shown
in Table~\ref{tab:deltaM}.  The reason is that the fluctuations
between the jackknife samples are unreasonably large. Lacking statistical 
control, we do not
consider these two strategies any further for $g_A$. In future higher
precision calculations, especially on $M_\pi \lesssim 200$~MeV
ensembles, we will continue to check whether estimates from the
$\{4^{},2^{\rm free}\}$ and $\{4^{N\pi},2^{\rm free}\}$ strategies
become more robust.
\item 
Overall, two- and $3^\ast$-state fits, irrespective of whether inputs of ground state 
parameters are from either the $\{4\}$ or the 
$\{4^{N\pi}\}$ fits to two-point functions, overlap on every
ensemble. The 3*-state fits are overparameterized with
respect to the two-state fits based on both the Akaike criteria and because the
uncertainty in the two additional fit parameters is $>100\%$ for the
following ensembles and strategies:\looseness-1
  \begin{itemize}
  \item $a094m270$: $\{4,3^\ast\}$, $\{4^{N\pi},3^\ast\}$
  \item $a091m170$: $\{4^{N\pi},3^\ast\}$
  \item $a091m170L$: $\{4,3^\ast\}$, $\{4^{N\pi},3^\ast\}$
  \end{itemize}
The values from $\{4^{N\pi},3^\ast\}$ agree with those from $\{4^{N\pi},2\}$ 
but have larger errors. To be conservative, we 
choose the $\{4^{N\pi},3^\ast\}$ results for all ensembles. 
\item 
There is a roughly $2\sigma$ difference between $\{4,3^\ast\}$ and
$\{4^{N\pi},3^\ast\}$ results on the $M_\pi \approx 170$~MeV
ensembles, $a091m170$, $a091m170L$ and $a071m170$, as shown in
Fig.~\ref{fig:diffcharges}. The $\{4^{N\pi},3^\ast\}$ values are
larger---a smaller mass gap implies a larger ESC and leads to a larger
$\tau \to \infty$ value since the convergence is from below as shown
in Fig.~\ref{fig:gAcomp}.  The difference is approximately $6\%$ at
$M_\pi = 170$~MeV, and becomes $\approx 8\%$ after the CCFV fits as
shown in Table~\ref{tab:gASTfinal} in Sec.~\ref{sec:CCFV}.
\item 
A similar difference of approximately $5\%$ is also present in the
axial form factor $G_A$ for the lowest nonzero momentum transfer,
${\vec q} = (1,0,0) 2 \pi /La$, data on the $M_\pi \approx 170$~MeV
ensembles between the $\{4,3^\ast\}$ and $\{4^{N\pi},3^\ast\}$
strategies as shown in Table~\ref{tab:GA-renormalized}.
\end{itemize}

The key issue to settle is whether the $N(1)\pi(-1)$ state, which 
is seen to contribute to the axial form factors at the lowest $Q^2$ and whose 
effect grows as $Q^2 \to 0$, also contributes at the approximately $5\%$  level
to the forward matrix element  as indicated by the data. We discuss this issue further 
in Sec.~\ref{sec:AFF}, and in Sec.~\ref{sec:CCFVcharges} where we
compare these estimates of $g_A$ to the second set of values obtained
by extrapolating $G_A(Q^2)$ to $Q^2=0$ using the dipole, Pad\'e and
$z$-expansion fits defined in Sec.~\ref{sec:AFFFinalFits}.


\begin{table*}      
\begin{ruledtabular}
\begin{tabular}{r|lll| lll|ll}
fit & $g_A^\text{bare}|_{{\bm q}=0}$ & $g_A|_{{\bm q}=0}^{{\rm Z}_1}$ & $g_A|_{{\bm q}=0}^{{\rm Z}_2}$ & $g_A^\text{bare}|_\text{dipole}$ & $g_A^\text{bare}|_{P_2}$ & $g_A^\text{bare}|_{z^2}$ & $g_A|_{z^2}^{{\rm Z}_1}$ & $g_A|_{z^2}^{{\rm Z}_2}$ \\ \hline
\hline  \multicolumn{9}{c}{$ a127m285$} \\ \hline  
$\{4,3^\ast\}$ & 1.433(13) & 1.264(22) & 1.238(19) & 1.424(13) & 1.423(14) & 1.424(13) & 1.255(21) & 1.230(19) \\ 
$\{4^{N\pi},3^\ast\}$ & 1.445(13) & 1.274(22) & 1.248(19) & 1.449(16) & 1.459(19) & 1.453(16) & 1.281(23) & 1.255(21) \\ 
$\{4^{N\pi},2^{A_4}\}$ & - & - & - & 1.458(18) & 1.488(26) & 1.465(20) & 1.291(25) & 1.266(23) \\ 
$\{4^{N\pi},2^\text{sim}\}$ & - & - & - & 1.429(20) & 1.432(32) & 1.421(22) & 1.252(26) & 1.227(24) \\ 
\hline  \multicolumn{9}{c}{$ a094m270$} \\ \hline  
$\{4,3^\ast\}$ & 1.431(51) & 1.263(48) & 1.256(45) & 1.360(27) & 1.390(52) & 1.386(33) & 1.224(34) & 1.216(30) \\ 
$\{4^{N\pi},3^\ast\}$ & 1.416(21) & 1.250(25) & 1.242(20) & 1.365(25) & 1.426(42) & 1.409(28) & 1.244(30) & 1.237(27) \\ 
$\{4^{N\pi},2^{A_4}\}$ & - & - & - &   &      &      &       &  \\ 
$\{4^{N\pi},2^\text{sim}\}$ & - & - & - & 1.350(25) & 1.379(49) & 1.375(33) & 1.213(33) & 1.206(31) \\ 
\hline  \multicolumn{9}{c}{$ a094m270L$} \\ \hline  
$\{4,3^\ast\}$ & 1.3892(96) & 1.231(21) & 1.236(14) & 1.387(9) & 1.393(10) & 1.392(9) & 1.234(21) & 1.239(14) \\ 
$\{4^{N\pi},3^\ast\}$ & 1.413(11) & 1.252(22) & 1.258(15) & 1.410(13) & 1.424(15) & 1.418(14) & 1.256(23) & 1.262(16) \\ 
$\{4^{N\pi},2^{A_4}\}$ & - & - & - & 1.412(10) & 1.434(13) & 1.426(11) & 1.264(22) & 1.269(15) \\ 
$\{4^{N\pi},2^\text{sim}\}$ & - & - & - & 1.397(12) & 1.414(17) & 1.406(14) & 1.246(23) & 1.251(17) \\ 
\hline  \multicolumn{9}{c}{$ a091m170$} \\ \hline  
$\{4,3^\ast\}$ & 1.419(20) & 1.251(25) & 1.244(21) & 1.399(15) & 1.402(19) & 1.413(19) & 1.247(25) & 1.240(21) \\ 
$\{4^{N\pi},3^\ast\}$ & 1.495(41) & 1.319(41) & 1.311(38) & 1.480(40) & 1.469(58) & 1.488(51) & 1.313(49) & 1.305(47) \\ 
$\{4^{N\pi},2^{A_4}\}$ & - & - & - & 1.412(21) & 1.504(36) & 1.504(31) & 1.327(34) & 1.319(31) \\ 
$\{4^{N\pi},2^\text{sim}\}$ & - & - & - & 1.421(25) & 1.442(41) & 1.451(37) & 1.280(38) & 1.273(35) \\ 
\hline  \multicolumn{9}{c}{$ a091m170L$} \\ \hline  
$\{4,3^\ast\}$ & 1.436(17) & 1.257(25) & 1.252(19) & 1.426(17) & 1.419(18) & 1.423(19) & 1.245(25) & 1.241(20) \\ 
$\{4^{N\pi},3^\ast\}$ & 1.521(41) & 1.331(42) & 1.327(39) & 1.502(44) & 1.487(51) & 1.496(49) & 1.309(47) & 1.305(45) \\ 
$\{4^{N\pi},2^{A_4}\}$ & - & - & - & 1.441(25) & 1.507(32) & 1.504(30) & 1.316(33) & 1.312(29) \\ 
$\{4^{N\pi},2^\text{sim}\}$ & - & - & - & 1.499(27) & 1.538(36) & 1.536(33) & 1.344(36) & 1.339(32) \\ 
\hline  \multicolumn{9}{c}{$ a073m270$} \\ \hline  
$\{4,3^\ast\}$ & 1.371(15) & 1.233(20) & 1.232(17) & 1.358(11) & 1.359(17) & 1.363(14) & 1.226(19) & 1.226(16) \\ 
$\{4^{N\pi},3^\ast\}$ & 1.384(12) & 1.245(18) & 1.244(15) & 1.361(11) & 1.402(18) & 1.392(13) & 1.251(19) & 1.251(16) \\ 
$\{4^{N\pi},2^{A_4}\}$ & - & - & - & 1.329(12) & 1.359(18) & 1.348(14) & 1.212(20) & 1.212(16) \\ 
$\{4^{N\pi},2^\text{sim}\}$ & - & - & - & 1.342(12) & 1.365(19) & 1.360(15) & 1.222(20) & 1.222(17) \\ 
\hline  \multicolumn{9}{c}{$ a071m170$} \\ \hline  
$\{4,3^\ast\}$ & 1.414(34) & 1.267(32) & 1.271(33) & 1.371(21) & 1.372(23) & 1.377(24) & 1.234(24) & 1.237(24) \\ 
$\{4^{N\pi},3^\ast\}$ & 1.479(38) & 1.325(36) & 1.329(36) & 1.448(37) & 1.476(49) & 1.484(46) & 1.329(42) & 1.333(43) \\ 
$\{4^{N\pi},2^{A_4}\}$ & - & - & - & 1.359(21) & 1.469(32) & 1.472(30) & 1.319(29) & 1.323(29) \\ 
$\{4^{N\pi},2^\text{sim}\}$ & - & - & - & 1.432(29) & 1.483(44) & 1.485(40) & 1.330(37) & 1.334(38) \\ 
\end{tabular}
\end{ruledtabular}
\caption{Results for $g_A$ from the seven ensembles and with the four
  strategies, specified in column one and defined in Appendix~\ref{sec:glossary}, used to control the excited state
  contamination.  The second column gives estimates from the forward
  matrix element ($q=0)$ for the two strategies $\{4,3^\ast\}$ and
  $\{4^{N\pi},3^\ast\}$ in which the excited state spectrum is taken
  from $\{4\}$ and $\{4^{N\pi}\}$ fits to ${\cal C}^{\rm
    2pt}$. Columns 5--7 give $g_A$ obtained by extrapolating $G_A(Q^2
  \neq 0)$ data using a dipole, $P_2$ Pad\'e and $z^2$ fits to all ten $Q^2
  \neq 0$ points. The fits to $\{4^{N\pi},2^{A_4}\}$ data on the
  $a094m270$ ensemble are not stable, so no results are presented.  The corresponding renormalized
  values using the two methods, ${\rm Z}_1 \equiv Z_A g_A^{\rm bare}$ and
  ${\rm Z}_2 \equiv (Z_A/Z_V) \times ( g_A^{\rm bare}/ g_V^{\rm bare})$, are
  given in columns 3-4 and 8-9.  }
\label{tab:gAdP2z2}
\end{table*}

  \begin{table*}        
  \begin{ruledtabular}
    \begin{tabular}{l |lll|lll}
fit & $g_S^\text{bare}$ & $g_S|^{{\rm Z}_1}$ & $g_S|^{{\rm Z}_2}$  &   $g_T^\text{bare}$ & $g_T|^{{\rm Z}_1}$ & $g_T|^{{\rm Z}_2}$  \\ 
\hline 
\multicolumn{7}{c}{$a127m285$} \\ \hline 
\{4,3*\} & 1.083(27)[0.94] & 0.897(28) & 0.874(41) & 1.173(10)[1.16] & 1.046(21) & 1.029(14) \\ 
\{$4^{N\pi}$,3*\} & 1.091(31)[0.96] & 0.904(30) & 0.880(43) & 1.169(12)[1.18] & 1.043(22) & 1.026(15) \\ 
\{$4,2^\text{free}$\} & 1.036(22)[1.16] & 0.858(24) & 0.836(37) & 1.1825(83)[1.10] & 1.055(20) & 1.038(13) \\ 
\{$4^{N\pi},2^\text{free}$\} & 1.041(21)[1.15] & 0.863(23) & 0.840(37) & 1.1839(92)[1.16] & 1.056(21) & 1.039(13) \\ 
\hline \multicolumn{7}{c}{$a094m270$} \\ \hline 
\{4,3*\} & 1.22(10)[1.21] & 0.965(83) & 0.953(84) & 1.102(24)[1.12] & 1.022(30) & 1.019(24) \\ 
\{$4^{N\pi}$,3*\} & 1.193(58)[1.22] & 0.942(48) & 0.930(51) & 1.108(19)[1.12] & 1.028(26) & 1.024(19) \\ 
\{$4,2^\text{free}$\} & 1.113(48)[1.19] & 0.878(40) & 0.867(44) & 1.140(25)[1.02] & 1.058(31) & 1.054(24) \\ 
\{$4^{N\pi},2^\text{free}$\} & 1.101(36)[1.21] & 0.869(31) & 0.858(36) & 1.133(10)[1.01] & 1.051(22) & 1.047(12) \\ 
\hline \multicolumn{7}{c}{$a094m270L$} \\ \hline 
\{4,3*\} & 1.195(24)[1.35] & 0.951(25) & 0.952(35) & 1.0923(86)[0.96] & 1.015(22) & 1.019(11) \\ 
\{$4^{N\pi}$,3*\} & 1.176(43)[1.33] & 0.936(38) & 0.937(44) & 1.095(13)[0.94] & 1.017(24) & 1.021(15) \\ 
\{$4,2^\text{free}$\} & 1.165(15)[1.44] & 0.927(20) & 0.928(30) & 1.1110(41)[1.03] & 1.032(22) & 1.0364(92) \\ 
\{$4^{N\pi},2^\text{free}$\} & 1.178(15)[1.44] & 0.938(20) & 0.939(31) & 1.1184(47)[1.11] & 1.039(22) & 1.0433(96) \\ 
\hline \multicolumn{7}{c}{$a091m170$} \\ \hline 
\{4,3*\} & 1.172(60)[0.96] & 0.926(51) & 0.918(54) & 1.054(14)[0.84] & 0.981(25) & 0.977(15) \\ 
\{$4^{N\pi}$,3*\} & 1.18(14)[0.95] & 0.93(11) & 0.92(11) & 1.063(39)[0.89] & 0.990(42) & 0.985(37) \\ 
\{$4,2^\text{free}$\} & 1.152(53)[0.98] & 0.910(45) & 0.902(48) & 1.083(12)[0.88] & 1.009(24) & 1.004(13) \\ 
\{$4^{N\pi},2^\text{free}$\} & 1.188(53)[1.00] & 0.938(45) & 0.930(49) & 1.107(16)[0.88] & 1.031(27) & 1.027(17) \\ 
\hline \multicolumn{7}{c}{$a091m170L$} \\ \hline 
\{4,3*\} & 1.145(73)[0.84] & 0.897(58) & 0.892(60) & 1.061(14)[0.96] & 0.983(20) & 0.982(15) \\ 
\{$4^{N\pi}$,3*\} & 1.17(14)[0.85] & 0.92(11) & 0.91(11) & 1.031(32)[1.01] & 0.955(34) & 0.954(31) \\ 
\{$4,2^\text{free}$\} & 1.132(43)[0.91] & 0.887(36) & 0.882(40) & 1.0977(91)[1.04] & 1.017(18) & 1.016(11) \\ 
\{$4^{N\pi},2^\text{free}$\} & 1.223(57)[0.95] & 0.958(47) & 0.952(50) & 1.149(26)[1.75] & 1.064(29) & 1.063(25) \\ 
\hline \multicolumn{7}{c}{$a073m270$} \\ \hline 
\{4,3*\} & 1.271(25)[1.13] & 0.989(23) & 0.989(37) & 1.0627(73)[0.87] & 1.021(21) & 1.0201(91) \\ 
\{$4^{N\pi}$,3*\} & 1.272(30)[1.09] & 0.990(26) & 0.989(40) & 1.0623(86)[0.88] & 1.020(21) & 1.020(10) \\ 
\{$4,2^\text{free}$\} & 1.230(14)[1.00] & 0.958(16) & 0.957(33) & 1.0823(51)[1.00] & 1.040(21) & 1.0389(78) \\ 
\{$4^{N\pi},2^\text{free}$\} & 1.235(14)[1.00] & 0.962(16) & 0.961(33) & 1.0853(46)[1.01] & 1.042(20) & 1.0418(76) \\ 
\hline \multicolumn{7}{c}{$a071m170$} \\ \hline 
\{4,3*\} & 1.22(13)[0.84] & 0.94(10) & 0.94(10) & 1.016(22)[0.92] & 0.980(26) & 0.983(22) \\ 
\{$4^{N\pi}$,3*\} & 1.24(21)[0.84] & 0.95(16) & 0.95(16) & 1.006(34)[0.89] & 0.971(36) & 0.974(33) \\ 
\{$4,2^\text{free}$\} & 1.182(72)[0.83] & 0.907(57) & 0.907(62) & 1.052(15)[0.89] & 1.016(21) & 1.019(16) \\ 
\{$4^{N\pi},2^\text{free}$\} & 1.230(72)[0.83] & 0.943(57) & 0.944(62) & 1.083(17)[0.96] & 1.045(23) & 1.049(18) \\ 
\end{tabular}
\end{ruledtabular}
\caption{Results for $g_S$ and $g_T$ on the seven ensembles and for the
  four strategies specified in column 1 and defined in
  Appendix~\ref{sec:glossary} that are used to control the excited
  state contamination.  The second and fifth columns give the bare
  values. The renormalized values using the two different methods,
  ${\rm Z}_1 \equiv Z_{S,T} g_{S,T}^{\rm bare}$ and ${\rm Z}_2 \equiv
  (Z_{S,T}/Z_V) \times ( g_{S,T}^{\rm bare}/ g_V^{\rm bare})$, are
  given in columns 3--4 and 6--7. The numbers within square brackets
  give the $\chi^2/$dof of the ESC fits. }
\label{tab:gSTs4e7}
\end{table*}


\subsection{\texorpdfstring{$g_S$}{gS}}

The data and fits to the largest four values of $\tau$ used to remove
ESC in $g_S$ are shown in Fig.~\ref{fig:gScomp}.  The statistical
errors in individual points are much larger compared to $g_A$ or
$g_T$, and are sizable for the $M_\pi \approx 170$~MeV ensembles. The
results after the ESC fits are collected together in Table~\ref{tab:gSTs4e7}. The notable
features in the data, fits, and results are the following:
\begin{itemize}
\item 
The $\{4,2^{\rm free}\}$ and $\{4^{N\pi},2^{\rm free}\}$ fits give
results with smaller errors compared to the $\{4^{N\pi},3^\ast\}$ and
$\{4,3^\ast\}$ fits. As shown in Table~\ref{tab:deltaM}, the $\Delta
{\widetilde M}_1 \approx 1$~GeV is, however, much larger than even
$\Delta {M}_1$, i.e., the result of the $\{4\}$-fits. Even accounting
for the fact that a two-state fit typically gives a larger $\Delta
{M}_1$ (this can be seen by comparing $a\Delta M_1^{\{2\}}$ with
$a\Delta M_1^{\{4\}}$ in Table~\ref{tab:deltaM}), the values from the
$\{2^{\rm free}\}$ fits are unexpectedly large.
\item 
Estimates from the four fit strategies are consistent on all ensembles
as shown in Fig.~\ref{fig:gScomp} and in Table~\ref{tab:gSTs4e7}. No
significant difference is observed between the $\{4^{N\pi},3^\ast\}$ and
$\{4,3^\ast\}$ values as shown in Fig.~\ref{fig:diffcharges}.  
\item 
The $\chi^2/$dof for the four fits are similar, so it cannot be used to 
distinguish between them.
\item 
The $3^\ast$-state fits are not overparameterized by the Akaike criteria.
\item 
On the $M_\pi \approx 170$~MeV ensembles, the $0 \leftrightarrow 2$
transition term is not well-determined in the $3^\ast$ fits.
\item 
The expected monotonic convergence in not yet realized for the
$\tau=19$ or 21 data on the $a071m170$ ensemble as shown in
Fig.~\ref{fig:gScomp}. However, as shown in Fig.~\ref{fig:UandDcomp}
in Appendix~\ref{sec:anatomy}, the data for the connected insertions
on $u$ and $d$ quarks do show it. On making the same ESC fits to each
of these to get the $\tau\to \infty$ values, and then constructing the
isovector combination $g_S^{u-d}$ gave overlapping values. The errors,
however, are larger, presumably because there is a cancellation of
fluctuations when fitting to the $u-d$ data.  The largest difference,
about $0.5 \sigma$, is in the $a091m170L$ and $a071m170$
ensembles. Based on an analysis of subsets of data, error reduction
comes mainly from the average over gauge configurations, i.e., the
average over multiple measurements on each configuration is less
effective as compared to that for $g_A$ and $g_T$.
\end{itemize}
Overall, we do not have an airtight criterion for picking one strategy
over the other. In Sec.~\ref{sec:CCFVcharges}, we perform the CCFV
extrapolation for all four cases, and the results, summarized in
Table~\ref{tab:gASTfinal}, show consistency within $1\, \sigma$.
Eventually in Sec.~\ref{sec:CCFVcharges}, we will invoke the fact that
the two $\{2^{\rm free}\}$ fits give an unexpectedly large $\Delta
{\widetilde M}_1$ to focus on the $\{4^{},3^\ast\}$ and
$\{4^{N\pi},3^\ast\}$ values, which give consistent results as shown
in Fig.~\ref{fig:diffcharges}.

\subsection{\texorpdfstring{$g_T$}{gT}}

The magnitude of the ESC and the errors in the data for $g_T$
are smaller than those in $g_A$ or $g_S$.  Nevertheless, we find that using a larger
$t_{\rm skip}$ improves the fits in many cases. Other features in the data are the following
\begin{itemize}
\item 
The $\chi^2/$dof of fits with all four strategies are, again,
reasonable and consistent as shown in Fig.~\ref{fig:gTcomp}.
\item
The $\Delta {\widetilde M}_1$ from $\{4,2^{\rm free}\}$ and
$\{4^{N\pi},2^{\rm free}\}$ strategies is determined with similar
precision (5--15\% error) as from the $\{4^{N\pi}\}$ and $\{4^{}\}$
fits to the two-point function. It is, however, much larger and
comparable to the values found in the $g_S$ analysis as shown in
Table~\ref{tab:deltaM}. Thus, the same argument made in the case of
$g_S$ for choosing results from $\{4,3^\ast\}$ or
$\{4^{N\pi},3^\ast\}$ applies.
\item
The $\{4,2^{\rm free}\}$ and $\{4^{N\pi},2^{\rm free}\}$ estimates are
systematically larger by 1--2$\sigma$ as can be seen in
Fig.~\ref{fig:gTcomp} and from Table~\ref{tab:gSTs4e7}.  This is
because a larger $\Delta {\widetilde M}_1$ leads to a smaller $\tau
\to \infty$ extrapolation and thus a larger $g_T$ because convergence is 
from above.
\item
We note a roughly $1\sigma$ difference between $\{4,3^\ast\}$ and
$\{4^{N\pi},3^\ast\}$ results on the $a091m170L$ and $a071m170$
ensembles, as shown in Fig.~\ref{fig:diffcharges}. While this $\approx
2\%$ difference is well within our error estimates, future
calculations, especially at $M_\pi \approx 135$~MeV, are needed to
confirm whether the low-lying multihadron states make a contribution
at the few percent level as $M_\pi \to 135$~MeV.
\item
For the $\{4^{N\pi},2^{\rm free}\}$ strategy, the gray band in
Fig.~\ref{fig:gTcomp} showing the $\tau = \infty$ value lies above the
largest $\tau$ data. This happens because the ratio data need not
converge monotonically for specific combinations of $\Delta
{\widetilde M}_1$ (or $\Delta { M}_1^{4^{\rm N\pi }}$) and the size of
the ESC in the three-point function. An example is when the
contribution of the excited states in the three-point function comes
with a positive sign (as for $g_T$ that converges from above) while
that from the two-point correlator always comes with a negative
sign. (The spectral decomposition of the two-point function in the
denominator is a sum of positive terms because our source and sink
interpolating operators are the same.)  We have checked that this
behavior describes our data, and leads to a nonmonotonic
convergence in the ratio for $g_T$, i.e., the ratio data go below the
gray band as $\tau$ is increased and then turn back up at values of
$\tau$ larger than accessible in current calculations. Our fits to the
three-point correlators, which show monotonic convergence in $\tau$, 
are, on the other hand, robust. 
\end{itemize}
Overall, as for $g_S$, the $\chi^2$/dof of the various fits to the
data do not help us select among the strategies.  We, therefore,
perform the CCFV extrapolation for all four strategies in
Sec.~\ref{sec:CCFVcharges} and then discuss our choice of the best
estimate.

\begin{figure*}[tbp] 
\subfigure
{
    \includegraphics[width=0.23\linewidth]{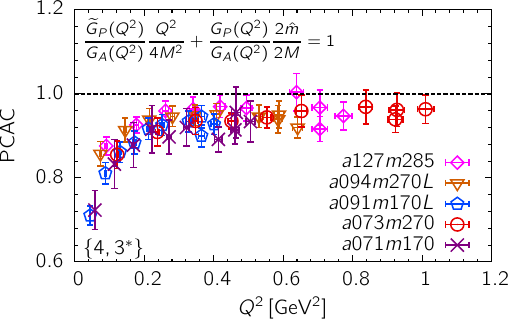}  
    \includegraphics[width=0.23\linewidth]{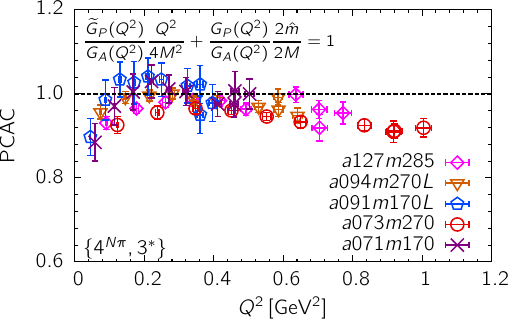}  
    \includegraphics[width=0.23\linewidth]{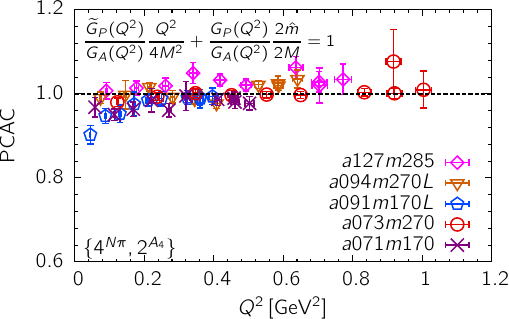}  
    \includegraphics[width=0.23\linewidth]{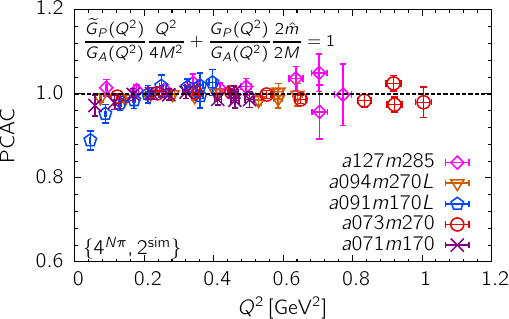}  
}
\subfigure
{
    \includegraphics[width=0.23\linewidth]{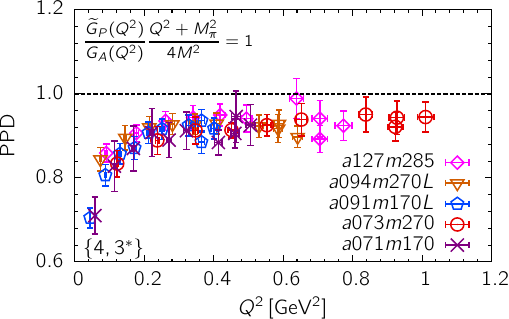}  
    \includegraphics[width=0.23\linewidth]{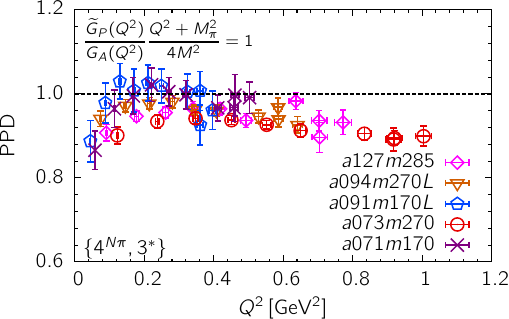}  
    \includegraphics[width=0.23\linewidth]{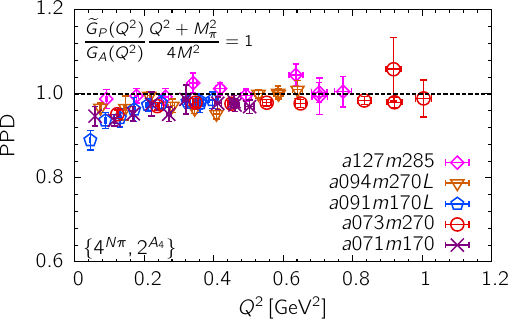}  
    \includegraphics[width=0.23\linewidth]{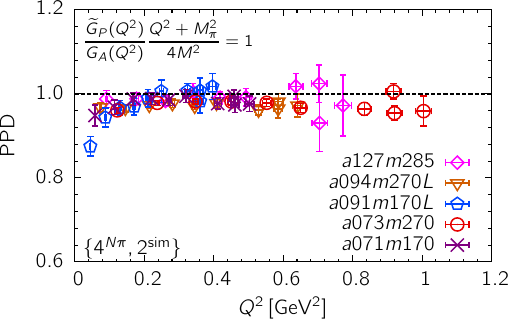}  
}
\subfigure
{
    \includegraphics[width=0.23\linewidth]{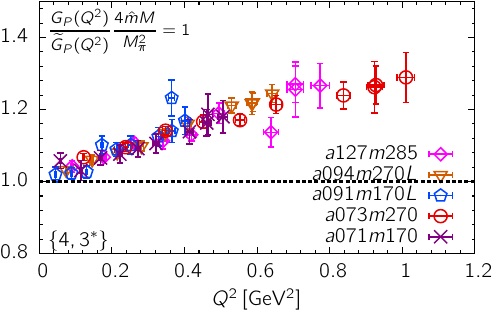}  
    \includegraphics[width=0.23\linewidth]{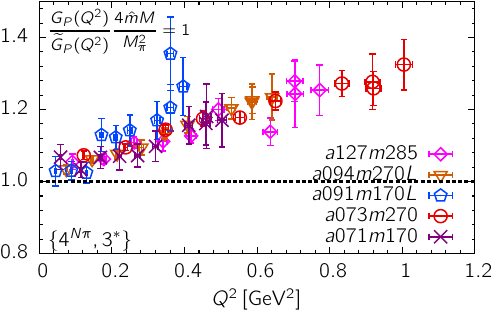}  
    \includegraphics[width=0.23\linewidth]{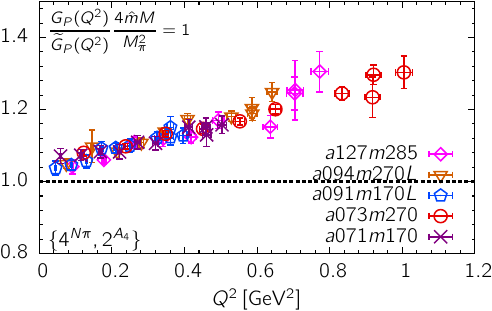}  
    \includegraphics[width=0.23\linewidth]{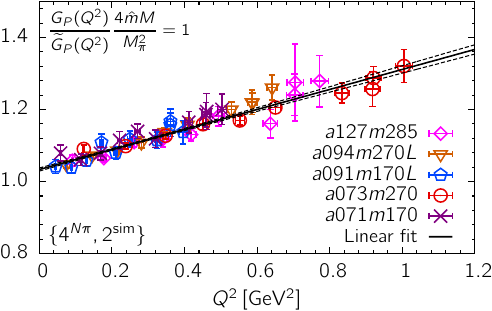}  
}
\caption{The top four panels show tests of the PCAC relation between
  the axial form factors for four analysis strategies specified at the
  bottom left corner. The middle panels show tests of the pion-pole
  dominance (PPD) hypothesis and the bottom panels show the quantity
  $4 M_N{\widehat m} G_P(Q^2) / M_\pi^2 {\widetilde G}_P(Q^2)$ that
  should equal unity for the PPD and the PCAC relation to be
  simultaneously satisfied.  A fit linear in $Q^2$ is shown in the bottom right panel. The symbols and color code used to show
  the data from the five larger volume ensembles are specified in the
  legends. Only data with $Q^2 \le 1$~GeV${}^2$ are shown as the
  errors above it are large in some cases.
\label{fig:PCAC}}
\end{figure*}
%

\section{The \texorpdfstring{$A_4$}{A4} three-point function at \texorpdfstring{$Q^2\neq 0$}{nonzero Q\suptwo} and Understanding ESC in \texorpdfstring{$G_A(Q^2)$}{GA(Q\suptwo)}}
\label{sec:A4}

In Ref.~\cite{Jang:2019vkm}, we showed that the first excited state
energies $M_1$ and $E_1$, obtained from the four-state fit $\{4\}$, are
much larger than those of the noninteracting multihadron
states relevant for extracting axial form factors: $N({\bm  q})
\pi({-\bm  q}) $ or $N({\bm  0}) \pi({-\bm  q})$, or $N \pi
\pi$ or even the $N(1440)$. The differences are striking at small
momentum transfers.  In fact, as illustrated in
Fig.~\ref{fig:2ptCOMP}, estimates of $E_1$ have large uncertainty, and only 
the ground state parameters are 
determined with few percent accuracy from fits to the two-point
functions. Even for $M_0$, in spite of the
seemingly long plateau in the effective-mass plots starting at
$\tau \sim 1$~fm, estimates from $\{4\}$- and $\{4^{N\pi}\}$-state fits
differ by 1--2\%.  In Ref.~\cite{Jang:2019vkm}, we also showed that
when ${\widetilde E}_1$ extracted from two-state fits to the $A_4$
three-point function $\langle {\cal N} (\tau,-\bm q) A_4(t,\bm
q) \overline{\cal N}(0,\bm 0)
\rangle$ is used to obtain $G_A$, $\widetilde{G}_P$ and $G_P$, the
PCAC relation between the three form factors is much better satisfied.  That strategy,
labeled $S_{\rm A4}$ in~\cite{Jang:2019vkm}, is called
$\{4^{},2^{\rm A_4}\}$ or $\{4^{N\pi},2^{\rm A_4}\}$ in this paper.

With high statistics data, we further explore the two- and three-state
fits to the $A_4$ correlator at nonzero momentum transfer.  We can
now make fits with the full covariance matrix and can take the first
excited state parameters from two-point correlators or leave $M_1$ and $E_1$
free along with the matrix elements, i.e., take only $M_0$, $E_0$,
${\cal A}_0$ and ${\cal A}_0^{\bm p}$ from one of the two four-state fits to
the two-point function.  To quantify the sensitivity of the form
factors to different choices for the mass gaps, we investigate
six strategies: $\{4^{},3^\ast\}$, $\{4^{N\pi},3^\ast\}$,
$\{4^{},2^{\rm A_4}\}$, $\{4^{N\pi},2^{\rm A_4}\}$, $\{4^{},2^{\rm
  sim}\}$ and $\{4^{N\pi},2^{\rm sim}\}$. The last two involve a
simultaneous fit, with common ${\widetilde M}_1$ and ${\widetilde
  E}_1$, to all four $A_\mu$ and the $P$ three-point functions as
discussed below. A more detailed discussion of the possible
excited states and the limitations of analyses is given in
Appendix~\ref{sec:AFFESC}. 

The first comparison of such fits to the three-point function $\langle
{\cal N} (\tau) A_4(t) \overline{\cal N}(0) \rangle$ is shown in
Fig.~\ref{fig:affA4COMP} for the four strategies $\{4^{},3^\ast\}$,
$\{4^{N\pi},3^\ast\}$, $\{4^{N\pi},2^{\rm A_4}\}$ and
$\{4^{N\pi},2^{\rm sim}\}$.  Data from six ensembles are shown for
momentum transfer ${\bm n} = (0,0,1)$ as these have large ESC.  For
the $\{4^{},3^\ast\}$ strategy, the $\chi^2/$dof of the fits, given in
the labels in Fig.~\ref{fig:affA4COMP}, are uniformly bad as was
pointed out in Ref.~\cite{Jang:2019vkm}. Also, as shown in
Fig.~\ref{fig:PCAC}, the form factors obtained with this strategy
do not satisfy the PCAC relation rewritten as
\begin{equation}
\frac{Q^2}{4 M_N^2} \frac{{\widetilde G}_P(Q^2)}{G_A(Q^2)} + 
\frac{2 {\widehat m}}{2M_N} \frac{G_P(Q^2)}{G_A(Q^2)} = 1 \,,
\label{eq:testPCAC}
\end{equation}
with $\widehat m$ given in Table~\ref{tab:gpiNN}.  Even though
$\{4^{},3^\ast\}$ data fail the PCAC test, we will continue to perform
a full analyses with it for the purpose of comparison.

The $\chi^2/$dof improves significantly with $\{4^{N\pi},3^\ast\}$ and
is the best with $\{4^{N\pi},2^{\rm A_4}\}$ as shown in
Fig.~\ref{fig:affA4COMP}.  The $\chi^2/$dof of the
$\{4^{N\pi},2^{\rm sim}\}$ fit is similar, however, recall it involves a
simultaneous fit to all five correlators. Also, estimates of
${\widetilde M}_1$ and ${\widetilde E}_1$ are similar in the two
cases.  The same is true with respect to satisfying PCAC as shown in
Fig.~\ref{fig:PCAC}.

Next, note that $\Delta M_1$ and $\Delta E_1$ decrease on going from
$\{4^{},3^\ast\}$ to $\{4^{N\pi},3^\ast\}$ to $\{4^{N\pi},2^{\rm
A_4}\}$; and the difference between $\Delta M_1$ and $\Delta E_1$ also
changes. Overall, the behavior using strategy $\{4^{N\pi},2^{\rm
A_4}\}$ is consistent with the results in Ref.~\cite{Jang:2019vkm}, i.e.,
(i) the $\chi^2/{\rm dof}$ of the fits are much reduced\footnote{The
$\chi^2$/dof is still large in many cases indicating that the fit
ansatz used to control ESC does not fully describe the data and
highlights the need for a more nuanced understanding of excited states
that contribute significantly. This caveat should be considered
implicit throughout the paper.}; (ii) the ${\widetilde M}_1$ and
${\widetilde E}_1$, which we label as ${\widetilde M}_1^{A_4}$ and
${\widetilde E}_{1}^{A_4}$, are much smaller than those obtained from
the $\{4\}$-fits to the two-point correlation function; and (iii)
${\widetilde M}_1^{A_4}$ and ${\widetilde E}_{1}^{A_4}$ are roughly
consistent with the noninteracting energies of $N({\bm q}) \pi({-\bm
q})$ and $N({\bm 0})
\pi({-\bm q})$ states, respectively, as shown in
Fig.~\ref{fig:AFF-deltaM}. These features are also consistent with the
effective field theory ($\chi$PT) result that, at leading (tree) order, the
axial current inserts a pion with momentum $\bm q$, i.e., the pion-pole
dominance (PPD) hypothesis~\cite{Bar:2018xyi,Bar:2019gfx}.

In contrast, fits to the $A_i$ correlators with ${\widetilde M}_1$ and
${\widetilde E}_1$ left as free parameters do not have good
$\chi^2/{\rm dof}$, i.e., these correlators do not constrain the
excited-state parameters. The reason is that the ground state dominates in the $A_i$ correlators, whereas the excited state is
dominant in $A_4$.\looseness-1

Using the ${\widetilde M}_1$ and ${\widetilde E}_1$ obtained from fits to $A_4$ to also analyze
$A_i$ and $P$ leads to form factors that are in much better agreement
with PCAC relation as shown in Fig.~\ref{fig:PCAC}.  This step, however,
assumes that the same combination of excited states provides the
dominant contribution to all five ($O = A_\mu$ and $P$) correlation
functions. If this is the case then, statistically, the more sound
method is to fit these five correlators simultaneously with common
${\widetilde M}_1$ and ${\widetilde E}_1$.  These strategies are labeled $\{4,2^{\rm sim}\}$ and 
$\{4^{N\pi},2^{\rm sim}\}$.  As expected, the resulting ${\widetilde M}_1$ and
${\widetilde E}_1$ from these simultaneous fits are similar to $M_1^{A_4}$ and
$E_1^{A_4}$ because these are mainly controlled by the $A_4$ correlator.

Figure~\ref{fig:PCAC} also shows tests of the pion-pole dominance
hypothesis, which, with the Goldberger-Treiman
relation~\cite{Goldberger:1958vp}, relates ${\widetilde G}_P(Q^2)$ to
$G_A(Q^2)$ as 
\begin{equation}
 \frac{Q^2+M_\pi^2}{4 M_N^2} \frac{{\widetilde G}_P(Q^2)}{G_A(Q^2)}   = 1 \,.
\label{eq:PPD} \\
\end{equation}
The behavior of the data for the combination in 
Eq.~\eqref{eq:testPCAC} (PCAC) and Eq.~\eqref{eq:PPDtest} (PPD) is very
similar and correlated, and $\{4^{N\pi},2^{\rm sim}\}$ gives the most
consistent outcome.  Noting this strong correlation, we examine the
relation 
\begin{equation}
2 {\widehat m} \frac{ 2 M_N}{M_\pi^2 } \frac{G_P(Q^2)}{ {\widetilde  G}_P(Q^2)} = 1 \,,
\label{eq:PPDtest}
\end{equation} 
which should hold for the PCAC relation, Eq.~\eqref{eq:testPCAC}, and
PPD, Eq.~\eqref{eq:PPD}, to be simultaneously satisfied. Following Ref.~\cite{Bernard:2001rs}, 
and working to first order in $\chi$PT in $M_\pi^2$ and $Q^2$, 
the left hand side of Eq.~\eqref{eq:PPDtest} can be expanded as 
\begin{equation}
 1 + \Delta + \frac{1}{6} \langle r^2_A \rangle M^2_\pi +
\frac{Q^2}{M^2_\pi} \left( \Delta + \frac{1}{6}\langle r^2_A \rangle M^2_\pi \right) \,,
\label{eq:PPDtestCPT} 
\end{equation}
where $\Delta \equiv 2 {\overline d}_{18} M_\pi^2/g_A$ is the
Goldberger-Treiman discrepancy, and ${\overline d}_{18} $ is an
unknown low-energy constant.  The data for the left hand side of
Eq.~\eqref{eq:PPDtest}, also presented in Fig.~\ref{fig:PCAC}, show
that the ratio is close to unity at $Q^2=0$ and has a significant,
essentially linear, increase with $Q^2$ on all seven ensembles. A
linear fit to the five larger volume ensembles, shown in the bottom
right panel in Fig.~\ref{fig:PCAC}, gives $1.033(5) + 0.272(16) Q^2$,
with $Q^2$ in GeV${}^{2}$. From Eq.~\eqref{eq:PPDtestCPT}, the 
quantity $ (\Delta + \frac{1}{6}\langle r^2_A \rangle M^2_\pi)$ should
equal the intercept minus one, and also the slope times
$M_\pi^2$. Using our result $\langle r^2_A \rangle = 0.43$~fm${}^2$
presented in Sec.~\ref{sec:CCFVrA}, we get $\Delta \sim 0$ from the
intercept and $\sim -0.02$ from the slope.  For comparison, using the
Goldberger-Treiman relation $g_A M_N = g_{\pi N N} F_\pi (1 + \Delta)$
and the experimental values $g_A = 1.27641$, $M_N=939$~MeV, $F_\pi =
92.2$~MeV and $g_{\pi N N } =
13.25$~\cite{Perez:2016aol,Reinert:2020mcu,Baru:2011bw} gives
$\Delta \sim -0.02$. In short, we show that the ratio defined in
Eq.~\eqref{eq:PPDtest} is not unity, and exhibits a linear dependence
on $Q^2$ that is consistent with the prediction of $\chi$PT.

The data in Fig.~\ref{fig:PCAC} also show that with the
$\{4^{N\pi},2^{\rm sim}\}$ strategy, the smallest $Q^2$ points on the
$a091m170L$ ensemble start to deviate away from unity for both the PCAC and
PPD relations but not those from the $a071m170$ ensemble.  In
contrast, for the $\{4,3^\ast\}$ strategy, the data from both
ensembles bend down at small $Q^2$, which we have shown is due to the
missed $N\pi$ states.  To investigate this difference between the
$a091m170L$ and $a071m170$ data with the $\{4^{N\pi},2^{\rm sim}\}$
strategy, we show $(Q^2 + M_\pi^2){\widetilde G}_P(Q^2)$ versus $Q^2$
in Fig.~\ref{fig:PPD} in Appendix~\ref{sec:compAFF}, and note that the
data move up as $a \to 0$ for all but the $\{4,3^\ast\}$ strategy, i.e.,
they indicate a dependence on $a$ when the $N\pi$ state is
included. Nevertheless, we cannot pinpoint whether the difference in
behavior is a discretization effect or a combination of statistical
and/or larger discretization effects in the $a091m170L$ data, or
indicates the need to include additional [multihadron] low energy excited states
in the fits. In the near future, we plan to double the statistics on
these two ensembles to better quantify the difference and explore
adding a third state, i.e., a $\{4^{N\pi},3^{\rm sim}\}$ fit.

\subsection{\texorpdfstring{$\{4^{N\pi},2^{\rm sim}\}$}{4N\textpi,2sim} is our preferred strategy for 
analyzing the axial form factors}
\label{sec:AFFconsistency}

Data from the two strategies $\{4^{N\pi},2^{A_4}\}$ and $\{4^{N\pi},2^{\rm
  sim}\}$ show much better agreement with the PCAC and PPD relations as
shown in Fig.~\ref{fig:PCAC}. To choose between them, we consider two
additional checks: First, the ground state matrix elements extracted
from the $A_4$ correlator with ${\bm q } \neq 0$ should satisfy the
relation $\partial_4 A_4 = (E_0-M_0) A_4$ for all $\bm
q$. Second, the value of the ground state matrix element $\langle {N}|
A_4 |N \rangle$ extracted from fits to $\langle {\cal N} A_4 {\cal N}
\rangle$ should agree with that reconstructed by inserting $G_A$ and
$\widetilde G_P$ calculated from the $A_i$ correlators into the right
hand side of Eq.~\eqref{eq:r2ff-GPGA4}.  The first condition is
satisfied by both strategies even though $\langle {N}| A_4 |N \rangle$
is very poorly determined with $\{4^{N\pi},2^{A_4}\}$. The second
check is satisfied within errors only by data from $\{4^{N\pi},2^{\rm
  sim}\}$. Based on these two consistency checks and the PCAC
relation, we select $\{4^{N\pi},2^{\rm sim}\}$ as our preferred
strategy for analyzing the axial form factors, however, we will
continue to examine all six strategies discussed above to exhibit the
spread.

The obvious next step is $\{4^{N\pi},3^{\rm sim}\}$ fits, i.e.,
leaving the first and second excited-state energy gaps as free
parameters (or using priors for them) in fits to the three-point
functions. With current data, we do not get meaningful results. Much
higher statistics are required.

\section{Axial vector form factors}
\label{sec:AFF}

As discussed in Sec.~\ref{sec:A4}, we compare six strategies to
extract the axial vector form factors, with our preferred one being
$\{4^{N\pi},2^{\rm sim}\}$. It makes the following assumption: the
excited-state contamination in all five channels, $A_\mu$ and $P$,
can, to a good approximation, be accounted for by a ``single low mass
effective excited state'' whose parameters can be determined from a
simultaneous two-state fit to the five three-point functions. Only the
ground state parameters are taken from fits to the two-point
functions.  

We find that the two sets of estimates using $\{4^{N\pi},2^{A_4}\}$ and
$\{4^{N \pi},2^{\rm sim}\}$, versus $\{4^{},2^{A_4}\}$ and
$\{4^{},2^{\rm sim}\}$ fits give overlapping results for the form factors, which
satisfy PCAC equally well. These two sets differ only in the $M_0$ and 
${\cal A}_0$ obtained from the $\{4^{N\pi}\}$- and $\{4\}$-state fits 
to the two-point functions, and these differences do 
not significantly impact the results for the form factors. It is the 
mass gap of the first excited state used in the fits to the 
three-point function 
that is important. In both the $\{2^{A_4}\}$ and $\{2^{\rm sim}\}$ fits, 
the output $\Delta {\widetilde E}_1$ is controlled by the $A_4$ correlator 
and corresponds to the $N \pi$ state as discussed in Sec.~\ref{sec:A4}. 
Thus, the impact of including the $N \pi$ state is far more significant in the
3-point functions, however, our approach is to consistently choose 
strategies in which the  mass gap in both the two- and three-point functions 
does or does not include the low-lying ($N \pi$) state. 
This is achieved with the $\{4,3^\ast\}$,
$\{4^{N \pi},3^\ast\}$, $\{4^{N\pi},2^{A_4}\}$ and $\{4^{N \pi},2^{\rm
sim}\}$ strategies (see Appendix~\ref{sec:glossary} for their definition), 
which are, therefore, used to present the final
results.  We do not discuss estimates from the $\{4^{},2^{A_4}\}$ and
$\{4^{},2^{\rm sim}\}$ strategies any further since all we can add
from their analysis is they give results consistent with 
$\{4^{N\pi},2^{A_4}\}$ and
$\{4^{N \pi},2^{\rm sim}\}$.\looseness-1

\begin{figure*}[tbp] 
\subfigure
{
    \includegraphics[width=0.47\linewidth]{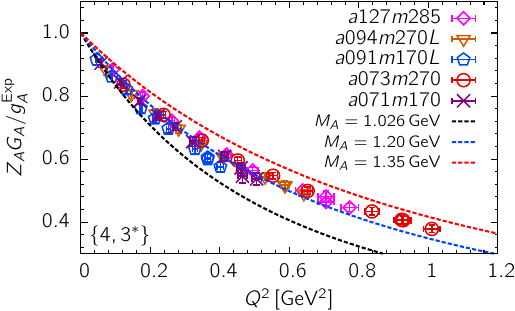} 
    \includegraphics[width=0.47\linewidth]{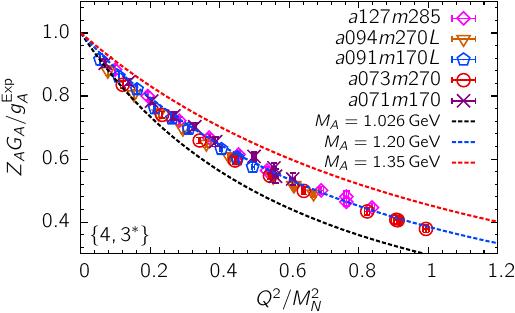} 
}
{
    \includegraphics[width=0.47\linewidth]{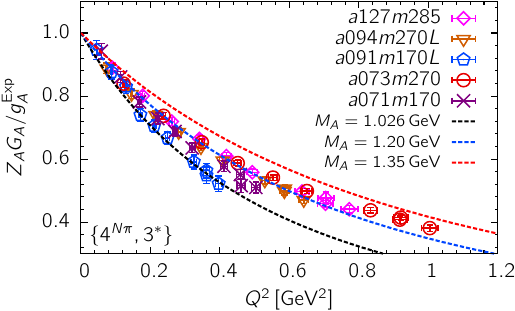} 
    \includegraphics[width=0.47\linewidth]{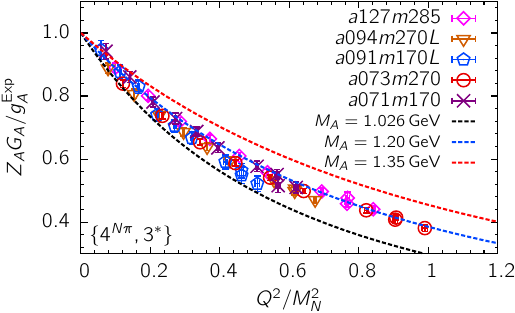} 
}
{
    \includegraphics[width=0.47\linewidth]{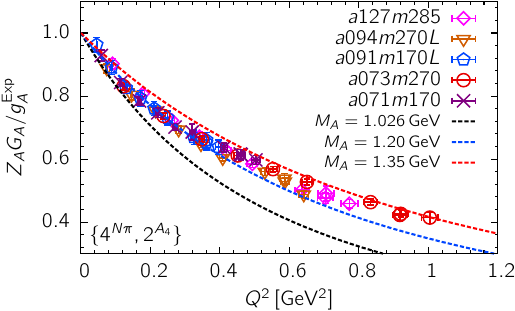} 
    \includegraphics[width=0.47\linewidth]{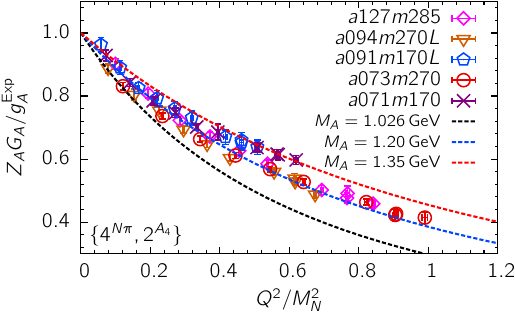} 
}
{
    \includegraphics[width=0.47\linewidth]{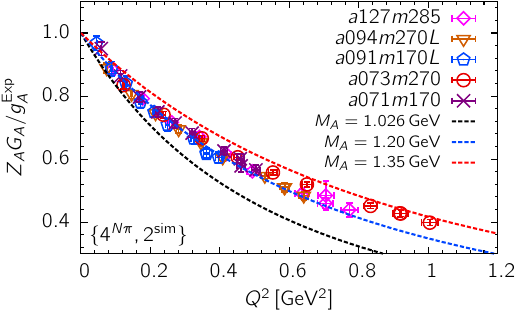} 
    \includegraphics[width=0.47\linewidth]{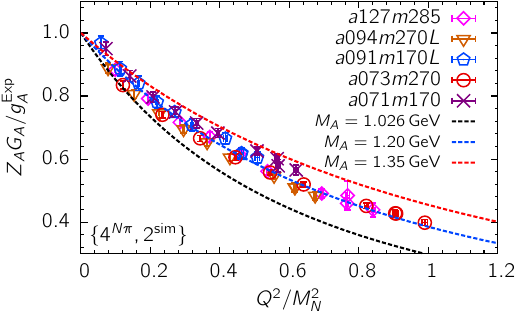} 
}
\caption{The data for the renormalized axial form factor $Z_A
  G_A(Q^2)/g_A^{\rm exp}$, with $g_A^{\rm exp}=1.277$,  are plotted versus $Q^2$ in GeV${}^2$ (left)
  and $Q^2/M_N^2$ (right).  Each panel shows the data from the five
  larger volume ensembles. The four rows show the results from four
  strategies, specified at the lower left corner of each panel that
  are used to control ESC. The three curves show the dipole ansatz
  with $M_A=1.026$, 1.2 and 1.35~GeV, and have been drawn only to
  guide the eye. The agreement of the three $\sim 270$ and the two 170~MeV
  data indicates that discretization errors are small. }
\label{fig:GA4s5e}
\end{figure*}

\begin{figure*}[tbp] 
\subfigure
{
    \includegraphics[width=0.47\linewidth]{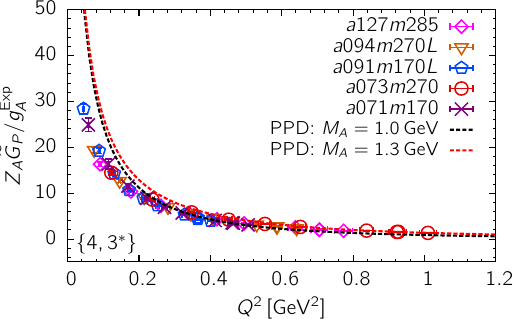} 
    \includegraphics[width=0.47\linewidth]{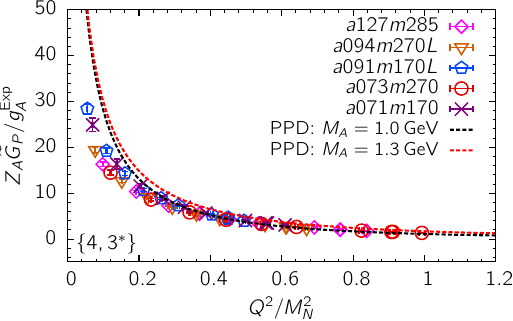} 
}
{
    \includegraphics[width=0.47\linewidth]{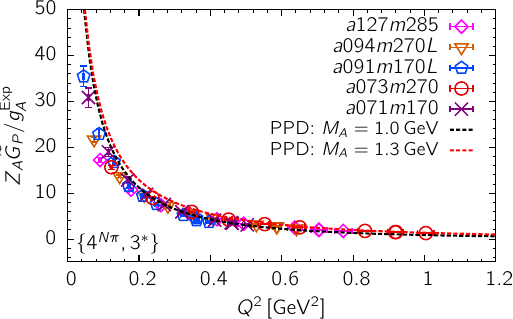} 
    \includegraphics[width=0.47\linewidth]{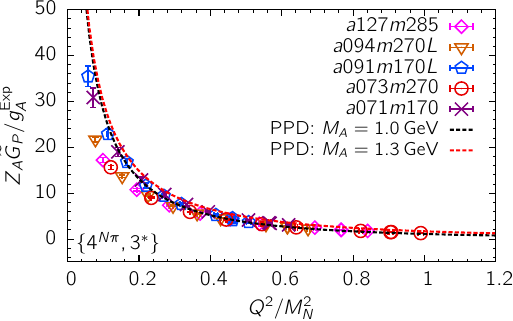} 
}
{
    \includegraphics[width=0.47\linewidth]{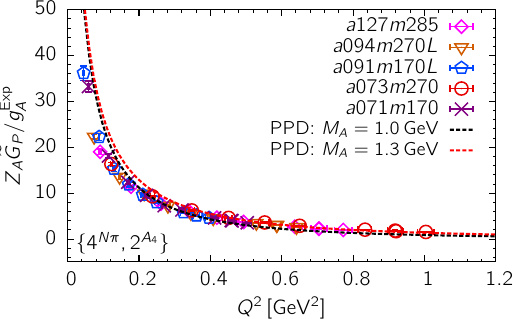} 
    \includegraphics[width=0.47\linewidth]{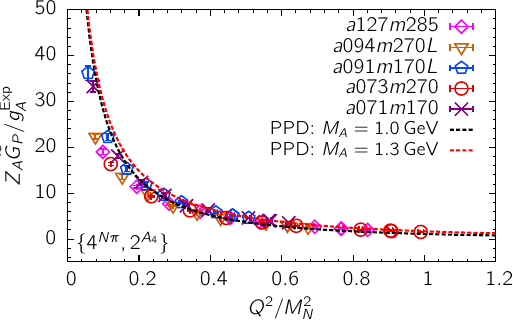} 
}
{
    \includegraphics[width=0.47\linewidth]{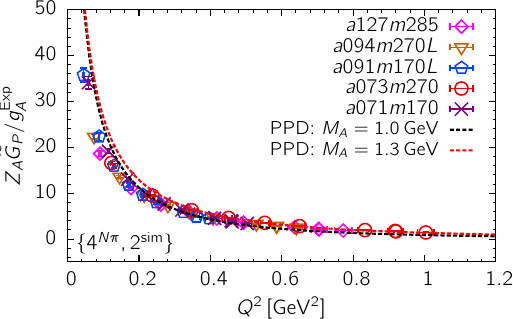} 
    \includegraphics[width=0.47\linewidth]{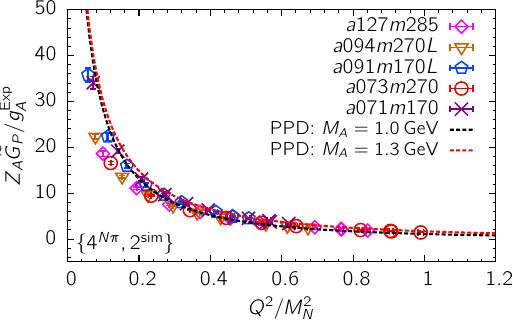} 
}
\caption{The data for the renormalized induced pseudoscalar form
  factor $Z_A {\widetilde G}_P(Q^2)/g_A^{\rm exp}$, with $g_A^{\rm exp}=1.277$, are plotted versus
  $Q^2$ in GeV${}^2$ (left) and $Q^2/M_N^2$ (right). Each panel shows
  the data from the five larger volume ensembles. The four rows show the
  results from four strategies for controlling ESC that are specified
  in the label at the bottom left corner. The difference among the
  three $\sim 270$ and the two 170~MeV data is more noticeable when plotted
  versus $Q^2/M_N^2$.  }
\label{fig:GPt4s5e}
\end{figure*}

\begin{figure*}[tbp] 
\subfigure
{
    \includegraphics[width=0.47\linewidth]{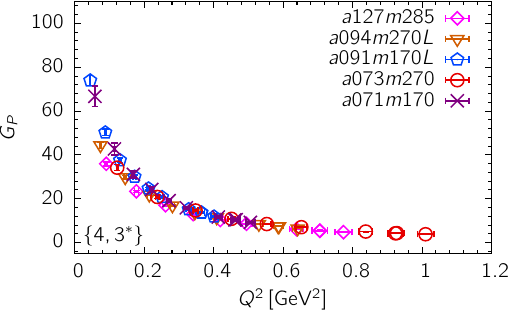} 
    \includegraphics[width=0.47\linewidth]{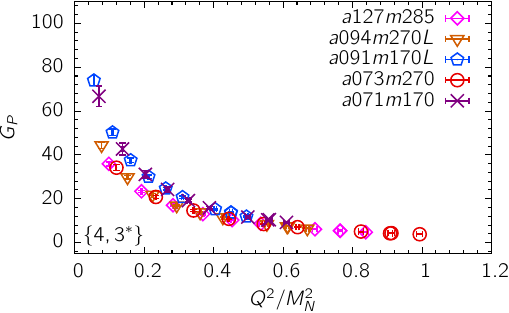} 
}
{
    \includegraphics[width=0.47\linewidth]{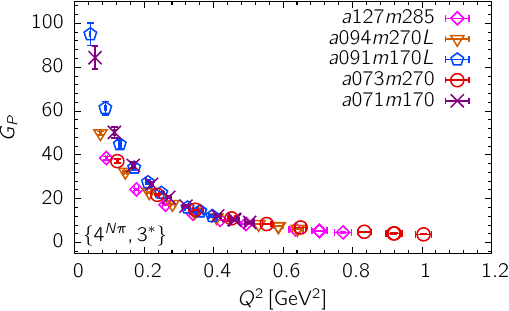} 
    \includegraphics[width=0.47\linewidth]{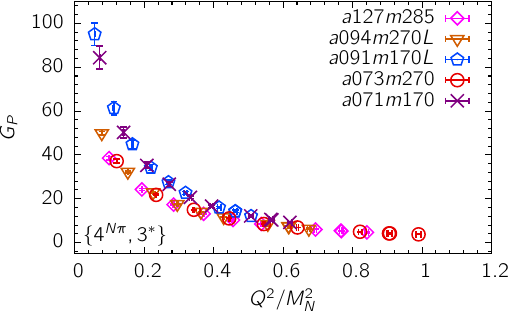} 
}
{
    \includegraphics[width=0.47\linewidth]{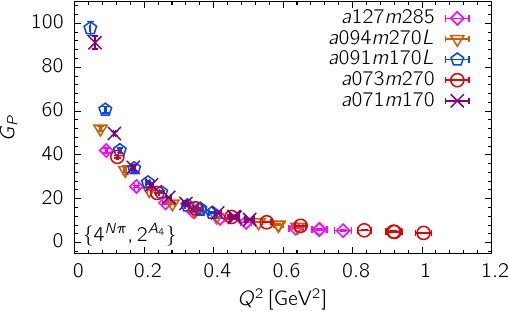} 
    \includegraphics[width=0.47\linewidth]{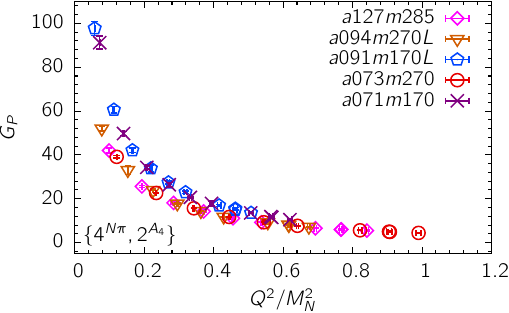} 
}
{
    \includegraphics[width=0.47\linewidth]{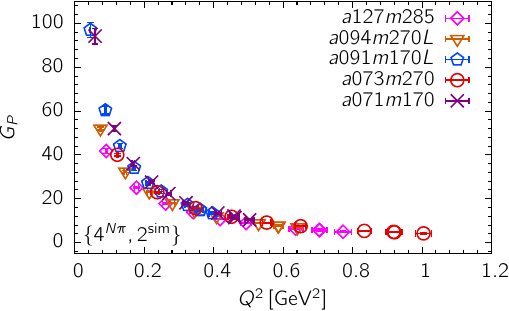} 
    \includegraphics[width=0.47\linewidth]{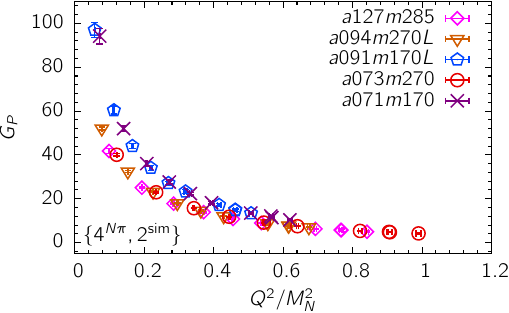} 
}
\caption{The unrenormalized pseudoscalar form factor ${
    G}_P(Q^2)$ is plotted versus $Q^2$ in GeV${}^2$ (left) and 
  $Q^2/M_N^2$ (right). Each panel shows the data from the five larger
  volume ensembles. The four rows show the results from four
  strategies for controlling ESC that are specified in the labels. The
  difference among the three $\sim 270$~MeV and the two 170~MeV data is more
  noticeable when plotted versus $Q^2/M_N^2$.  }
\label{fig:GPS4s5e}
\end{figure*}

The data for $ Z_A {G}_A(Q^2)$ and $Z_A{\widetilde G}_P(Q^2)$ for the four
remaining strategies are given in Tables~\ref{tab:GA-renormalized}
and~\ref{tab:GPt-renormalized} and plotted in Figs.~\ref{fig:GA4s5e}
and~\ref{fig:GPt4s5e}, where we divide them by $g_A^{\rm exp}
= 1.277$ so that the value should equal unity at \(Q^2=0\) in the CCFV limit.  Similarly, the
unrenormalized ${ G}_P(Q^2)$ is given in
Table~\ref{tab:GP-unrenormalized} and plotted in
Fig.~\ref{fig:GPS4s5e}. The latter is used primarily to check the
PCAC and PPD relations as shown in Fig.~\ref{fig:PCAC}. 

\begin{table}      
\begin{ruledtabular}
\begin{tabular}{r|lll}
fit & $\langle r_A^2 \rangle |_\text{dipole}$ & $\langle r_A^2 \rangle |_{P_2}$ & $\langle r_A^2 \rangle |_{z^2}$ \\ \hline
\hline  \multicolumn{4}{c}{$ a127m285$} \\ \hline  
$\{4,3^\ast\}$ & 0.293(13) & 0.293(20) & 0.297(15) \\ 
$\{4^{N\pi},3^\ast\}$ & 0.315(13) & 0.333(22) & 0.323(15) \\ 
$\{4^{N\pi},2^{A_4}\}$ & 0.302(15) & 0.349(32) & 0.315(18) \\ 
$\{4^{N\pi},2^\text{sim}\}$ & 0.304(15) & 0.310(42) & 0.297(21) \\ 
\hline  \multicolumn{4}{c}{$ a094m270$} \\ \hline  
$\{4,3^\ast\}$ & 0.255(18) & 0.293(65) & 0.291(29) \\ 
$\{4^{N\pi},3^\ast\}$ & 0.265(14) & 0.340(43) & 0.314(18) \\ 
$\{4^{N\pi},2^{A_4}\}$ &    &        &       \\ 
$\{4^{N\pi},2^\text{sim}\}$ & 0.247(11) & 0.280(48) & 0.278(23) \\ 
\hline  \multicolumn{4}{c}{$ a094m270L$} \\ \hline  
$\{4,3^\ast\}$ & 0.290(11) & 0.305(18) & 0.305(13) \\ 
$\{4^{N\pi},3^\ast\}$ & 0.317(11) & 0.348(20) & 0.336(13) \\ 
$\{4^{N\pi},2^{A_4}\}$ & 0.312(9) & 0.358(19) & 0.339(12) \\ 
$\{4^{N\pi},2^\text{sim}\}$ & 0.298(10) & 0.333(26) & 0.317(16) \\ 
\hline  \multicolumn{4}{c}{$ a091m170$} \\ \hline  
$\{4,3^\ast\}$ & 0.301(15) & 0.307(30) & 0.340(29) \\ 
$\{4^{N\pi},3^\ast\}$ & 0.376(33) & 0.355(92) & 0.411(69) \\ 
$\{4^{N\pi},2^{A_4}\}$ & 0.292(16) & 0.459(52) & 0.466(41) \\ 
$\{4^{N\pi},2^\text{sim}\}$ & 0.306(16) & 0.350(59) & 0.378(53) \\ 
\hline  \multicolumn{4}{c}{$ a091m170L$} \\ \hline  
$\{4,3^\ast\}$ & 0.341(19) & 0.323(30) & 0.342(30) \\ 
$\{4^{N\pi},3^\ast\}$ & 0.449(45) & 0.426(74) & 0.462(63) \\ 
$\{4^{N\pi},2^{A_4}\}$ & 0.311(20) & 0.486(48) & 0.484(40) \\ 
$\{4^{N\pi},2^\text{sim}\}$ & 0.369(19) & 0.478(61) & 0.479(51) \\ 
\hline  \multicolumn{4}{c}{$ a073m270$} \\ \hline  
$\{4,3^\ast\}$ & 0.269(12) & 0.270(24) & 0.280(17) \\ 
$\{4^{N\pi},3^\ast\}$ & 0.271(9) & 0.330(22) & 0.312(12) \\ 
$\{4^{N\pi},2^{A_4}\}$ & 0.242(9) & 0.287(21) & 0.271(14) \\ 
$\{4^{N\pi},2^\text{sim}\}$ & 0.253(8) & 0.285(22) & 0.278(13) \\ 
\hline  \multicolumn{4}{c}{$ a071m170$} \\ \hline  
$\{4,3^\ast\}$ & 0.284(22) & 0.288(36) & 0.306(38) \\ 
$\{4^{N\pi},3^\ast\}$ & 0.368(29) & 0.428(66) & 0.455(56) \\ 
$\{4^{N\pi},2^{A_4}\}$ & 0.271(15) & 0.494(47) & 0.507(39) \\ 
$\{4^{N\pi},2^\text{sim}\}$ & 0.308(18) & 0.424(67) & 0.438(59) \\ 
\end{tabular}
\end{ruledtabular}
\caption{Results for $\langle r_A^2 \rangle $ from a dipole, $P_2$ Pad\'e and
  $z^2$ fits to all ten $Q^2 \neq 0$ points for the seven ensembles and
  the four strategies  in column 1 (see Appendix~\ref{sec:glossary}), used to control excited state
  contamination. The fits to $\{4^{N\pi},2^{A_4}\}$ data on the
  $a094m270$ ensemble are not stable so no results are given.}
\label{tab:rAdP2z2}
\end{table}

\subsection{Parameterizing the \texorpdfstring{$Q^2$}{Q\suptwo} behavior of \texorpdfstring{$G_A(Q^2)$}{GA(Q\suptwo)} and the extraction of \texorpdfstring{$g_A$}{gA} and \texorpdfstring{$\langle r_A^2 \rangle$}{<rA2>}}
\label{sec:AFFFinalFits}

Our primary goal is to calculate the axial form factors, $G_A$ and
${\widetilde G}_P$, as a function of $Q^2$ as these are needed in the
calculation of the neutrino-nucleus cross-sections.  These results are
shown in Figs.~\ref{fig:GA4s5e} and~\ref{fig:GPt4s5e}. 

In most current lattice QCD calculations, the smallest nonzero lattice
momentum, which is also the gap between the discrete momenta, is
large, $|\bm q_{\rm min} | \ge 200$~MeV.  Consequently, it is
important to keep in mind that obtaining the slope and the value at
$Q^2=0$ from fits to lattice data with $Q^2 \gtrsim 0.04$~GeV${}^2$
have an associated systematic uncertainty.  This can be estimated
by comparing $g_A$ obtained directly at $Q^2=0$ from the forward
matrix element with the extrapolated value $G_A(Q^2 \to 0)$. In this
work, we perform this extrapolation using three parameterizations,
dipole, Pad\'e and $z$-expansion, as discussed below and in
Sec.~\ref{sec:CCFVcharges}.

Historically, the dipole ($D$) ansatz has been used to parameterize the
$Q^2$ behavior of $G_A(Q^2)$: 
\begin{equation}
G_A(Q^2)|_D = \frac{G_A(0)}{(1+Q^2/{\cal M}_A^2)^{2}} \ \Longrightarrow \ \langle r_A^2\rangle = \frac{12}{{\cal M}_A^2} \,. 
\label{eq:dipole}
\end{equation}
It is the Fourier transform of a distribution exponentially falling in space, and appealing for phenomenological analyses because it
has only one unknown parameter, the axial mass ${\cal M}_A$ since
$g_A$ is known accurately from experiments.  
Also, it goes to zero as $Q^4$ for large 
$Q^2 $ as predicted by QCD perturbation 
theory~\cite{Smith:1972xh,Lepage:1980fj}.

The second parameterization used is the model-independent
$z$-expansion~\cite{Hill:2010yb,Bhattacharya:2011ah}:
\begin{equation}
 \frac{G_A(Q^2)}{G_A(0)} = \sum_{k=0}^{\infty} a_k z(Q^2)^k  \,,
\label{eq:Zexpansion}
\end{equation}
where the $a_k$ are fit parameters and $z$ is defined to be 
\begin{equation}
  z = \frac{\sqrt{t_\text{cut}+Q^2}-\sqrt{t_\text{cut}+{\overline t}_0}}
           {\sqrt{t_\text{cut}+Q^2}+\sqrt{t_\text{cut}+{\overline t}_0}} \,.
\label{eq:Zdef}
\end{equation}
In terms of $z$, the form factors are analytical within the unit
circle with the nearest singularity, a branch cut, at $Q^2 = -
t_\text{cut} = -9M_\pi^2$ (or $-4 M_\pi^2$ in the vector channel). We
choose the parameter ${\overline t}_0$, which is the value of $-Q^2$
that is mapped to $z=0$, to be the midpoint of the range of $Q^2$
values on each ensemble to minimize the maximum value of $|z|$ as
discussed in Ref.~\cite{Jang:2019vkm}. For the seven ensembles listed
in Table~\ref{tab:Ensembles}, this corresponds to ${\overline t}_0 =
\{0.4,\ 0.6,\ 0.3,\ 0.3,\ 0.2,\ 0.5,\ 0.25\}$~GeV${}^2$,
respectively. We find no significant difference in the results on
using ${\overline t}_0 = 0$.

\begin{figure}[tbp] 
    \includegraphics[width=0.94\linewidth]{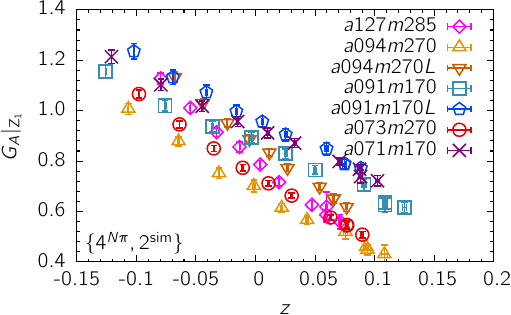} 
\caption{The renormalized $G_A$ plotted versus $z$ for the seven
  ensembles. The data are from the $\{4^{N\pi},2^{\rm sim}\}$
  strategy.
\label{fig:GAversusz}}
\end{figure}

\begin{figure*}[tbp] 
{
    \includegraphics[width=0.32\linewidth]{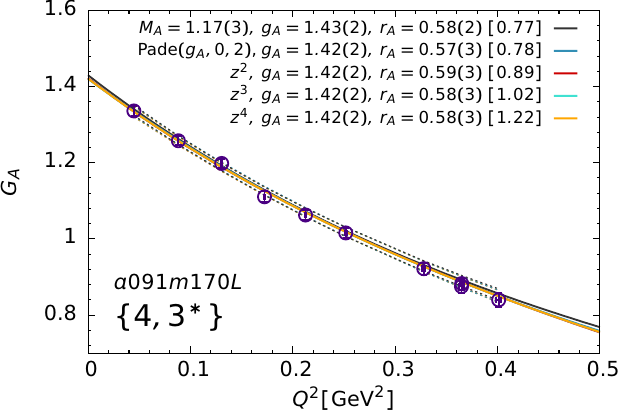} 
    \includegraphics[width=0.32\linewidth]{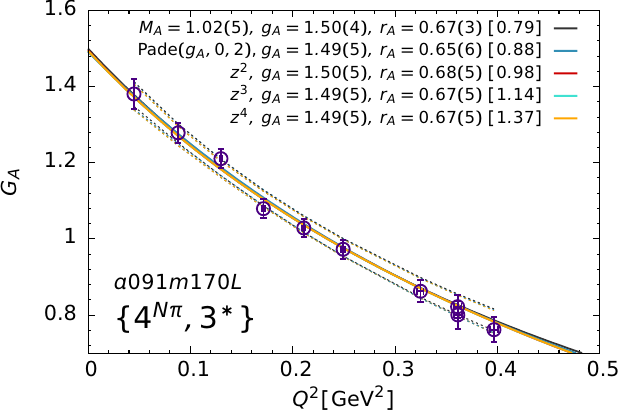} 
    \includegraphics[width=0.32\linewidth]{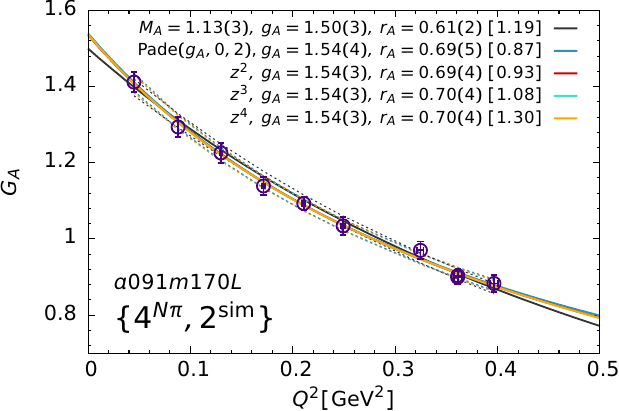} 
}
{
    \includegraphics[width=0.32\linewidth]{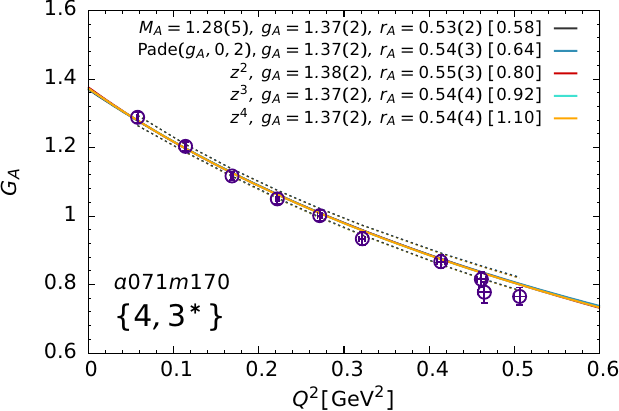} 
    \includegraphics[width=0.32\linewidth]{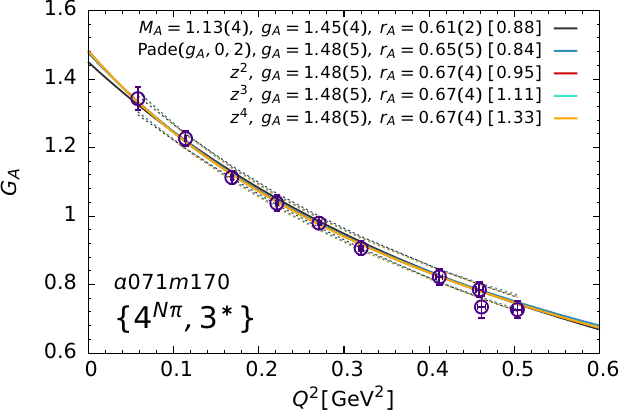} 
    \includegraphics[width=0.32\linewidth]{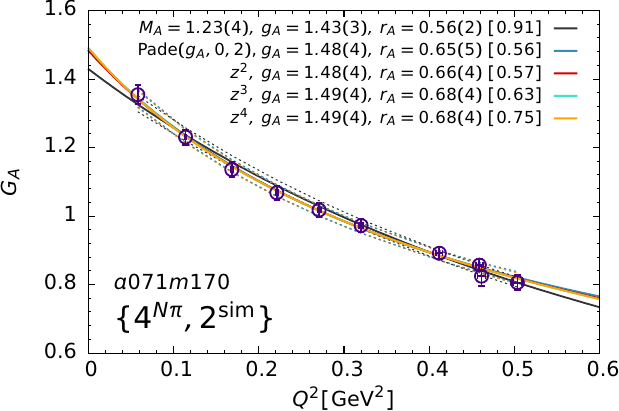} 
}
\caption{Plot of $G_A$ versus $Q^2$ for the $a091m170L$ (top row) and
  $a071m170$ (bottom row) ensembles. Also shown are the dipole, Pad\'e
  ($g_A$,0,2) and $z^{\{2,3,4\}}$ fits to the ten $Q^2$ points. The
  results for unrenormalized $g_A$ and $r_A$ (in fm) are given in the legends:
  dipole (top line), Pad\'e (second line) and $z^{2,3,4}$ (lines
  3-5). The $\chi^2$/dof of the fits are given within square brackets.
  The error bands of the fits are shown by dotted lines of the same
  color only over the range of the data for clarity.
\label{fig:GAcompQ2}}
\end{figure*}

The data for $Z_A G_A(Q^2)/1.277$ are plotted versus $z$ in
Fig.~\ref{fig:GAversusz} for the $\{4^{N\pi},2^{\rm sim}\}$
strategy and show only small deviations from linearity. As a result, 
$z$-expansion fits with $z^{\{2,3,4\}}$ truncations give essentially
identical results for both $g_A$ and $\langle r_A^2 \rangle$. As shown
in Fig.~\ref{fig:GAcompQ2}, the augmented $\chi^2$ does not decrease by two
units on increasing the order of truncation from $z^2 \to z^3 \to
z^4$. Therefore the $z^{\{3,4\}}$ fits are considered overparameterized by the
Akaike information criteria~\cite{1100705}. In
Ref.~\cite{Jang:2019jkn}, we had observed that fitting the precise
experimental data for the electric and magnetic form factors
stabilizes for $z^k$ truncated at $k \ge 4$. Our current lattice
data with ten points are well fit by the $z^2$ ($z^3$)
truncation for the axial (vector) form factors 
as discussed further in Sec.~\ref{sec:VFF}. 

\begin{table*}      
\begin{ruledtabular}
\begin{tabular}{r|llllrllll}
ESC strategy & $c_0$ & $c_1$ & $c_2$ & $c_3$ & [$\chi^2/dof$] & $g_P^\ast|_{{\rm Z}_1}  $ & $g_{\pi NN}|_{{\rm Z}_1}$ & $g_P^\ast|_{{\rm Z}_2}  $ & $g_{\pi NN}|_{{\rm Z}_2}$\\ 
\hline
\hline
\multicolumn{10}{c}{$Q^2$ fits to the $a091m170L$ data} \\
\hline
\hline
$\{4,3^\ast\}$&0.0356(16) & 0.136(37) & -2.13(30) & - & 
    [29.38/7] & 3.89(15) & 
7.52(39) & 
3.87(15) & 
7.50(38) 
\\ 
$\{4,3^\ast\}$&0.0312(18) & 0.545(93) & -11.8(2.0) & 65(13) & 
  [6.22/6] & 3.76(16) & 
6.59(42) & 
3.75(15) & 
6.57(41) 
\\ 
$\{4^{N\pi},3^\ast\}$&0.0501(41) & -0.13(10) & -0.99(74) & - & 
  [15.91/7] & 5.19(36) & 
10.59(90) & 
5.17(36) & 
10.55(90) 
\\ 
$\{4^{N\pi},3^\ast\}$&0.0425(49) & 0.43(23) & -14.0(4.7) & 87(31) & 
  [8.04/6] & 4.86(38) & 
9.0(1.1) & 
4.84(38) & 
8.9(1.1) 
\\ 
$\{4^{N\pi},2^{A_4}\}$&0.0548(23) & -0.287(64) & 0.86(48) & - & 
  [3.59/7] & 5.55(21) & 
11.56(57) & 
5.53(19) & 
11.53(54) 
\\ 
$\{4^{N\pi},2^{A_4}\}$&0.0530(33) & -0.18(15) & -1.2(2.9) & 13(18) & 
  [3.06/6] & 5.45(25) & 
11.20(75) & 
5.43(23) & 
11.17(73) 
\\ 
$\{4^{N\pi},2^\text{sim}\}$&0.0529(25) & -0.196(83) & -0.05(71) & - & 
  [4.02/7] & 5.43(21) & 
11.17(60) & 
5.41(20) & 
11.13(58) 
\\ 
$\{4^{N\pi},2^\text{sim}\}$&0.0516(40) & -0.11(21) & -1.8(4.3) & 12(28) & 
  [3.85/6] & 5.36(27) & 
10.90(89) & 
5.34(26) & 
10.86(88) 
\\ 
\hline
\hline
\multicolumn{10}{c}{$Q^2$ fits to the $a071m170$ data } \\
\hline
\hline
$\{4,3^\ast\}$&0.0192(17) & 0.116(67) & -2.04(70) & - & 
    [10.98/7] & 3.69(27) & 
6.89(63) & 
3.71(27) & 
6.91(63) 
\\ 
$\{4,3^\ast\}$&0.0174(18) & 0.47(14) & -13.7(4.1) & 102(36) & 
  [2.81/6] & 3.66(27) & 
6.22(67) & 
3.67(27) & 
6.24(67) 
\\ 
$\{4^{N\pi},3^\ast\}$&0.0318(27) & -0.231(94) & 0.32(92) & - & 
  [7.43/7] & 5.73(42) & 
11.38(99) & 
5.75(43) & 
11.4(1.0) 
\\ 
$\{4^{N\pi},3^\ast\}$&0.0271(34) & 0.21(23) & -12.0(5.8) & 104(48) & 
  [2.82/6] & 5.24(48) & 
9.7(1.3) & 
5.25(49) & 
9.7(1.3) 
\\ 
$\{4^{N\pi},2^{A_4}\}$&0.0325(11) & -0.295(49) & 1.83(59) & - & 
  [7.24/7] & 5.81(17) & 
11.64(48) & 
5.83(17) & 
11.68(48) 
\\ 
$\{4^{N\pi},2^{A_4}\}$&0.0359(20) & -0.60(16) & 10.0(4.2) & -67(34) & 
  [3.34/6] & 6.19(26) & 
12.87(80) & 
6.21(26) & 
12.91(80) 
\\ 
$\{4^{N\pi},2^\text{sim}\}$&0.0342(15) & -0.295(66) & 1.22(76) & - & 
  [2.54/7] & 6.13(23) & 
12.24(61) & 
6.15(24) & 
12.28(62) 
\\ 
$\{4^{N\pi},2^\text{sim}\}$&0.0354(26) & -0.40(19) & 4.1(5.0) & -24(41) & 
  [2.20/6] & 6.27(34) & 
12.69(98) & 
6.29(34) & 
12.73(98) 
\\ 
\end{tabular}
\end{ruledtabular}
\caption{Results of fits to $\frac{m_\mu}{2M_N}\widetilde G_P(Q^2)$
  versus $Q^2$ using the ansatz and the parameters $c_i$ defined in
  Eq.~\eqref{eq:GPt_fit}. The strategies used to remove ESC in column
  1 are defined in Appendix~\ref{sec:glossary}. Fits to $a091m170L$
  data (top) and $a071m170$ (bottom), with and without the finite
  volume ($c_3$) term, are compared for the four strategies listed in
  column one.  All ten values of $Q^2$ are used in the fits and
  results are given for both renormalization methods.  }
\label{tab:GPt_Q2fit_D7E7}
\end{table*}

\begin{table*}[t]       
\begin{ruledtabular}
\begin{tabular}{r|ll|ll||ll|ll}
Ensemble & \multicolumn{2}{c|}{$\{4,3^\ast\}$} & \multicolumn{2}{c||}{$\{4^{N\pi},2^{\rm sim}\}$} & \multicolumn{2}{c|}{$\{4,3^\ast\}$} & \multicolumn{2}{c}{$\{4^{N\pi},2^{\rm sim}\}$} \\
         & $g_P^\ast|_{{\rm Z}_1}$ & $g_P^\ast|_{{\rm Z}_2}$ & $g_P^\ast|_{{\rm Z}_1}$ & $g_P^\ast|_{{\rm Z}_2}$  & $g_{\pi NN}|_{{\rm Z}_1}$ &  $g_{\pi NN}|_{{\rm Z}_2}$ & $g_{\pi NN}|_{{\rm Z}_1}$ &  $g_{\pi NN}|_{{\rm Z}_2}$ \\ 
\hline
$a127m285$  & 2.266(66) & 2.221(61)  & 2.655(81) & 2.602(78)  & 11.30(53) & 11.08(51)  & 13.64(67) & 13.37(65)  \\ 
$a094m270$  & 2.52(16) & 2.50(16)  & 2.87(10) & 2.851(96)  & 11.27(89) & 11.20(87)  & 12.97(59) & 12.90(57)  \\ 
$a094m270L$  & 2.455(94) & 2.465(89)  & 2.919(68) & 2.931(55)  & 10.89(56) & 10.94(54)  & 13.46(46) & 13.51(43)  \\ 
$a091m170$  & 3.93(14) & 3.91(14)  & 5.53(22) & 5.50(21)  & 7.77(37) & 7.73(36)  & 11.30(56) & 11.24(55)  \\ 
$a091m170L$  & 3.89(15) & 3.87(15)  & 5.43(21) & 5.41(20)  & 7.52(39) & 7.50(38)  & 11.17(60) & 11.13(58)  \\ 
$a073m270$  & 2.45(11) & 2.45(10)  & 2.883(54) & 2.883(48)  & 11.11(62) & 11.11(62)  & 13.30(40) & 13.30(39)  \\ 
$a071m170$  & 3.69(27) & 3.71(27)  & 6.13(23) & 6.15(24)  & 6.89(63) & 6.91(63)  & 12.24(61) & 12.28(62)  \\ 
\end{tabular}
\end{ruledtabular}
\caption{Results for $g_P^\ast \equiv \frac{m_\mu}{2M_N} \widetilde
  G_P(0.88m_\mu^2)$ and $g_{\pi NN} \equiv \frac{c_0}{2a^2 m_\mu
    F_\pi}$ from fits to $\frac{m_\mu}{2M_N}\widetilde G_P(Q^2) $
  using Eq.~\protect\eqref{eq:GPt_fit} with the term proportional to
  $c_3$ set to zero.  Estimates from the two renormalization methods
  and the two strategies $\{4,3^\ast\}$ and $\{4^{N\pi},2^{\rm sim}\}$
  are compared.}
\label{tab:gPstargpiNN}
\end{table*}


We also examine $z$-expansion fits with sum rules that ensure that 
$G_A(Q^2)$ falls as $ Q^{-4}$ with $Q^2 \to \infty$ as 
predicted by perturbation theory~\cite{Lepage:1980fj} following the
procedure described in Ref.~\cite{Jang:2019jkn}. Results of 
analyses with and without sum rules overlap. Our
final results for $g_A$ (Table~\ref{tab:gAdP2z2}) and $\langle r_A^2
\rangle$ (Table~\ref{tab:rAdP2z2}) are taken from fits without sum
rules as these quantities characterize the behavior at $Q^2=0$.  To
stabilize all these $z$-expansion fits, we use Gaussian priors for all the 
$a_k$ with central value zero and width five.

Last, we make two Pad\'e fits, $P_2 \equiv P(g, 0, 2)$ and $P_3
\equiv P(g, 1, 3)$, defined as 
\begin{align}
P(g, 0, 2) &= \frac{g}{1 + b_1 Q^2 + b_2 Q^4} 
\label{eq:defPade2} \,, \\
P(g, 1, 3) &= \frac{g(1+a_1 Q^2)}{1 + b_1 Q^2 + b_2 Q^4 + b_3 Q^6} \,.
\label{eq:defPade3} 
\end{align}
These also incorporate the $1/Q^4$ behavior expected at large
$Q^2$. Since the calculation is done for spacelike $Q^2$ and at
values sufficiently far from the physical poles and cuts, their
influence is expected to be small. Therefore, these Pad\'e fits should provide
an equally good parameterization as the $z$-expansion.

We find that $P_2$ gives results consistent with the $z^{2,3,4}$
fits, and has the virtue of being easier to visualize in terms of
powers of $Q^2$.  In Sec.~\ref{sec:FFfits} we will also present a
$P(g, 0, 2)$ and $z^2$ (or $z^3$) parameterization of the
axial, electric and magnetic form factors ignoring lattice artifacts,
with results given in Eqs.~\eqref{eq:GAPade},~\eqref{eq:GAz2}
and~\eqref{eq:GEMPade}.

To explore systematic errors due to the limited range of $Q^2$ points
fit, we compare results obtained by fitting all ten $Q^2\neq 0$ points
versus the six with the smallest $Q^2$ values. This cut, based on the
number of points rather than a value of $Q^2$ in physical units, is
chosen because, in the problematic cases in the vector channel, the
errors are large in the four largest $Q^2$ points as can be seen in
Figs.~\ref{fig:GEsummary}--\ref{fig:GMsummary}. Based on this comparison, we
selected ten-point fits for the axial form factors and six-point for the
vector.

Results for $g_A$ and $\langle r_A^2 \rangle$ depend on both the
strategy used to obtain the ground state matrix element and on the
fits (dipole, or the $z$-expansion or the Pad\'e) used to
parameterize the $Q^2$ behavior of $G_A(Q^2)$.  In particular, the
value of the low $Q^2$ points in $G_A(Q^2)$ vary between the
strategies as shown in Fig.~\ref{fig:FFcompare}, which in turn leads
to differences in the $Q^2$ parameterization, i.e., in  $g_A$ 
and $\langle r_A^2 \rangle$. These differences can be inferred 
from the labels in Fig.~\ref{fig:GAcompQ2}, where the three panels 
give $Q^2$ fits to $G_A(Q^2
\neq 0)$ data for the $\{4,3^\ast\}$,
$\{4^{N\pi},3^\ast\}$ and $\{4^{N\pi},2^{\rm sim}\}$ strategies for
the $a091m170L$ and $a071m170$ ensembles. Recall that the difference in $g_A$, 
obtained from the forward matrix element, between the $\{4,3^\ast\}$
and $\{4^{N\pi},3^\ast\}$ strategies was shown in 
Fig.~\ref{fig:diffcharges}.

Comparing results for $g_A$ and
$\langle r_A^2 \rangle$ from the seven ensembles, summarized in
Tables~\ref{tab:gAdP2z2} and~\ref{tab:rAdP2z2}, we note the following points:
\begin{itemize}
\item
With the $\{4,3^\ast\}$ strategy, results for $g_A$ from
dipole, $z^{\{2,3,4\}}$ and Pad\'e fits agree with those measured directly from the
forward matrix element on all ensembles. The fits have reasonable
$\chi^2$/dof. 
\item
For the $\{4^{N\pi},3^\ast\}$ strategy, similar agreement is seen
between results from the dipole, $z^{\{2,3,4\}}$ and Pad\'e fits, and from
the forward matrix element.  However, these estimates are larger than
those with the $\{4,3^\ast\}$ strategy, especially for the
$M_\pi \approx 170$~MeV ensembles (see Fig.~\ref{fig:diffcharges}).
\item
With the $\{4^{N\pi},2^{A_4}\}$ and the preferred $\{4^{N\pi},2^{\rm
  sim}\}$ strategies, (i) the dipole estimates are smaller than the
$z^{\{2,3,4\}}$ or the Pad\'e values on all three $M_\pi = 170$~MeV
ensembles, and (ii) the $\chi^2$/dof becomes larger for the dipole fit
to the data from all three $\{4^{N\pi}\}$ strategies, mainly because
it misses the low $Q^2$ points.
\end{itemize}

A key point is that the differences observed on the $M_\pi \approx 170$~MeV
ensembles are not evident at $M_\pi \sim 270$~MeV. This is consistent
with the earlier discussion that the difference in the mass gaps
between the $\{4\}$ and $\{4^{N\pi}\}$ fits become larger as $M_\pi$
decreases, i.e., the mass gap of the $N\pi$ state decreases. In
short, the data shown in Tables~\ref{tab:gAdP2z2} and~\ref{tab:rAdP2z2} indicate that
estimates of $g_A$ and $\langle r_A^2 \rangle$ become increasingly
sensitive to the ESC strategy as $M_\pi \to 135$~MeV. Also, the dipole
fit starts to fail.  This $M_\pi$ dependent behavior has a significant
impact on the final estimates obtained from the CCFV fits as discussed
in Sec.~\ref{sec:CCFV} and shown in Fig.~\ref{fig:CCFVgA}.

\section{The induced pseudoscalar form factor \texorpdfstring{${\widetilde G}_P(Q^2)$}{G\tildeabove P(Q\suptwo)} and the extraction of \texorpdfstring{$g_P^\ast$}{gP*} and \texorpdfstring{$g_{\pi N N}$}{g\textpi NN}}
\label{sec:GP}

The data for the renormalized induced pseudoscalar form factor $Z_A
{\widetilde G}_P(Q^2)/(g_A^{\rm exp})$ from the five larger volume
ensembles are plotted versus $Q^2$ and $Q^2/M_N^2$ in
Fig.~\ref{fig:GPt4s5e}. Overall, the data show dependence on the pion
mass, i.e., data fall into two bands for ensembles with $M_\pi \approx
270$ and 170~MeV. This dependence is more evident when plotted versus
$Q^2/M_N^2$. On the other hand, we do not observe a significant $a$
dependence.

The $Q^2$ dependence of ${\widetilde G}_P(Q^2)$, given in Table~\ref{tab:GPt-renormalized}, is analyzed 
using the small $Q^2$ expansion of the pion-pole dominance ansatz  given in Eq.~\eqref{eq:PPD}: 
\begin{equation}
\frac{m_\mu}{2 M_N} {\widetilde G}_P(Q^2) = \frac{c_0}{a^2(M_\pi^2 + Q^2)} + c_1 +c_2 a^2  Q^2 + c_3 a^4 Q^ 4\,,
\label{eq:GPt_fit}
\end{equation}
where the leading term is the pion-pole term and the polynomial
approximates the dependence coming from the small $Q^2$ behavior of
$G_A(Q^2)$.  It is also the behavior predicted for small $Q^2$ and $
M_\pi^2$ by the leading order chiral perturbation
theory~\cite{Bernard:2001rs}. In practice, this ansatz fits the data
over a large range of $Q^2$, 2.5$M_\pi^{2}$ -- 50$M_\pi^{2}$ in units
of $M_\pi=135$~MeV, as given in Table~\protect\ref{tab:Q2}.

From these fits, we extract $g_P^\ast$ and the pion-nucleon coupling,
$\gpNN$, using the following expressions:
\begin{align}
g_P^\ast \equiv &\; \frac{m_\mu}{2M_N}  \widetilde G_P(0.88m_\mu^2) 
\label{eq:gpstardef} \,, \\
g_{\pi NN} \equiv &\;  \lim_{Q^2 \to -M_\pi^2} \frac{M_\pi^2 + Q^2}{4M_N F_\pi} {\widetilde G}_P(Q^2) = \frac{c_0}{2a^2 m_\mu F_\pi} \,, 
\label{eq:gpiNNdef}
\end{align}
where $\gpNN$ is defined as the residue of ${\widetilde G}_P(Q^2)$ at
the pion pole at $Q^2 = -M_\pi^2$, and $m_\mu = 105.7$~MeV is the muon
mass and $F_\pi=92.2$~MeV is the pion decay constant.

We carried out fits to ${\widetilde G}_P(Q^2)$ versus $Q^2$ to get
$g_P^\ast$ and $\gpNN$ using (i) just the
sma
llest six $Q^2$ points and (ii) to all ten. On all seven ensembles
and for all four strategies (except for the four highest momenta points
with the $\{4^{N\pi},2^{A_4}\}$ strategy on the $a094m270$ ensemble
that could not be fit reliably) the estimates from these two fits are
consistent at $< 1\sigma$ level.  For our final results, we choose the
ten-point fits.

A second issue is whether the $Q^4$ term in Eq.~\eqref{eq:GPt_fit} is
needed or is an overparameterization. Results of the fits with
and without the $Q^4$ term are given in Table~\ref{tab:GPt_Q2fit_D7E7}
for the $a091m170L$ and $a071m170$ ensembles.  We note a significant
difference between the $\{4,3^{\ast}\}$ and $\{4^{N\pi},3^{\ast}\}$
strategies and in both cases there is a large reduction in the total $\chi^2$, which 
justifies including the $Q^4$ term by the Akaike information
criteria~\cite{1100705}. The errors on $c_0$ are, however, about a
factor of 2 larger with the $\{4^{N\pi},3^{\ast}\}$ strategy.
Estimates from the $\{4^{N\pi},2^{A_4}\}$ and the $\{4^{N\pi},2^{\rm
  sim}\}$ strategies are consistent and larger than  those from even
$\{4^{N\pi},3^{\ast}\}$. In these two cases, the $Q^4$ term is an
overparameterization by AIC and in fits including it, even the $c_1$ are poorly determined. 

Data from all seven ensembles obtained using strategies $\{4,3^\ast\}$
and $\{4^{N\pi},2^{\rm sim}\}$ are given for the two ways of
renormalizing the axial current in Table~\ref{tab:gPstargpiNN}. It
shows clearly that the main difference in the estimates comes from
whether the $N\pi$ state is included in the analysis.

Our final results are presented with the $\{4^{N\pi},2^{\rm sim}\}$
strategy based on the discussion in Sec.~\ref{sec:AFFconsistency} and
with the term proportional to $c_3$ set to zero. The CCFV fits to the
data in Table~\ref{tab:gPstargpiNN} are discussed in
Sec.~\ref{sec:CCFVgPstar} where we also compare our final results for
$\gpNN$ with the phenomenological Goldberger-Treiman relation and the
experimental value from the $\pi N$ scattering length.


\begin{figure*}[tbp] 
\subfigure
{
    \includegraphics[width=0.235\linewidth]{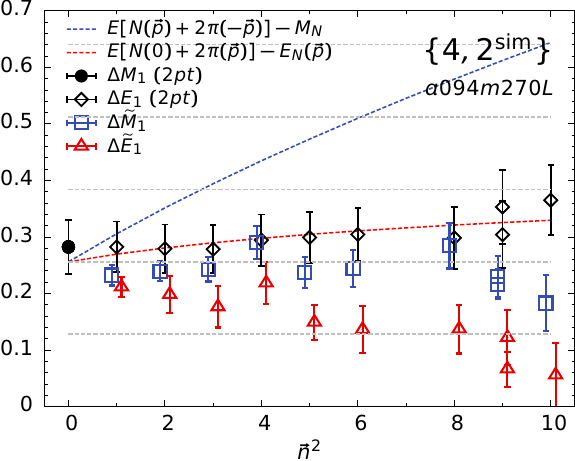}  
    \includegraphics[width=0.235\linewidth]{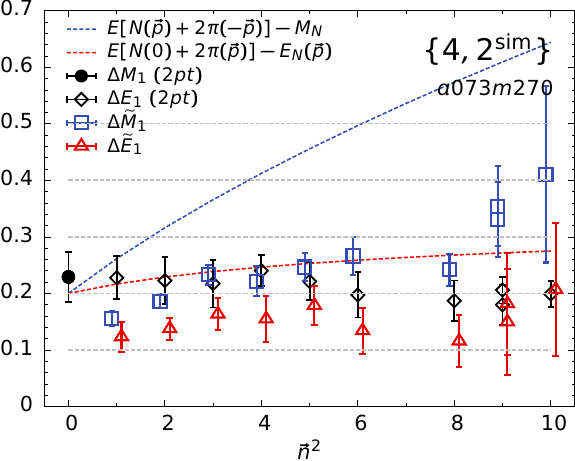}  
    \includegraphics[width=0.235\linewidth]{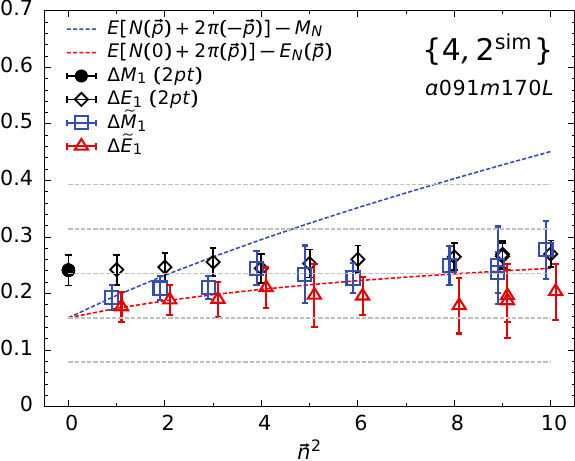}  
    \includegraphics[width=0.235\linewidth]{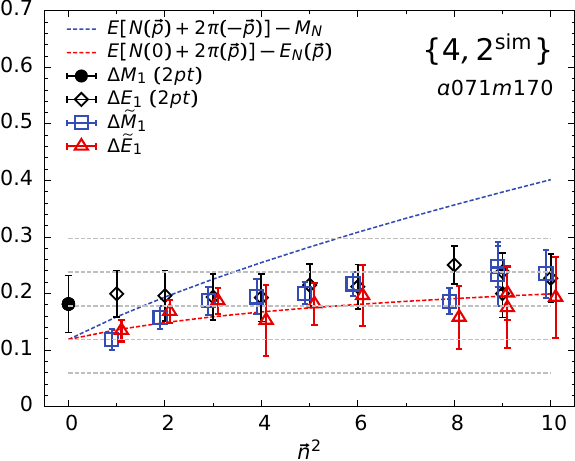}  
}
{
    \includegraphics[width=0.235\linewidth]{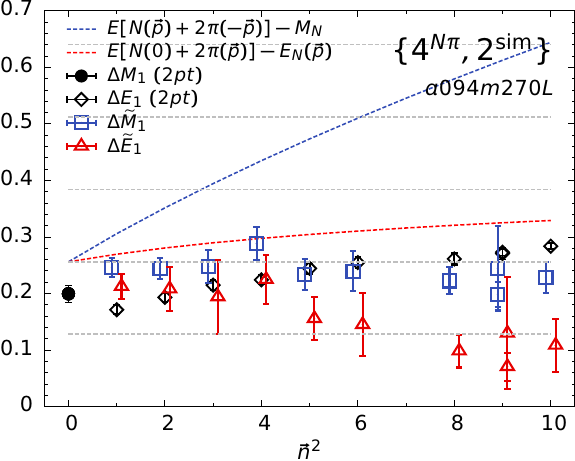}  
    \includegraphics[width=0.235\linewidth]{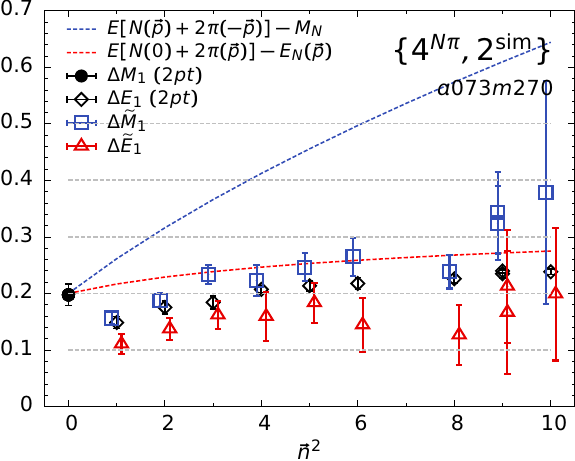}  
    \includegraphics[width=0.235\linewidth]{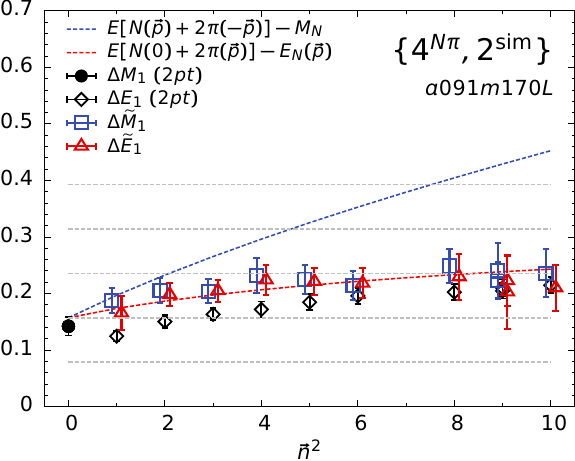}  
    \includegraphics[width=0.235\linewidth]{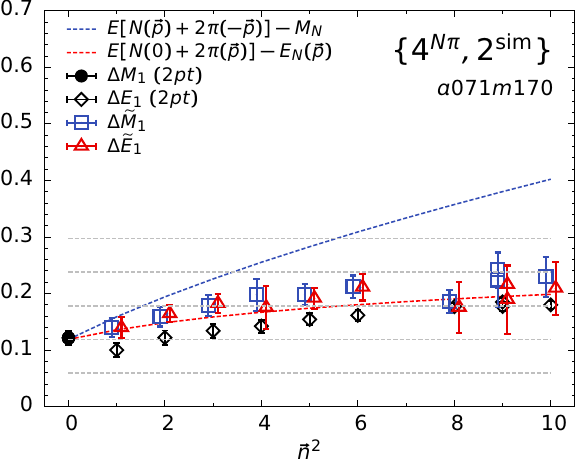}  
}
\caption{Estimates, in lattice units, of $\Delta M_1$ (black filled
  circles) and $\Delta E_1$ (open black diamonds) from fits to the
  two-point function for four ensembles in the order $a094m270L$,
  $a073m270$ , $a091m170L$ and $a071m170$ in each row.  Each panel
  also shows the values of $\Delta {\widetilde M}_1$ (open blue
  squares) and $\Delta {\widetilde E}_1$ (open red triangles) from the
  $\{4,2^{\rm sim}\}$ (top row) and the $\{4^{N\pi},2^{\rm sim}\}$
  (bottom row) fits to the vector three-point functions.  The mass
  gaps of the noninteracting $N(\bm q) 2\pi(-\bm q)$ and $N(\bm
  0) 2\pi(-\bm q)$ states are shown by the dotted blue and red 
  lines. The horizontal dotted black lines show the masses of $1, 2,
  \ldots$ pions.
  \label{fig:VFF-deltaM}}
\end{figure*}
%

\section{Electric and Magnetic Form Factors}
\label{sec:VFF}

To obtain the electric and magnetic form factors, we analyze the three
sets of correlators, $\Re V_4$, $\Im V_i$ and $\Re V_i$ defined in
Eqs.~\eqref{eq:GM1}--\eqref{eq:GE4} using four strategies
$\{4,3^\ast\}$, $\{4^{N\pi},3^{\ast}\}$, $\{4,2^{\rm sim}\}$ and
$\{4^{N\pi},2^{\rm sim}\}$ to remove ESC. In the $\{2^{\rm sim}\}$
fits to the three-point functions, all three correlators are fit
simultaneously with common $\Delta {\widetilde M}_1$ and
$\Delta {\widetilde E}_1$.  Only the ground state
parameters are taken from the two-point function. Fits with different
strategies are illustrated in Figs.~\ref{fig:GECOMP}, \ref{fig:ViCOMP}
and~\ref{fig:GMCOMP} using the lowest momentum transfer (${\bm n}^2 =
1$) data that have significant ESC and a good statistical signal, and the
fits are stable with respect to variations in $\tau$ and $t_{\rm
  skip}$.  The $\chi^2/$dof of the fits and the values of $\Delta M_1$
and $\Delta E_1$ entering in the fits to the three-point functions (or
$\Delta {\widetilde M}_1$ and $\Delta {\widetilde E}_1$ that are
outcomes in the $\{2^{\rm sim}\}$ fits) are given in the legends. Note
that for each strategy, the mass gaps in fits to the three correlation
functions are the same since they either are taken from fits to two-point
functions for the first two strategies or are outputs of simultaneous
fits in the two $\{2^{\rm sim}\}$ cases.

The first issue we investigate is whether the excited states that
contribute to these three correlators can be identified.  The analog
of the pion-pole dominance in the axial channel is vector-meson
dominance, i.e., the vector current, $V_\mu (\bm q)$, couples to the
$\rho$-meson, the lowest excitation in the vector channel, and thus to
the $2\pi(\bm q)$ state.  In this case, the dominant excited state
contributing to $\Delta {\widetilde M}_1$ and $\Delta {\widetilde
  E}_1$ should be $N(\bm q) 2\pi(-\bm q)$ (and/or $N(\bm 0) 2\pi(\bm
0))$ and $N(\bm 0) 2\pi(\bm q)$, respectively, where
$2\pi(\bm q)$ is a two pion state with total momentum $\bm q$. 

In Fig.~\ref{fig:VFF-deltaM}, the $\Delta {\widetilde M}_1$ and
$\Delta {\widetilde E}_1$ from simultaneous $\{2^{\rm sim}\}$ fits are
compared to the $\Delta M_1$ and $\Delta E_1$ from the $\{4\}$- and
$\{4^{N\pi}\}$-state fits to the two-point functions and to the mass
gaps expected for a specified state (dotted lines).  Our criteria for
identification of a state is when the $\Delta {\widetilde M}_1$ or
$\Delta {\widetilde E}_1$ agree with the corresponding dotted line.  We remind the
readers that $\{4^{N\pi}\}$-state fits are also relevant for the
vector channel because the mass gap of the $N(\bm 0)\pi(\bm 0)\pi(\bm
0)$ state is close to that for the $N(\bm 1)\pi(-\bm 1)$ state for our
ensembles.  The data exhibit the following features:
\begin{itemize} 
\item
The $\Delta {\widetilde E}_1$ (open red triangles) for the $170$~MeV
ensembles are consistent with the energy of a noninteracting $N(\bm
0) 2\pi(\bm q)$ state shown by the red dotted line. This agreement is seen for 
both the $\{4^{},2^{\rm sim}\}$ and $\{4^{N\pi},2^{\rm sim}\}$ strategies. 
\item
The $\Delta {\widetilde E}_1$ for the $270$~MeV ensembles lie between 1 and 2 times 
$M_\pi$. The closest association would be $N(\bm q) \pi(\bm 0)$ or
$N(\bm 0) \pi(\bm q)$ or $N(\bm q) 2\pi(\bm 0)$ states but not the $N(\bm 0) 2\pi(\bm q)$
state  shown by the red dotted line.
\item
The values of $\Delta {\widetilde M}_1$ (blue squares) lie much below
the $N(\bm q) 2\pi(-\bm q)$ state shown by the blue dotted line for
the $270$~MeV ensembles, however, the difference decreases
significantly in the data from the $170$~MeV ensembles. The increase
with $\bm q$ also becomes similar in shape to that for $N(\bm q) 2\pi(-\bm q)$.
\item
The $\Delta {\widetilde M}_1 $ are similar to $ \Delta {\widetilde E}_1$ for the
$170$~MeV ensembles while they lie about $M_\pi/2$ above for the $270$~MeV
ensembles. This behavior is very different from the axial case shown in
Fig.~\ref{fig:AFF-deltaM}.
\item
With  $\{4^{N\pi},2^{\rm sim}\}$, the mass gap $\Delta {\widetilde
M}_1 \approx \Delta {\widetilde E}_1$ and comes close to $\Delta {E}_1$ used in 
$\{4^{N\pi},3^\ast\}$ for both $170$~MeV ensembles. Such
an agreement in the mass gaps in the $\{4^{N\pi},2^{\rm sim}\}$ and
$\{4^{N\pi},3^\ast\}$ strategies implies that they should give similar results.
\end{itemize}
These trends in $\Delta {\widetilde M}_1$ and $\Delta {\widetilde
  E}_1$ support vector meson dominance, i.e., the insertion of $2\pi(\bm
  q)$ by the current, which we anticipate will become even more
  apparent on physical $M_\pi = 135$~MeV ensembles.  This is in
  analogy with pion-pole dominance with the axial current inserting
  $\pi(\bm q)$ as inferred from Fig.~\ref{fig:AFF-deltaM}. The values of
  $\Delta {\widetilde M}_1$ from the $M_\pi \approx 270$~MeV ensembles
  lying close to $2 M_\pi$ suggest that the $N(\bm 0) 2\pi(\bm 0)$
  state and its tower also contribute on the ${\bm p} = 0$ side of
  the operator.

Next, we investigated whether the data for $G_E$ from $\Im V_i$, which
show large ESC as illustrated in Fig.~\ref{fig:ViCOMP} and similar to
that seen in $\langle {\cal N}^\dagger A_4 {\cal N} \rangle$, provide
further insight on the identity of the excited states. We find that
the $\chi^2/$dof of even the $\{4,3^\ast\}$ fits is not unreasonably
large compared to the other strategies even though the values of
$\Delta M_1$ and $\Delta E_1$ are significantly different.  Overall,
current data for $G_E^{V_i}$ do not help us decide which excited states give the
dominant contribution.

\begin{figure*}[tbp] 
\subfigure
{
    \includegraphics[width=0.24\linewidth]{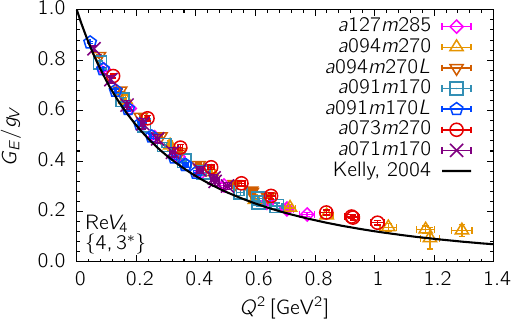}  
    \includegraphics[width=0.24\linewidth]{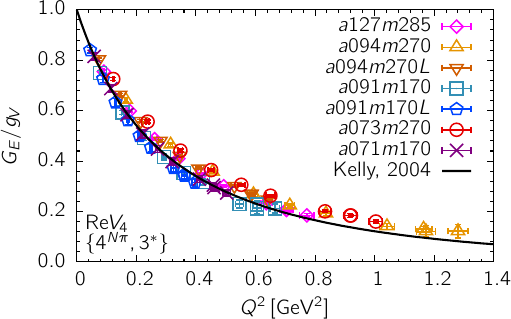} 
    \includegraphics[width=0.24\linewidth]{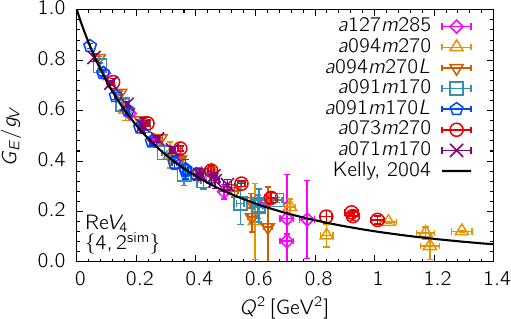}   
    \includegraphics[width=0.24\linewidth]{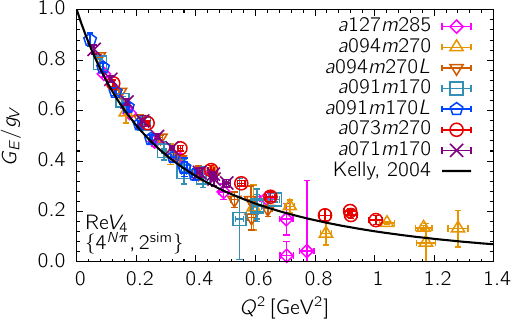}   
}
{
    \includegraphics[width=0.24\linewidth]{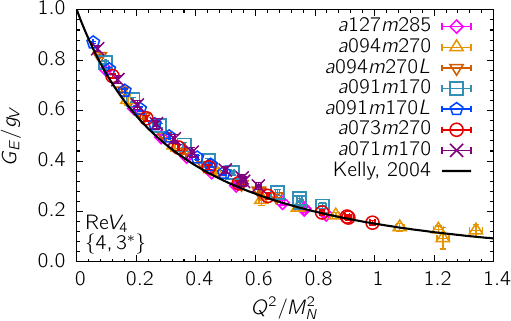}  
    \includegraphics[width=0.24\linewidth]{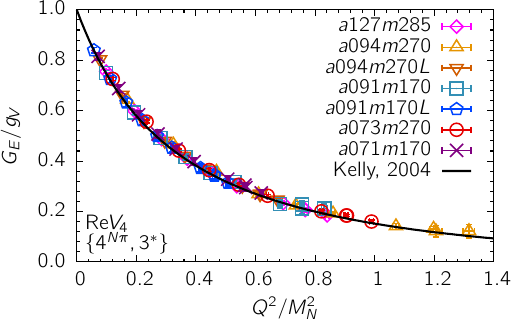} 
    \includegraphics[width=0.24\linewidth]{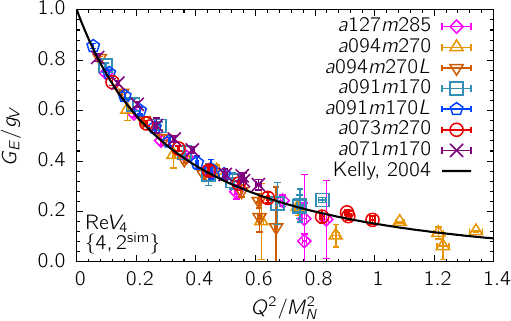}   
    \includegraphics[width=0.24\linewidth]{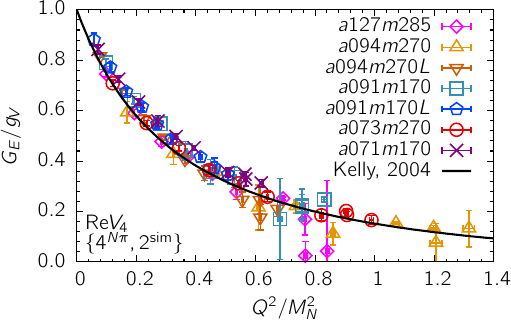}   
}
\vspace{-0.1in}
\caption{$G_E(Q^2)$ from $\Re V_4$ plotted versus $Q^2$ in GeV${}^2$
  (top panels) and versus $Q^2/M_N^2$ (bottom panels). Each panel
  shows the data for the seven ensembles, and each row compares the
  four strategies used to remove ESC.
  \label{fig:GEsummary}}
\end{figure*}

\begin{figure*}[tbp] 
\subfigure
{
    \includegraphics[width=0.24\linewidth]{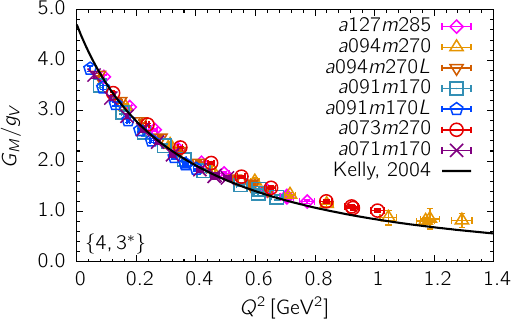}  
    \includegraphics[width=0.24\linewidth]{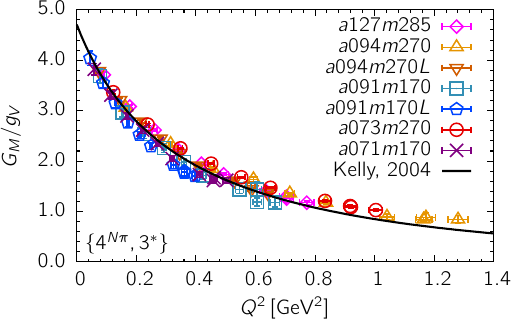} 
    \includegraphics[width=0.24\linewidth]{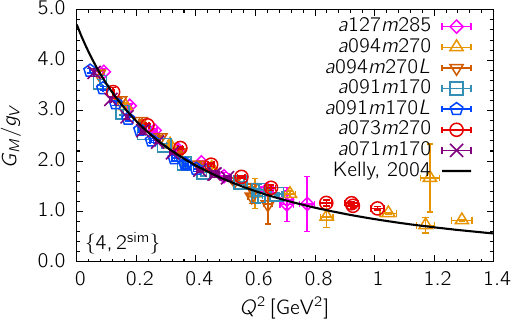}   
    \includegraphics[width=0.24\linewidth]{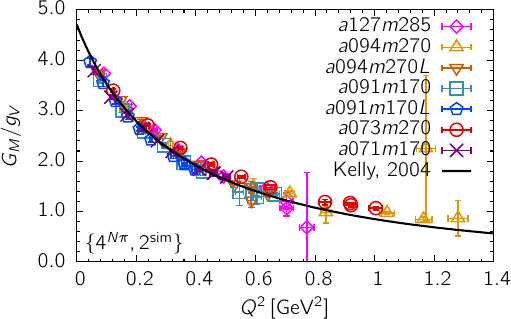}   
}
{
    \includegraphics[width=0.24\linewidth]{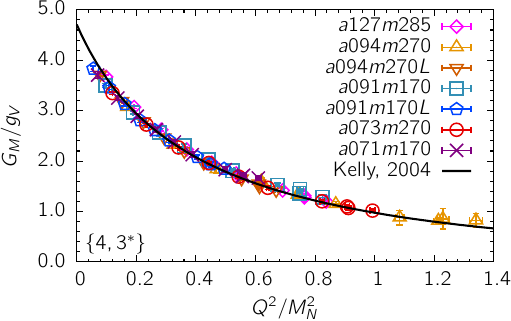}  
    \includegraphics[width=0.24\linewidth]{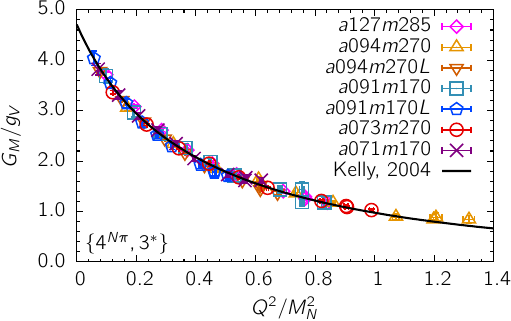} 
    \includegraphics[width=0.24\linewidth]{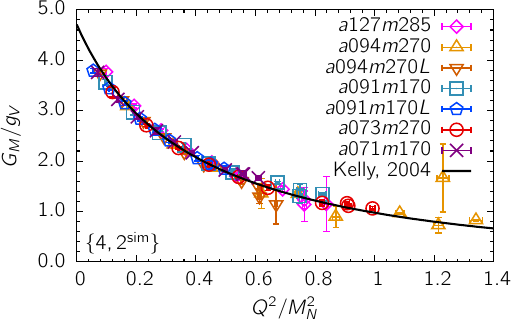}   
    \includegraphics[width=0.24\linewidth]{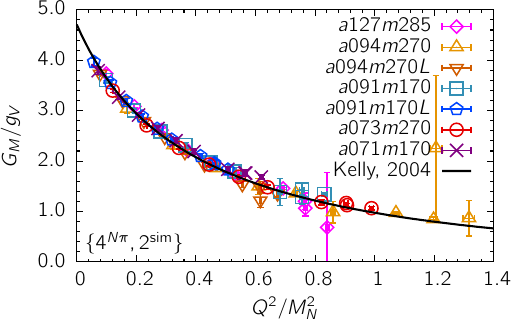}   
}
\vspace{-0.1in}
\caption{$G_M(Q^2)$ from $\Re V_i$ plotted versus $Q^2$ in
  GeV${}^2$ (top panels) and versus $Q^2/M_N^2$ (bottom panels). Each panel
  shows the data for the seven ensembles, and each row compares the
  four strategies used to remove ESC.
  \label{fig:GMsummary}}
\end{figure*}

An important feature in the ESC fits shown in Figs.~\ref{fig:GECOMP}
and~\ref{fig:GMCOMP} in Appendix~\ref{sec:TcompVFF} is that while the
differences in the mass gaps between the four strategies are large,
the variation in results for $G_E^{V_4}$ and $G_M^{V_i}$ is $\lesssim
5\%$. The smallness of the variation is further highlighted in
Figs.~\ref{fig:GEsummary} and~\ref{fig:GMsummary}---all four estimates
of the form factors are consistent within errors with the Kelly
parameterization of the experimental data. 

We base our choice of which strategy to choose for presenting the
final results on the trends in the mass gaps illustrated in
Fig.~\ref{fig:VFF-deltaM}. The first is the growing agreement between
$\Delta {\widetilde M}_1$ and $\Delta {\widetilde E}_1$ in the
$\{4^{N\pi},2^{\rm sim}\}$ data. Next, their agreement with the
$\Delta {M}_1$ and $\Delta {E}_1$ from the $\{4^{N\pi}\}$ fits.
Last, $\Delta {\widetilde M}_1 \approx 2M_\pi$ suggests that the
lowest excited state $N(0) \pi (0) \pi(0)$ also contributes.  These
trends suggest that the $\{4^{N\pi},2^{\rm sim}\}$ and
$\{4^{N\pi},3^{\ast}\}$ strategies should give similar results for the
form factors. Thus we will choose between these when presenting the
final results. \looseness-1

Results for the renormalized form factors from the four strategies
are given in
Tables~\ref{tab:FF-GE4strategies},~\ref{tab:FF-GEi4strategies},
and~\ref{tab:FF-GM4strategies}. The $\chi^2/$dof of the fits used to
remove the ESC are reasonable in most cases.  The errors are the
smallest in the $\{4,3^\ast\}$ data, and are large for many of the large
$Q^2$ points from the $\{4,2^{\rm sim}\}$ and the $\{4^{N\pi},2^{\rm
  sim}\}$ fits.  For this reason, we choose fits to the smallest six $Q^2$ points 
for calculating the charge radii.

\begin{figure*}[tbp] 
\subfigure
{
    \includegraphics[width=0.325\linewidth]{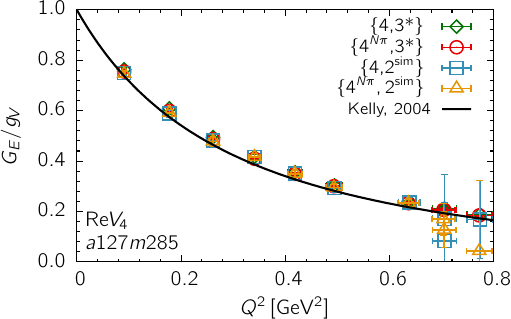} 
    \includegraphics[width=0.325\linewidth]{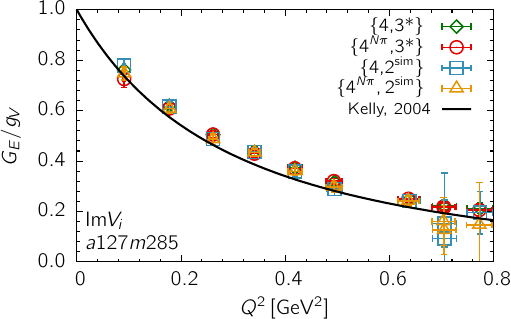} 
    \includegraphics[width=0.325\linewidth]{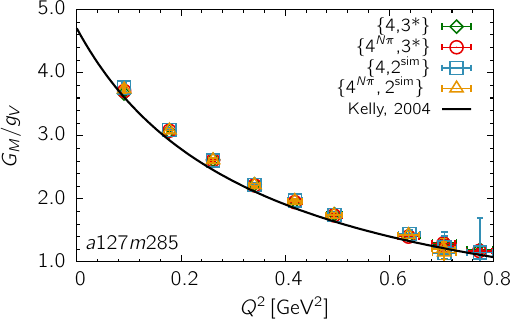} 
}
{
    \includegraphics[width=0.325\linewidth]{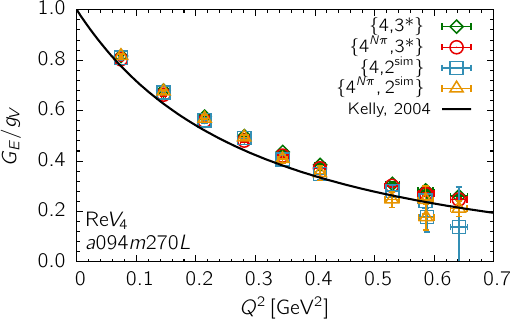} 
    \includegraphics[width=0.325\linewidth]{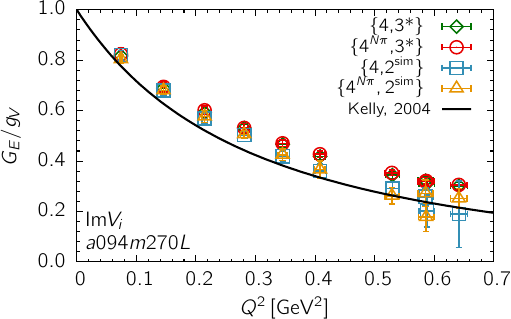} 
    \includegraphics[width=0.325\linewidth]{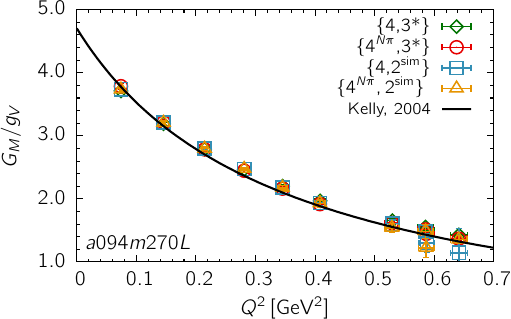} 
}
{
    \includegraphics[width=0.325\linewidth]{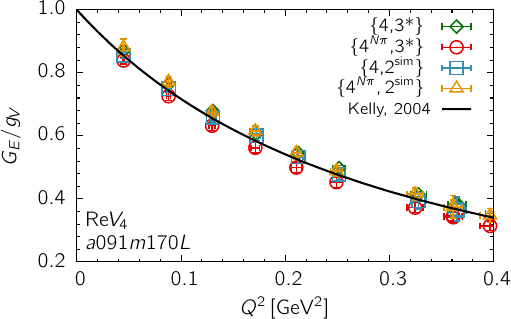} 
    \includegraphics[width=0.325\linewidth]{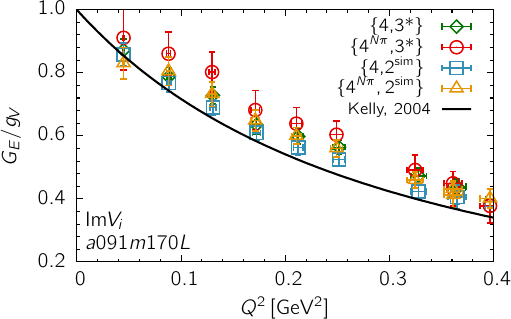} 
    \includegraphics[width=0.325\linewidth]{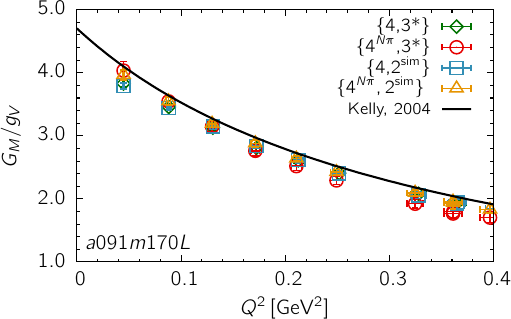} 
}
{
    \includegraphics[width=0.325\linewidth]{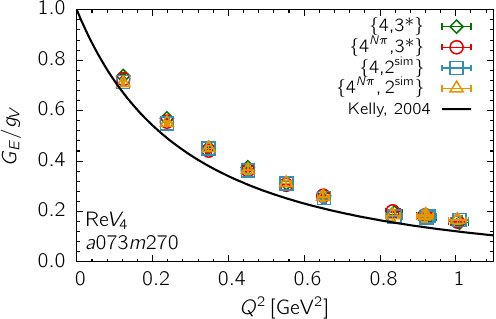} 
    \includegraphics[width=0.325\linewidth]{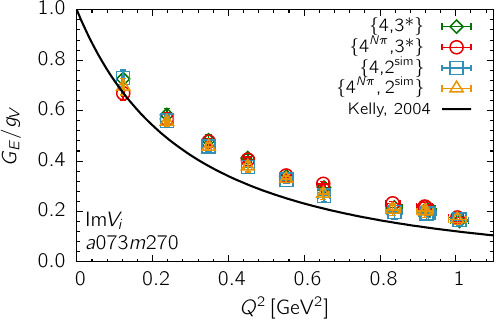} 
    \includegraphics[width=0.325\linewidth]{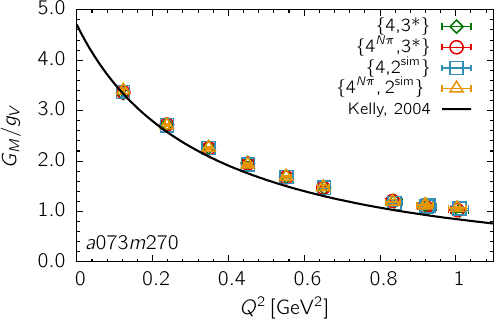} 
}
{
    \includegraphics[width=0.325\linewidth]{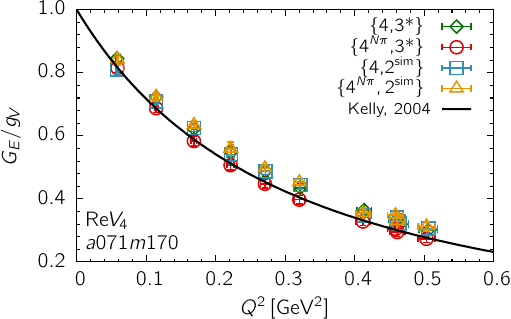} 
    \includegraphics[width=0.325\linewidth]{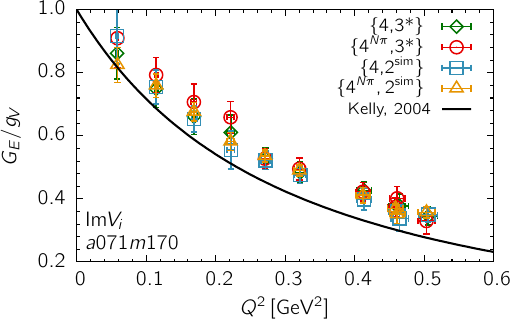} 
    \includegraphics[width=0.325\linewidth]{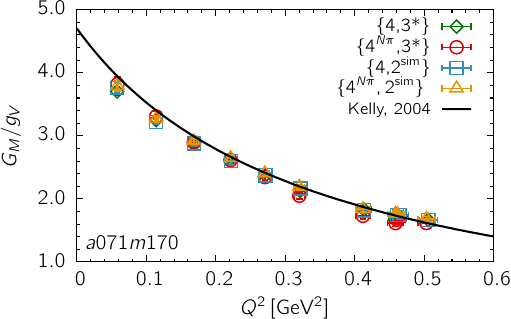} 
}
\caption{Each panel shows a comparison between the renormalized form factors $G_E^{\Re V_4}$ (left), $G_E^{\Im V_i}$
  (middle), and $G_M^{\Re V_i}$ (right) obtained using four strategies and plotted versus $Q^2$ in
  GeV${}^2$. The labels specify the strategy used to remove the ESC, and
  the ensemble ID. The solid black line shows the Kelly
  fit to the experimental data. \looseness-1
  \label{fig:GEGM-strategy}}
\end{figure*}

\begin{figure*}[tbp] 
\subfigure
{
    \includegraphics[width=0.24\linewidth]{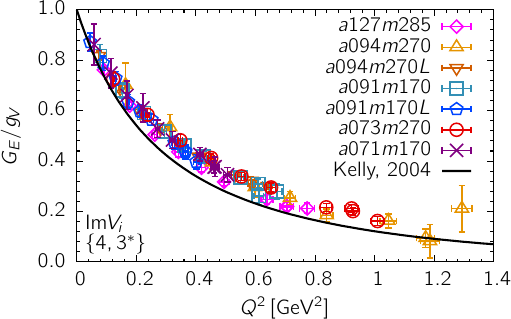}  
    \includegraphics[width=0.24\linewidth]{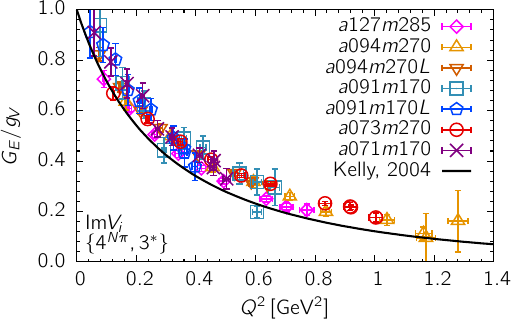} 
    \includegraphics[width=0.24\linewidth]{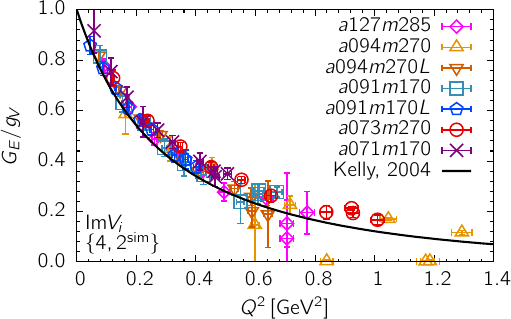}   
    \includegraphics[width=0.24\linewidth]{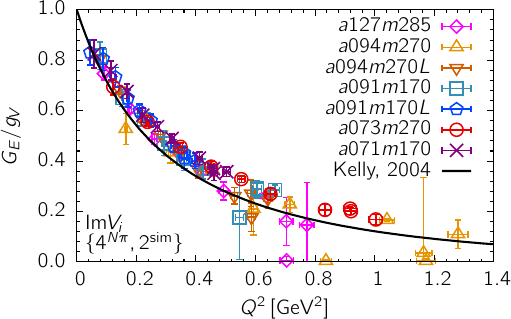}   
}
{
    \includegraphics[width=0.24\linewidth]{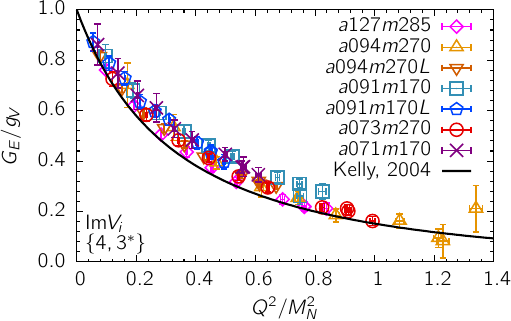}  
    \includegraphics[width=0.24\linewidth]{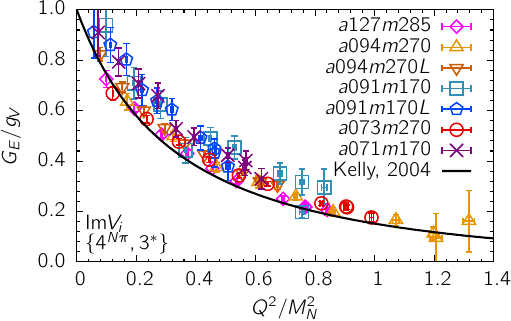} 
    \includegraphics[width=0.24\linewidth]{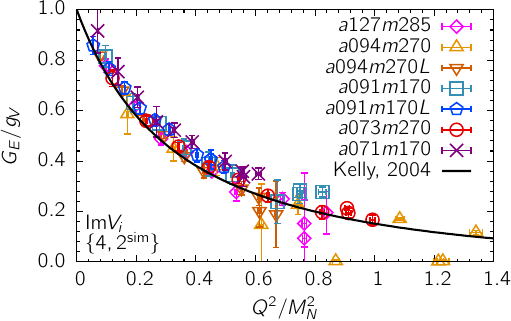}   
    \includegraphics[width=0.24\linewidth]{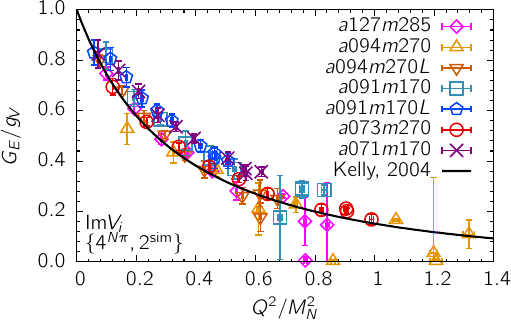}   
}
\vspace{-0.1in}
\caption{$G_E(Q^2)$ from $\Im V_i$ plotted versus $Q^2$ in GeV${}^2$ (top panels) and versus 
  $Q^2/M_N^2$ (bottom panels). Each panel
  shows the data for the seven ensembles, and each row compares the
  four strategies used to remove ESC.  
  \label{fig:GEisummary}}
\end{figure*}

A comparison of the form factors, and the errors in them, among the
four strategies is shown in Fig.~\ref{fig:GEGM-strategy} for the five
large volume ensembles. For each strategy, the full data from the
seven ensembles are shown in
Figs.~\ref{fig:GEsummary},~\ref{fig:GEisummary}
and~\ref{fig:GMsummary}.  The $G_E^{V_i}$ show significant variation
between the strategies, with the $\{4^{},2^{\rm sim}\}$ data being
closest to the Kelly curve. Part of this observed variation is a
result of a poorer statistical signal and part due to less control over
ESC. For these reasons, we do not include $G_E^{V_i}$ in our final
analysis; however, this channel influences the extraction of $\Delta
{\widetilde M}_1$ and $\Delta {\widetilde E}_1$ from the simultaneous
$\{2^{\rm sim}\}$ fits.

For the two cases with the best signal, $G_E$ from $\Re V_4$ and $G_M$
from $\Re V_i$, we make the following observations from
Fig.~\ref{fig:GEGM-strategy} using the Kelly curve as a benchmark and
to guide the eye:
\begin{itemize}
\item
No significant difference is observed between the data from the two
simultaneous fits, $\{4,2^{\rm sim}\}$ versus $\{4^{N\pi},2^{\rm
  sim}\}$, i.e., the differences in the ground state parameters used do not significantly 
affect the results.  On the four largest $Q^2$ points, the errors are large in
many cases, but the overall shape of the data is similar for all four
strategies.
\item
Results for $G_E^{V_4}$ and $G_M^{V_i}$ lie close to the
Kelly parameterization for all four strategies, with the
$\{4^{N\pi},3^{\ast}\}$ data plotted versus $Q^2/M_N^2$ showing the
best agreement. 
\item
All four strategies give consistent results on the $M_\pi \approx 270$~MeV 
ensembles. 
\item
In Fig.~\ref{fig:GEGM-strategy}, one can notice (i) a small spread
among the four strategies in $G_E^{V_4}$ on the $M_\pi \approx
170$~MeV ensembles, (ii) a small upward movement of data from
$a091m170L$ to $a071m170$, and (iii) the $\{4,3^\ast\}$ and both
$\{2^{\rm sim}\}$ data on $a
\approx 0.07$~fm ensembles lie above the Kelly curve.
\item 
The $G_M^{V_i}$ data also move upwards from $a091m170L$ to
$a071m170$. The $\{4,3^\ast\}$ strategy data lie below others on the
two smallest $Q^2$ points.
\item
The data plotted versus $Q^2$ show some dependence on $a$ and/or
$M_\pi^2$, whereas when plotted versus $Q^2/M_N^2$, no significant
dependence on either $a$ or $M_\pi^2$ is observed, and the agreement
with the Kelly curve is better. The size of the observed difference 
among the data plotted versus $Q^2$ or
$Q^2/M_N^2$ can be accounted for by discretization errors. 
 Assuming that there is a cancellation of these in the
analysis versus the dimensionless quantity $Q^2/M_N^2$, 
we choose it for presenting our final results.
\end{itemize}


\begin{figure*}[tbp] 
\subfigure
{
    \includegraphics[width=0.48\linewidth]{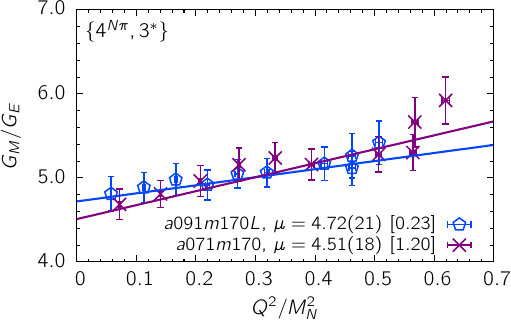} \hspace{0.2cm} 
    \includegraphics[width=0.48\linewidth]{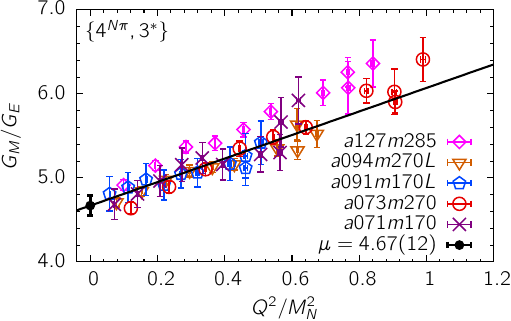}  
}
\caption{A linear fit to the smallest six $Q^2$ points for $G_M/G_E $
  from the $a091m170L$ and $a071m170$ ensembles obtained with the
  $\{4^{N\pi},3^\ast\}$ strategy. The intercept at $Q^2=0$ gives
  $\mu^{u-d}$.  The left panel shows separate fits to the two
  ensembles and the right to the combined data. Also shown, for
  comparison, in the right panel are the data from the other three
  larger volume $M_\pi \sim 270$~MeV ensembles.
  \label{fig:VFF-G_EoverG_M}}
\end{figure*}
%

\begin{table*}   
  \begin{ruledtabular}
    \begin{tabular}{r |llll | llll }
 & \multicolumn{4}{c|}{$\langle r_E^2\rangle|_{\rm dipole} $} & \multicolumn{4}{c}{$ \langle r_E^2\rangle|_{z^3} $} \\
Ensemble & $\{4,3^\ast\}$ & $\{4^{N\pi},3^\ast\}$ & $\{4,2^\text{sim}\}$ & $\{4^{N\pi},2^\text{sim}\}$ & $\{4,3^\ast\}$ & $\{4^{N\pi},3^\ast\}$ & $\{4,2^\text{sim}\}$ & $\{4^{N\pi},2^\text{sim}\}$ \\ \hline
$a127m285$  & 0.738(28)  & 0.773(27)  & 0.778(36)  & 0.777(38)  & 0.734(30)  & 0.768(30)  & 0.782(39)  & 0.778(43)  \\ 
$a094m270$  & 0.698(37)  & 0.704(20)  & 0.705(49)  & 0.706(48)  & 0.656(52)  & 0.699(32)  & 0.692(62)  & 0.711(63)  \\ 
$a094m270L$  & 0.682(22)  & 0.734(19)  & 0.698(22)  & 0.684(23)  & 0.669(24)  & 0.737(25)  & 0.701(25)  & 0.674(26)  \\ 
$a091m170$  & 0.740(27)  & 0.891(32)  & 0.767(36)  & 0.728(32)  & 0.726(41)  & 0.969(77)  & 0.847(86)  & 0.772(98)  \\ 
$a091m170L$  & 0.768(28)  & 0.902(54)  & 0.809(40)  & 0.784(38)  & 0.737(43)  & 0.893(79)  & 0.880(83)  & 0.76(10)  \\ 
$a073m270$  & 0.643(23)  & 0.681(19)  & 0.667(25)  & 0.664(24)  & 0.625(26)  & 0.662(25)  & 0.712(33)  & 0.710(33)  \\ 
$a071m170$  & 0.747(42)  & 0.854(43)  & 0.737(29)  & 0.712(25)  & 0.666(76)  & 0.834(96)  & 0.883(96)  & 0.72(11)  \\ 
\hline
 & \multicolumn{4}{c|}{$\langle r_M^2\rangle|_{\rm dipole} $} & \multicolumn{4}{c}{$ \langle r_M^2\rangle|_{z^3} $} \\
Ensemble & $\{4,3^\ast\}$ & $\{4^{N\pi},3^\ast\}$ & $\{4,2^\text{sim}\}$ & $\{4^{N\pi},2^\text{sim}\}$ & $\{4,3^\ast\}$ & $\{4^{N\pi},3^\ast\}$ & $\{4,2^\text{sim}\}$ & $\{4^{N\pi},2^\text{sim}\}$ \\ \hline
$a127m285$  & 0.582(22)  & 0.613(23)  & 0.627(29)  & 0.624(29)  & 0.569(33)  & 0.627(34)  & 0.672(34)  & 0.654(35)  \\ 
$a094m270$  & 0.507(25)  & 0.505(19)  & 0.544(29)  & 0.536(26)  & 0.565(36)  & 0.623(36)  & 0.634(31)  & 0.657(37)  \\ 
$a094m270L$  & 0.544(19)  & 0.613(19)  & 0.564(18)  & 0.558(17)  & 0.592(34)  & 0.642(34)  & 0.576(35)  & 0.568(34)  \\ 
$a091m170$  & 0.562(23)  & 0.691(39)  & 0.592(26)  & 0.615(29)  & 0.77(11)  & 1.00(11)  & 0.765(86)  & 0.743(95)  \\ 
$a091m170L$  & 0.630(29)  & 0.817(52)  & 0.610(27)  & 0.678(30)  & 0.61(11)  & 0.88(11)  & 0.55(10)  & 0.66(11)  \\ 
$a073m270$  & 0.495(18)  & 0.514(16)  & 0.509(20)  & 0.522(18)  & 0.527(40)  & 0.545(40)  & 0.613(26)  & 0.636(36)  \\ 
$a071m170$  & 0.562(31)  & 0.679(37)  & 0.581(25)  & 0.582(23)  & 0.71(12)  & 0.85(11)  & 0.89(10)  & 0.83(11)  \\ 
\hline
 & \multicolumn{4}{c|}{$ \langle \mu \rangle|_{\rm dipole} $} & \multicolumn{4}{c}{$ \langle \mu \rangle|_{z^3} $} \\
Ensemble & $\{4,3^\ast\}$ & $\{4^{N\pi},3^\ast\}$ & $\{4,2^\text{sim}\}$ & $\{4^{N\pi},2^\text{sim}\}$ & $\{4,3^\ast\}$ & $\{4^{N\pi},3^\ast\}$ & $\{4,2^\text{sim}\}$ & $\{4^{N\pi},2^\text{sim}\}$ \\ \hline
$a127m285$  & 4.558(51)  & 4.696(64)  & 4.753(84)  & 4.730(82)  & 4.538(56)  & 4.712(71)  & 4.823(89)  & 4.771(86)  \\ 
$a094m270$  & 4.252(84)  & 4.249(76)  & 4.421(94)  & 4.421(93)  & 4.343(67)  & 4.452(72)  & 4.542(73)  & 4.558(75)  \\ 
$a094m270L$  & 4.369(41)  & 4.571(57)  & 4.444(44)  & 4.422(41)  & 4.419(47)  & 4.578(61)  & 4.441(53)  & 4.426(47)  \\ 
$a091m170$  & 4.177(55)  & 4.598(95)  & 4.303(71)  & 4.359(72)  & 4.321(83)  & 4.749(54)  & 4.445(64)  & 4.474(77)  \\ 
$a091m170L$  & 4.323(64)  & 4.717(99)  & 4.275(57)  & 4.494(83)  & 4.311(78)  & 4.735(85)  & 4.224(72)  & 4.484(84)  \\ 
$a073m270$  & 4.273(52)  & 4.332(52)  & 4.307(65)  & 4.371(58)  & 4.301(71)  & 4.374(75)  & 4.487(70)  & 4.550(72)  \\ 
$a071m170$  & 4.200(78)  & 4.526(96)  & 4.230(70)  & 4.286(74)  & 4.281(82)  & 4.560(75)  & 4.455(79)  & 4.469(80)  \\ 
\end{tabular}
    \end{ruledtabular} \caption{Results for the isovector electric
      charge radius squared $\langle r_E^2 \rangle$ (top); magnetic
      charge radius squared, $\langle r_M^2 \rangle$ (middle); and 
      magnetic moment, $\mu^p - \mu^n$ (bottom), for the seven
      ensembles obtained using the dipole and the $z^3$
      parameterization of the $Q^2$ behavior. These fits were made
      keeping the smallest six $Q^2 \neq 0$ points. In fits to $G_M$,
      we included the point $G_M(0)/g_V$, obtained by linearly
      extrapolating $G_E/G_M$ to $Q^2 = 0$, as a prior with width
      $0.2$.  Data are compared for the four strategies ($\{4,3^*\}$,
      $\{4^{N\pi},3^\ast\}$, $\{4^{},2^\text{sim}\}$ and
      $\{4^{N\pi},2^\text{sim}\}$) for controlling ESC (see Appendix~\ref{sec:glossary}). Dipole
      estimates are not included in the final results as explained in
      the text.}  
\label{tab:rErMmu} 
\end{table*}

As mentioned above, the analog of the PCAC relation for the
electromagnetic form factors is the conserved vector charge, i.e.,
$\lim_{Q^2 \to 0} G_E(Q^2) \equiv g_V = 1/Z_V$. Since $g_V$ from the
forward matrix element has $O(1\%)$ excited-state effect as shown in
Fig.~\ref{fig:gVd7}, one could use it to pick the best strategy, i.e.,
the one for which the extrapolation of $G_E(Q^2)$ to $Q^2 = 0$ using
the $z^2$ or Pad\'e fit is most consistent with $g_V$. However, data
from all four strategies shown in Figs.~\ref{fig:GEsummary}
and~\ref{fig:GMsummary}, are consistent within expected lattice
artifacts with the Kelly parameterization, so this check does not help
in picking among the strategies.

The reduction in scatter in the form factors under variation in $a$
and $M_\pi$ when plotted versus $Q^2/M_N^2$ is consistent with the
analysis of clover-on-HISQ data presented in Ref.~\cite{Jang:2019jkn},
where results by other collaborations carried out at or near the
physical pion mass were also reviewed.  On the other hand, the
improvement in agreement with the Kelly curve of the clover-on-clover
data presented here is striking.  (See in particular the
$\{4^{N\pi},3^\ast\}$ strategy data plotted versus $Q^2/M_N^2$ in
Figs.~\ref{fig:GEsummary} and~\ref{fig:GMsummary}.)  Beyond the fact
that the clover-on-clover formulation is unitary, the only substantial
change in the lattice methodology we have made over the clover-on-HISQ
calculations is the random parity transformation (see
Eq.~\eqref{eq:parity} in Sec.~\ref{sec:setup}) on all the
lattices~\cite{Yoon:2016dij,Yoon:2016jzj}. Symmetry under parity plays
an important role in constraining the excited states that should
contribute; for example it disallows the $N(0) \pi (0) $ state. So,
while we expect improvement in the precision with which correlation
functions or contributions that should be zero under parity
transformation are indeed zero, the level of improvement in agreement
with the Kelly parameterization calls for further study.

For the $\{4,3^\ast\}$ strategy, the data in
Fig.~\ref{fig:GEGM-strategy} for $G_E(Q^2)$ lie above the Kelly curve
and the low $Q^2$ points of $G_M(Q^2)$ lie below. This behavior is in
accord with the deviations pointed out in
Ref.~\cite{Jang:2019jkn}. The data with the $\{4^{N\pi},3^\ast\}$
strategy are more consistent with the Kelly result. We hypothesize on
the basis of the observed improvement with the Kelly curve, the
behavior of the mass gaps shown in Fig.~\ref{fig:VFF-deltaM}, and the
vector meson dominance model that the low lying multihadron
excitations contribute.  While significantly more data, especially on
$M_\pi \approx 135$~MeV ensembles, are needed to validate this
conjecture, we will select between $\{4^{N\pi},3^\ast\}$ and $\{4^{N\pi},2^{\rm
sim}\}$ strategies for presenting results in this paper.  Of these two
strategies, the statistical precision of the current data is better
for $\{4^{N\pi},3^{\ast}\}$ and it has the advantage of including
three states in the fit.  On the other hand $\{2^{\rm sim}\}$ is,
statistically, better motivated if the same set of states contribute
to the three correlation functions.  For the time being, we will
continue to analyze all four strategies since it is instructive to
explore the differences.

The errors in the current lattice data are much larger than in the
Kelly parameterization of the experimental data and cover a smaller
range in $Q^2$. It will be some time before lattice data reach the
precision of experiments even in the range $0.04 < Q^2 <
1$~GeV${}^2$. Nevertheless, we regard the consistency of our results
with the Kelly curve an important and necessary step in demonstrating
control over all systematic uncertainties in the calculations of form
factors. The main thrust of future improvements will be on increasing
the statistics, designing better nucleon interpolating operators to
further control ESC, extending the calculation to more values of
$a$ and $M_\pi$ to confirm the observed lack of dependence on them, and 
obtaining data at smaller values of $Q^2$.

Having obtained $G_E(Q^2)$ and $G_M(Q^2)$ from the four strategies to
control ESC, we again parameterize the $Q^2$ dependence using the
dipole, $z$-expansion and Pad\'e fits. From these fits, we extract the
electric and magnetic isovector charge radii squared, $\langle r_E^2
\rangle$ and $\langle r_M^2 \rangle$, and the magnetic moment $\mu$.
These data are given in Table~\ref{tab:rErMmu} and exhibit two
noteworthy features: (i) the estimates with $\{4^{N\pi},3^{\ast}\}$
are larger, and (ii) the intercept at $Q^2=0$ of fits to $G_M/g_V$
shows the beginning of a flare-out, especially for $z$-expansion fits
with sum rules. This second feature suggests that $Q^2=0$ is already
at the edge of reliability of extrapolation of the fits to our data,
which have $Q^2_{\rm min} \gtrsim 0.04$~GeV${}^2$.

In Ref.~\cite{Jang:2019jkn}, we had shown that the ratio $G_E/G_M$
exhibits a linear behavior versus $Q^2$ and had used it to get an
estimate of $G_M(Q^2=0)=\mu$. The clover-on-clover data presented in
this study confirms this behavior as illustrated in
Fig.~\ref{fig:VFF-G_EoverG_M} for the $a091m170L$ and $a071m170$
ensembles. So we use this value of $G_M(Q^2=0)/g_V$ as a prior in the
fits to $G_M(Q^2)/g_V$. The error in it is  $\lesssim 0.2$ for
all ensembles, so we select 0.2 for the width. Setting the width to
0.3 changes the estimates by $\lesssim \sigma/3$ for both $\rMsq$ and
$\mu$. Overall, the use of the prior stabilizes the fits near $Q^2 =
0$, but does not change the results for $\rMsq$ or $\mu$
significantly.  The dipole, Pad\'e and $z$-expansion fits for the four
strategies are illustrated in Figs.~\ref{fig:VFF-Q2D7}
and~\ref{fig:VFF-Q2E7}) in Appendix~\ref{sec:TcompVFF} for the
$a091m170L$ and $a071m170$ ensembles, respectively. The values of
$\rEsq,\ \rMsq, \ \mu$ obtained, and the prior used, are given in the
labels.  These fits are made to the six smallest $Q^2 $ points since
the errors are large in some of the higher $Q^2$ data. For
completeness, we state that the results of fits to all ten points are
essentially the same.

Two important points: first, the current data (six or ten values of
$Q^2$) can be fit by the {$z^{2}$} and  $z^{3}$ truncations and $z^{4}$ is an
overparameterization. We note a change between $z^{2}$ and $z^{3}$ and
reasonable stability between $z^{3}$ and $z^{4}$. Thus all subsequent
results are with fits using the $z^3$ truncation. Second, the two
Pad\'e fits give overlapping results, and the $P(g,1,3)$ is again an
overparameterization.

To obtain the continuum limit values for $\rEsq$, $\rMsq$ and $\mu$,
the CCFV fits to the data given in Table~\ref{tab:rErMmu} are
discussed in Sec.~\ref{sec:CCFVEMradii}.

\section{Final results from the chiral-continuum-finite-volume fits}
\label{sec:CCFV}

In this section, we examine the dependence of the isovector charges,
$g_{A,S,T}^{u-d}$, the axial charge radius $\langle r_A^2 \rangle$,
the induced pseudoscalar charge $g_P^\ast$, the pion-nucleon coupling
$g_{\pi N N}$, the electric and magnetic charge radii, $\langle r_E^2
\rangle$ and $\langle r_M^2 \rangle$, and the magnetic moment
$\mu^{u-d}$ on the lattice spacing $a$, pion mass $M_\pi$, and the
lattice size parameter $M_\pi L$. The data are shown in
Figs.~\ref{fig:CCFVgA}--\ref{fig:CCFV-VFF-mu} in
Appendix~\ref{sec:appendixCCFV} along with the CCFV fit results as pink
bands.  In cases for which the largest variation is versus $M_\pi^2$,
we also show, for comparison, the result of just a chiral fit by a
gray band. The more these two bands overlap, the more dominant is the 
chiral correction. 

The overall framework of the CCFV analysis is as follows. A
simultaneous CCFV fit in the three variables is made to get the
results at the physical point defined as $M_\pi = 135$~MeV, $a=0$ and
$M_\pi L = \infty$.  With seven data points, we can only include
leading order corrections in each variable to avoid
overparameterization.  Keeping just the leading terms, we cannot
directly assess a systematic error associated with possible higher-order 
corrections to the CCFV ansatz. What we do evaluate is whether
the final error estimate from the simultaneous CCFV fit is
conservative in comparison to the observed change under extrapolation
in each parameter. In particular, for each quantity, we compare the
change between the data from the ensemble closest to the physical
point and the extrapolated value. For example, when discretization
errors are dominant, we compare the difference between data at
$a071m170$ and the extrapolated value with the error estimate from the
CCFV fit to determine if the latter is conservative enough.

In all cases, the discretization corrections are taken to be linear in
$a$ as our calculation (lattice action and operators) is not fully
$O(a)$ improved.

To evaluate possible finite volume corrections in a given observable,
we compare the data on the two pairs of ensembles
$\{a094m270,a094m270L\}$ and $\{a091m170,a091m170L\}$. Second, we also
compare outputs of chiral-continuum (CC) fits to the five larger
volume data with CCFV fits to the seven points, and check for
overparameterization.  Differences between the two fits, if
significant in comparison to the quoted error, are evaluated for
whether an additional systematic uncertainty should be assigned.
Overall, finite-volume corrections are observed to be small for $M_\pi
L > 4$.

The analysis so far has been carried out with a number of strategies
for removing ESC in the various quantities.  As already discussed, the
overriding uncertainty in the final analysis comes from whether the
low-lying $N\pi$ or $N\pi\pi$ states are relevant and included.
Including them significantly impacts the estimates from the $M_\pi
\approx 170$~MeV ensembles and thus the chiral extrapolation. In many
cases the errors in the $\approx 170$~MeV data are much larger than in
the $M_\pi \approx 270$~MeV points. Thus, their weight in the CCFV
fits is small. This is a serious limitation.  In subsequent sections,
we will discuss this and other issues on a case-by-case basis, and
provide our reasons for picking the strategy used to present the final
results and the assessment of the need for an additional systematic
uncertainty.

\subsection{The CCFV extrapolation for \texorpdfstring{$g_{A,S,T}^{u-d}$}{g(A,S,T)(u-d)}}
\label{sec:CCFVcharges}

The leading order CCFV fit ansatz used for all three isovector charges is 
\begin{equation}
g(a,M_\pi,M_\pi L) = c_1+ c_2 a + c_3 M_\pi^2 +c_4\frac{M_\pi^2e^{-M_\pi L}}{\sqrt{M_\pi L}} \,.
\label{eq:CCFVcharges}
\end{equation}
Results from these CCFV and CC ($c_4$ set to zero) fits to
$g_{A,S,T}^{u-d}$ are summarized in Table~\ref{tab:gASTfinal} and the
CCFV fits are shown in Figs.~\ref{fig:CCFVgA},~\ref{fig:CCFVgS}
and~\ref{fig:CCFVgT} in Appendix~\ref{sec:appendixCCFV}.  Overall,
these data indicate possible finite-volume corrections in
$g_{A}^{u-d}$, but no significant effect is observed in $g_{S}^{u-d}$
or $g_{T}^{u-d}$. Below, just before Eq.~\eqref{eq:finalcharges}, we
also discuss the change in results (i) on assuming that the
discretization errors begin at $O(a^2)$, i.e., replacing the term $c_2
a$ by $c_2 a^2$, and (ii) without any discretization error term.

All averages presented in this section are averages weighted by the inverse square of the errors. In most
cases the $\chi^2$ of the different fits whose results we average are
very similar, so the averages constructed using AIC weights are also
the same. Furthermore, both of these are also consistent with unweighted
averages.  We caution the reader that, for brevity, we use the
term average to denote averages weighted by the inverse square of the errors.

We note a number of systematic shifts of $O(0.03)$ in results
summarized in Table~\ref{tab:gASTfinal}, which, while smaller than the
individual total analysis errors in most cases, need to be
addressed. These are (i) between the two renormalization methods ${\rm
Z}_1$ and ${\rm Z}_2$, (ii) between the CC and CCFV fits and (iii) the
variation between the various strategies. 

The two methods of renormalization, ${\rm Z}_1$ and ${\rm Z}_2$, are
equally well motivated, however, as discussed in Sec.~\ref{sec:gV}, and 
the errors in the renormalization constants are better controlled with
${\rm Z}_1$ for $g_S$ and with ${\rm Z}_2 $ for $g_A$ and $g_T$. We,
therefore average the $g_S$ values obtained with ${\rm Z}_1$, 
given in Table~\ref{tab:gASTfinal} and specified below,  
 and $g_A$ and $g_T$ with ${\rm Z}_2$.  To
account for the difference in results obtained using ${\rm Z}_1$ versus 
${\rm Z}_2$, we assign an additional systematic uncertainty 
for all three charges.

Second, comparing the CCFV and CC estimates, there is a notable
difference only in $g_{A}^{u-d}$, which we discuss below.  For
$g_{S}^{u-d}$ and $g_{T}^{u-d}$, the CCFV fits have slightly larger
errors but in most cases the reduction in $\chi^2$ is not sufficient
to warrant including the finite volume correction term by the Akaike
criteria. As they are consistent, we present the average of
the CC and CCFV results.

On the third issue, for $g_{S}^{u-d}$ and $g_{T}^{u-d}$, the two
$\{2^{\rm free}\}$ strategies yield an unexpectedly large $\Delta
{\widetilde M}_1$. A larger value is expected in a two-state fit, i.e.,
it constitutes an effective mass gap representing the contribution of
all excited states. Including a third state improves the estimate for
$\Delta M_1$.  Therefore, as discussed in Sec.~\ref{sec:charges}, we
will choose the final results from the strategies that use a
three-state fit, $\{4^{},3^\ast\}$ and $\{4^{N\pi},3^\ast\}$. The
axial charge $g_{A}^{u-d}$ requires a more extensive analysis with
respect to ESC that is presented below.\looseness-1

{$\bm {g_{A}^{u-d}}$:} The axial charges, summarized in
Table~\ref{tab:gASTfinal} for the various strategies considered, are
obtained in two different ways: (i) from the forward matrix element,
which for the $\{4^{},3^\ast\}$ and $\{4^{N\pi},3^\ast\}$ strategies
are given in rows one and eight, and (ii) by extrapolating the form
factor $G_A(Q^2)$ to $Q^2=0$. To specify the parameterization used in
the second case, we introduce a third symbol,
$\{D\}$/$\{z^2\}$/$\{P_2\}$, to represent a dipole/$z^2$/$P(g, 0, 2)$
fit. For example $\{4^{N\pi},3^\ast,{z}^2\}$ means form factors obtained using
the $\{4^{N\pi},3^\ast\}$ strategy and extrapolated using the $z^2$
fit (the glossary in Appendix~\ref{sec:glossary} describes the various fits). In many of
the CCFV fits, the data show no significant finite volume correction,
especially above $M_\pi L > 4.0$. The effect is much smaller than the
overall analysis error from the CCFV fit shown in
Fig.~\ref{fig:CCFVgA} in Appendix~\ref{sec:appendixCCFV}. So we also
performed CC fits to data neglecting the two small volume ensembles,
$a094m270$ and $a091m170$. These are labeled as $\{{\widehat D}\}$ or
$\{{\widehat z}^2\}$ or $\{{\widehat P}_2\}$. Overall, the main issue
that needs to be resolved in both ways is whether the $N ({\bm 1}) \pi
({-\bm 1})$ state should be included in the analysis.

With the $\{4^{},3^\ast\}$ strategy (first seven rows in
Table~\ref{tab:gASTfinal}), the $\Delta M_1$ from a four-state fit is
large, about $600$~MeV, and the $\tau \to \infty$ value for
$g_{A}^{u-d}$ is smaller, about 5\% below the experimental value. 
In this case, estimates from the forward matrix element
(first row) and those using the dipole or $z^2$ or Pad\'e
parameterization of the form factors give consistent results.
Comparison of these estimates from the $Q^2$ fits is shown in the two
left panels in Fig.~\ref{fig:GAcompQ2} for the two $M_\pi \sim
170$~MeV ensembles.

With the $\{4^{N\pi},3^\ast\}$ strategy [uses the $N ({\bm 1}) \pi ({-\bm
  1})$ as the lowest excited state as discussed in
Sec.~\ref{sec:spectrum}], we find that the finite-volume correction
term is negligible as shown by the CCFV fit to the
$\{4^{N\pi},3^\ast\}$ data in Fig.~\ref{fig:CCFVgA}.  Comparing the
results in rows 9-14, we note that the estimates with the dipole fit,
$\{4^{N\pi},3^\ast, D\}$, are smaller. The reason is that the dipole fit misses the
lowest $Q^2$ point on the $M_\pi \approx 170$~MeV ensembles as
illustrated in the middle panels in Fig.~\ref{fig:GAcompQ2}.

With the preferred $\{4^{N\pi},2^\text{sim}\}$ strategy, selected on the basis of 
satisfying the PCAC relation, only results from the extrapolation of
the form factor are possible. Within errors, the estimates in each of
the four columns in Table~\ref{tab:gASTfinal} are consistent, but
two of the three $O(0.03)$ shifts discussed above (renormalization,
and finite volume indicated by CCFV versus CC estimates) are manifest.  We derive our best estimate as
follows. The finite volume systematic is not well controlled, so we
average the larger volume, $M_\pi L > 4$, CC-fit values $\{4^{N\pi},2^\text{sim},{\widehat z}^2\}$ and the
$\{4^{N\pi},2^\text{sim}, {\widehat P}_2 \}$. For renormalization, 
we choose the ${\rm Z}_2$ estimates as discussed in Sec.~\ref{sec:gV}. With these 
choices, our result 
is $g_{A}^{u-d} = 1.32(5)$.  The
same selection procedure applied to the $\{4,3^\ast\}$ strategy gives
$g_{A}^{u-d} = 1.23(4)$. The large difference, $\sim 0.09$, makes it
clear that establishing whether the low-mass $N \pi$ state(s)
contribute is essential to the extraction of $g_{A}^{u-d}$.

Three systematic uncertainties, summarized in
Eq.~\eqref{eq:finalcharges}, are added to the above estimate. These
are taken to be half the spread in the data in
Table~\ref{tab:gASTfinal} as follows: For renormalization it is half
the difference between the ${\rm Z}_1$ and ${\rm Z}_2$ values, i.e.,
0.02. Half the spread in results between the strategies that include
the $N\pi$ state when removing ESC gives 0.04. For finite volume
corrections, half the difference between the CC and CCFV fit values
gives 0.02. In all these averages and error estimates, we do not
consider the dipole fit values since these fits miss the lowest $Q^2$
point on the $M_\pi \approx 170$~MeV ensembles. This is illustrated in
the right panels in Fig.~\ref{fig:GAcompQ2}.\looseness-1

Overall, in the CCFV fits we note (i) there is tiny if any $a$
dependence in data from any of the strategies investigated; (ii) there
is almost no dependence on $M_\pi^2$ for $\{4,3^\ast\}$ but a
significant one in the strategies that include the $N\pi$ state; and
(iii) there is an indication of a finite volume correction with the
$\{4^{N\pi},2^\text{sim},z^2\}$ and $\{4^{N\pi},2^\text{sim},P_2\}$
strategies.  Of these three changes, the largest effect  is in 
the slope versus $M_\pi^2$ on including
the $N\pi$ state. The contribution of the $N \pi$
state grows as $Q^2 \to 0$ and $M_\pi \to 135$~MeV. Since
$G_A(Q^2)$ is analytical and monotonic in $Q^2$, we expect the
influence of the $N\pi$ state to persist at $Q^2 = 0$ in the sense
that the value of $g_A$ obtained directly at $Q^2 = 0$ from the
forward matrix element calculated using the $A_3$ correlator must
agree in the continuum limit with that extracted from a $z$-expansion
(or Pad\'e) fit to the form factor.  Even though our data satisfy this
check individually for both $\{4,3^\ast\}$ and $\{4^{N\pi},3^\ast\}$
strategies as shown in Table~\ref{tab:gASTfinal}, the value of $g_A$,
however, is different. The estimate from $\{4^{N\pi},3^\ast\}$ varies
between $1.28(5)$ and $ 1.33(5)$. This is consistent with our final
result, $g_{A}^{u-d} = 1.32(6)$, and the error covers the 
$\{4^{N\pi},2^\text{sim},{\widehat z}^2\}$ and the
$\{4^{N\pi},2^\text{sim}, {\widehat P}_2\}$ estimates.

We consider $\{4^{N\pi},2^\text{sim},{\widehat z}^2\}$ and
$\{4^{N\pi},2^\text{sim}, {\widehat P}_2\}$ as two models because, up
to some reasonably small $Q^2$, both the fixed order $z$-expansion and
the Pad\'e should give the same intercept in the limit of perfect
data.  The reason we take the weighted average and do not include the
AIC weight is because the $\chi^2$ of both is abnormally small 
as discussed below.

$\bm {g_{S}^{u-d}}$: We neglect the results from the two $\{2^{\rm
  free}\}$ strategies, which are somewhat larger, because the
associated $\Delta {\widetilde M}_1$ is larger than even that from the
$\{4\}$ fit as discussed in Sec.~\ref{sec:charges}. Results from
$\{4,3^\ast\}$ and $\{4^{N\pi},3^\ast\}$ overlap (see
Fig.~\ref{fig:diffcharges}) and no significant finite-volume correction
is observed. Thus we average estimates from the latter two strategies 
and the two fits, CCFV and CC, all with the ${\rm Z}_1$ 
renormalization method (see Sec.~\ref{sec:gV}). The result is 
q$g_{S}^{u-d} = 1.06(9)$. Note that the error estimate covers the
larger but neglected $\{2^{\rm free}\}$ values.

The most significant variation in the CCFV fits shown in
Fig.~\ref{fig:CCFVgS} in Appendix~\ref{sec:appendixCCFV} is versus
$a$. The difference between the $a=0.071$~fm and the $a=0$ value is
$\sim 0.12$, so we assign, in Eq.~\eqref{eq:finalcharges}, an
additional systematic uncertainty of 0.06 for possible incomplete
accounting of the discretization error in the CCFV or CC fits.
Estimates from the two renormalization methods show a difference of
$\sim 0.04$, so we assign an additional systematic uncertainty of
$0.02$.

$\bm {g_{T}^{u-d}}$: We again neglect the results from the two
$\{2^{\rm free}\}$ fits for the same reason as for $g_{S}^{u-d}$.
Similarly, we take the weighted average of the remaining four estimates in
Table~\ref{tab:gASTfinal} with ${\rm Z}_2$ renormalization and get $g_{T}^{u-d} = 0.97(3)$.  The
largest variation in the CCFV fits shown in Fig.~\ref{fig:CCFVgT} in
Appendix~\ref{sec:appendixCCFV} is versus $M_\pi^2$, with a possible
$\sim 0.02$ difference between $M_\pi =170$ and the extrapolated
135~MeV value. This difference is covered by the overall analysis (CC or CCFV) error.
There is also a $\approx 0.02$ difference between the two ES strategies (see
Fig.~\ref{fig:diffcharges}), so we assign a $0.01$ 
uncertainty for possible additional ES effects. Last, the two
renormalization methods give estimates that differ by $\sim 0.02$, so
we assign an additional $0.01$ uncertainty due to it.

\textbf{Remarks on discretization errors}: The
discretization correction in the CC and CCFV fit ansatz,
Eq.~\eqref{eq:CCFVcharges}, is taken to be linear in $a$ since our
action and the axial operator are not fully $O(a)$ improved. 
We have also carried out the analysis with the errors starting
at $O(a^2)$, i.e., using $c_2 a^2$ instead of $c_2 a$ in
Eq.~\eqref{eq:CCFVcharges} and assuming 
the linear in $a$ correction is negligible. The $\chi^2$ of the two sets of CCFV fits
are essentially the same for all three charges. The corresponding estimates for the
charges change to $g_{A}^{u-d} = 1.34(4)$, $g_{S}^{u-d} = 0.97(6)$
and $g_{T}^{u-d} = 0.97(2)$. The reason for the smaller CCFV fit
errors is that the range of extrapolation to the continuum limit is
smaller in $a^2$. We keep the larger error estimates from fits with
$c_2 a$ but assign an additional discretization uncertainty of $0.02$
and $0.01$ for $g_{A}^{u-d} $ and $g_{T}^{u-d} $, respectively. The
largest change in $g_{S}^{u-d}$ is with respect to $a$ and the error already assigned
covers the variation between $c_2 a$ and $c_2 a^2$ fits.

We also show chiral fits (gray bands) for $g_A$ and $g_T$ in the middle
panels of Figs~\ref{fig:CCFVgA} and~\ref{fig:CCFVgT}. The reason for neglecting
discretization and finite volume corrections is the
observation that the data on the five large volume lattices do not
show a significant dependence on $a$ or $M_\pi L$. In all cases, these
results overlap with the CCFV values but have smaller errors. The
similar $\chi^2$ suggests that the CCFV fits are
overparameterized. Nevertheless, as discussed above, for the final 
results we quote the CC values and errors for $g_A$ and CCFV for $g_S$
and $g_T$.

\textbf{Remarks on low $\chi^2$ values in CCFV fits}: The $\chi^2$ of
the two fits $\{ 4^{N\pi},2^\text{sim},{\widehat z^2 } \}$ and
$\{4^{N\pi},2^\text{sim},{\widehat P}_2 \}$ used to get $g_A^{u-d}$
are essentially zero as given in Table~\ref{tab:gASTfinal}. The
following two factors could explain such $\chi^2 \ll 1$: (i) the
errors assigned to the data points are overestimated, and (ii) the
fits are overparameterized. The first because the error in the
multiplicative renormalization factor $Z_A$ is of the same size as
the statistical error in $g_A^{\rm bare}$ (see Tables~\ref{tab:gVZV}
and~\ref{tab:gAdP2z2}) and is neither normally distributed nor
independent.  The second because the discretization errors are small
and including the $c_2$ term is an overparameterization. We have
chosen to include it (CC fit) but do not construct 
an AIC weighted average due to the small $\chi^2$. 

Within this framework, our final results are
\begin{equation}
\setlength{\tabcolsep}{6pt}
\begin{tabular}{l|ccccc}
{\rm Charge} & {\rm Value} & $\delta${\rm ES} & $\delta$Z & $\delta a$       & $\delta${\rm FV}   \\
\hline
$g_{A}^{u-d}$ & 1.32(6)   &   (4)           & (2)      &   (2)           & (2)               \\
$g_{S}^{u-d}$ & 1.06(9)   &                 & (2)      &   (6)           &                    \\
$g_{T}^{u-d}$ & 0.97(3)   &   (1)           & (1)      &   (1)           &                    \\
\end{tabular}
\label{eq:finalcharges}
\end{equation}
where the first error is the overall analysis uncertainty and $\delta$ES, $\delta$Z, $\delta a$, and $\delta$FV are the
additional systematic uncertainties due to excited states,
renormalization, discretization and finite volume artifacts.  
Combining these systematic errors in quadrature, our results are: 
\begin{eqnarray}
g_{A}^{u-d} &=&  1.32(6)(5)_{\rm sys} \,,  \nonumber \\
g_{S}^{u-d} &=&  1.06(9)(6)_{\rm sys}  \,,  \nonumber \\
g_{T}^{u-d} &=&  0.97(3)(2)_{\rm sys} \,.
\label{eq:finalcharges1}
\end{eqnarray}
Even with our high statistics data, the errors in $g_A^{u-d}$ are much
larger than in the experimental value $g_{A}^{u-d} =
1.2764(1)$~\cite{Mendenhall:2012tz,Brown:2017mhw,Mund:2012fq}. Estimates
for $g_S^{u-d}$ and $g_T^{u-d}$ are consistent with results 
in Ref.~~\cite{Gupta:2018qil} obtained using
the clover-on-HISQ formulation.

\begin{table*}   
\begin{ruledtabular}
\begin{tabular}{r|cc|cc}
Strategy & $g_A^{u-d}|_{{\rm Z}_1}$ ($c_4=0$) & $g_A^{u-d}|_{{\rm Z}_1}$ & $g_A^{u-d}|_{{\rm Z}_2}$ ($c_4=0$) & $g_A^{u-d}|_{{\rm Z}_2}$\\ \hline
\{4,3*\} & 1.215(48) [0.26] & 1.203(59) [0.31] & 1.250(42) [0.18] & 1.250(51) [0.24] \\
$\{ 4,3^*,z^2 \}$ & 1.194(44) [0.04] & 1.200(52) [0.05] & 1.230(39) [0.12] & 1.242(46) [0.05] \\
$\{ 4,3^*,{\widehat z^2 } \}$ & 1.194(44) [0.02] &  & 1.230(40) [0.14] &  \\
$\{ 4,3^*,P_2 \}$ & 1.184(46) [0.02] & 1.191(56) [0.01] & 1.221(41) [0.16] & 1.239(49) [0.06] \\
$\{ 4,3^*,{\widehat P_2 } \}$ & 1.185(46) [0.00] &  & 1.222(41) [0.23] &  \\
$\{ 4,3^*,D \}$ & 1.183(42) [0.35] & 1.206(48) [0.15] & 1.217(36) [0.59] & 1.248(42) [0.03] \\
$\{ 4,3^*,{\widehat D } \}$ & 1.184(42) [0.05] &  & 1.219(37) [0.13] &  \\
\{$4^{N\pi}$,3*\} & 1.280(48) [0.11] & 1.288(55) [0.12] & 1.317(42) [0.14] & 1.331(47) [0.03] \\
$\{ 4^{N\pi},3^*,z^2 \}$ & 1.274(52) [0.24] & 1.289(61) [0.24] & 1.307(48) [0.24] & 1.328(55) [0.13] \\
$\{ 4^{N\pi},3^*,{\widehat z^2 } \}$ & 1.277(54) [0.24] &  & 1.312(49) [0.15] &  \\
$\{ 4^{N\pi},3^*,P_2 \}$ & 1.272(57) [0.14] & 1.273(69) [0.18] & 1.308(53) [0.10] & 1.316(62) [0.12] \\
$\{ 4^{N\pi},3^*,{\widehat P_2 } \}$ & 1.277(58) [0.20] &  & 1.313(54) [0.10] &  \\
$\{ 4^{N\pi},3^*,D \}$ & 1.222(49) [0.61] & 1.262(56) [0.14] & 1.256(43) [0.98] & 1.303(50) [0.03] \\
$\{ 4^{N\pi},3^*,{\widehat D } \}$ & 1.225(50) [0.01] &  & 1.260(45) [0.18] &  \\
$\{ 4,2^\text{sim},z^2 \}$ & 1.248(55) [0.84] & 1.295(66) [0.56] & 1.276(51) [1.07] & 1.332(61) [0.52] \\
$\{ 4,2^\text{sim},{\widehat z^2 } \}$ & 1.263(56) [0.05] &  & 1.296(52) [0.04] &  \\
$\{ 4,2^\text{sim},P_2 \}$ & 1.239(64) [0.91] & 1.290(78) [0.78] & 1.269(59) [1.12] & 1.332(73) [0.75] \\
$\{ 4,2^\text{sim},{\widehat P_2 } \}$ & 1.257(65) [0.07] &  & 1.293(60) [0.05] &  \\
$\{ 4,2^\text{sim},D \}$ & 1.160(47) [1.50] & 1.219(54) [0.35] & 1.193(43) [2.33] & 1.261(49) [0.27] \\
$\{ 4,2^\text{sim},{\widehat D } \}$ & 1.159(47) [0.06] &  & 1.192(43) [0.49] &  \\
$\{ 4^{N\pi},2^\text{sim},z^2 \}$ & 1.279(54) [0.68] & 1.320(62) [0.34] & 1.308(50) [1.04] & 1.357(57) [0.37] \\
$\{ 4^{N\pi},2^\text{sim},{\widehat z^2 } \}$ & 1.290(54) [0.00] &  & 1.322(50) [0.26] &  \\
$\{ 4^{N\pi},2^\text{sim},P_2 \}$ & 1.273(63) [0.73] & 1.326(75) [0.38] & 1.303(59) [1.07] & 1.368(69) [0.38] \\
$\{ 4^{N\pi},2^\text{sim},{\widehat P_2 } \}$ & 1.283(64) [0.00] &  & 1.316(59) [0.23] &  \\
$\{ 4^{N\pi},2^\text{sim},D \}$ & 1.210(48) [1.14] & 1.259(55) [0.27] & 1.242(44) [1.90] & 1.299(49) [0.27] \\
$\{ 4^{N\pi},2^\text{sim},{\widehat D } \}$ & 1.215(49) [0.02] &  & 1.250(44) [0.43] &  \\
\hline
Strategy & $g_S^{u-d}|_{{\rm Z}_1}$ ($c_4=0$) & $g_S^{u-d}|_{{\rm Z}_1}$ & $g_S^{u-d}|_{{\rm Z}_2}$ ($c_4=0$) & $g_S^{u-d}|_{{\rm Z}_2}$\\ \hline
$\{4,3^\ast\}$ & 1.068(68) [0.05] & 1.052(92) [0.04] & 1.101(96) [0.05] & 1.09(12) [0.07] \\
$\{4^{N\pi},3^\ast\}$ & 1.062(93) [0.05] & 1.06(11) [0.06] & 1.10(11) [0.02] & 1.10(13) [0.02] \\
\{$4,2^\text{free}$\} & 1.056(52) [0.39] & 1.086(63) [0.28] & 1.081(82) [0.40] & 1.118(92) [0.27] \\
\{$4^{N\pi},2^\text{free}$\} & 1.100(52) [1.01] & 1.157(61) [0.25] & 1.120(82) [0.85] & 1.186(91) [0.21] \\
\hline
Strategy & $g_T^{u-d}|_{{\rm Z}_1}$ ($c_4=0$) & $g_T^{u-d}|_{{\rm Z}_1}$ & $g_T^{u-d}|_{{\rm Z}_2}$ ($c_4=0$) & $g_T^{u-d}|_{{\rm Z}_2}$\\ \hline
$\{4,3^\ast\}$ & 0.944(46) [0.06] & 0.942(53) [0.08] & 0.968(27) [0.03] & 0.971(34) [0.03] \\
$\{4^{N\pi},3^\ast\}$ & 0.938(50) [0.14] & 0.926(57) [0.13] & 0.962(33) [0.15] & 0.955(38) [0.17] \\
\{$4,2^\text{free}$\} & 0.995(43) [0.15] & 0.985(50) [0.15] & 1.017(24) [0.26] & 1.017(29) [0.35] \\
\{$4^{N\pi},2^\text{free}$\} & 1.027(44) [0.22] & 1.027(50) [0.29] & 1.047(25) [0.46] & 1.047(28) [0.61] \\
\end{tabular}
\end{ruledtabular}
\caption{ Results for the renormalized $g_{A,S,T}^{u-d}$ after CC ($c_4=0$) and 
  CCFV extrapolations using Eq.~\protect\eqref{eq:CCFVcharges} for the
  various strategies used to remove the ESC listed in column one that
  are discussed in Secs.~\protect\ref{sec:charges}
  and~\protect\ref{sec:AFF}, as well as in Appendix~\ref{sec:glossary}.  The
  results for $g_A^{u-d}$ labeled with additional $z^2$/$P_2$/$D$ use
  $g_A = G_A(Q^2=0)$ obtained by extrapolating $G_A(Q^2)$ to $Q^2=0$
  using these fits to all ten $Q^2 \ne 0$ points (see glossary in
  Appendix~\ref{sec:glossary}). The results in rows with ${\widehat
  D}$/${\widehat z}^2$/${\widehat P}_2$ are from CC fits to data
  excluding the small volume $a094m270$ and $a091m170$ ensembles.  The
  $\chi^2$/dof of the CC and CCFV fits are given within the square
  brackets.\looseness-1
\label{tab:gASTfinal}}
\end{table*}

\subsection{The CCFV extrapolation for the axial charge radius squared 
\texorpdfstring{$\langle r_A^2 \rangle$}{<rA\suptwo>}}
\label{sec:CCFVrA}

The data given in Table~\ref{tab:rAdP2z2} show no significant
difference between the $\{4,3^\ast\}$ and $\{4^{N\pi},2^{\rm sim}\}$
strategies on the $M_\pi \approx 270$~MeV ensembles. However, there is
a difference on the $M_\pi \approx 170$~MeV ensembles due to the
inclusion of the $N\pi$ state.  We have summarized our reasons for
choosing the $\{4^{N\pi},2^{\rm sim}\}$ strategy for the analysis of
the axial form factors $G_A$ and ${\widetilde G}_P$ in
Sec.~\ref{sec:AFFconsistency}, and we will use it to obtain the
quantities derived from them, $\langle r_A^2 \rangle$, $ g_P^\ast$, and
$g_{\pi N N}$.

The CCFV ansatz used to fit  $\langle r_A^2 \rangle$, 
\begin{align}
  r_{A}^2 (a,M_\pi,L) &=& c_1 + c_2a + c_3 M_\pi^2 +  c_4 M_\pi^2 \frac{e^{-M_\pi L}}{\sqrt{M_\pi L}} \,, 
\label{eq:CCFV-rA} 
\end{align}
is the same as for the isovector charges given in
Eq~\eqref{eq:CCFVcharges}.  Fits with the $\{4^{N\pi},2^{\rm
  sim},z^2\}$ strategy are shown in Fig.~\ref{fig:CCFV-AFF} and the
results summarized in Table~\ref{tab:rAfinal}.  We note a strong
dependence on $M_\pi^2$ and a slight increase with both $M_\pi L$ and
$a$.  Most of the increase with $M_\pi L$ takes place for $M_\pi L <
4$; therefore, we take the final result from the $\{4^{N\pi},2^{\rm
  sim},{\widehat z}^2\}$ analysis:
\begin{equation}
r^2_A|_{z^{2}}  =  0.428(53)(30)\ {\rm fm}^2 \ \Rightarrow r_A|_{z^{2}}  =  0.65(4)(2)\ {\rm fm}\,,
\label{eq:rAfinal}
\end{equation}
where the second, systematic, uncertainty is the difference from the
$\{4^{N\pi},2^{\rm sim},{\widehat P}_2\}$ value. This result is
consistent with the $\{4^{N\pi},2^{\rm sim},z^2\}$ and
$\{4^{N\pi},2^{\rm sim},P_2\}$ values, and the quoted error also covers
the spread in the CCFV estimates from the $\{4^{N\pi},3^{\ast}\}$,
$\{4,2^{\rm sim}\}$, $\{4^{N\pi},2^{\rm sim}\}$ strategies and both
$z^2$ and $P_2$ fits.  

Results for $\langle r_A^2 \rangle$ using the dipole parameterization
of the $Q^2$ behavior are significantly smaller than those from the
$z^2$ or $P_2$ fits, and the
$\chi^2$/dof is large in many cases. More important, these fits miss
the low $Q^2$ points as illustrated in Fig.~\ref{fig:GAcompQ2}. So we
do not include the dipole estimates in deriving the final results.

Our result, $ r_A = 0.65(4)(2)$~fm, is consistent with the three
phenomenological/experimental values: (i) a
weighted world average of (quasi)elastic neutrino and antineutrino
scattering data~\cite{Bernard:2001rs}, (ii) charged pion
electroproduction experiments~\cite{Bernard:2001rs}, and (iii) a
reanalysis of the deuterium target data~\cite{Meyer:2016oeg}: 
\begin{align}
r_A&= 0.666(17){\rm fm} \   ({\cal M}_A = 1.03(2){\rm GeV} )   \ [\nu,{\overline \nu}\ {\rm scattering}]    \,,  \nonumber \\
r_A&= 0.639(10){\rm fm} \  ({\cal M}_A = 1.07(2){\rm GeV} )   \ [{\rm Electroprod.}] \,,  \nonumber \\
r_A&= 0.68(16){\rm fm}  \ \  ({\cal M}_A  = 1.00(24){\rm GeV}) \ [{\rm Deuterium}] \,. 
\label{eq:rA_expt}
\end{align}
In this list, we do not quote the MiniBooNE value ${\cal
M}_A=1.35(17)$~GeV ($r_A = 0.506$~fm)~\cite{AguilarArevalo:2010zc} as it is not the outcome of
an analysis, but the best value that reproduces the double
differential cross section for charged current quasielastic neutrino
and antineutrino scattering data off carbon analyzed with a dipole
ansatz and a relativistic Fermi gas model of nuclear
interactions~\cite{Smith:1972xh}. It will be interesting to see an
update of the MiniBooNE analysis with our parameterization of
$G_A(Q^2)$ given in Eq.~\eqref{eq:GAPade} and a more realistic model
of nuclear interactions~\cite{Lynn:2019rdt,Carlson:2014vla}.

\begin{table*}   
\begin{ruledtabular}
\begin{tabular}{r|cc|cc|cc}
ESC Strategy                       & $z^2$ ($c_4=0$)   & $z^2$             & $P_2$ ($c_4=0$)   & $P_2$             & dipole ($c_4=0$)  & dipole \\ \hline
$\{4,3^\ast\}$                     & 0.307(38) [0.27] & 0.319(45) [0.29] & 0.276(48) [0.36] & 0.298(58) [0.33] & 0.262(29) [1.58] & 0.297(34) [0.78] \\
$\{ 4,3^\ast \} \dag$              & 0.306(40) [0.51] &  & 0.277(48) [0.70] &  & 0.270(29) [1.41] &  \\
$\{ 4^{N\pi},3^\ast \}$            & 0.424(45) [0.34] & 0.446(49) [0.10] & 0.408(62) [0.17] & 0.412(72) [0.23] & 0.315(30) [2.75] & 0.362(34) [0.46] \\
$\{ 4^{N\pi},3^\ast \} \dag$       & 0.441(49) [0.07] &  & 0.421(65) [0.06] &  & 0.327(33) [1.53] &  \\
$\{ 4,2^\text{sim} \}$             & 0.413(47) [1.95] & 0.450(53) [1.89] & 0.375(75) [1.99] & 0.434(90) [2.20] & 0.228(25) [6.06] & 0.281(28) [0.31] \\
$\{ 4,2^\text{sim} \} \dag$        & 0.465(51) [0.42] &  & 0.445(80) [0.57] &  & 0.224(26) [2.01] &  \\
$\{ 4^{N\pi},2^\text{sim} \}$      & 0.399(49) [1.01] & 0.439(55) [0.47] & 0.366(80) [0.90] & 0.437(93) [0.48] & 0.244(27) [4.71] & 0.283(29) [0.98] \\
$\{ 4^{N\pi},2^\text{sim} \} \dag$ & 0.428(53) [0.33] &  & 0.398(83) [0.31] &  & 0.243(28) [2.04] &  \\
\end{tabular}
\end{ruledtabular}
\caption{ Results for the axial charge radius $\langle r_A^2 \rangle$
  from (i) different strategies for removing ESC listed in column one (see Appendix~\ref{sec:glossary})
  and (ii) fits to the ten $Q^2 \ne 0$ points for the axial form
  factor $G_A(Q^2)$ using the $z^2$, $P_2$ Pad\'e and the dipole
  parameterizations. The additional $\dag$ in column one denotes results from 
  CC fits with $c_4 = 0$, i.e., neglecting the small volume ($a094m270$ and $a091m170$) points. 
  The $\chi^2$/dof of the $Q^2$ fits are given within the square brackets.\looseness-1
\label{tab:rAfinal}}
\end{table*}

\subsection{The CCFV extrapolation for \texorpdfstring{$ g_P^\ast$}{gP*} and \texorpdfstring{$g_{\pi N N}$}{g\textpi NN} }
\label{sec:CCFVgPstar}

To perform the CCFV fit for $g_P^\ast$ given in
Table~\ref{tab:gPstargpiNN}, we use the ansatz
\begin{align}
  g_P^\ast (a,M_\pi,M_\pi L)/g_A =&  d_1 + d_2 a + \frac{d_4}{ M_\pi^2 + 0.88 m_\mu^2}   \nonumber \\
      +&  d_3 M_\pi^2 + \frac{d_5 M_\pi^2}{\sqrt{M_\pi L}} e^{-M_\pi L}\,,
\label{eq:extrap-gP} 
\end{align}
where the leading behavior in $M_\pi^2$ is taken to be the pion-pole
term evaluated at the momentum scale of the muon capture
experiment~\cite{Andreev:2012fj,Andreev:2015evt}.  The data and fit in
Fig.~\ref{fig:CCFV-AFF} in Appendix~\ref{sec:appendixCCFV} 
show no significant dependence on either $a$
or $M_\pi L$ but a strong dependence on $M_\pi^2$. The
result of the CCFV fit to the $\{4^{N\pi},2^{\rm sim}\}$ data is
\begin{eqnarray}
g_P^\ast     &=& 7.9(7)(9)_{\rm sys} \,, 
\label{eq:GPstar_final}
\end{eqnarray} 
where the second systematic uncertainty is half the change from 
the $a071m170$ point in the chiral extrapolation. 
The two methods for renormalization give overlapping results, so we do
not assess an additional systematic uncertainty due to it. To underscore the
importance of including the $N\pi$ state in the analysis of ESC, note
that the analogous result with the $\{4,3^{\ast}\}$ strategy is $ 3.9(1.1)$.

Experimentally, ${\widetilde G}_P(Q^2=0.88m_\mu^2)$ is determined from  muon capture by a
proton, $\mu^{-} + p \to \nu_\mu + n$~\cite{Andreev:2012fj,Andreev:2015evt}. 
Current estimates from the MuCap
experiment~\cite{Andreev:2012fj,Andreev:2015evt}, and from chiral
perturbation theory~\cite{Schindler:2006it,Bernard:2001rs} are 
\begin{eqnarray}
g_P^\ast|_{\rm MuCap}     &=& 8.06(55) \,,  \nonumber \\
g_P^\ast|_{\chi{\rm PT}}  &=& 8.29^{+0.24}_{-0.13} \pm 0.52 \,,
\label{eq:GPstar_pheno}
\end{eqnarray} 
respectively. 

The CCFV fit to the pion-nucleon coupling $g_{\pi N N }$
data, also given in Table~\ref{tab:gPstargpiNN}, was carried out using
the ansatz given in the right-hand side of Eq.~\eqref{eq:CCFV-rA}. The
result of the fit, shown in Fig.~\ref{fig:CCFV-AFF} in
Appendix~\ref{sec:appendixCCFV}, is
\begin{eqnarray}
g_{\pi NN}     &=& 12.4(1.2) \,.
\label{eq:GpiNN_final}
\end{eqnarray}
Again, the dominant dependence of the data is on $M_\pi^2$ but there
is no significant change from the $a071m170$ value. The variation with
the renormalization method is $\sim 0.3\sigma$. These are much
smaller than the quoted $1\sigma$ error, so we do not assign an
additional systematic uncertainty. For comparison, the result with the
$\{4,3^{\ast}\}$ strategy that does not include the $N\pi$ state is $
6.8(1.3)$.

To summarize, results for all three quantities, $\rAsq$, $g_P^\ast$ and $g_{\pi NN}$
given in Eqs.~\eqref{eq:CCFV-rA},~\eqref{eq:GPstar_final}
and~\eqref{eq:GpiNN_final} come in reasonable agreement with
phenomenological values with the $\{4^{N \pi},2^{\rm sim}\}$ strategy
that is singled out on the basis of the axial form factors satisfying
the PCAC relation.
%


\subsection{Goldberger-Treiman relation and \texorpdfstring{$F_\pi$}{f\textpi}}
\label{sec:CCFVgoldberger}

The Goldberger-Treiman (GT) relation predicts $g_{\pi N N} (1 + \Delta) = M_N
g_A/F_\pi $ as discussed in Sec.~\ref{sec:AFFconsistency}.
Three of these quantities, $ M_N$ (Table~\ref{tab:Ensembles}), $g_A$
(Table~\ref{tab:gAdP2z2}) and $F_\pi$ (Table~\ref{tab:gpiNN}), are
calculated in this work. Data for the product $M_N g_A/F_\pi $,
which is independent of the renormalization constant $Z_A$ and the lattice scale, are also given in 
Table~\ref{tab:gpiNN} for each ensemble.
The CCFV fits to these data for $M_N g_A/F_\pi $ and $F_\pi$ using the
ansatz given in Eq.~\eqref{eq:CCFV-rA} are shown in
Fig.~\ref{fig:gpiNN} in Appendix~\ref{sec:appendixCCFV}.  The result
for $M_N g_A/F_\pi $ depends, as expected, on the strategy used to
determine $g_A$, and for the two extreme values for $g_A$ obtained from $\{4,3^{\ast}\}$
and $\{4^{N\pi},2^{\rm sim}, z^2\}$ fits discussed in
Sec.~\ref{sec:CCFVcharges}, it is $ = 12.65(38)$ and $13.58(49)$,
respectively. We also show the CCFV fit for $F_\pi$ in the bottom row
of Fig.~\ref{fig:gpiNN} and find $F_\pi = 93.0(3.9)$ (96.1(3.6))~MeV
with ${\rm Z}_1$ (${\rm Z}_2$) renormalization. These CCFV fits to the
$F_\pi$ and $M_N g_A/F_\pi $ data show significant variation with $a$
and $M_\pi$.  Thus, to improve precision more $\{a, M_\pi, M_\pi L\}$
points are needed.

For comparison, using the experimental values, $g_A=1.2764$,
$M_N=939$~MeV and $F_\pi=92.2$~MeV and ignoring the Goldberger-Treiman
discrepancy $\Delta$ (see discussion in Sec.~\ref{sec:A4}) give $\gpNN =
M_N g_A/F_\pi = 13$.  The phenomenological estimate obtained from the
$\pi N$ scattering length analysis is $13.25(5)$~\cite{Perez:2016aol,Reinert:2020mcu,Baru:2011bw}.

\begin{table*}[t]   
\begin{ruledtabular}
\begin{tabular}{lc|ll|ll|ll}
 & & \multicolumn{2}{c|}{$\langle r_E^2 \rangle$} & \multicolumn{2}{c|}{$\langle r_M^2 \rangle$} & \multicolumn{2}{c}{$\mu$} \\ 
ESCfit & $Q^2$fit & CC & CCFV & CC & CCFV & CC & CCFV  \\ 
\hline
$\{4,3^*\}$                 & $D$   & 0.633(60)[0.26] & 0.658(75)[0.25] & 0.479(48)[1.24] & 0.579(67)[0.04] & 3.78(12)[1.36] & 3.95(15)[0.54] \\  
                            & $P_2$ & 0.589(74)[0.07] & 0.613(98)[0.04] & 0.491(93)[0.26] & 0.49(14)[0.34] & 3.81(14)[0.30] & 3.86(17)[0.30] \\   
                            & $z^3$ & 0.562(71)[0.05] & 0.577(98)[0.05] & 0.73(12)[0.50] & 0.71(17)[0.66] & 3.95(13)[0.28] & 4.03(16)[0.13] \\    
\hline                                                                                                                                            
$\{4^{N\pi},3^*\}$          & $D$   & 0.792(58)[0.36] & 0.843(77)[0.16] & 0.651(54)[5.28] & 0.879(77)[1.19] & 4.07(14)[3.17] & 4.43(17)[0.20] \\  
                            & $P_2$ & 0.792(81)[0.69] & 0.85(12)[0.76] & 0.64(13)[0.85] & 0.48(19)[0.70] & 4.03(18)[0.25] & 4.06(22)[0.32] \\     
                            & $z^3$ & 0.803(87)[0.43] & 0.84(12)[0.52] & 0.97(12)[0.56] & 0.94(16)[0.72] & 4.15(15)[0.54] & 4.24(18)[0.44] \\     
\hline                                                                                                                                            
$\{4,2^\text{sim}\}$        & $D$   & 0.621(64)[0.26] & 0.646(85)[0.28] & 0.457(52)[0.27] & 0.494(68)[0.12] & 3.64(15)[0.45] & 3.64(18)[0.59] \\
                            & $P_2$ & 0.65(11)[0.53] & 0.65(16)[0.71] & 0.52(13)[2.03] & 0.31(17)[1.47] & 3.80(20)[2.89] & 3.62(23)[2.68] \\
                            & $z^3$ & 0.80(10)[0.57] & 0.80(14)[0.76] & 0.73(11)[2.17] & 0.60(15)[2.37] & 4.00(17)[3.85] & 3.85(19)[4.15] \\
\hline
$\{4^{N\pi},2^\text{sim}\}$ & $D$   & 0.590(63)[0.52] & 0.623(81)[0.54] & 0.497(51)[1.04] & 0.564(67)[0.62] & 3.83(15)[0.89] & 3.86(18)[1.17] \\
                            & $P_2$ & 0.59(12)[0.97] & 0.49(18)[1.10] & 0.67(12)[1.39] & 0.49(18)[1.25] & 4.04(20)[2.51] & 3.86(22)[2.10] \\
                            & $z^3$ & 0.66(11)[0.79] & 0.55(16)[0.79] & 0.77(12)[1.41] & 0.60(17)[1.18] & 4.17(17)[2.28] & 4.05(19)[2.45] \\
\end{tabular}
\end{ruledtabular}
\caption{Results for $\langle r_E^2 \rangle$, $\langle r_M^2 \rangle$
  and $\mu$ from CC and CCFV fits to data from the four strategies,
  $\{4,3^*\}$, $\{4^{N\pi},3^\ast\}$, $\{4,2^\text{sim}\}$ and
  $\{4^{N\pi},2^\text{sim}\}$, used to control ESC (see
  Appendix~\ref{sec:glossary}). The $Q^2$ behavior of the data from
  each strategy is parameterized using the dipole ($D$), Pad\'e ($P_2$)
  and the $z^3$ fits.  The $\chi^2$/dof of the CC/CCFV fits are given
  within the square brackets.}
\label{tab:CCFV-EM-prior}
\end{table*}

\subsection{CCFV fits to the electric and magnetic radii, 
\texorpdfstring{$\langle r_E^2 \rangle$}{<rE\suptwo>} and \texorpdfstring{$\langle r_M^2 \rangle$}{<rM\suptwo>}, and the magnetic moment \texorpdfstring{$\mu$}{mu}}
\label{sec:CCFVEMradii}

The CCFV fits to each of these three quantities 
have four free parameters denoted by $c_i^{\{E,M,\mu\}}$. The fit ansatz for the electric
mean-square charge radius used is
\begin{equation}
  \expv{r_E^2}(a,M_\pi,L) = c_1^E + c_2^E a + c_3^E \ln \frac{M_\pi^2}{\lambda^2} + 
   c_4^E \ln \frac{M_\pi^2}{\lambda^2} e^{-M_\pi L} \,,
  \label{eq:rEsq-extrap}
\end{equation}
%
where the mass scale $\lambda$ is chosen to be $M_\rho=775\,\MeV$ and
the form of the chiral and FV corrections are taken from
Refs.~\cite{Bernard:1998gv,Kubis:2000zd,Gockeler:2003ay}.  For the
magnetic mean charge radius squared, we use
\begin{equation}
  \expv{r_M^2}(a,M_\pi,L) = c_1^M + c_2^M a + \frac{c_3^M}{M_\pi} + \frac{c_4^M}{M_\pi} e^{-M_\pi L} \,, 
  \label{eq:rMsq-extrap}
\end{equation}
where the leading dependence on $M_\pi$ is taken from Refs.~\cite{Bernard:1998gv,Kubis:2000zd}. 
Last, the  CCFV ansatz used for the magnetic moment is 
\begin{align}
  \mu(a,M_\pi,L) = &c_1^\mu + c_2^\mu a + c_3^\mu M_\pi + \nonumber \\
   &c_4^\mu M_\pi\left(1-\frac{2}{M_\pi L}\right) e^{-M_\pi L} \,. 
  \label{eq:mu-extrap}
\end{align}
where the forms of the chiral and finite-volume correction terms are
taken from Refs.~\cite{Kubis:2000zd,Beane:2004tw}.  All masses are 
expressed in units of GeV and the lattice spacing in fm.

In all three CCFV fit ansatz,
Eqs.~\eqref{eq:rEsq-extrap}--\eqref{eq:mu-extrap}, results from the
heavy baryon chiral perturbation theory ($\chi$PT) have been used only
to determine the form of the leading order chiral correction. For
example, for $\mu$, $\chi$PT predicts the slope, $c_3^\mu$, of the
linear dependence on $M_\pi$ as $M_N g_A^2/(4 \pi F_\pi^2)$~\cite{Beane:2003xv} with
$F_\pi=92.2$~MeV~\cite{PhysRevD.98.030001}; however, we leave $c_3^\mu$ a
free parameter.  Also, we include only the leading nonanalytical term
in Eqs.~\eqref{eq:rEsq-extrap} and~\eqref{eq:rMsq-extrap}.

Data for $\expv{r_E^2}$, $\expv{r_M^2}$ and $\mu$ from the four
strategies and the CCFV fits to them are shown in
Figs.~\ref{fig:CCFV-VFF-rE}, ~\ref{fig:CCFV-VFF-rM},
and~\ref{fig:CCFV-VFF-mu}.  The results are collected together in
Table~\ref{tab:CCFV-EM-prior}. We remind the reader that a prior for
$G_M(0)/g_V \equiv \mu$, obtained from the linear extrapolation of $G_E/G_M$, is
included in the $Q^2$ fits to $G_M$ to get $\langle r_M^2 \rangle$ and
$\mu$ on each ensemble. \looseness-1

In Sec.~\ref{sec:VFF}, we had presented evidence that the low-lying
multihadron $N \pi \pi$ state is relevant, and as $M_\pi \to 135$~MeV,
estimates from the $\{4^{N\pi},3^\ast\}$ and $\{4^{N\pi},2^{\rm
  sim}\}$ strategies should agree.  This is not manifest in
Table~\ref{tab:CCFV-EM-prior} for $\expv{r_E^2}$ or $\expv{r_M^2}$ and
estimates from $\{4^{N\pi},2^{\rm sim}\}$ are  smaller.
Furthermore, the data, and therefore the CCFV fits, have three
additional weaknesses:
\begin{itemize}
\item
The errors in $\expv{r_E^2}$ and $\expv{r_M^2}$ at $M_\pi \approx
170$~MeV and with the $z^3$ and Pad\'e fits are larger by a factor of
2--3 compared to $M_\pi \approx 270$~MeV points as can be seen
from the data in Table~\ref{tab:rErMmu} for all four strategies, and
from Figs.~\ref{fig:CCFV-VFF-rE} and~\ref{fig:CCFV-VFF-rM}. The CCFV
fits are therefore dominated by the smaller error $M_\pi \approx
270$~MeV points.
\item
To a lesser extent, the same is true for the data with the dipole fit and the 
$\{4^{N\pi},3^\ast\}$ strategy. 
\item
The dipole fits to the $a071m170$ data with the $\{4^{N\pi},2^{\rm
  sim}\}$ strategy shown in Fig.~\ref{fig:VFF-Q2E7} miss the low $Q^2$
points, and the results differ from those from the $z^3$ or the $P_2$ analyses.
\end{itemize}
In short, these CCFV fits are not yet robust. For our best estimate,
we take the average of the $z^3$ and $P_2$ fits to the
$\{4^{N\pi},3^\ast\}$ strategy data and the larger of the two analyses error. The
same is done for $\mu^{u-d} \equiv \mu^{p-n}$ even though errors in it
at the two values of $M_\pi$ are comparable and the CCFV fits are
reasonable. In both cases we use half the spread between the
$\{4^{N\pi},3^\ast\}$ and the $\{4^{N\pi},2^\text{sim}\}$ values as an
additional systematic uncertainty for possible residual ESC and $Q^2$
fit ansatz dependence.

With the above selections, our final results are 
\begin{eqnarray}
\expv{r_E^2}^{u-d}  &=& 0.85(12)(19)_{\rm sys} \ {\rm fm}^2 \ \Rightarrow r_E  =  0.92(12)\ {\rm fm}\,\,,  \nonumber \\
\expv{r_M^2}^{u-d}  &=& 0.71(19)(23)_{\rm sys} \ {\rm fm}^2 \ \Rightarrow r_M  =  0.84(18)\ {\rm fm}\,\,,  \nonumber \\
\mu^{u-d}           &=& 4.15(22)(10)_{\rm sys} \,.
\label{eq:VFF_final}
\end{eqnarray}
These radii are consistent with values obtained from the Kelly
parameterization~\cite{Kelly:2004hm} of the experimental data given in
Eq.~\eqref{eq:isovectorradiiKelly} (see our review in appendix D in
Ref.~\cite{Jang:2019jkn}), and the more precise value of the proton
charge radius $r_p = 0.831\pm 0.007_{\rm stat} \pm 0.012_{\rm sys}$
from the PRad experiment at Jefferson Lab~\cite{Xiong:2019umf} that
{claims to resolve} the ``proton radius puzzle'' by reconciling the values from
$e-p$ scattering with those from muonic hydrogen.  The errors in the
lattice results are, of course, much larger and do not provide independent input
on the ``proton radius puzzle''.  The $\mu^{p-n}$ is about $2\sigma$
smaller than the precisely measured value $\mu^{p-n}|_{\rm exp} =
4.7059$.

\begin{figure}[tb] 
\subfigure 
{
    \includegraphics[width=0.97\linewidth]{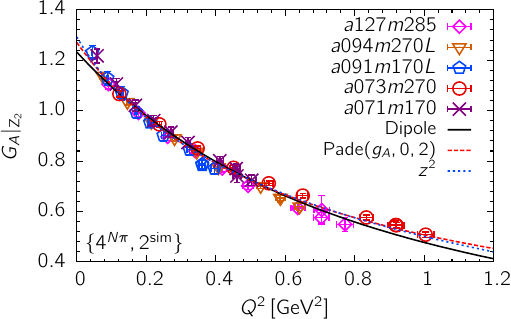}   
}
{
    \includegraphics[width=0.97\linewidth]{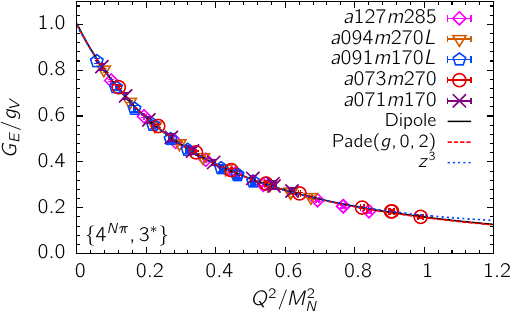}   
}
{
    \includegraphics[width=0.97\linewidth]{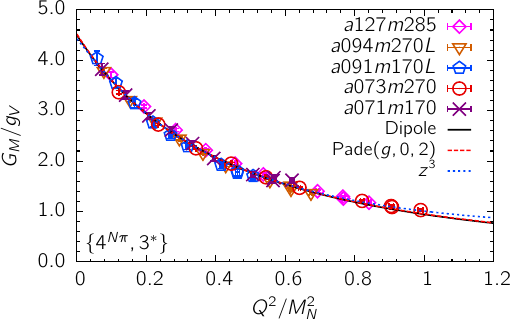} 
}
\caption{Comparison of the dipole, $P_2$ Pad\'e, and 
  $z$-expansion fits to the combined data from the five larger volume
  ensembles. We selected $\{4^{N\pi},2^{\rm sim}\}$ data for $G_A$ and
  $\{4^{N\pi},3^\ast\}$ for $G_E$ and $G_M$ as they show the least dependence on
  $a$ and $M_\pi$, which is neglected in these fits.  Result of the $P_2$ fit
  to $G_A$ is given in Eq.~\protect\eqref{eq:GAPade}, and to $G_E$ and
  $G_M$ in Eq.~\protect\eqref{eq:GEMPade}.}
\label{fig:FFfits}
\end{figure}
%
\section{Parameterizing the form factors \texorpdfstring{$G_A$}{GA}, \texorpdfstring{$G_E$}{GE} and \texorpdfstring{$G_M$}{GM} using Pad\'e and \texorpdfstring{$z$}{z}-expansion fits}
\label{sec:FFfits}
%

The Pad\'e and $z$-expansion fits to form factors presented in this section should be
considered a good heuristic, i.e., they serve our primary goal to provide
a good but simple parameterization of the lattice data. This is in the
same spirit as the phenomenologically useful Kelly parameterization~\cite{Kelly:2004hm} 
of $G_E$ and $G_M$ that are well
measured in electron scattering experiments, or the
rational function fit used in a recent analysis of the PRad experiment at
Jefferson Lab~\cite{Xiong:2019umf}. Note that the 
improvement in the precision with which the proton radius is extracted
and the likely resolution of the ``proton radius puzzle'' in Ref.~\cite{Xiong:2019umf} 
has come from increasing the range of $Q^2$ and the accuracy of the data and not
from the parameterization.

On the other hand, the axial form factors of the nucleon, 
$G_A$ and ${\widetilde G}_P$, that are important inputs in the
analysis of neutrino-nucleus scattering are not well measured due to
safety concerns with liquid hydrogen targets. Traditionally, $G_A$ has
been parameterized using the dipole ansatz, Eq.~\eqref{eq:dipole},
with estimates of the axial mass, $M_A$, ranging from 1 to 1.35~GeV,
and ${\widetilde G}_P$ obtained from $G_A$ using the PPD
hypothesis~\cite{Kronfeld:2019nfb}. Our analysis shows that a dipole
ansatz does not have enough free parameters to fit the data over the
range $0.04 < Q^2 < 1$~GeV${}^2$, nevertheless, we include it in this
section for comparison.  Furthermore, as discussed in
Secs.~\ref{sec:CCFVcharges},~\ref{sec:CCFVrA}
and~\ref{sec:CCFVEMradii}, while the data for the form factors have
small errors, the CCFV fits to charges and charge radii derived from
them are not yet robust, a consequence of having only seven ensembles
and the relatively larger errors in the $M_\pi \approx 170$~MeV data.
Thus, we did not present a $\{a \to 0, M_\pi=135{\rm MeV}, M_\pi
L \to \infty\}$ limit parameterization of the form factors in those
sections. On a positive note, the small dependence of $G_A$, $G_E$ and
$G_M$ on $\{a, M_\pi, M_\pi L\}$ observed in
Figs.~\ref{fig:GA4s5e},~\ref{fig:GEsummary} and~\ref{fig:GMsummary}
motivated the following heuristic analysis.

This simple parameterization assumes that the dependence on
$a$, $M_\pi$ and $M_\pi L$ can be neglected, with the intent to
subsequently include $a$ and $M_\pi$ dependent corrections as data get
better. (This assumption is the least-well motivated for $G_A$.)  To
reduce the impact of the neglected finite volume corrections, we do not
include data from the two small volume ensembles, $a094m270$ and
$a091m170$ with $M_\pi L \lesssim 4$, which show some evidence of
finite volume corrections. With the remaining data from five ensembles
(a total of fifty $Q^2 \ne 0$ points for $G_A$ and thirty for
$G_{E,M}$), we compare six parameterizations for each of the three
form factors: the dipole, two Pad\'e, $P(g,0,2)$ and $P(g,1,3)$, and
three $z$-expansion fits, $z^{2,3,4}$.  For $G_A$, we use the
preferred $\{4^{N\pi},2^{\rm sim}\}$ data with ${\rm Z}_2$
renormalization and remark that ${\rm Z}_1$ gives overlapping
results. For $G_E$ and $G_M$, we use the $\{4^{N\pi},3^\ast\}$ data.

The data and three of the six fits are compared in Fig.~\ref{fig:FFfits}. The 
results are summarized in Table~\ref{tab:FFfits}.  We observe the following:
\begin{itemize}
\item
The two $P(g,0,2)$ and $P(g,1,3)$ Pad\'e results are essentially
identical and stable for all three form factors. On the basis of the Akaike criteria, 
$P(g,1,3)$ is an overparameterization.
\item
The dipole fit to $G_A$ is poor and shows deviations near $Q^2=0$ and at large $Q^2$.
Similar, but smaller, deviations are seen for $G_M$. The dipole is a reasonable fit 
only for $G_E$.
\item
The $z^n$-expansion fits do not show convergence:
Table~\ref{tab:gASTfinal} shows variation between the $z^{2,3,4}$
estimates, and an increase in errors.  Furthermore, these estimates
now depend on the choice of $t_0$ [see Eq.~\eqref{eq:Zdef}] with the
overall midpoint value $t_0=0.5~\text{GeV}^2$ giving the smallest
$\chi^2$. As in Sec.~\ref{sec:CCFV}, our best choice based on the
Akaike criteria is again $z^2$ for $G_A$ and $z^3$ for $G_E$ and
$G_M$.
\end{itemize}

\begin{table*}  
\begin{ruledtabular}
\begin{tabular}{c |lll|lll|lll l}
fit             & $\rAsq~[\text{fm}^2]$ & $g_A$ & $\chi^2/dof$ & $\rEsq~[\text{fm}^2]$ & $g_V$ & $\chi^2/dof$ & $\rMsq~[\text{fm}^2]$ & $\mu$ & $\chi^2/dof$ \\ \hline
dipole           & 0.283(04) & 1.232(09) & 1.98 [95/48] & 0.799(08) & 1.003(04) & 0.46 [13/28] & 0.628(08) & 4.499(35) & 1.60 [45/28]\\ 
Pad\'e$(g,0,2)$  & 0.356(13) & 1.270(11) & 1.27 [60/47] & 0.778(19) & 0.999(05) & 0.42 [11/27] & 0.642(23) & 4.520(48) & 1.64 [44/27]\\ 
Pad\'e$(g,1,3)$  & 0.356(13) & 1.271(11) & 1.17 [53/45] & 0.778(19) & 0.999(05) & 0.45 [11/25] & 0.652(24) & 4.532(48) & 1.54 [38/25]\\ 
${\widehat z}^2$ & 0.426(15) & 1.292(11) & 1.15 [54/47] & 1.081(16) & 1.048(05) & 2.25 [61/27] & 0.919(24) & 4.750(49) & 2.42 [65/27]\\ 
${\widehat z}^3$ & 0.454(43) & 1.301(17) & 1.17 [54/46] & 0.743(54) & 0.996(9)  & 0.43 [11/26] & 0.48(10)  & 4.424(85) & 1.66 [43/26]\\ 
${\widehat z}^4$ & 0.67(11)  & 1.349(29) & 1.11 [50/45] & 0.66(18)  & 0.987(20) & 0.44 [11/25] & 1.04(33)  & 4.72(20)  & 1.61 [40/25]\\ 
\end{tabular}
\end{ruledtabular}
\caption{Results for the charge radii and charges obtained using the dipole, Pad\'e, and $z$-expansion fits
  to the renormalized form factors $G_A(Q^2)|_{{\rm Z}_2}$,
  $G_E(\frac{Q^2}{M_N^2})/g_V$ and $G_M(\frac{Q^2}{M_N^2})/g_V$.  The
  fits are made to the combined data from the five larger volume
  ensembles. The value $t_0=0.5~\text{GeV}^2$ (midpoint of the $Q^2$
  range) is used in the $z$-expansion fits for all data. The
  $\{4^{N\pi},2^\text{sim}\}$ data for $G_A$ and the
  $\{4^{N\pi},3^*\}$ data for $G_E$ and $G_M$ have been selected for
  this analysis as they exhibit the least dependence on $a$, $M_\pi$, 
  and $M_\pi L$ as shown in
  Figs.~\protect\ref{fig:GA4s5e},~\protect\ref{fig:GEsummary}
  and~\protect\ref{fig:GMsummary}.}
\label{tab:FFfits}
\end{table*}

Incorporating these observations and bearing in mind the caveats, our best
parameterizations of $G_A$, neglecting $\{a, M_\pi, M_\pi L\}$ dependent lattice artifacts, are (i)  the 
$\{4^{N\pi},2^{\rm sim},{\widehat P}_2\}$ fit:
\begin{align}
G_A(Q^2) &\equiv  \frac{g_A}{1 + b_0 \frac{Q^2}{4M_N^2}  + b_1(\frac{Q^2}{4M_N^2})^2 } \nonumber \\
         &=  \frac{1.270(11)}{1 + 5.36(20) \frac{Q^2}{4M_N^2} - 0.22(81) (\frac{Q^2}{4M_N^2})^2 } \,,
\label{eq:GAPade}
\end{align}
with $\chi^2$/dof=1.27 and $M_N=939$~MeV, and (ii) the $\{4^{N\pi},2^{\rm sim},{\widehat z}^2\}$ fit that gives
\begin{equation}
G_A(Q^2) =  0.725(5) - 1.63(3) z + 0.17(13) z^2 \,,
\label{eq:GAz2}
\end{equation}
with $\chi^2$/dof=1.15, and $z$ defined in Eq.~\eqref{eq:Zdef} with $t_0 =0.5$~GeV${}^2$. 
For our best results, we take the average of these  
$\{4^{N\pi},2^{\rm sim},{\widehat P}_2\}$ and $\{4^{N\pi},2^{\rm sim},{\widehat z}^2\}$ 
values given in Table~\ref{tab:FFfits} to get 
\begin{eqnarray}
g_A^{u-d}           &=& 1.281(11)(22)_{\rm sys} \,, \nonumber \\
\expv{r_A^2}^{u-d}  &=& 0.391(15)(70)_{\rm sys} \ {\rm fm}^2 \ \,,  
\label{eq:AFF_final2}
\end{eqnarray}
which are slightly
smaller than the values in Eqs.~\eqref{eq:finalcharges} and~\eqref{eq:rAfinal}.  The 
second, systematic, error is taken to be the difference between the two 
estimates averaged.

Similarly, the results of the $\{4^{N\pi},3^\ast,{\widehat P}_2\}$ 
and $\{4^{N\pi},3^\ast,{\widehat z}^3\}$ fits to  $G_E$ and $G_M$ are
\begin{align}
G_E(Q^2) &=  \frac{0.999(5)}{1 + 11.72(29) \frac{Q^2}{4M_N^2} + 38.5(1.9) (\frac{Q^2}{4M_N^2})^2 } \,, {\rm or} \nonumber \\
         &=  0.290(3) - 1.23(3) z + 1.72(19) z^2 \nonumber\\
         &\qquad\qquad\qquad\qquad\qquad\qquad{}+ 2.48(35) z^3  \,, \nonumber \\
G_M(Q^2) &=  \frac{4.52(5)}{1 + 9.68(35) \frac{Q^2}{4M_N^2} + 21.3(1.8) (\frac{Q^2}{4M_N^2})^2 } \,, {\rm or} \nonumber \\
         &=  1.613(11) - 5.74(14) z + 6.1(1.2) z^2\nonumber\\
         &\qquad\qquad\qquad\qquad\qquad\qquad{}+ 11.9(2.5) z^3  \,.
\label{eq:GEMPade}
\end{align}
Both sets of fits have very similar $\chi^2$/dof: $\approx 0.43$ and $\approx 1.65$ for $G_E$ and $G_M$, respectively. The variance-covariance matrices of the above six fits are given in Appendix~\ref{sec:covar}. 
The results are: 
\begin{equation}
\setlength{\tabcolsep}{6pt}
\begin{tabular}{l|cc}
                                 & $\{4^{N\pi},3^\ast,{\widehat P}_2\}$ & $\{4^{N\pi},3^\ast,{\widehat z}^3\}$   \\
\hline
$\expv{r_E^2}^{u-d}$ fm${}^2$  & $0.778(19)(50)_{\rm sys}$ & $0.743(54)(50)_{\rm sys}$      \\
$\expv{r_M^2}^{u-d}$ fm${}^2$  & $0.642(23)(80)_{\rm sys}$ & $0.48(10)(8)_{\rm sys}$      \\
$\mu^{u-d}$                    & $4.52(5)(10)_{\rm sys}  $ & $4.42(9)(10)_{\rm sys}  $  \,.    \\
\end{tabular}
\label{eq:VFF_final2}
\end{equation}
The second, systematic, error in both cases is taken to be half the spread between the 
$\{4^{N\pi},3^\ast,{\widehat P}_2\}$, $\{4^{N\pi},3^\ast,{\widehat z}^3\}$,
$\{4^{N\pi},2^{\rm sim},{\widehat P}_2\}$ and $\{4^{N\pi},2^{\rm sim},{\widehat z}^3\}$ estimates.

Next, we explored adding corrections due to $\{a,M_\pi\}$ in these
combined fits by expanding all parameters in them, for example
$b_0 \to (b_0^0 + b_0^a a + b_0^m {\cal O}(M_\pi^2))$, where ${\cal
O}(M_\pi^2))$ is $\log (M_\pi^2)$ for $\rEsq$ and $ 1/M_\pi$ for
$\rMsq$.  The result is that the $\chi^2$ is reduced only marginally
but the errors in the observables jump by a factor of 6 or more with
any (even one) additional parameter.  Also, in most cases the extra
parameter(s) are essentially undetermined indicating
overparameterization. Our conclusion again is that much higher
precision data on more ensembles are needed to include $\{a,M_\pi\}$
dependent corrections in this approach.

Another estimate of $\mu^{u-d}$ is obtained from a linear fit to
the $G_M/G_E$ data as shown in Fig.~\ref{fig:VFF-G_EoverG_M}.  The left
panel shows separate fits to the $a091m170L$ and $a071m170$ data with
the $\{4^{N\pi},3^\ast\}$ strategy.  The right panel shows the fit to
the combined data from these two ensembles. (Data from the other three
larger volume $M_\pi \sim 270$~MeV ensembles are included only for
comparison.)  The result from the fit to the two $M_\pi \approx
170$~MeV ensembles, $\mu^{u-d} = 4.67(12)$, is consistent with that
in Eq.~\eqref{eq:VFF_final2}.

This heuristic analysis has the advantage of evading the
two-step process used to get results given in Sec.~\ref{sec:CCFV}:
first a parameterization of the $Q^2$ behavior and then CCFV fits to
the observables with just leading order corrections in $\{a,M_\pi,
M_\pi L\}$.  The disadvantage is assuming that the $\{a,M_\pi, M_\pi
L\}$ corrections can be neglected, even though the data in
Figs.~\ref{fig:GA4s5e},~\ref{fig:GEsummary} and~\ref{fig:GMsummary}
suggest it. The remarkable outcome is that the estimates from the heuristic 
analysis are consistent with those given in
Eqs.~\eqref{eq:finalcharges},~\eqref{eq:rAfinal},
and~\eqref{eq:VFF_final} but with much smaller errors in all
cases. Also note that these fits give $G_E(Q^2=0) = 0.999(5)$ and
$G_M(Q^2=0) = 4.52(5)$, i.e., a necessary consistency check against the
precisely known values for the electric charge and the magnetic moment.

To understand why the dipole fit does not work for $G_A$ in this case
also, we note that the errors on points at small $Q^2$ grow as $Q^2
\to 0$ because the extrapolation in $\tau$ to remove ESC in the
$\{4^{N\pi},2^{\rm sim}\}$ fits is large on the $170$~MeV ensembles as
can be seen from Fig.~\ref{fig:aff4STcomp}. Similarly, the errors grow
as $Q^2$ increases because the statistical signal-to-noise degrades. Thus, the
dipole fit in Fig.~\ref{fig:FFfits} with $g_A$ and $M_A$ left as free
parameters is anchored by the smaller error points in the middle and
fails at both ends as it does not have enough degrees of freedom to
fully capture the curvature. The Pad\'e $\{g_A,0,2\}$, with one additional
degree of freedom, is sufficient.


\begin{table*}[p]  
\rotatebox{90}{%
\vbox{\scriptsize \setlength{\tabcolsep}{3.9pt}
  \renewcommand{\arraystretch}{1.2}
\hbox{\begin{tabular}{|c|c|c|c|c|c|c|c|c|c|c|c| }
\hline
Collab.      &Ref.&$g_A^{u-d}$ & $g_S^{u-d}$  & $g_T^{u-d}$  & $\rAsq^{u-d}$  &  $g_P^\ast$  & $g_{\pi NN}$ & $\rEsq^{u-d}$  & $\rMsq^{u-d}$  & $\mu^{p-n}$ & Action \\
             &    &            &              &              &  [fm${}^2$]    &              &              &  [fm${}^2$]    &  [fm${}^2$]    &             & \\
\hline
\hline
NME   & This   &$1.32(6)$&$1.06(9)$&$0.97(3)$& 0.428(53)  &7.9(7)  &  12.4(1.2)   &0.85(12) & 0.71(19)& 4.15(22) & $N_f=2+1$ \\
&Work A&$(4)_{\rm ES}(2)_Z (2)_a(2)_{\rm FV} $&$(2)_Z(6)_a$&$(1)_{\rm ES}(1)_Z (1)_{a}$&$(30)_{\rm ES}$&$(9)_{\chi}$  & &$(19)_{\rm ES}$&$(23)_{\rm ES}$&$(10)_{\rm ES}$& clover-on-clover\\
\hline
NME   & This   &$1.270(11)$&$$&$$& 0.356(13)   &  &   & 0.778(19) &0.642(23) & 4.52(5)& $N_f=2+1$ \\
&Work B&$(22)_{\rm sys}$&&&$(70)_{\rm sys}$&  &  &$(50)_{\rm sys}$   & $(80)_{\rm sys}$&$(10)_{\rm sys}$& clover-on-clover\\
\hline
 PNDME &\cite{Gupta:2018qil,Rajan:2017lxk,Jang:2019jkn,Jang:2019vkm} 
   &$1.218(25)(30)$&$1.022(80)(60)$ &$0.989(32)(10)$&  0.24(7) &  & & 0.591(62)  &  0.45(12)&  3.94(16) & $N_f=2+1+1$ \\
 &&& & & 0.24(3) &  &  & 0.586(21) & 0.495(50) & 3.98(15)& Clover-on-HISQ\\
\hline
ETMC  &\cite{Alexandrou:2020okk,Alexandrou:2019brg,Alexandrou:2018sjm} 
    &  $1.286(23)$  & $1.35(17)$  &  $0.936(25)(16)$ &0.343(42)  &  &   &  0.634(26)  & 0.51(13)& 3.97(16) &   $N_f=2+1+1$ \\
&&&&&&& & &&& Twisted Mass\\
\hline
RBC-  &\cite{Abramczyk:2019fnf}
      &  $1.15(5)$  & $0.9(3)$  &  $1.04(5)$ &     & &  &  & & &   $N_f=2+1$ \\
UKQCD &&&&& & &&&&& Domain Wall\\
\hline
CalLat  &\cite{Walker-Loud:2019cif,Chang:2018uxx} 
    &  $1.2642(93)$  &       &   &  &  &  &  & & &   $N_f=2+1+1$ \\
&&&&& & &&&&& DW-on-HISQ\\
\hline
PACS  &\cite{Ishikawa:2018rew,Shintani:2018ozy,Tsukamoto:2019dmi} 
   &$1.273(24)$&    &    & 0.419(28)    &     &    & 0.766(26) & 0.648(58) &  4.42(14)   &   $N_f=2+1$ \\
   &&$(9)_Z (5)_{\rm ES}$&  &  &$(49)_{\rm sys}$ &  &&$(49)_{\rm ES}$&$(441)_{\rm ES}$&$(32)_{\rm ES}$& clover-on-clover\\
\hline
JLQCD&\cite{Ishikawa:2018rew}&1.123(28)&0.88(8)&1.08(3) &   &  &&&&&  Overlap\\
      &&$(29)_\chi (90)_a$&$(3)_\chi (7)_a$&$(3)_\chi (9)_a$& &  &&&&& \\
\hline
RQCD  &\cite{Bali:2019yiy} 
  &$1.302(45)$&    &     & 0.449(88) &8.68(45) & 14.78(1.81)  &   & & & $N_f=2+1$ \\
($!z^{4+3}$)&   & $(42)_{\rm ES}(38)_m(46)_a$&    &     & $(42)_{\rm ES}(42)_m(49)_a$ & $(18)_{\rm ES}(23)_m(16)_a$ & $(72)_{\rm ES}(98)_m(67)_a$  &   & & &    \\
&&1.229(24)&&& 0.272(33) & 8.30(24)&  12.93(80) &  &  && clover-on-clover\\
(dipole)&&$(6)_{\rm ES}(3)_m(17)_a$&&& $(6)_{\rm ES}(7)_m(24)_a$  & $(14)_{\rm ES}(6)_m(8)_a$ & $(44)_{\rm ES}(20)_m(44)_a$ &  &  && \\
\hline
Mainz &\cite{Harris:2019bih,Djukanovic:2021cgp}&$1.242(25)_{\rm stat}$&$1.13(11)_{\rm stat}$&$0.965(38)_{\rm stat}$&   &  &  &0.800(25)   &0.661(30)  & 4.71(11) &  $N_f=2+1$ \\
      &          &$(+00,-31)_{\rm sys}$  &$(+07,-06)_{\rm sys}$  & $(+13,-41)_{\rm sys}$ &          &  &  &$(22)_{\rm sys}$   &$(11)_{\rm sys}$  &$(13)_{\rm sys}$  &  clover-on-clover \\
\hline
LHPC    &\cite{Hasan:2019noy,Hasan:2017wwt} 
  &$1.265(49)$  &  0.927(303)   & 0.972(41)   & 0.249(12)   &  &  & 0.608(15) &  & 3.899(38)& $N_f=2+1$ \\
&&&&& 0.295(68) &  &  & 0.787(87)&& 4.75(15)&  clover-on-clover \\
\hline
$\chi$QCD &\cite{Liang:2018pis} 
   &$1.254(16)(30)$&        &          &    &  &  &  & & & $N_f=2+1$ \\
&&&&&   &  &&&&&  Overlap-on-DW\\
\hline
\end{tabular}}
\hsize0.99\textheight
\caption{Comparison of lattice QCD results from simulations
  with $N_f=2+1+1$ and $2+1$ flavors that satisfy the criteria on
  chiral and continuum extrapolation defined in the text. Only the
  latest (best) results for each quantity from a given
  collaboration/calculation are quoted. We present our (NME) results
  from two sets of analyses: (A) the analysis presented in
  Sec.~\ref{sec:CCFV}, and (B) in Sec.~\ref{sec:FFfits}.  The PNDME
  estimates on the top are from a $z^{3,4}$ analysis and at the bottom
  from a dipole fit.  The two RQCD~\cite{Bali:2019yiy} results are
  from the $z$-expansion (top) and the dipole (below) fits to the
  $Q^2$ behavior of $G_A$. The notation for systematic errors is:
  FV=finite volume, Z=renormalization, ES=excited states,
  $a$=discretization, $\chi$=$m$=chiral, and sys=systematic (or when
  the second error has no label) }
\label{tab:LatCompare}
}%
}
\end{table*}

\section{Comparison with Previous Lattice QCD Calculations}
\label{sec:CompLQCD}

In this section, we compare with results from other recent lattice
calculations done with either 2+1+1 or 2+1 dynamical flavors. 
We assume that a dynamical charm in the lattice
generation does not significantly impact the quantities composed of
light quarks that are investigated here; i.e., the two formulations give
the same results. For a more extensive review of the calculation of
the charges, we direct the reader to the Flavor Lattice Averaging
Group (FLAG) Reviews 2019~\cite{Aoki:2019cca} and 2021~\cite{Aoki:2021kgd}.

It is important to note that all of these isovector quantities
from different calculations are expected to only agree at the physical
point, and thus a CCFV extrapolation is necessary.  We have therefore
applied the following criteria in selecting the calculations to
compare.  We require that (i) the results are either obtained at
$M_\pi \approx 135$~MeV or have been extrapolated to it, and similarly
(ii) they include ensembles with $a < 0.1$~fm or a continuum extrapolation
has been performed. (iii) We find that so far no other calculation has
carried out the extensive high statistics analysis of excited
states presented in this work, so we do not apply an excited-state
criterion for inclusion, but will comment on the method used to control
ESC and the outcome.  \looseness-1

The results compared are summarized in Table~\ref{tab:LatCompare}.
For each collaboration, we quote the latest (or best in the words of
the authors) value for each observable, which is often given in
different publications.  Overall, it is evident that a complete
control over systematic uncertainties, especially excited-state
effects, is still work under progress.

The PNDME results~\cite{Gupta:2018qil,Rajan:2017lxk,Jang:2019jkn} are
from a clover-on-HISQ formulation using eleven 2+1+1-flavor HISQ
ensembles, including two at the physical pion mass. All the quoted
results are from CCFV fits to data with the $\{4,3^\ast\}$ strategy,
i.e., they represent the status~\cite{Aoki:2019cca} before $N \pi$ (or $N \pi \pi$) states
were included in fits to remove the ESC.

The ETM
Collaboration~\cite{Alexandrou:2020okk,Alexandrou:2019brg,Alexandrou:2018sjm}
has presented results for most of the quantities analyzed in this
work. Their latest results are from one 2+1+1-flavor twisted mass
clover-improved ensemble with $a = 0.0801(4)$~fm, $M_\pi =
139(1)$~MeV, $M_\pi L= 3.62$, so issues of continuum extrapolation,
finite volume corrections and chiral behavior are not addressed.
(Our $a071m170$  ensemble provides data at similar values of $Q^2$.)
Their statistical sample is 750 lattices separated by 4 trajectories
each, and results for the isovector
charges~\cite{Alexandrou:2019brg} are taken from a
two-state fit, $\{2,2\}$ in our notation. Their axial form
factors~\cite{Alexandrou:2020okk} do not satisfy the PCAC relation,
and their estimates presented for ${\widetilde G}_P$ and $G_P$ are not
the calculated values but those obtained from $G_A$ using the
pion-pole dominance relation.  Consequently, we do not quote their
estimates for $g_P^\ast$ and $g_{\pi N N}$. Both the dipole and
$z$-expansion fits to $G_A(Q^2)$ obtained from $\{2,2\}$ strategy work
well and give $ \rAsq=0.343(42)(16)$~fm${}^2$, which is consistent with
our $\{4, 3^\ast\}$ value. The electric and magnetic form factors,
presented in Ref.~\cite{Alexandrou:2018sjm}, are well fit by a dipole
ansatz, however, they differ from the Kelly parameterization at small
$Q^2$, as also seen in the PNDME results in Ref.~\cite{Jang:2019jkn}.

The RBC-UKQCD Collaboration has analyzed two ensembles of 2+1-flavor
domain wall (DW) fermions with Iwasaki plus
dislocation-suppressing-determinant-ratio (DSDR) gauge action at $a =
1.378(7)$~fm and with $M_\pi = 249.4(3)$ and $172.3(3)$~MeV. They
report issues of long autocorrelations in a statistical sample of
only 700 trajectories, which may explain an underestimate of $g_A$
and a large uncertainty in $g_S$.

The CalLat Collaboration~\cite{Walker-Loud:2019cif,Chang:2018uxx}
report $g_A^{u-d}$ with percent level accuracy using the
domain-wall-on-HISQ formulation. In their calculation, the operator
is already summed over all insertion times $t$ during generation; therefore, 
they can only analyze their data versus $\tau$. They use
two-state fits to data starting at much smaller source-sink
separations, $0.2 \lesssim \tau \lesssim 0.8$~fm, where many higher
excited states contribute and sensitivity to contributions from $N\pi$ states would
be small. They do not explicitly include an $N\pi$ state in their
analysis.  Thus, the balance between control over statistical versus
systematic errors, especially the impact of the inclusion of the
$N\pi$ state(s), remains to be addressed.  The CCFV fits are made to
data from 16 ensembles at three values of $a \approx 0.09,\ 0.12,
0.15$~fm and five values of $M_\pi \approx 400, 350, 310, 220,
130$~MeV.

The PACS
Collaboration~\cite{Ishikawa:2018rew,Shintani:2018ozy,Tsukamoto:2019dmi}
uses a single $128^4$ ensemble generated with 2+1-flavor
stout-smeared $O(a)$ improved Wilson-clover fermions and Iwasaki gauge
action at $a = 0.0846(7)$~fm and $M_\pi = 135(9)$~MeV. While the
lattice volume is large, $M_\pi L = 7.4$, results have been
presented from only 20 lattices, each separated by 10 trajectories.  The
JLQCD~\cite{Yamanaka:2018uud} use 2+1-flavor overlap formulations with
a single value of $a=0.11$~fm, 4 values of $M_\pi =
293,379,453,540$~MeV and 50 gauge configurations. In both
calculations, even though some of their estimates are reasonable, the
control over the statistical and various systematic uncertainties we
have discussed is limited. For example, on the key issue of excited
states, in Ref.~\cite{Shintani:2018ozy} they find no significant
excited-state effects over the range $0.84 < \tau < 1.35$~fm, in
contradiction to all other calculations. Also, estimates from
$96^4$~\cite{Ishikawa:2018rew} and $128^4$~\cite{Shintani:2018ozy}
lattices with the same lattice spacing but with $M_\pi=146$ versus
$135$~MeV show much larger differences than expected, presumably due
to low statistics in both calculations. \looseness-1

The RQCD Collaboration~\cite{Bali:2019yiy} has presented results for
the axial form factors on 37 ensembles with 2+1 flavors of
nonperturbatively $O(a)$ improved Wilson-clover fermions with a tree-
level Symanzik improved gauge action generated by the CLS
Collaboration~\cite{Bruno:2014jqa}.  These ensembles cover five values of the
lattice spacing and include two physical pion mass ensembles. To
remove excited states they use a strategy similar to $\{3,2\}$ for the
axial charge and $G_A$, and to $\{4^{N\pi},3^\ast\}$ for ${\widetilde
  G}_P$ and $G_P$ form factors.  The resulting form factors 
satisfy the PCAC relation at a level similar to that presented in this
work.  They find that both the dipole and the $z$-expansion ansatz fit the
$Q^2$ behavior of $G_A(Q^2)$; however, results for $g_A^{u-d}$,
$\rAsq^{u-d}$, and $g_P^\ast$ are different as can be seen from the summary in
Table~\ref{tab:LatCompare}.

The Mainz Collaboration~\cite{Harris:2019bih} analyzed 11 CLS
ensembles~\cite{Bruno:2014jqa} that are common with the RQCD work
described above. On these ensembles, the pion mass ranges between $203$
and $353$~MeV. To control ESC they explore the summation method and
two-state fits with a common value for $\Delta M_1$ for six
quantities, the three charges and three Mellin moments that give the
momentum fraction, helicity, and transversity. Note that our data for
$\Delta M_1$ (or $\Delta {\widetilde M}_1$) for the three charges
given in Table~\ref{tab:deltaM} and for the three moments given in
Ref.~\cite{Mondal:2020ela} do not support using a common value for
$\Delta M_1$ in the analysis of all six quantities. Their final
results are also obtained with the CCFV ansatz given in
Eq.~\eqref{eq:CCFVcharges}, which they call ABDE.  For the vector form
factors~\cite{Djukanovic:2021cgp}, they analyze 10 CLS ensembles
including one with the physical pion mass; however, errors in data from
it are large.  The  ESC is controlled using the summation and two-state fit
methods ($\{2,2\}$ in our notation), which give consistent values.
The errors in the data with the summation method, especially at the
larger $Q^2$, are much larger. They employ dipole and $z$-expansion
parameterization of the $Q^2$ behavior, and the chiral-continuum
extrapolation using heavy baryon chiral perturbation theory (HBChPT)
supplemented with leading order corrections for lattice discretization
and finite volume. Their final estimates are obtained from a
model-agnostic average (summation, two-state fits, dipole,
$z$-expansion, HBChPT, and cuts on $Q^2$ and $M_\pi^2$ values) with
weights given by the Akaike information criteria.

The LHPC Collaboration~\cite{Hasan:2019noy,Hasan:2017wwt} analyzed two
physical pion mass 2+1-flavor ensembles generated with 2-HEX-smeared
Wilson-clover action (the Budapest-Marseille-Wuppertal ensembles).
One of their main observations from the study of
charges~\cite{Hasan:2019noy} is a significant variation in $Z_S$
between the RI${}^\prime$-MOM and RI-sMOM renormalization schemes
which, along with the statistical errors and extrapolation in $a$
uncertainty, accounts for the large error in $g_S$. In
Ref.~\cite{Hasan:2017wwt}, they present results for $\mu^{p-n}$ and
charge radii from two methods: traditional ($z$-expansion) and
derivative. In Table~\ref{tab:LatCompare}, we quote their results 
from the traditional method as
recommended by them, and from the two analyses for handling ESC:
$\tau/a=10$ ratio data (top) and summation (bottom), which
differ. Systematic uncertainties were not evaluated in either set of
estimates.

The $\chi$QCD Collaboration~\cite{Liang:2018pis} used the
overlap-on-domain-wall formulation on three 2+1-flavor domain-wall
ensembles generated by the RBC/UKQCD Collaboration. On each of these
ensembles, data with 5--6 values of the valence pion mass are
generated. They obtain $g_A^{u-d}$ using a CC fit to these partially
quenched data.

From the summary of results in Table~\ref{tab:LatCompare}, we conclude
that, overall, results for $g_T^{u-d}$ are consistent within
$5\%$, and for $g_S^{u-d}$ within $10\%$, and sensitivity to excited
states in their extraction is small. For all other quantities such as
$g_A^{u-d}$, the charge radii, $g_P^\ast$ and $g_{\pi NN}$, results
from analyses that do not include the $N\pi$ states give smaller
values compared to phenomenology.

\section{Conclusions}
\label{sec:conclusions}

We have presented an analysis of isovector charges and axial and
electromagnetic form factors on seven 2+1-flavor Wilson-clover
ensembles generated by the JLab/W\&M/LANL/MIT
Collaborations~\cite{JLAB:2016} and described in
Table~\ref{tab:Ensembles}. This unitary clover-on-clover calculation
is an improvement over our previous work using the nonunitary
clover-on-HISQ
formulation~\cite{Gupta:2018qil,Rajan:2017lxk,Jang:2019vkm,Jang:2019jkn}.
In addition, high-statistics data have allowed us to make significant
progress in understanding key issues in controlling other systematic
uncertainties including excited state contamination in various nucleon
matrix elements. 

The excited-state contributions to each observable
are analyzed using a number of possible values of the energy of the
first excited state, which is assumed to provide the dominant
contamination. The axial form factors extracted including the
low-lying multihadron $N \pi$ state satisfy the PCAC relation between
them and are consistent with the pion-pole dominance hypothesis.  We
also find evidence that the $N \pi \pi$ state, theoretically supported
by the vector-meson dominance hypothesis, contributes to the electric
and magnetic form factors. They show much less sensitivity to the
excited state mass gap, and the results agree with the experimental data 
parameterized using the Kelly result~\cite{Kelly:2004hm}.

Results of the pseudoscalar decay constant, $F_\pi$, after CCFV fits
to data with two methods for renormalization are $F_\pi|_{{\rm Z}_1}=
93.0(3.9)$ (this CCFV fit is shown in Fig.~\ref{fig:gpiNN}) and
$F_\pi|_{{\rm Z}_2}= 95.9(3.5)$. These estimates agree with the
experimental value to within a few percent. Noting that $F_\pi$ data points have
small statistical errors, the difference and the size of the errors after CCFV fits,
$\approx 4\%$, should be regarded as a measure of the overall accuracy
of the CCFV fits with seven data points, especially in observables
that show significant variations with respect to $\{a,M_\pi, M_\pi
L\}$.

The results for the three isovector charges obtained from the forward
matrix elements [see Eq.~\eqref{eq:finalcharges1}] are $g_{A}^{u-d} =
1.32(6)(5)_{\rm sys}$ (this estimate includes input from the extrapolation of the 
$Q^2 \neq 0$ data), $g_{S}^{u-d} = 1.06(9)(6)_{\rm sys}$, and
$g_{T}^{u-d} = 0.97(3)(2)_{\rm sys}$. The first overall
analysis error is conservative with respect to the variation
observed under CCFV extrapolations. Estimation of 
systematic uncertainties are discussed in
Sec.~\ref{sec:CCFVcharges}. The scalar and tensor charges
$g_{S,T}^{u-d}$ do not show a significant dependence on the value of the
first excited state mass, so we consider their estimate robust.

The value of $g_A^{u-d}$ has been extracted in two ways, one from the
forward matrix element and the second from an extrapolation of the
axial form factor to $Q^2=0$. These two ways must give the same result
in the continuum limit that should agree with the experimental
value. We find that $g_{A}^{u-d}$ is sensitive to the inclusion of the
$N\pi$ state.  Our results have a $\sim 10\%$ spread depending on the
ESC strategy and the $Q^2$ fits used as discussed in
Sec.~\ref{sec:CCFVcharges}.  A snapshot of the spread is given in
Table~\ref{tab:gASTfinal}. The change in $G_A(Q^2)$ on including the
$N\pi$ state is, in most cases, a few percent (see
Table~\ref{tab:GA-renormalized}): the largest change ($3$--$5$\%) is
in the smallest $Q^2$ point (${\bm n}^2=1$) on the $M_\pi \approx
170$~MeV ensembles, however, it is precisely the change in the low
$Q^2$ points that has the largest impact on the extraction of
$g_A^{u-d}$ from fits to $G_A$. Similarly, the change in the forward
matrix element is about $6\%$ (see Fig.~\ref{fig:diffcharges}).  These
changes are of the same size as our overall analysis error estimate, $\sim 5\%$, and
the additional systematic uncertainty included in the final result
$g_{A}^{u-d} = 1.32(06)(05)_{\rm sys}$.  Thus, this level of the possible
contribution of the $N\pi$ state in extracting $G_A$, and its impact
on the improvement of the PCAC relation, is just at the level of our
current resolution. Our conclusion, therefore, is as follows: to fully resolve the
issue of the size of the contributions of the $N\pi$ states in the
extraction of $G_A$ and to improve the precision of lattice estimates
of $g_A^{u-d}$ requires more extensive data.

To fit the $Q^2$ dependence of the form factors, we explore the
$z$-expansion, the dipole ansatz, and Pad\'e fits. Estimates from the
$z^{2,3,4}$ truncation of the $z$-expansion give consistent results
for the axial form factors and we take final values from the $z^2$
fits to avoid overparameterization.  For the vector form factors we
use the $z^3$ truncation.  The second order Pad\'e, $P(g,0,2)$, with
three free parameters, is found to provide an equally good
parameterization.  The dipole ansatz does not provide a good fit to
$G_A(Q^2)$ obtained including the $N \pi$ state when removing the
ESC. It provides a reasonable fit to the electric form factor, and
less so to the magnetic.

We have carried out two analyses to get charge radii from the form
factors, and both sets of  results are summarized in Table~\ref{tab:LatCompare}.
In the first, the $Q^2$ dependence of data from each ensemble is
parameterized using the dipole, Pad\'e and $z$-expansion, and the
lattice artifacts in the resulting values of the charges and the charge radii
due to discretization, finite volume effects, and heavier than
physical values of quark masses are then removed by  simultaneous
CCFV fits keeping leading order corrections in the three variables
$\{a, M_\pi, M_\pi L \}$.  The results are the following:
(i) the axial charge radius squared, $\rAsq=0.428(53)(30)_{\rm sys}$~fm${}^2$, (ii)
the induced pseudoscalar charge, $g_P^\ast=7.9(7)(9)_{\rm sys}$, (iii) the
pion-nucleon coupling $\gpNN = 12.4(1.2)$, (iv) the electric charge
radius squared, $\expv{r_E^2}^{u-d} = 0.85(12)(19)_{\rm sys} \ {\rm fm}^2$, (v)
the magnetic charge radius squared, $\expv{r_M^2}^{u-d} = 0.71(19)(23)_{\rm sys}
\ {\rm fm}^2$, and (vi) the magnetic moment $\mu^{u-d} =
4.15(22)(10)_{\rm sys}$. At this point, we do not consider deviations from 
phenomenological/experimental results significant. In the axial channel, to obtain this
improved consistency of results and for the form factors to satisfy
the PCAC relation between them, it was crucial to include the $N\pi$
state in the removal of ESC.\looseness-1

The electric and magnetic form factors $G_E$ and $G_M$, shown in
Figs.~\ref{fig:GEsummary} and~\ref{fig:GMsummary}, exhibit much less
sensitivity to the value of the mass gap of the excited state. Our
results agree with the Kelly parameterization of the experimental data
over the range $0.04 \lesssim Q^2 \lesssim 1.2$~GeV${}^2$ when plotted
as a function of $Q^2/M_N^2$, and show no significant variation with
respect to either $a$ or $M_\pi^2$. This agreement is a major improvement over
our previous work using the clover-on-HISQ formulation presented in
Ref.~\cite{Jang:2019jkn}.

A {second, heuristic,} analysis of form factors, presented in
Sec.~\ref{sec:FFfits}, explores the same set of parameterizations
(see Table~\ref{tab:FFfits}) but makes a single fit to data from all
five larger volume ensembles as shown in
Figs.~\ref{fig:GA4s5e},~\ref{fig:GEsummary} and~\ref{fig:GMsummary},
i.e., ignoring $\{a,M_\pi, M_\pi L\}$ dependent artifacts. The
$P(g,0,2)$ Pad\'e does a good job of parameterizing the $Q^2$ behavior
and the results are given in Eqs.~\eqref{eq:GAPade}
and~\eqref{eq:GEMPade}. 

The results for $g_A^{u-d}$, $\rAsq^{u-d}$, $\rEsq^{u-d}$, $\rMsq^{u-d}$, and $\mu^{u-d}$
from these two sets of analyses, summarized in
Table~\ref{tab:LatCompare}, are consistent but the errors
from the second set are smaller; a consequence of the analysis becoming simpler 
on ignoring $\{a,M_\pi, M_\pi L\}$ dependent artifacts. 

Our goal is to provide a parameterization of the form factors themselves
versus $Q^2$ for input into phenomenological analyses. 
In the analysis method ``A,'' we do not have
a robust theoretical guide or adequate data for performing CCFV extrapolations of each
of the coefficients of the $z$-expansion or Pad\'e fits (the $a_k$ in
Eq.~\eqref{eq:Zexpansion} or the $b_i$ in Eq.~\eqref{eq:defPade2})
determined from fits to individual ensembles. In method ``B,'' we 
make the assumption that the $\{a,M_\pi, M_\pi L\}$ dependent artifacts can be
ignored (see the data in Fig.~\ref{fig:FFfits}). Under this assumption, 
Eqs.~\eqref{eq:GAPade},~\eqref{eq:GAz2}, and~\eqref{eq:GEMPade} 
give our continuum limit parameterization of the form factors. \looseness-1

Overall, our results for the form factors are consistent with
phenomenological/experimental values.  For this agreement, it was
essential to include the low-energy $N\pi$ ($N\pi\pi$) excited state
in the analysis of the axial form factors, and to a smaller extent in
the vector channel.  Motivation for including these states
comes from $\chi$PT, pion-pole dominance for axial, and vector meson
dominance for vector channels. Our data support these hypotheses, and
the estimates of $\Delta M_1$ are in rough agreement with those
expected with $N\pi$ (or $N\pi\pi$ for vector) states (see
Figs.~\ref{fig:AFF-deltaM} and~\ref{fig:VFF-deltaM} for the axial and
vector cases, respectively).  The change in the axial form factors is
only a few percent; however, it is large, $\sim 35\%$, in both the
induced pseudoscalar, ${\widetilde G}_P$, and the pseudoscalar, $G_P$,
form factors. With these changes, the resulting form factors satisfy
the PCAC relation between them. Furthermore, the estimates
of the induced pseudoscalar charge, $g_P^\ast = 7.9(7)(9)$, and of the
pion-nucleon coupling $\gpNN = 12.4(1.2)$ become consistent with
phenomenology.\looseness-1

The change in the electric and magnetic form factors between the four ESC 
 strategies is small as shown in Figs.~\ref{fig:GEsummary}
 and~\ref{fig:GMsummary}. A significant reduction in the 
 dependence on $\{a,M_\pi\}$ of both form factors is 
 observed when plotted versus $Q^2/M_N^2$.  This provided motivation
 for the Pad\'e and $z$-expansion parameterization presented in Sec.~\ref{sec:FFfits} and the  results in 
 Eqs.~\eqref{eq:GAPade}--\eqref{eq:VFF_final2}.

To increase precision, address the issue of the spread in results due
to different estimates of the relevant mass gap, and to resolve whether
additional $N\pi$ state[s] should be included in the analysis, higher
statistics data at more values of $\{a,M_\pi, M_\pi L\}$ are
needed. The benchmarks for improvement will continue to be satisfying
the PCAC relation between the axial form factors, the agreement with
the experimental value $g_A^{u-d}=1.2764(1)$, and the well
measured vector form factors $G_E$ and $G_M$.

\begin{acknowledgments}

We thank V. Cirigliano and E. Mereghetti for many helpful discussions, and E. Ruiz Arriola for 
informing us of the current results for the pion-nucleon coupling. 
The calculations used the Chroma software
suite~\cite{Edwards:2004sx}, the multigrid invertor for generating quark 
propagators~\cite{Brannick:2007ue,Babich:2010qb,Osborn:2010mb}, the 
QUDA library~\cite{Clark:2009wm,Babich:2010mu,Babich:2011np},  
and the mutigrid with QUDA~\cite{Clark:2016rdz}. This research used resources at (i) the
National Energy Research Scientific Computing Center, a DOE Office of
Science User Facility supported by the Office of Science of the
U.S. Department of Energy under Contract No. DE-AC02-05CH11231; (ii)
the Oak Ridge Leadership Computing Facility, which is a DOE Office of
Science User Facility supported under Contract DE-AC05-00OR22725, and
was awarded through the INCITE Program Projects No. PHY138 and No. HEP133, and ALCC Program Project No. LGT107; 
(iii) the USQCD Collaboration, which is funded by the Office of Science of
the U.S. Department of Energy, and (iv) Institutional Computing at Los
Alamos National Laboratory.  T. Bhattacharya and R. Gupta were partly
supported by the U.S. Department of Energy, Office of Science, Office
of High Energy Physics under Contract No.~DE-AC52-06NA25396.  B. Jo\'o
is supported by the U.S. Department of Energy, Office of Science,
under Contract DE-AC05-06OR22725. F. Winter is supported by the
U.S. Department of Energy, Office of Science, Office of Nuclear
Physics under Contract DE-AC05-06OR23177.  We acknowledge support from
the U.S. Department of Energy, Office of Science, Office of Advanced
Scientific Computing Research and Office of Nuclear Physics,
Scientific Discovery through Advanced Computing (SciDAC) program, and
of the U.S. Department of Energy Exascale Computing Project.
T. Bhattacharya, R. Gupta, S. Mondal, S. Park and B.Yoon were partly
supported by the LANL LDRD program, and S. Park by the Center for
Nonlinear Studies.
\end{acknowledgments}

\appendix

\section{Glossary of labels used to describe the various fits made}
\label{sec:glossary}

A summary of the abbreviations used to describe the various analysis strategies and fits 
is given below in order of the three entries such as in $\{4^{N\pi},3^\ast,{\widehat P}_2\}$. 

The first entry specifies the fits to the two-point function used to extract the 
spectrum. It has two possibilities:
\begin{itemize}
\item
$\{4\}$ denotes that four-state fits are made to the two-point
function. Empirical Bayesian priors with
wide widths for excited state energies and amplitudes are used 
only to stabilize the fits. These fits are
illustrated in the left-hand column in Fig.~\ref{fig:2ptCOMP}.
\item
$\{4^{N\pi}\}$ denotes that a prior for $\Delta E_1$ with a narrow
width centered about the energy of the noninteracting $N \pi$ (or of 
$N \pi \pi$ that is essentially degenerate) state is used in four-state fits
to the two-point function. Priors on higher states are similar to
those in $\{4\}$ fits. These fits are illustrated in the right-hand
column in Fig.~\ref{fig:2ptCOMP}.
\end{itemize}
The second entry specifies the four different fits made to the three-point functions:
\begin{itemize}
\item
$\{3^\ast\}$ specifies that three-state fits are made to the
three-point functions with the spectrum taken from either $\{4\}$ or
$\{4^{N\pi}\}$ fits to two-point functions, and the $\langle 2 | O | 2 \rangle$ term in
Eq.~\eqref{eq:3pt} is set to zero.
\item
$\{2^{A_4}\}$: This is a two-state fit to the three spatial axial
vector and the pseudoscalar three-point functions with a common
$\Delta E_1$ determined from fits to the $A_4$ correlator. The ground
state parameters are taken from either $\{4\}$ or $\{4^{N\pi}\}$ fits.
\item
$\{2^{\rm sim}\}$: This is a two-state fit to a set of three-point
functions. In the axial channel it denotes that a simultaneous fit to
the four axial vector and the pseudoscalar channels is made with a common
$\Delta E_1$. In the vector channel it denotes a simultaneous fit to
the three distinct correlation functions described in
Sec.~\ref{sec:VFF}.  The ground state parameters are taken from either
$\{4\}$ or $\{4^{N\pi}\}$ fits.  In both cases, the output mass gap is
called $\Delta {\widetilde E}_1$.
\item
$\{2^{\rm free}\}$: This is a two-state fit to an individual
three-point function with $\Delta E_1$ left as a free parameter. The
output mass gap is called $\Delta {\widetilde E}_1$.  The ground state
parameters are taken from either $\{4\}$ or $\{4^{N\pi}\}$ fits.
\end{itemize}
The third entry specifies the six  different fits made to the form factors to 
parameterize the $Q^2$ behavior:
\begin{itemize}
\item
$\{D\}$ and $\{\widehat D\}$: ``D'' stands for a dipole fit.  The hat
in $\{\widehat D\}$ specifies that subsequent CCFV fits to quantities
such as $g_A$, $\rEsq$, $\rMsq$ and $\mu$ have been carried out neglecting
the two small volume 
{points, $a094m270$ and $a091m170$, and the finite-volume correction term have 
been neglected, i.e., only a CC fit is performed. }
\item
$\{z^k\}$ and $\{{\widehat z}^k\}$: These are z-expansion fits 
truncated at power $k$. The hat in the
label $\{\widehat z\}$ again specifies that subsequent CC fits have
been done neglecting the two small volume 
{points, $a094m270$ and $a091m170$, and the finite-volume correction term have 
been neglected, i.e., only a CC fit is performed. }
\item
$\{P_n\}$ and $\{{\widehat P}_n\}$: ``P'' stands for a Pad\'e fit.
The subscript $n$ specifies the order of the Pad\'e as discussed in
Sec.~\ref{sec:AFF}.  The hat {in $\{\widehat P\}$} again
specifies that the two small volume 
{points, $a094m270$ and $a091m170$, and the finite-volume correction term have
been neglected, i.e.,  only a CC fit is performed. }
\end{itemize}

\section{Lattice parameters and the values of \texorpdfstring{$Q^2$}{Q\suptwo} from the two four-state fits, \texorpdfstring{$\{4\}$}{4} and \texorpdfstring{$\{4^{N\pi}\}$}{4Npi}}
\label{sec:compQ2}

In this appendix, we give the parameters of the seven ensembles in 
Table~\ref{tab:Ensembles} and the corresponding parameters used in the calculation
of the clover propagators in Table~\ref{tab:cloverparams}. The values
of momentum transfer squared, $Q^2$, obtained from the two four-state
fits, $\{4\}$ and $\{4^{N\pi}\}$, to the two-point correlation
function are given in Table~\ref{tab:Q2}.

\leavevmode

\onecolumngrid

\begin{table*}[h]    
\setlength{\tabcolsep}{3pt}
\begin{ruledtabular}
\begin{tabular}{l|c|c|c|cc|cc|c||c|c|c|c}
ID          & $\beta$ & $a$        &$M_\pi$ &$M_N^{\{4\}}$ &$M_N^{\{4^{N\pi}\}}$ & \multicolumn{2}{c|}{Size} & $M_\pi L$ & Lattices     &  $N_{\rm HP}$ & $N_{\rm LP}$   & $\tau$ \\ 
            &         & [fm]       &  [MeV] &  [MeV] &  [MeV] &  $L/a$    & $T/a$    &           &              &               &                &        \\ 
\hline
$a127m285$ & 6.1 & 0.127(2) & 285(5) & 961(15) & 958(15) & 32 & 96 & 5.87 & 2,002 & 8,008 & 256,256      & \{8,{{10,12,14}}\} \\      
$a094m270$ & 6.3 & 0.094(1) & 269(3) & 982(15) & 986(11) & 32 & 64 & 4.09 & 2,469 & 7,407 & 237,024      & \{8,10,{{12,14,16}}\} \\   
$a094m270L$ & 6.3 & 0.094(1) & 269(3) & 979(11) & 976(11) & 48 & 128 & 6.15 & 4,510 & 18,040 & 577,280   & \{8,10,12,{{14,16,18}}\} \\
$a091m170$ & 6.3 & 0.091(1) & 169(2) & 903(11) & 895(12) & 48 & 96 & 3.75 & 4,012 & 16,048 & 513,536     & \{8,10,{{12,14,16}}\} \\   
$a091m170L$ & 6.3 & 0.091(1) & 170(2) & 901(11) & 884(13) & 64 & 128 & 5.03 & 2,002 & 10,010 & 320,320   & \{8,10,{{12,14,16}}\} \\   
$a073m270$ & 6.5 & 0.0728(8) & 272(3) & 1008(11) & 1007(11) & 48 & 128 & 4.81 & 4,720 & 18,880 & 604,160 & \{11,13,{{15,17,19}}\} \\  
$a071m170$ & 6.5 & 0.0707(8) & 166(2) & 911(13) & 901(12) & 72 & 192 & 4.28 & 2,500 & 15,000 & 240,000   & \{13,{{15,17,19}},21\} \\  
\end{tabular}
\end{ruledtabular}
\caption{\label{tab:Ensembles}Parameters of the seven isotropic clover
  ensembles being generated by the JLab/W\&M/LANL/MIT Collaboration
  using the highly tuned CHROMA code.  Each row gives the ensemble ID
  and parameters, the number of lattices analyzed, the number of high
  precision, $N_{\rm HP}$, and low precision, $N_{\rm LP}$,
  measurements of isovector quantities made and the values of
  source-sink separation $\tau$ simulated. Each lattice is separated
  by 4--6 trajectories with $\approx 92\%$ acceptance rate in the {Hybrid Monte Carlo (HMC)}
  algorithm. The nucleon mass, $M_N$, is given for the two fit
  strategies $\{4\}$ and $\{4^{N\pi}\}$ defined in the text. The
  lattice spacing $a$ is determined from the Wilson flow parameter
  $w_0$ using the method proposed in
  Ref.~\protect\cite{Borsanyi:2012zs}.  }
\end{table*}

\vspace*{-0.3cm}
\begin{table*}[h]      
\centering
\begin{ruledtabular}
\begin{tabular}{l|ccc|c|c}
           ID       &  $m_l$     &  $m_s$     & $c_{\text{SW}}$ & Smearing parameters  & RMS smearing radius       \\
\hline                                        
$a127m285$          & $-0.2850$  & $-0.2450$  & 1.24931 & \{5, 50\}    & 5.79(1)  \\
$a094m270$          & $-0.2390$  & $-0.2050$  & 1.20537 & \{7, 91\}    & 7.72(3)  \\
$a094m270L$         & $-0.2390$  & $-0.2050$  & 1.20537 & \{7, 91\}    & 7.76(4)  \\
$a091m170$          & $-0.2416$  & $-0.2050$  & 1.20537 & \{7, 91\}    & 7.64(3)  \\
$a091m170L$         & $-0.2416$  & $-0.2050$  & 1.20537 & \{7, 91\}    & 7.76(4)  \\
$a073m270$          & $-0.2070$  & $-0.1750$  & 1.17008 & \{9, 150\}   & 9.84(1)  \\
$a071m170$          & $-0.2091$  & $-0.1778$  & 1.17008 & \{10, 185\}  & 10.71(2)  \\
\end{tabular}
\end{ruledtabular}
\caption{The parameters used in the calculation of the clover
  propagators.  The hopping parameter for the light/strange quarks,
  $\kappa_{l,s}$, in the clover action is given by $2\kappa_{l,s} =
  1/(m_{l,s}+4)$.  $c_{\rm SW}$ is the Sheikholeslami-Wohlert improvement coefficient in 
the clover action.  The parameters used to construct Gaussian smeared
  sources~\protect\cite{Gusken:1989ad}, $\{\sigma, N_{\text{KG}}\}$,
  are given in the fifth column where $N_{\text{KG}}$ is the number
  of applications of the Klein-Gordon operator and the width of the
  smearing is controlled by the coefficient $\sigma$, both in Chroma
  convention~\cite{Edwards:2004sx}.  The resulting root-mean-square
  radius of the smearing in lattice units, defined as $\sqrt{\int dr
    \, r^4 S^\dag S /\int dr \, r^2 S^\dag S} $ with $S(r)$ the value
  of the smeared source at radial distance $r$, is given in the last
  column.\looseness-1  }
\label{tab:cloverparams}
\end{table*}


\cleardoublepage
\begin{table*}[h]  
  \begin{ruledtabular}
    \begin{tabular}{c |lllllll}
\multicolumn{8}{c}{$Q^2$ values with  strategy $\{4^{}\}$ } \\
\hline
$\bm {n}$ & $a127m285$ & $a094m270$ & $a094m270L$ & $a091m170$ & $a091m170L$ & $a073m270$ & $a071m170$\\ \hline
$(1,0,0)$ & 0.091(03) & 0.164(04) & 0.074(02) & 0.078(02) & 0.045(01) & 0.122(03) & 0.058(01)\\
$(1,1,0)$ & 0.178(06) & 0.314(07) & 0.146(03) & 0.154(03) & 0.088(02) & 0.238(05) & 0.114(03)\\
$(1,1,1)$ & 0.262(08) & 0.453(11) & 0.215(05) & 0.226(05) & 0.131(03) & 0.348(08) & 0.169(04)\\
$(2,0,0)$ & 0.341(11) & 0.598(15) & 0.281(06) & 0.294(07) & 0.172(04) & 0.451(10) & 0.222(05)\\
$(2,1,0)$ & 0.419(13) & 0.716(18) & 0.346(07) & 0.361(08) & 0.213(05) & 0.553(12) & 0.272(07)\\
$(2,1,1)$ & 0.495(16) & 0.839(21) & 0.409(09) & 0.426(10) & 0.252(06) & 0.652(15) & 0.322(08)\\
$(2,2,0)$ & 0.638(21) & 1.046(28) & 0.530(12) & 0.549(13) & 0.328(07) & 0.838(20) & 0.413(10)\\
$(2,2,1)$ & 0.705(23) & 1.172(30) & 0.588(13) & 0.609(15) & 0.365(08) & 0.927(22) & 0.461(12)\\
$(3,0,0)$ & 0.706(23) & 1.186(32) & 0.586(13) & 0.611(16) & 0.365(08) & 0.923(22) & 0.465(13)\\
$(3,1,0)$ & 0.774(25) & 1.293(34) & 0.642(14) & 0.672(17) & 0.401(09) & 1.010(24) & 0.506(13)\\
\hline
\multicolumn{8}{c}{$Q^2$ values with  strategy $\{4^{N\pi}\}$ } \\
\hline
$\bm {n}$ & $a127m285$ & $a094m270$ & $a094m270L$ & $a091m170$ & $a091m170L$ & $a073m270$ & $a071m170$\\ \hline
$(1,0,0)$ & 0.091(03) & 0.165(04) & 0.074(02) & 0.078(02) & 0.045(01) & 0.122(03) & 0.058(01)\\
$(1,1,0)$ & 0.178(06) & 0.315(07) & 0.146(03) & 0.154(03) & 0.088(02) & 0.238(05) & 0.114(03)\\
$(1,1,1)$ & 0.261(08) & 0.456(10) & 0.215(05) & 0.225(05) & 0.130(03) & 0.348(08) & 0.168(04)\\
$(2,0,0)$ & 0.341(11) & 0.593(13) & 0.281(06) & 0.293(07) & 0.171(04) & 0.451(10) & 0.221(05)\\
$(2,1,0)$ & 0.418(13) & 0.715(16) & 0.345(07) & 0.359(08) & 0.211(05) & 0.552(12) & 0.271(06)\\
$(2,1,1)$ & 0.493(16) & 0.837(19) & 0.408(09) & 0.424(10) & 0.249(06) & 0.650(15) & 0.320(08)\\
$(2,2,0)$ & 0.636(20) & 1.042(25) & 0.529(11) & 0.546(13) & 0.325(08) & 0.833(19) & 0.412(10)\\
$(2,2,1)$ & 0.704(22) & 1.165(28) & 0.587(13) & 0.606(14) & 0.361(09) & 0.921(21) & 0.459(11)\\
$(3,0,0)$ & 0.705(23) & 1.173(32) & 0.586(13) & 0.605(15) & 0.361(09) & 0.918(21) & 0.461(11)\\
$(3,1,0)$ & 0.772(25) & 1.280(33) & 0.641(14) & 0.665(16) & 0.397(10) & 1.004(23) & 0.503(12)\\
\end{tabular} \vspace{5mm}
\end{ruledtabular}
 \caption{Data for the momentum transfer squared, $Q^2={\bm q}^2-(E-M_N)^2$, in units of
   $\text{GeV}^2$, for the two strategies $\{4\}$ (top) and $\{4^{N\pi}\}$
   (bottom) used in the analysis of the form factors. }
\label{tab:Q2}
\end{table*}

\twocolumngrid
\section{Comparison of charges extracted using 4 strategies}
\label{sec:compgAST}

\begin{figure*}[p] 
\subfigure
{
    \includegraphics[width=0.24\linewidth]{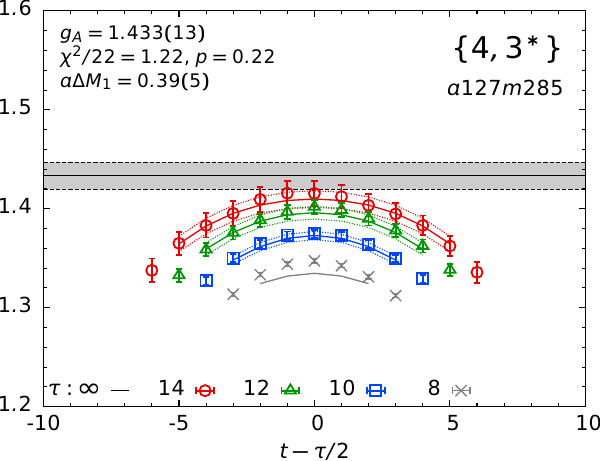}    
    \includegraphics[width=0.24\linewidth]{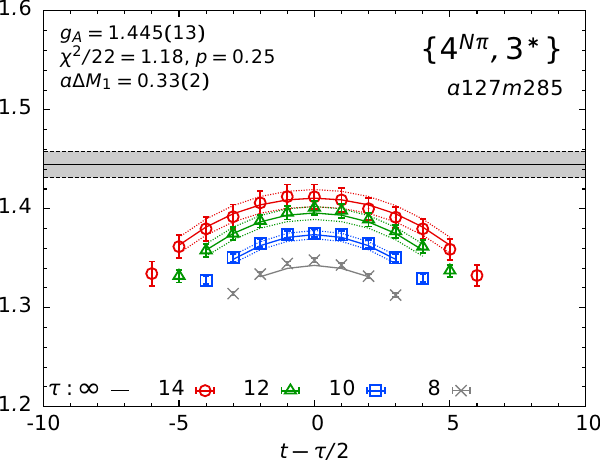}  
    \includegraphics[width=0.24\linewidth]{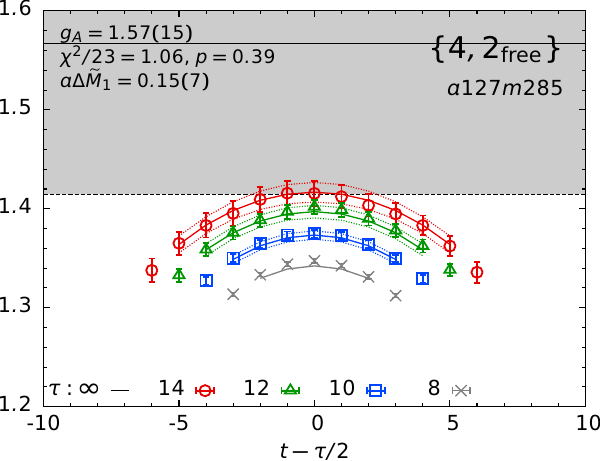}    
    \includegraphics[width=0.24\linewidth]{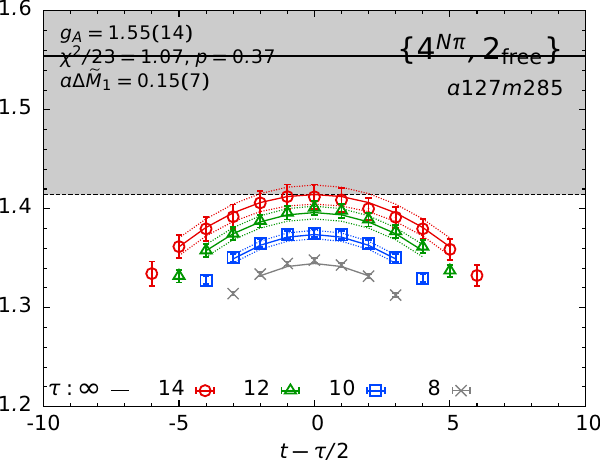} 
}
{
    \includegraphics[width=0.24\linewidth]{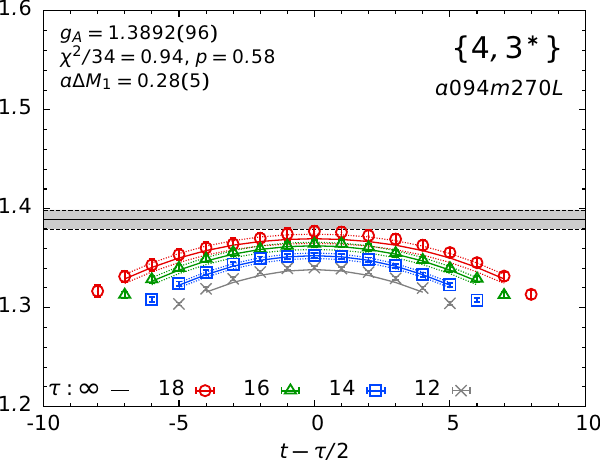}    
    \includegraphics[width=0.24\linewidth]{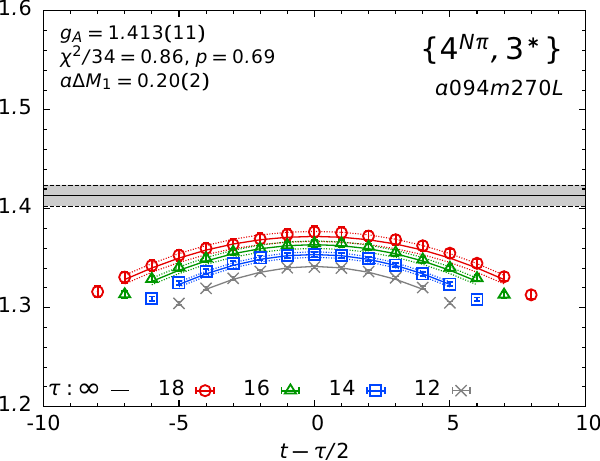}  
    \includegraphics[width=0.24\linewidth]{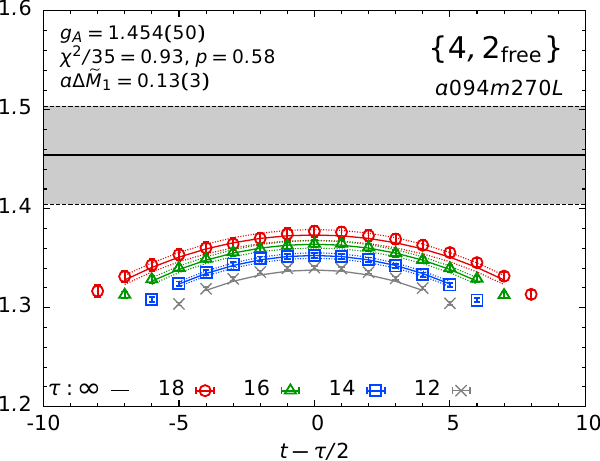}    
    \includegraphics[width=0.24\linewidth]{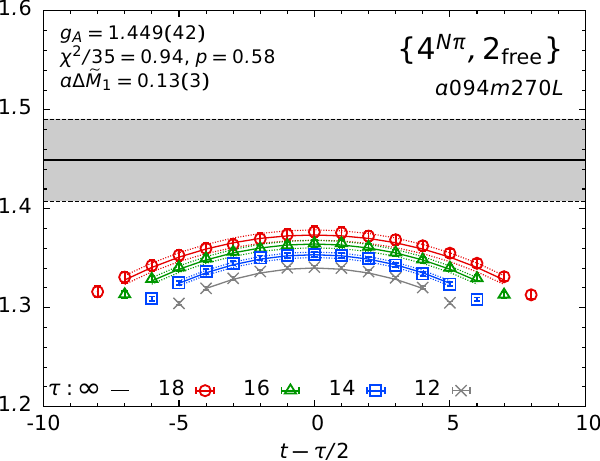} 
}
{
    \includegraphics[width=0.24\linewidth]{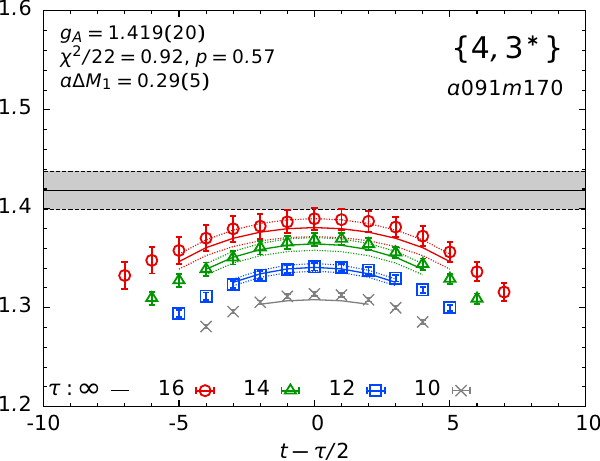}    
    \includegraphics[width=0.24\linewidth]{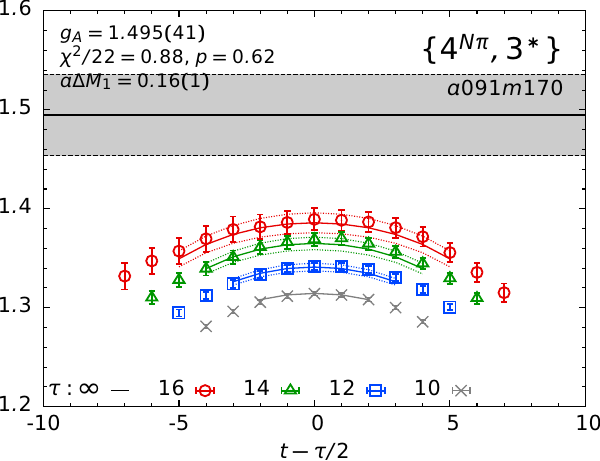}  
    \includegraphics[width=0.24\linewidth]{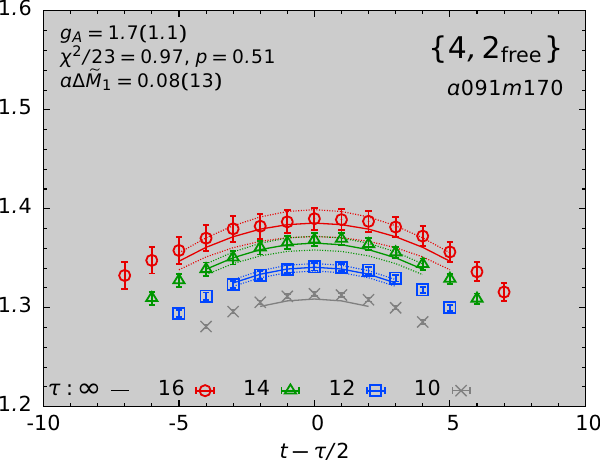}    
    \includegraphics[width=0.24\linewidth]{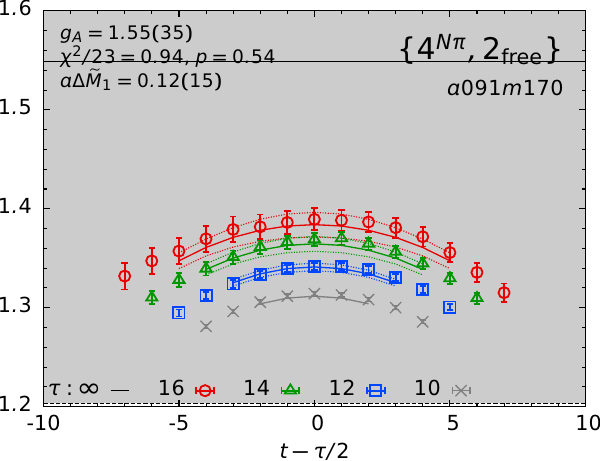} 
}
{
    \includegraphics[width=0.24\linewidth]{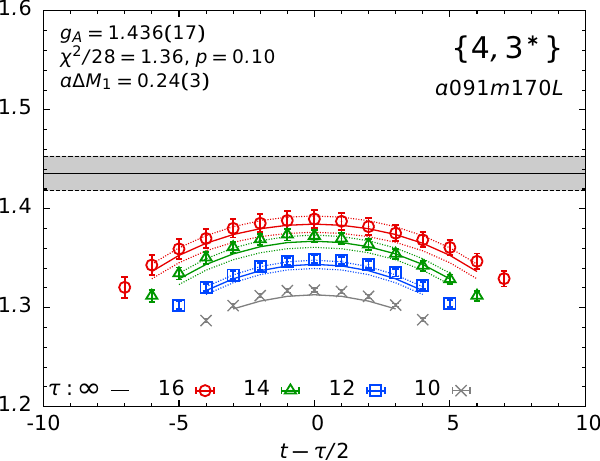}    
    \includegraphics[width=0.24\linewidth]{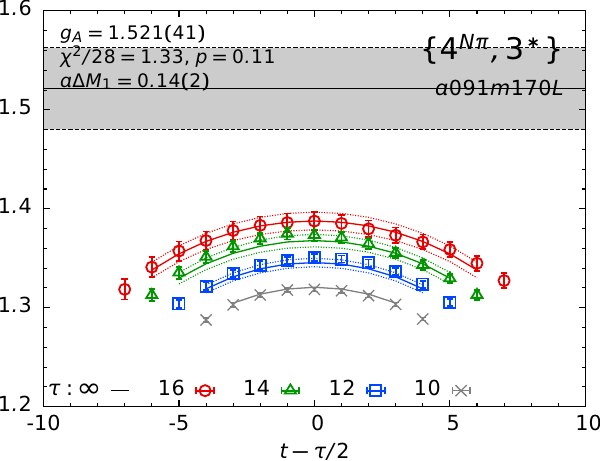}  
    \includegraphics[width=0.24\linewidth]{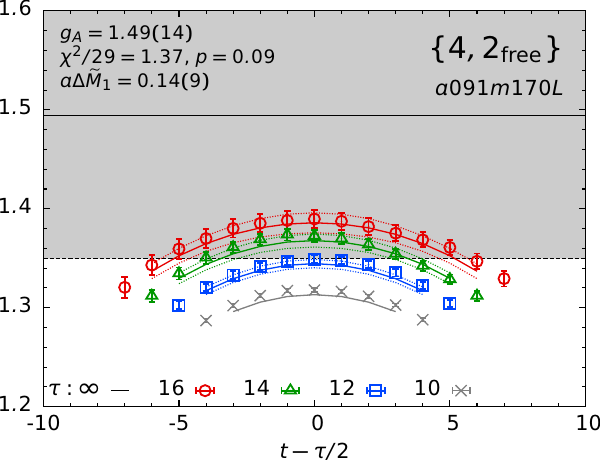}    
    \includegraphics[width=0.24\linewidth]{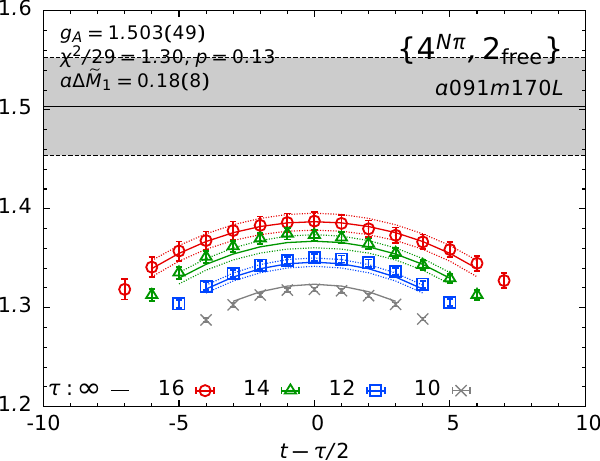} 
}
{
    \includegraphics[width=0.24\linewidth]{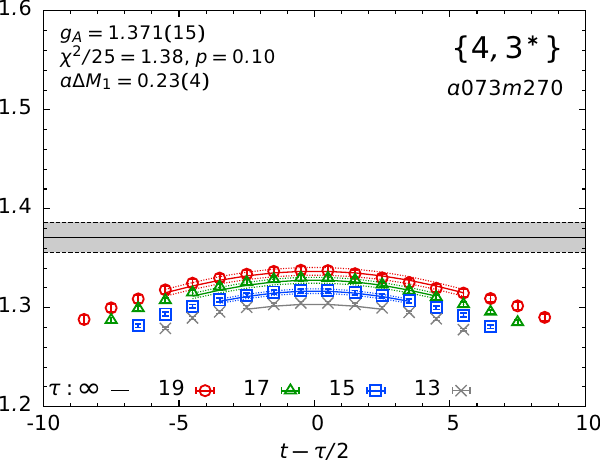}    
    \includegraphics[width=0.24\linewidth]{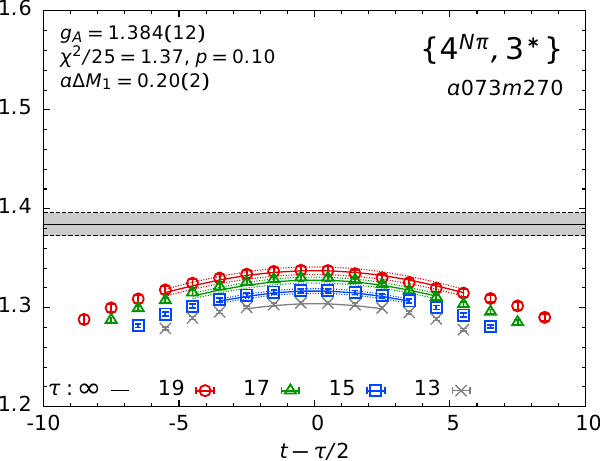}  
    \includegraphics[width=0.24\linewidth]{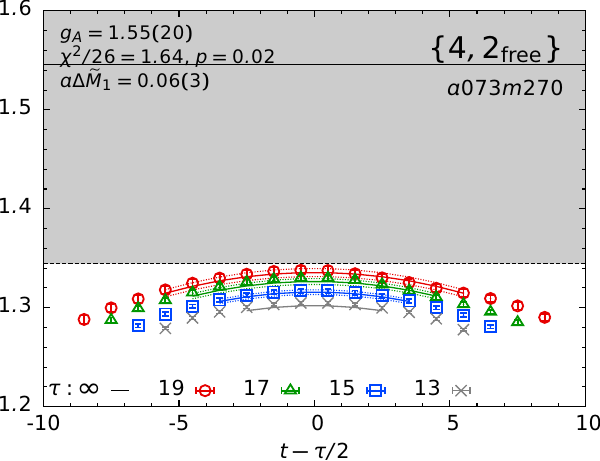}    
    \includegraphics[width=0.24\linewidth]{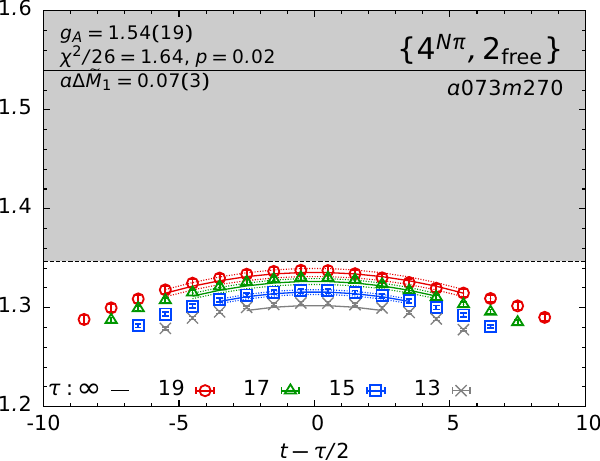} 
}
{
    \includegraphics[width=0.24\linewidth]{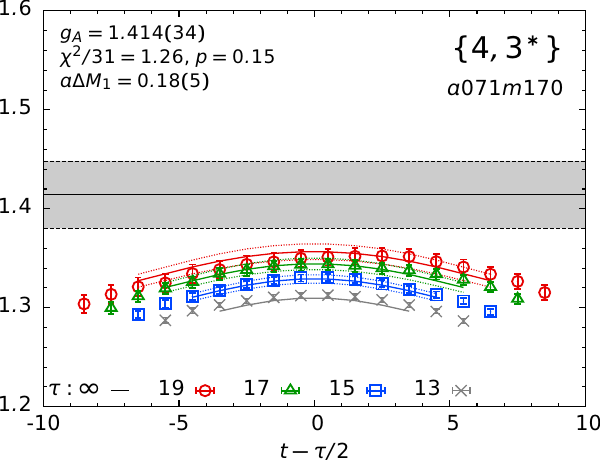}    
    \includegraphics[width=0.24\linewidth]{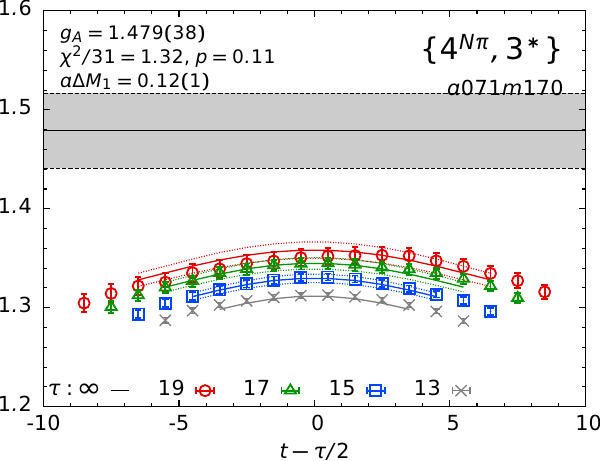}  
    \includegraphics[width=0.24\linewidth]{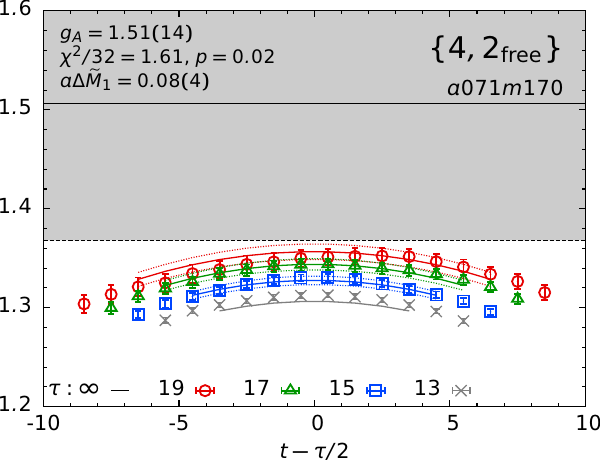}    
    \includegraphics[width=0.24\linewidth]{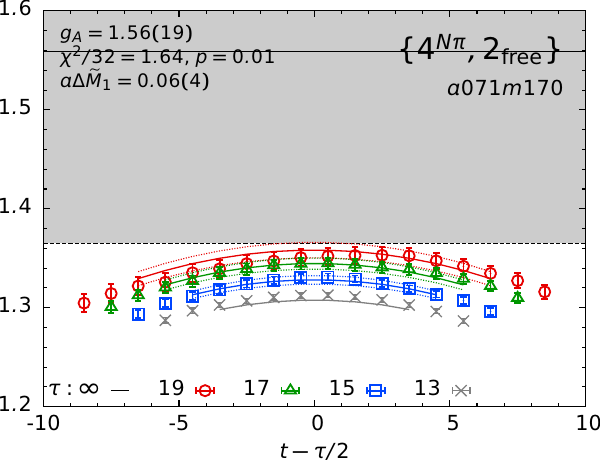} 
}
\caption{Each panel gives the data for the ratio 
${\cal R}_{\mathcal{A}}(t, \tau) =
C^{3\text{pt}}_{\mathcal{A}}(t,\tau;\bm{0},\bm{0})/C^{2\text{pt}}(\tau,\bm{0})$ 
defined in Eq.~\eqref{eq:ratio}, which 
gives the unrenormalized axial charge $g_{A}^{u-d}$ in the limit
$\tau \to \infty$, plotted as a function of $t -\tau/2$ for the
four largest values of $\tau$. The data connected by lines of the same
color for the three largest $\tau$ are used in the fit to get the
$\tau \to \infty$ value given by the black line with its gray error
band. Data at the smallest $\tau$, shown as gray crosses, are not used
in the fit. The four panels in each row show the excited-state fits to the same data but 
with the four strategies, $\{4^{},3^\ast\}$ (left column),
$\{4^{N\pi},3^\ast\}$ (second column), $\{4^{},2^{\rm free}\}$ (third
column), and $\{4^{N\pi},2^{\rm free}\}$ (right column). The labels
give the bare charge $g_A$, the $\chi^2/$dof of the fit, the mass gap
$a \Delta M_1$ (or $a \Delta {\widetilde M}_1$) and the ensemble ID.}  
\label{fig:gAcomp}
\end{figure*}

\begin{figure*}[tbp] 
\subfigure
{
    \includegraphics[width=0.24\linewidth]{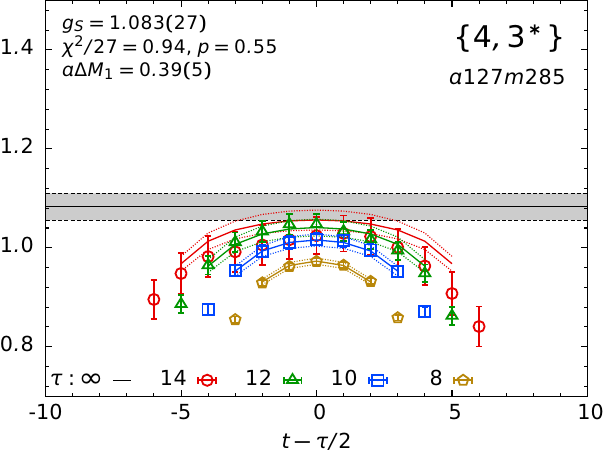}    
    \includegraphics[width=0.24\linewidth]{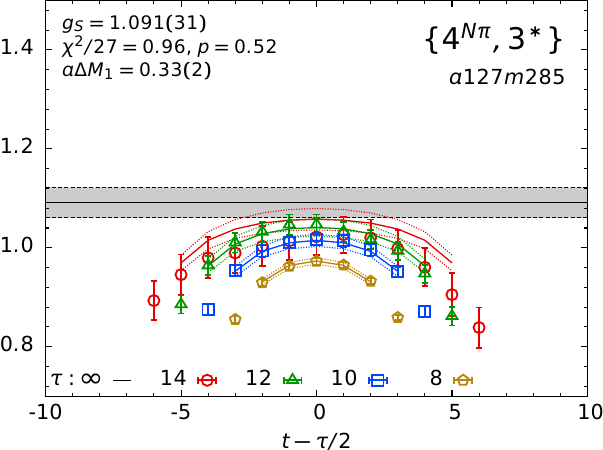}  
    \includegraphics[width=0.24\linewidth]{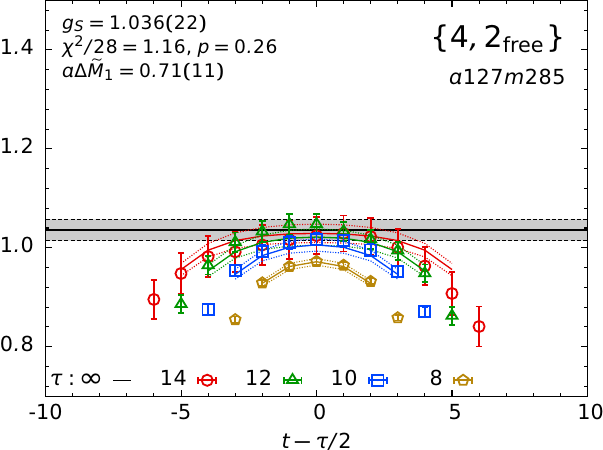}    
    \includegraphics[width=0.24\linewidth]{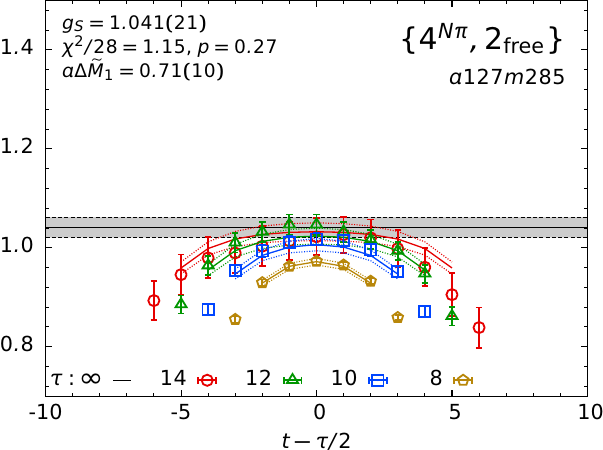} 
}
{
    \includegraphics[width=0.24\linewidth]{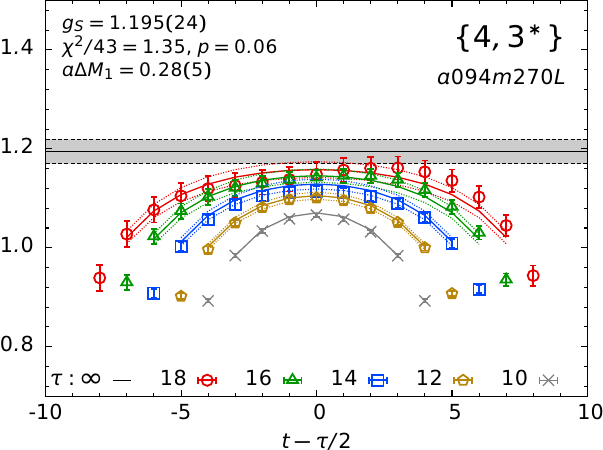}    
    \includegraphics[width=0.24\linewidth]{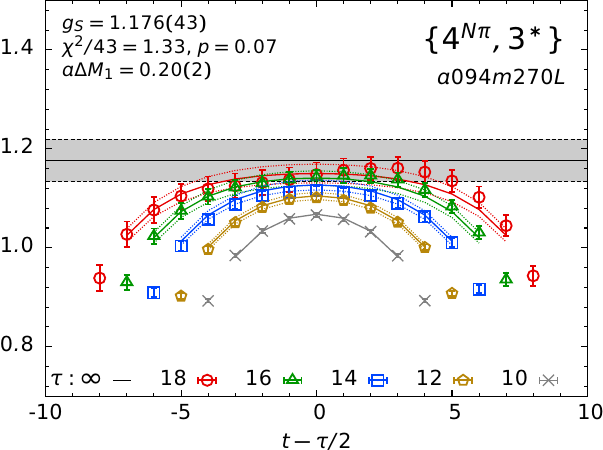}  
    \includegraphics[width=0.24\linewidth]{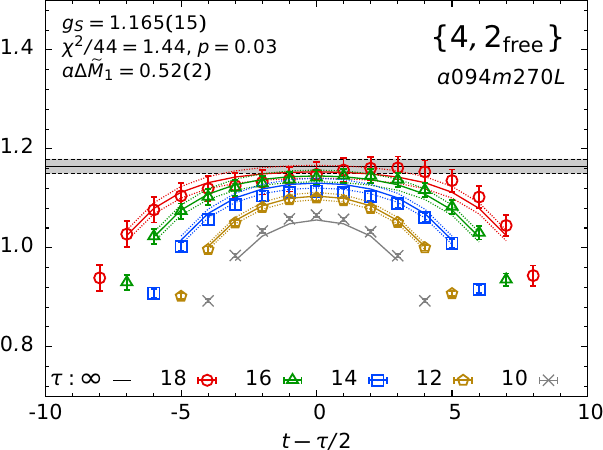}    
    \includegraphics[width=0.24\linewidth]{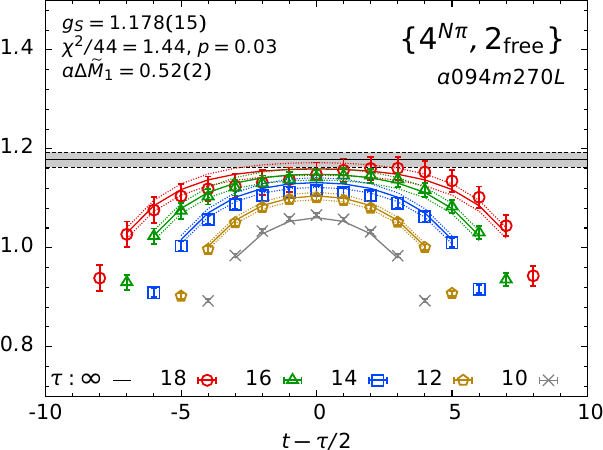} 
}
{
    \includegraphics[width=0.24\linewidth]{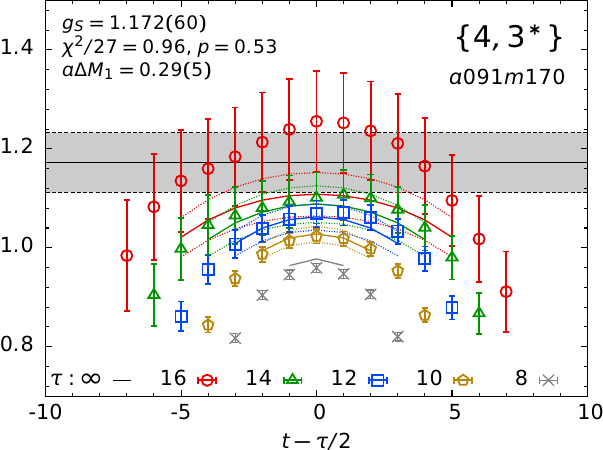}    
    \includegraphics[width=0.24\linewidth]{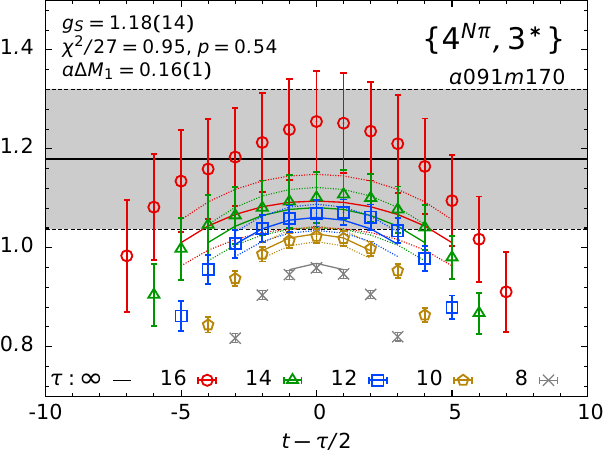}  
    \includegraphics[width=0.24\linewidth]{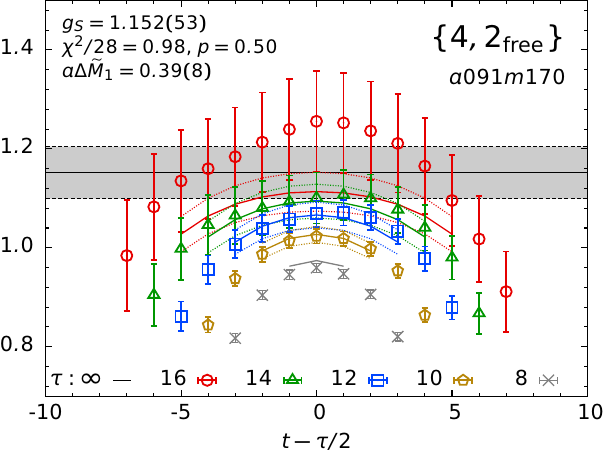}    
    \includegraphics[width=0.24\linewidth]{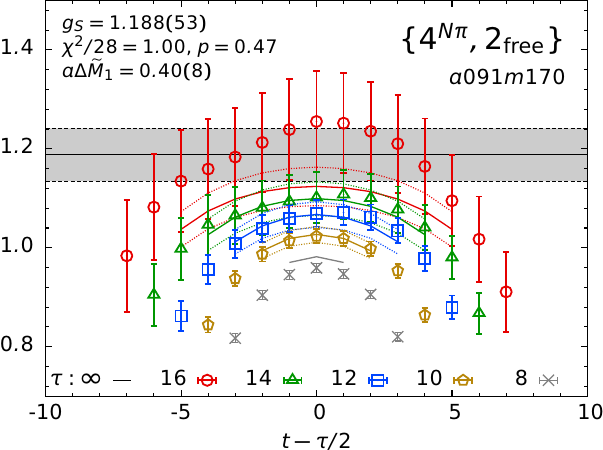} 
}
{
    \includegraphics[width=0.24\linewidth]{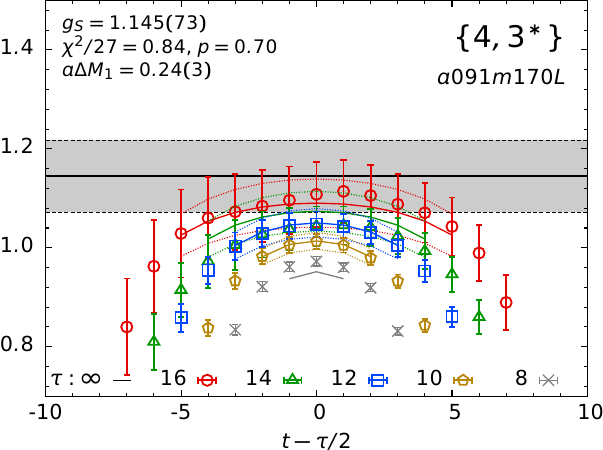}    
    \includegraphics[width=0.24\linewidth]{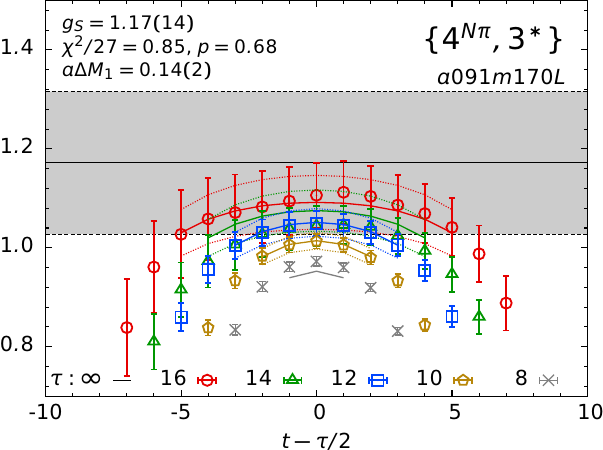}  
    \includegraphics[width=0.24\linewidth]{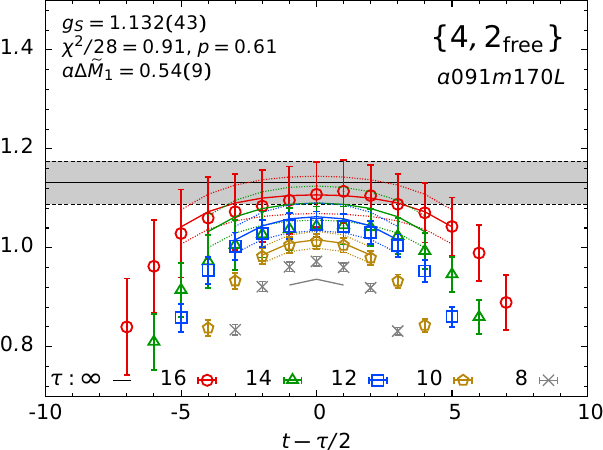}    
    \includegraphics[width=0.24\linewidth]{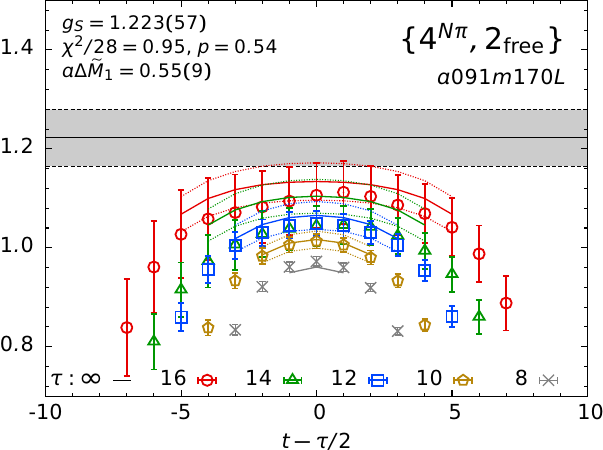} 
}
{
    \includegraphics[width=0.24\linewidth]{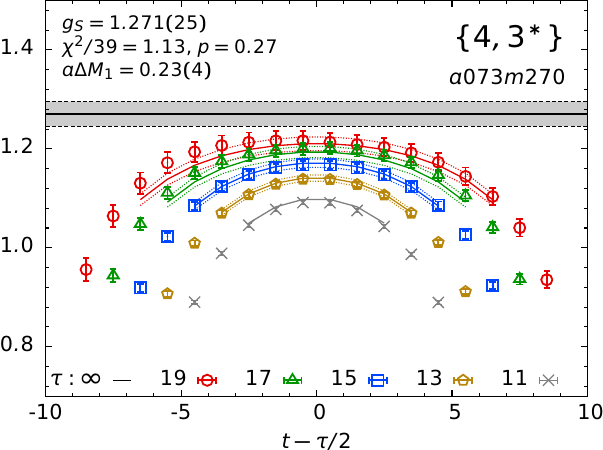}    
    \includegraphics[width=0.24\linewidth]{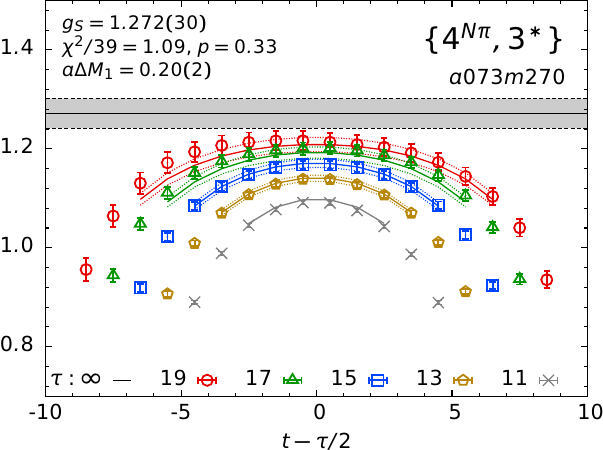}  
    \includegraphics[width=0.24\linewidth]{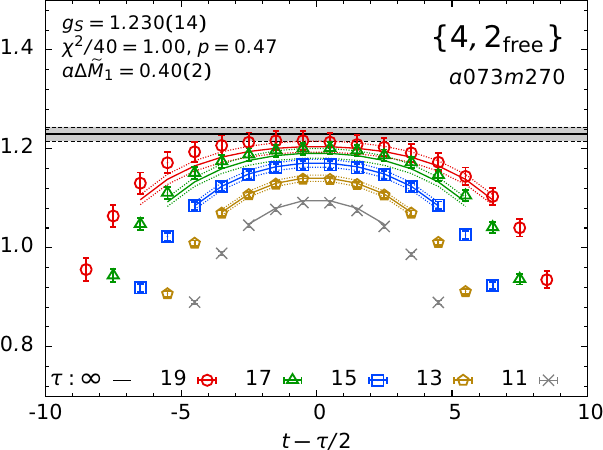}    
    \includegraphics[width=0.24\linewidth]{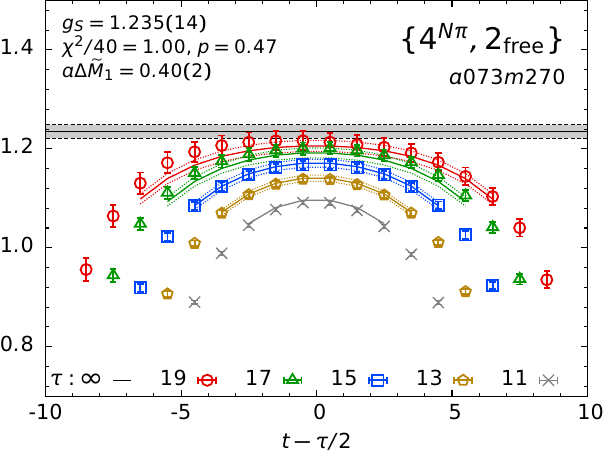} 
}
{
    \includegraphics[width=0.24\linewidth]{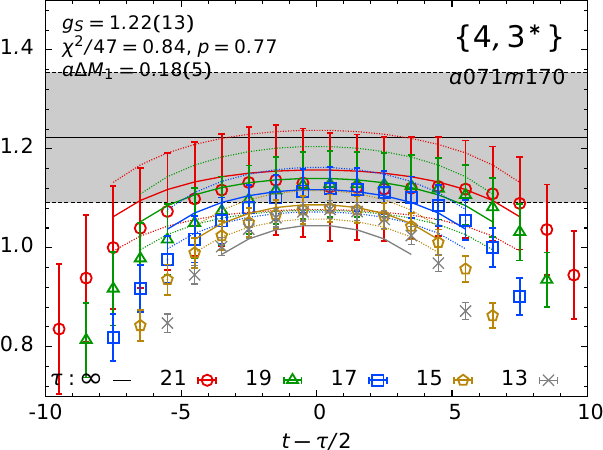}    
    \includegraphics[width=0.24\linewidth]{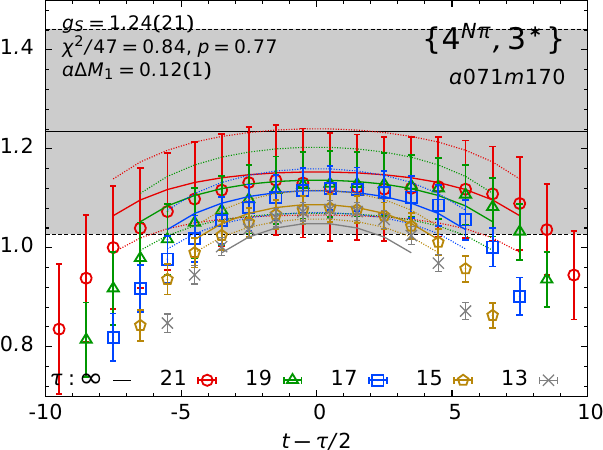}  
    \includegraphics[width=0.24\linewidth]{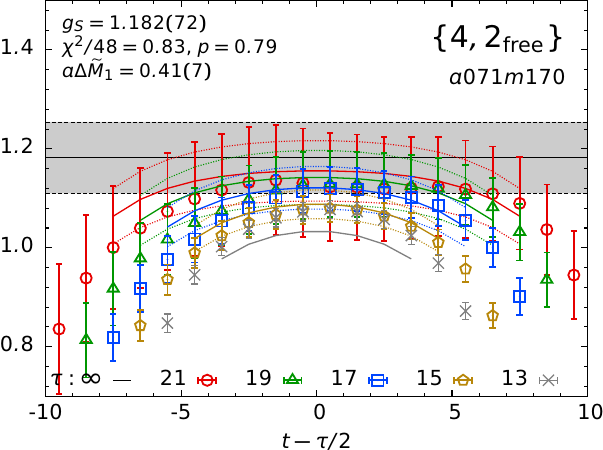}    
    \includegraphics[width=0.24\linewidth]{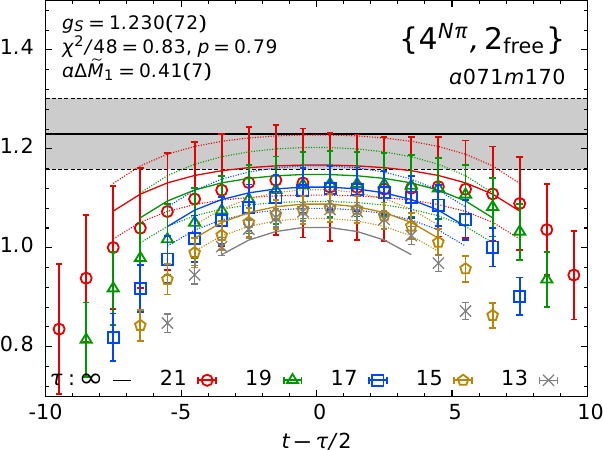} 
}
\caption{Each panel shows the data for the ratio defined in Eq.~\protect\eqref{eq:ratio} that
  gives the unrenormalized scalar charge $g_{S}^{u-d}$ in the limit
  $\tau \to \infty$, and plotted as a function of $t -\tau/2$ for the
  for the five largest values of $\tau$ (four for $a127m285$). In each panel, the data
  with the four largest $\tau$ and connected by lines of the same color are
  used in the fit to get the $\tau \to \infty$ value (gray band).  The
  rest is the same as in Fig.~\protect\ref{fig:gAcomp}.
  \label{fig:gScomp}}
\end{figure*}

\begin{figure*}[tbp] 
\subfigure
{
    \includegraphics[width=0.24\linewidth]{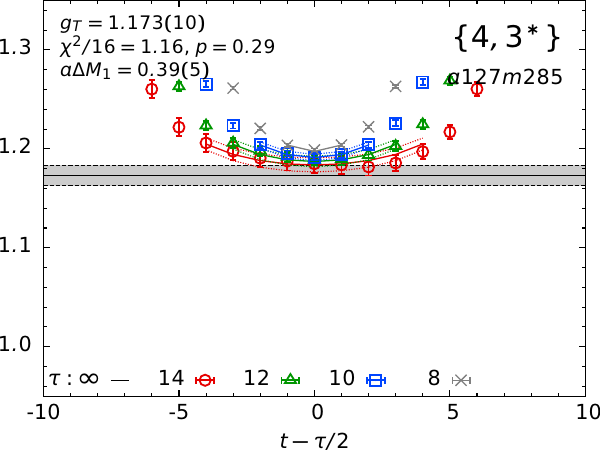}    
    \includegraphics[width=0.24\linewidth]{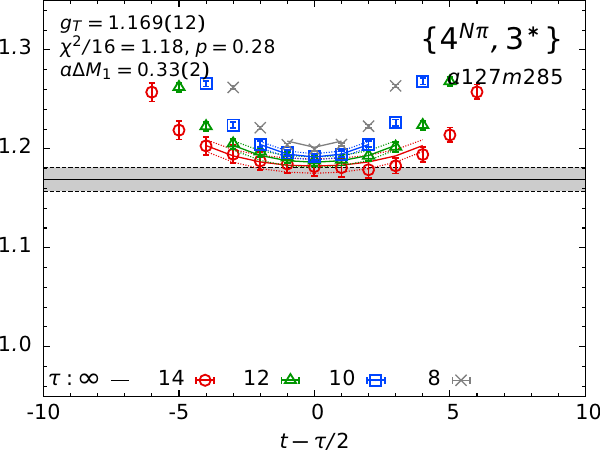}  
    \includegraphics[width=0.24\linewidth]{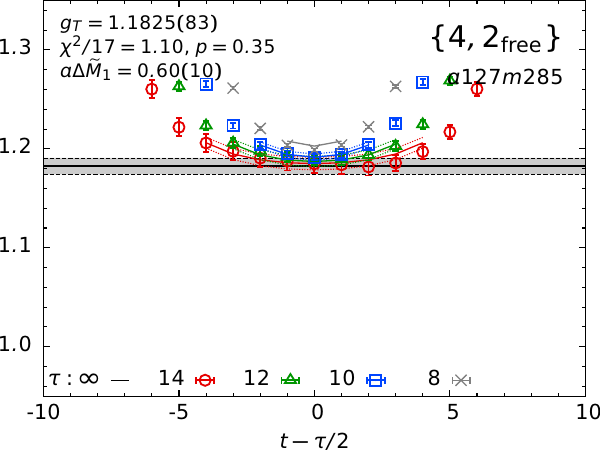}    
    \includegraphics[width=0.24\linewidth]{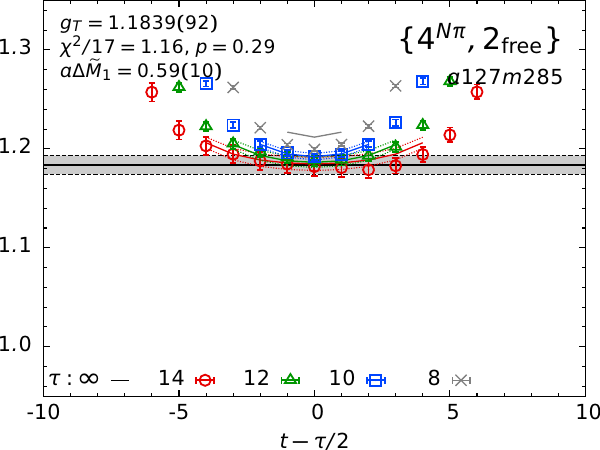} 
}
{
    \includegraphics[width=0.24\linewidth]{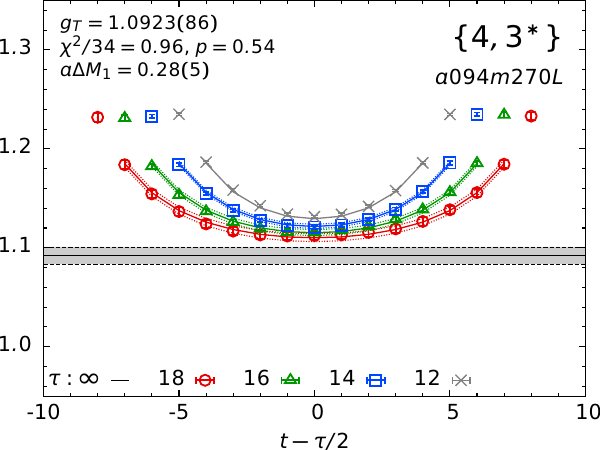}    
    \includegraphics[width=0.24\linewidth]{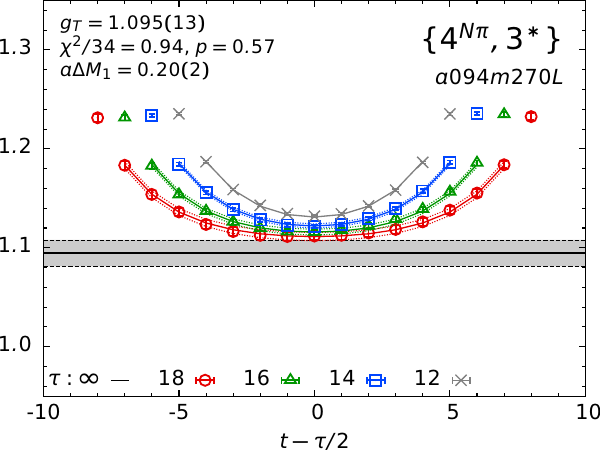}  
    \includegraphics[width=0.24\linewidth]{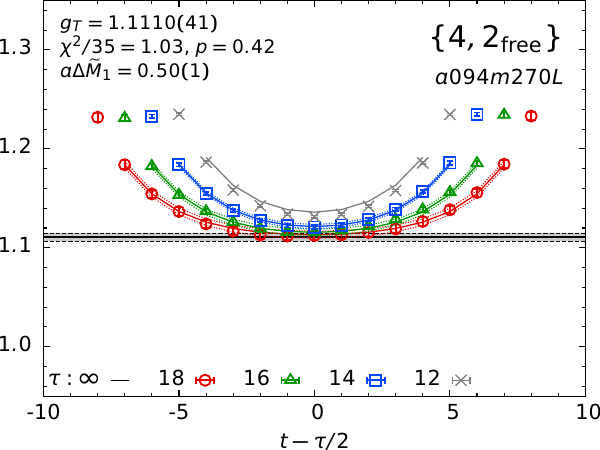}    
    \includegraphics[width=0.24\linewidth]{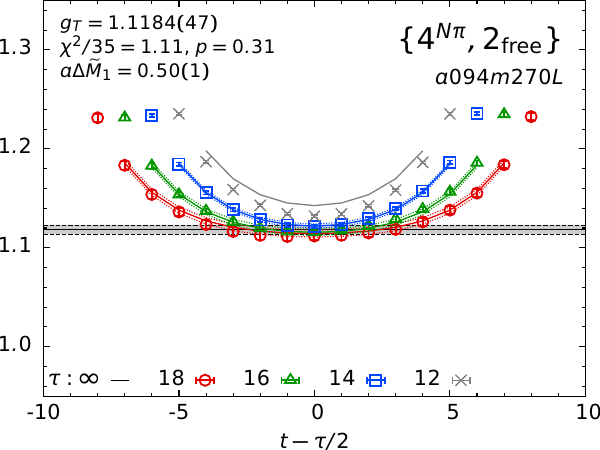} 
}
{
    \includegraphics[width=0.24\linewidth]{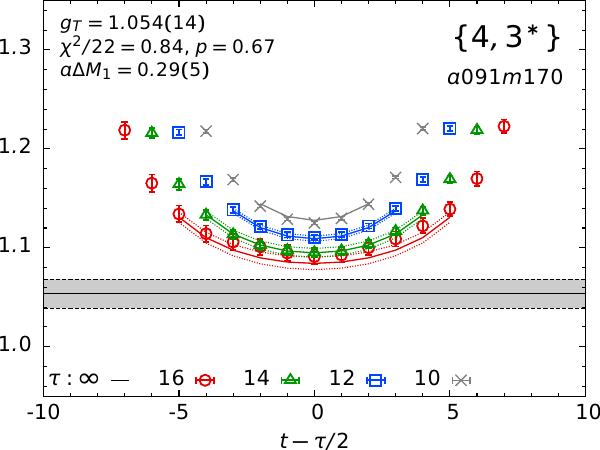}    
    \includegraphics[width=0.24\linewidth]{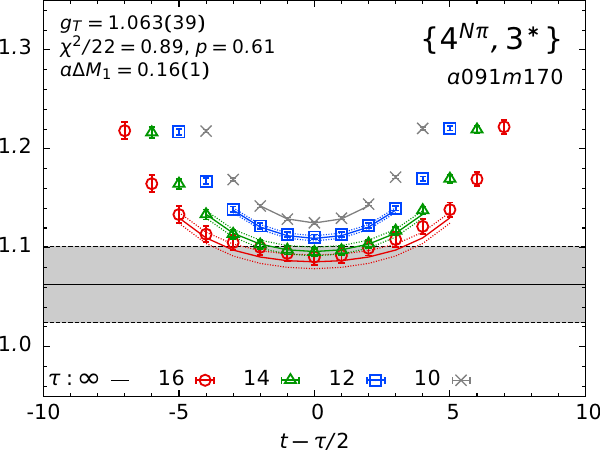}  
    \includegraphics[width=0.24\linewidth]{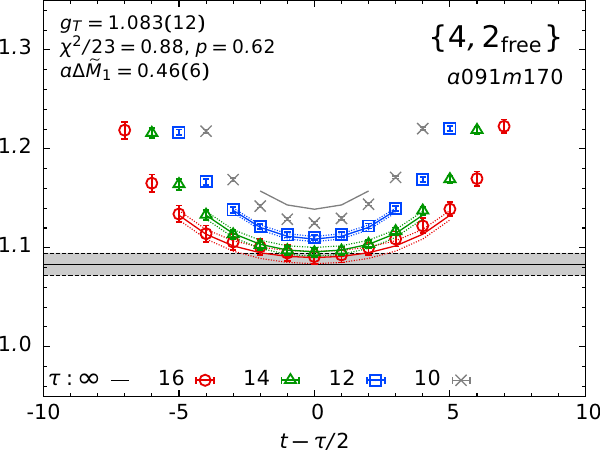}    
    \includegraphics[width=0.24\linewidth]{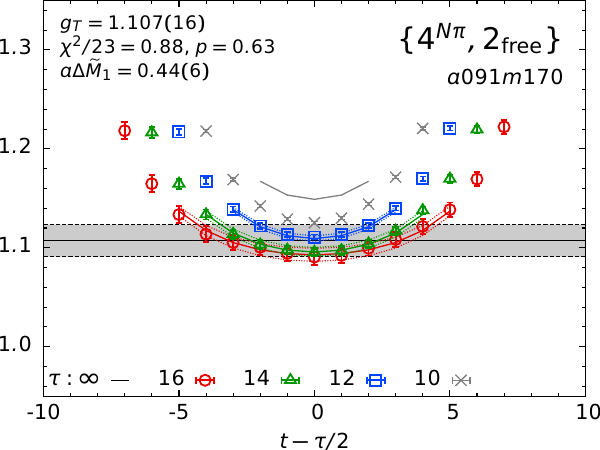} 
}
{
    \includegraphics[width=0.24\linewidth]{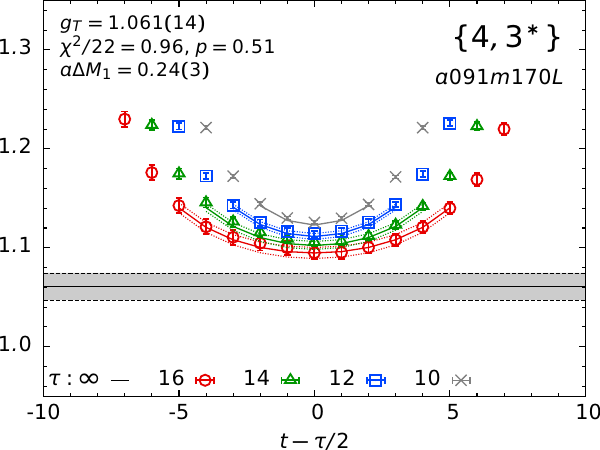}    
    \includegraphics[width=0.24\linewidth]{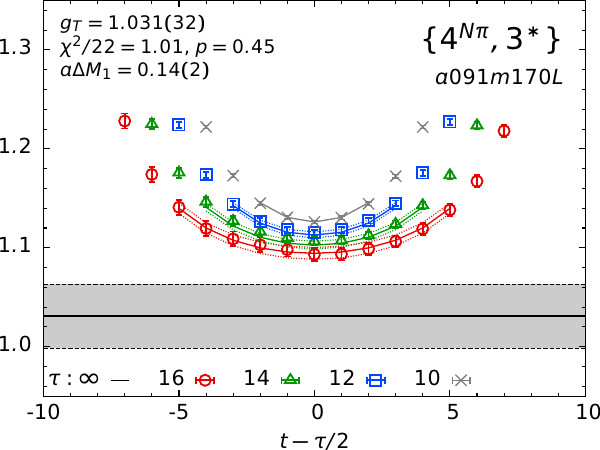}  
    \includegraphics[width=0.24\linewidth]{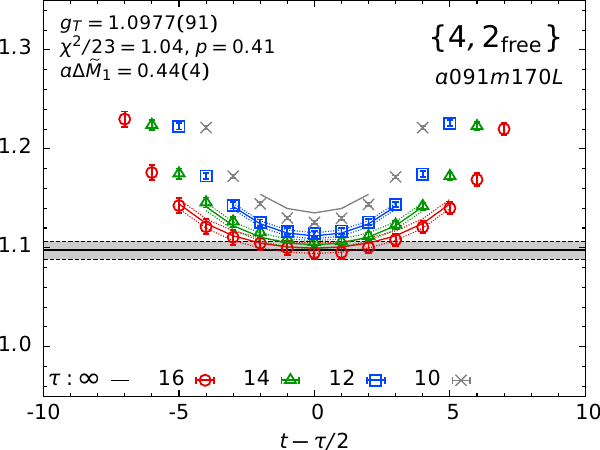}    
    \includegraphics[width=0.24\linewidth]{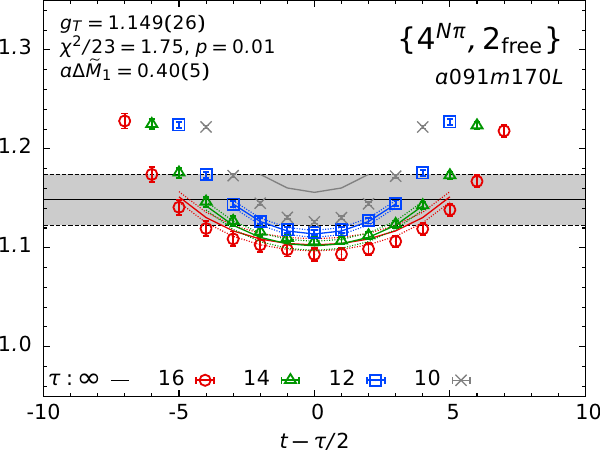} 
}
{
    \includegraphics[width=0.24\linewidth]{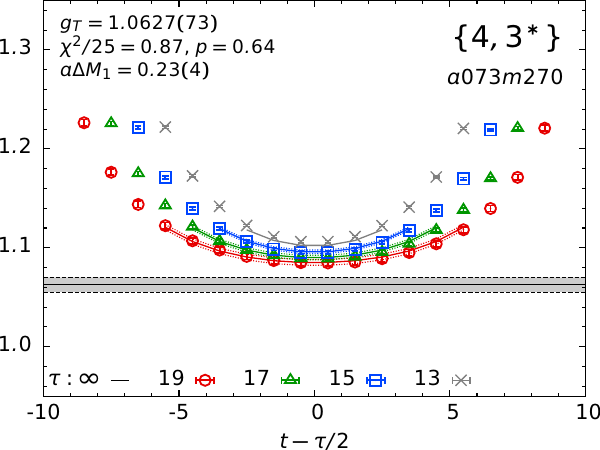}    
    \includegraphics[width=0.24\linewidth]{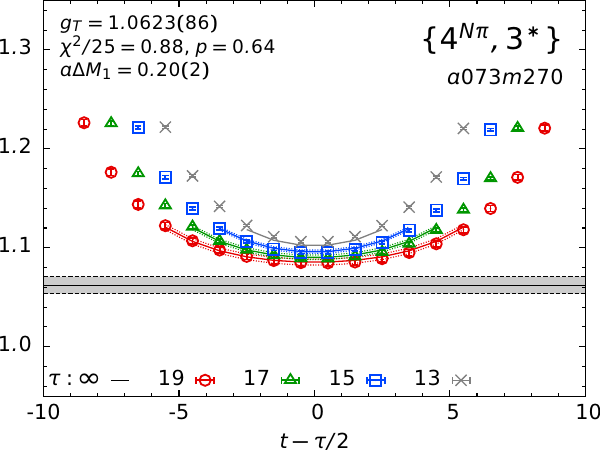}  
    \includegraphics[width=0.24\linewidth]{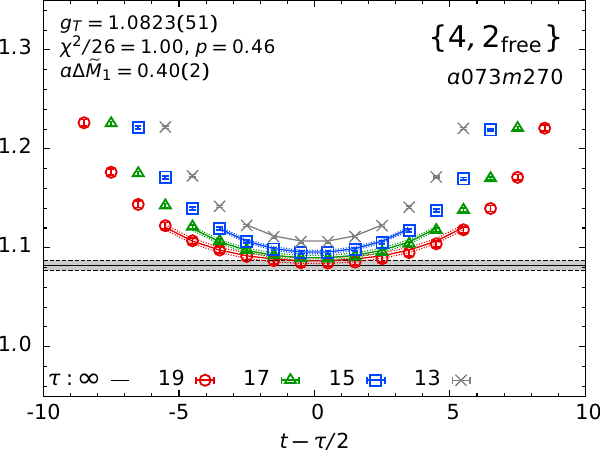}    
    \includegraphics[width=0.24\linewidth]{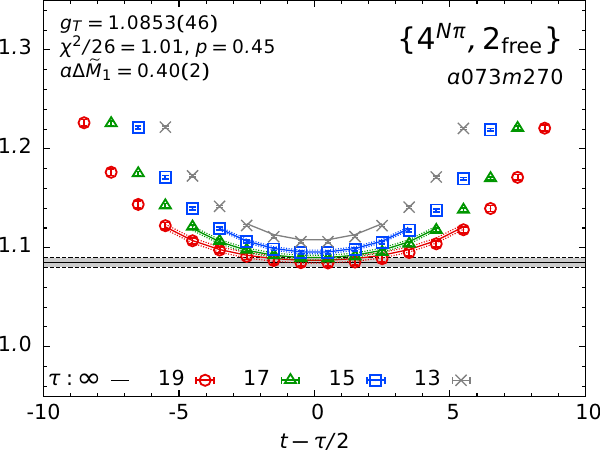} 
}
{
    \includegraphics[width=0.24\linewidth]{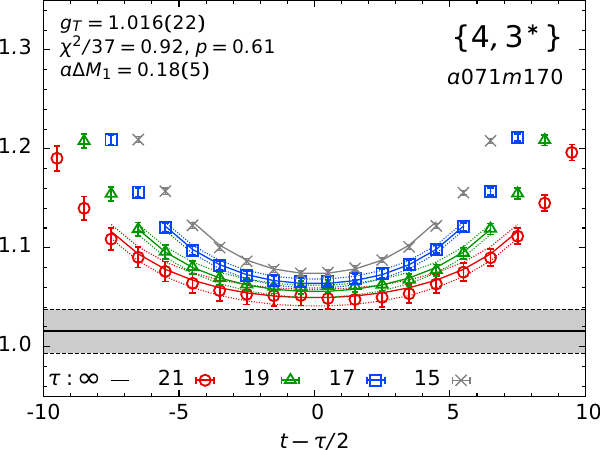}    
    \includegraphics[width=0.24\linewidth]{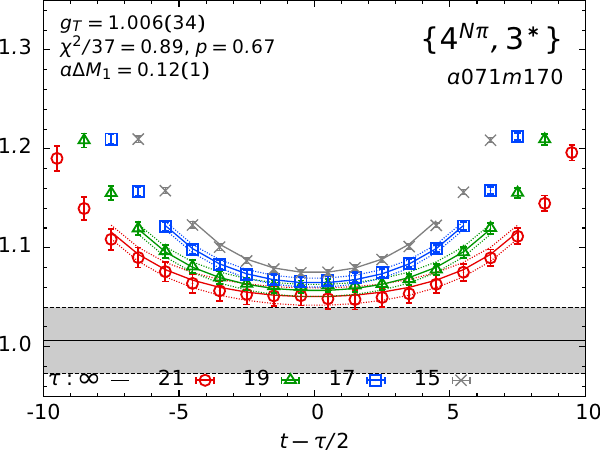}  
    \includegraphics[width=0.24\linewidth]{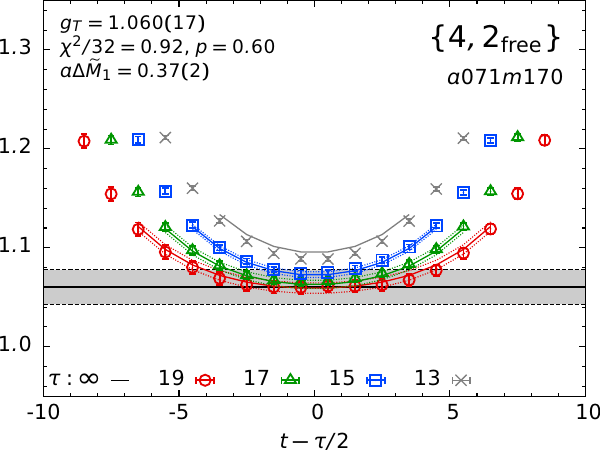}    
    \includegraphics[width=0.24\linewidth]{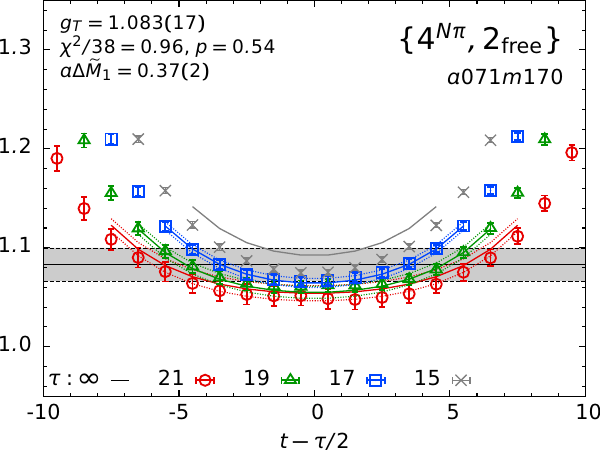} 
}
\caption{Each panel shows the data for the ratio defined in Eq.~\protect\eqref{eq:ratio} that
  gives the unrenormalized tensor charge $g_{T}^{u-d}$ in the limit
  $\tau \to \infty$, and plotted as a function of $t -\tau/2$ for the
  four largest values of $\tau$. The rest is the same as in
  Fig.~\protect\ref{fig:gAcomp}.
  \label{fig:gTcomp}}
\end{figure*}

In this Appendix, we show the data and the fits made to control ESC in
$g_A$, $g_S$ and $g_T$ in Figs.~\ref{fig:gAcomp},~\ref{fig:gScomp}
and~\ref{fig:gTcomp}, respectively, using the four strategies,
$\{4^{},3^\ast\}$, $\{4^{N\pi},3^\ast\}$, $\{4^{},2^{\rm free}\}$, and
$\{4^{N\pi},2^{\rm free}\}$ discussed in Sec.~\ref{sec:charges}.  The
results for the charges are summarized in Tables~\ref{tab:gAdP2z2}
and~\ref{tab:gSTs4e7}.

\vspace*{\baselineskip}\leavevmode

\onecolumngrid\hrule width 0pt\twocolumngrid\leavevmode\cleardoublepage
\section{Anatomy of the excited-state contamination in the charges}
\label{sec:anatomy}

\begin{figure*}[tbp] 
\subfigure
{
    \includegraphics[width=0.24\linewidth]{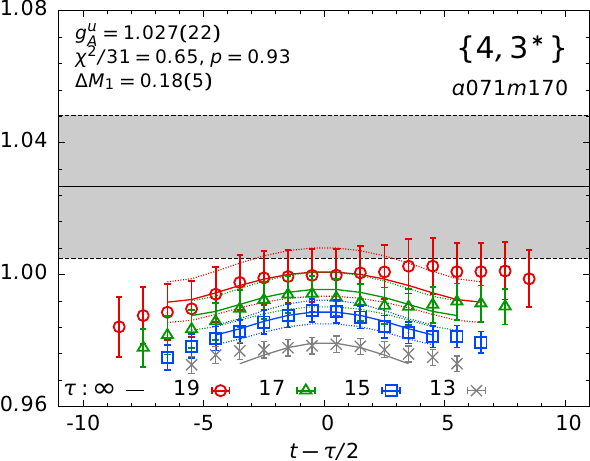}    
    \includegraphics[width=0.24\linewidth]{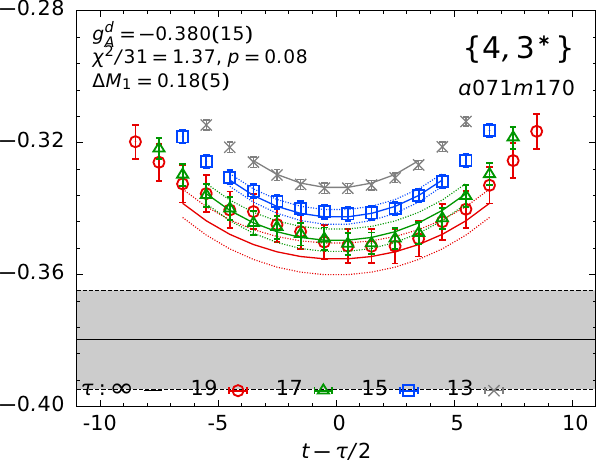}    
    \includegraphics[width=0.24\linewidth]{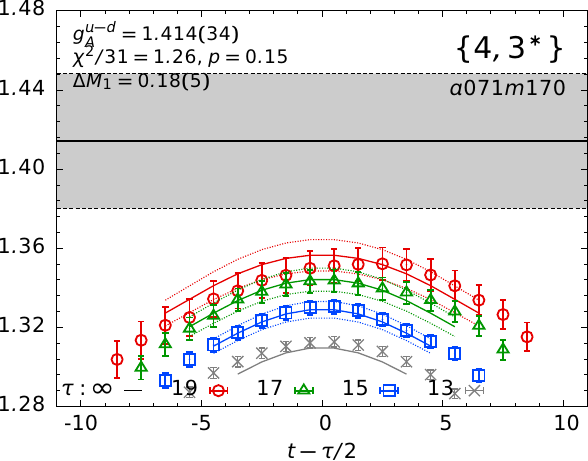}    
    \includegraphics[width=0.24\linewidth]{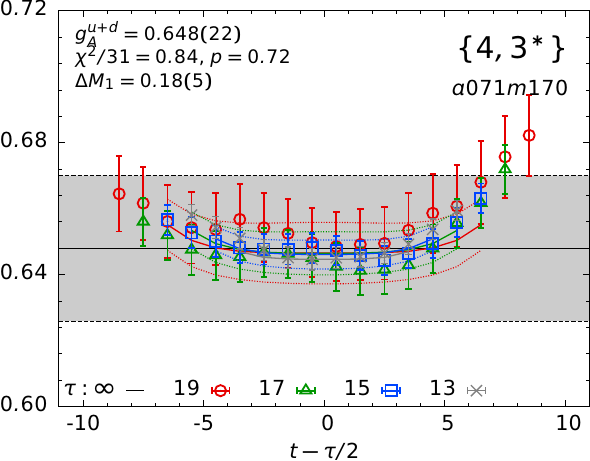}    
}
{
    \includegraphics[width=0.24\linewidth]{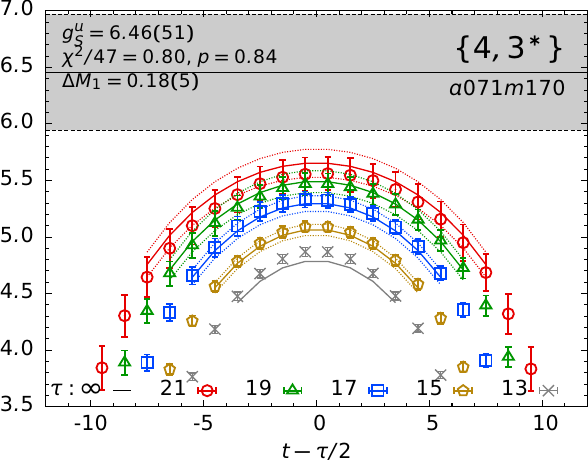}    
    \includegraphics[width=0.24\linewidth]{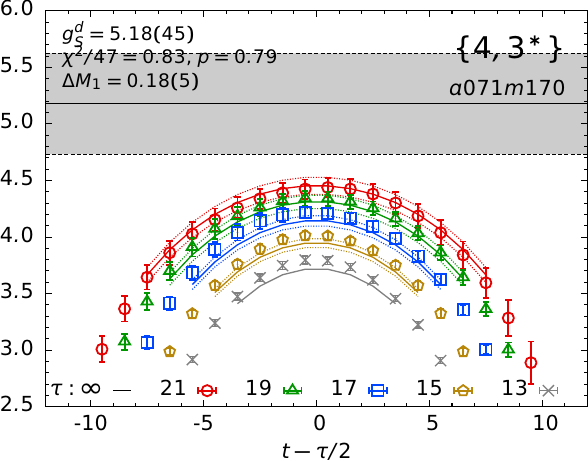}    
    \includegraphics[width=0.24\linewidth]{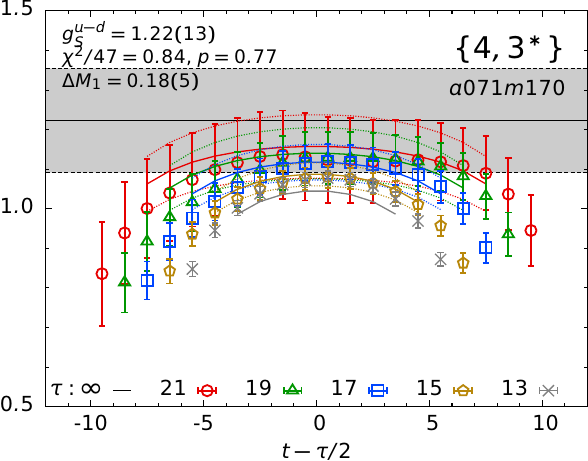}    
    \includegraphics[width=0.24\linewidth]{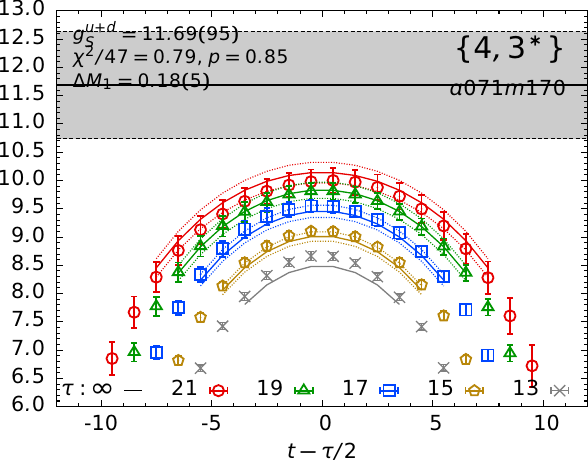}    
}
{
    \includegraphics[width=0.24\linewidth]{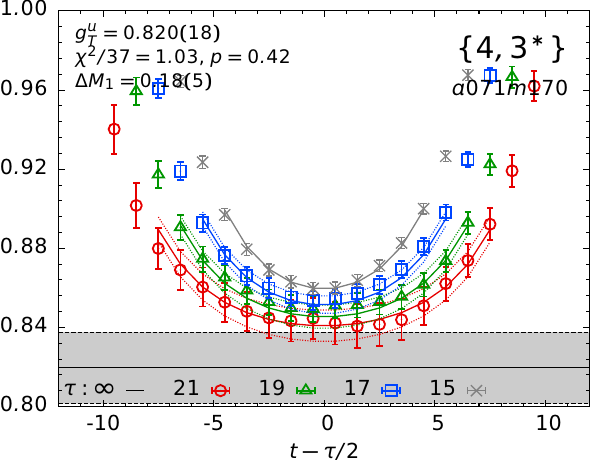}    
    \includegraphics[width=0.24\linewidth]{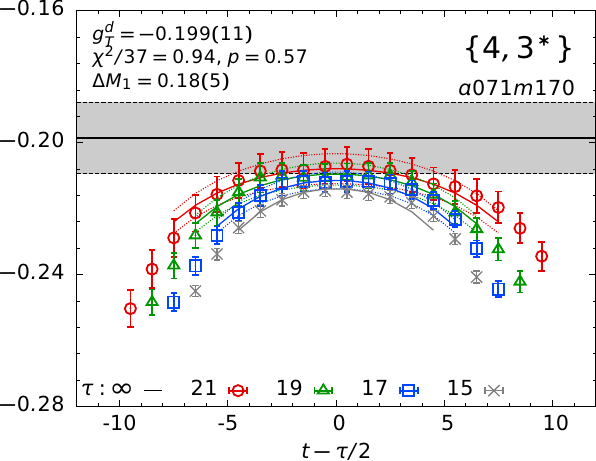}    
    \includegraphics[width=0.24\linewidth]{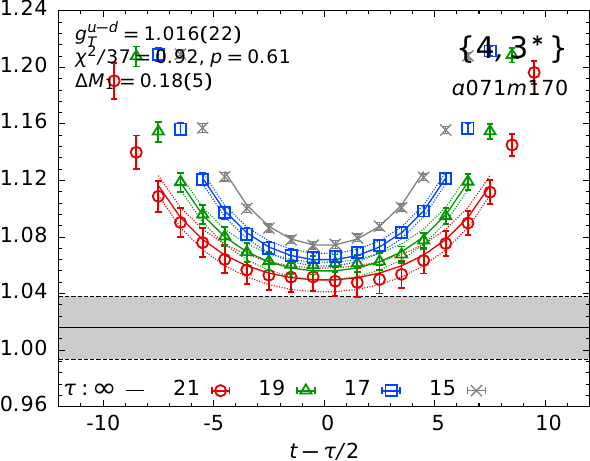}    
    \includegraphics[width=0.24\linewidth]{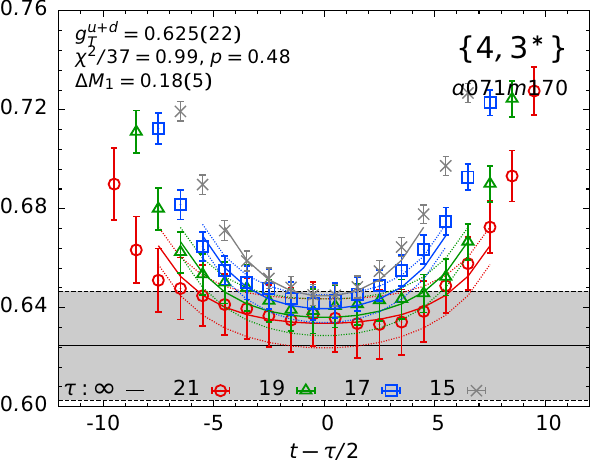}    
}
\caption{Data for the ratio defined in Eq.~\protect\eqref{eq:ratio}
  for different operator insertions---on the $u$ quark
  (left column), $d$ quark (second column), $u-d$ combination (third
  column) and the connected part of the $u+d$ combination (right
  column)---are shown for the $a071m170$ ensemble. Data for $g_{A}$ (top
  row), $g_{S}$ (middle row) and $g_{T}$ (bottom row), are plotted as
  a function of $t -\tau/2$ for the values of $\tau$ specified in the
  labels. All the fits to get the $\tau \to \infty$ values are with
  the $\{4,3^\ast\}$ strategy.  \label{fig:UandDcomp}}
\end{figure*}

In this Appendix, we compare fits to the data for the three charges,
$g_{A,S,T}$, in Fig.~\ref{fig:UandDcomp} to highlight (i) the
differences in ESC for the $u$, $d$, $u-d$ and $u+d$ quark bilinear
operator insertions and (ii) how these ESC patterns impact the
extraction of the isovector and isoscalar (connected only)
combinations.  Data are presented for the $a071m170$ ensemble, which
have the largest statistical errors.  The fits are made using the
$\{4,3^\ast\}$ strategy.  We also examine the data for symmetry about
$(t-\tau/2)$, monotonic convergence versus $\tau$ and the size of
errors, and how these impact our ability to remove ESC.

The ESC in the axial channel is equally large in magnitude for
insertion in the $u$ and $d$ quarks. It adds in the $u-d$ combination
as the data have opposite signs, but cancel in $u+d$.  In the case of
the scalar charge, the ESC in both the $u$ and $d$ insertions are a
similar fraction of the value. Thus, it adds in $u+d$. In the $u-d$ combination, 
there is a large cancellation; however, significant ESC
remains as shown in Fig.~\ref{fig:gScomp}.  In the case of the tensor
charge, the value and the ESC in the insertion in the $u$ quark is
much larger, and it dominates in both the $u-d$ and the $u+d$
combinations.  Overall, in the $u+d$ axial and $u-d$ scalar cases,
where there is a cancellation, much higher statistical precision in
the $\tau > 1.5$~fm data is needed to demonstrate monotonic convergence and
improve the reliability of $n$-state fits.

Given these patterns, we made fits with the same set of ESC strategies
to data with separate insertions of $u$ and $d$ quark operators. The
goal was to see whether these fits, especially in the scalar channel,
are more stable and the $g_S^{u-d}$ combination constructed from 
individual ESC fits has better precision. What we found, on all seven
ensembles and for all three charges, is that direct fits to the $u-d$ data gave values
and errors consistent with those obtained by combining results from
separate fits to data with $u$ and $d$ insertions. The largest
differences are in $g_S^{u-d}$ for the $a091m170L$ (about $1 \sigma$)
and $a071m170$ (about $0.5 \sigma$) ensembles. This check shows that
our error estimates are reasonable even in the worst cases.  In short,
examining the separate fits did provide a better understanding of the
ESC and of the statistical precision of the fits but did not improve
the estimates for the isovector charges.

\onecolumngrid\hrule width 0pt\twocolumngrid\cleardoublepage
\section{Excited states in the axial three-point functions}
\label{sec:AFFESC}

On a finite lattice, one has towers of eigenstates of the transfer
matrix labeled by their quantum numbers.  A strict identification with
physical states such as $N(\bm 0) \pi({\bm q})$ and $N(-\bm q)
\pi({\bm q})$ can only be done in infinite volume and in the continuum
limit.  As mentioned in the text, both $N(\bm p) \pi(-{\bm p})$ and
$N(\bm 0) \pi({\bm 0}) \pi({\bm 0})$ have the right quantum numbers
(spin, parity, $G$-parity) to contribute to the axial channel. It is the
magnitude of their couplings that decides the size of their
contributions. These need to be determined nonperturbatively from fits
to the three-point functions for high precision results. In such
analyses, for example in the axial channel, $\chi$PT is a good guide.

In a series of papers, B\"ar has presented the predictions of
$\chi$PT~\cite{Bar:2016uoj,Bar:2018xyi,Bar:2019gfx} keeping one
excited state, $N\pi$, in the analysis. At the tree level,
consistent with the pion-pole dominance hypothesis, the axial current
$A_\mu(\bm q)$ couples through a pion with momentum $q_\mu$. In our
setup, for the matrix elements of the three spatial $A_i$, the interaction with $\pi({\bm q})$ 
causes the transitions to the excited states $N(0) \to N(0) \pi (\bm
q)$ and $N(-\bm q) \to N(-\bm q) \pi (\bm q)$ in addition to the
desired ground state transitions $N(0) \to N(\bm q)$ and $N(-\bm q)
\to N(\bm 0)$. These ESC arise at tree-level, depend on $\bm q$ and are expected
to be large in the ${\widetilde G}_P$ and $G_P$ form factors.  In
addition, at the loop-level, all states with the right quantum numbers
such as $N(\bm q) \pi(\bm 0)$, $N(\bm 0) \pi(\bm q)$, $N(\bm 0) \pi
(\bm 0) \pi (\bm 0)$ and the full tower of $N(-\bm p) \pi (\bm p)$
states with all allowed values of $\bm p$ on the $\bm p =0$ side of
the three-point function, can contribute to all three form
factors. These loop-level contributions are estimated to be a few
percent effect and show only a mild dependence on $\bm p$.

\begin{figure*}[tbp] 
\subfigure
{
    \includegraphics[width=0.325\linewidth]{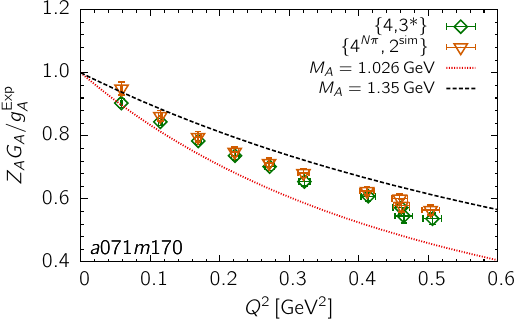} 
    \includegraphics[width=0.325\linewidth]{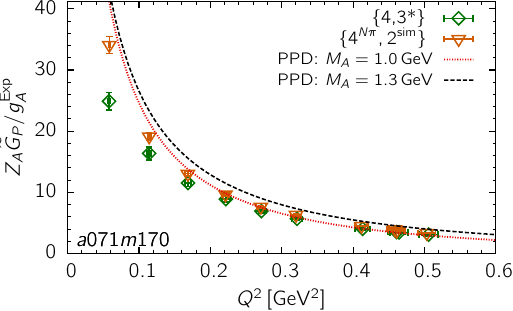} 
    \includegraphics[width=0.325\linewidth]{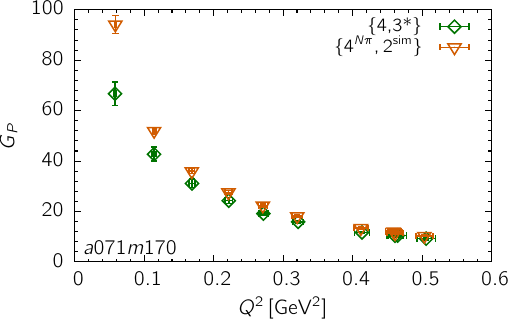} 
}
{
    \includegraphics[width=0.325\linewidth]{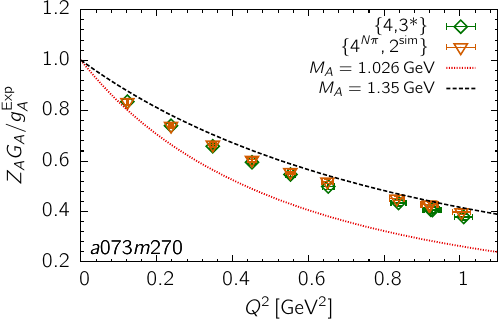} 
    \includegraphics[width=0.325\linewidth]{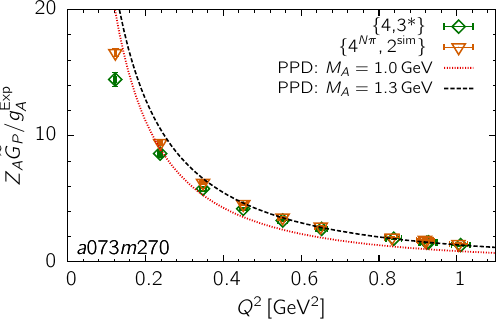} 
    \includegraphics[width=0.325\linewidth]{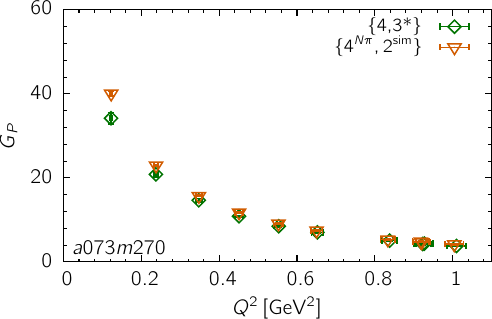} 
}
\caption{The form factors $Z_AG_A/g_A^{\rm exp}$, $Z_A {\widetilde G}_P/g_A^{\rm exp}$ and
  $G_P$ from the two strategies $\{4,3^\ast\}$ and $\{4^{N\pi},2^{\rm
  sim}\}$ are compared in each panel for the ensembles
  $a071m170$ (top row) and $a073m270$ (bottom row). We also show two dipole
  fits with $M_A = 1.026$ and $1.35$ to $G_A$, and a pion-pole
  dominance fit to ${\widetilde G}_P$ with $G_A$ given by the dipole
  ansatz to guide the eye.  \label{fig:FFcompare}}
\end{figure*}

The $\{4^{N\pi},2^{\rm sim}\}$ strategy analysis of the axial form
factors includes the $N\pi$ state predicted by tree-level $\chi$PT
analysis but neglects the contribution of all other states that can
contribute at loop-level. Compared to $\{4,3^\ast\}$, this changes
${\widetilde G}_P$ and $G_P$ by $\sim 35\%$ and $G_A$ by $\sim 5\%$ at
the smallest $Q^2$ point on the $a071m170$ ensemble as shown in
Fig.~\ref{fig:FFcompare}. The difference is much smaller on the $M_\pi
\approx 270$~MeV ensembles as shown for the $a073m270$ ensemble, i.e.,
the effect of the $N\pi$ state increases as $Q^2 \to 0$ and $M_\pi \to
0$. For the axial charge $g_A$ obtained from $A_3$, there is no
tree-level contribution due to the kinematic constraint. Our analysis
in Sec.~\ref{sec:charges}, including only the lowest, $N(- 1) \pi (1)$
(or the approximately degenerate $N(\bm 0) \pi (\bm 0) \pi (\bm 0)$)
state that can contribute at loop-level indicates that the effect
could be $\sim 8\%$ for $M_\pi = 135$~MeV. The impact of the remaining
tower of excited states in either case is unknown. In this appendix,
we discuss these effects and how best to proceed to remove all ESC.

First, we discuss the evidence that multihadron states
contribute. Next, we point out why it will be difficult to resolve all
relevant states from fits to the two-point function.  Last, we
provide some thoughts on how the analysis presented in this work can
be extended.

\begin{figure*}[tbp] 
{
    \includegraphics[width=0.24\linewidth]{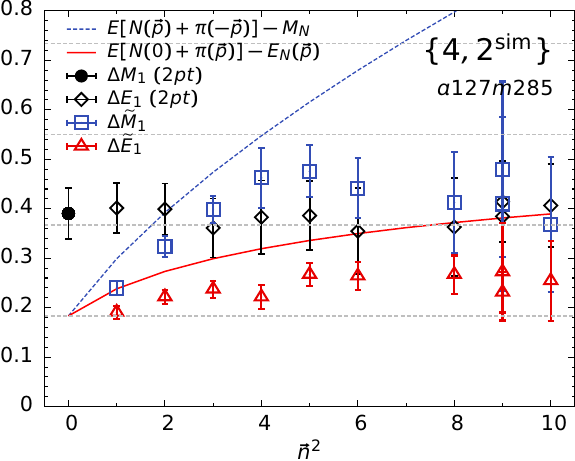} \hspace{0.2in} 
    \includegraphics[width=0.24\linewidth]{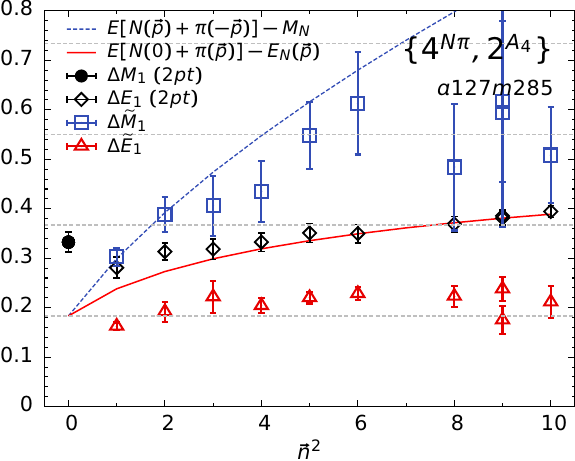}\hspace{0.2in} 
    \includegraphics[width=0.24\linewidth]{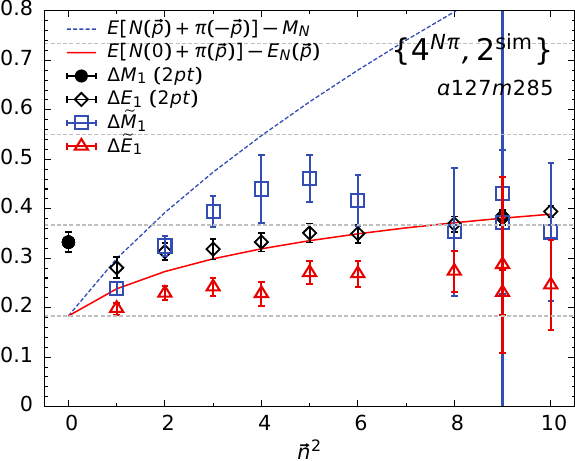} 
}
{
}
{
    \includegraphics[width=0.24\linewidth]{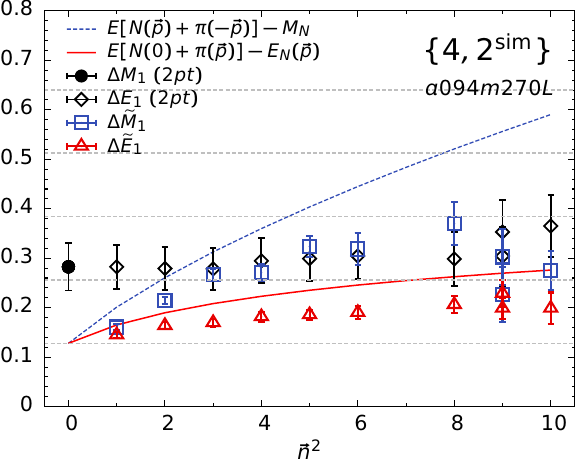}  \hspace{0.2in} 
    \includegraphics[width=0.24\linewidth]{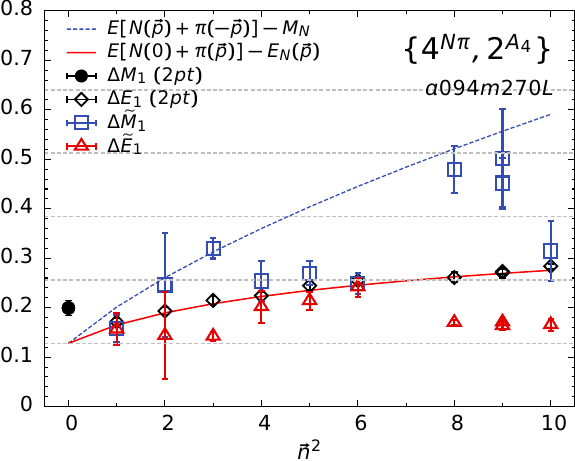}\hspace{0.2in} 
    \includegraphics[width=0.24\linewidth]{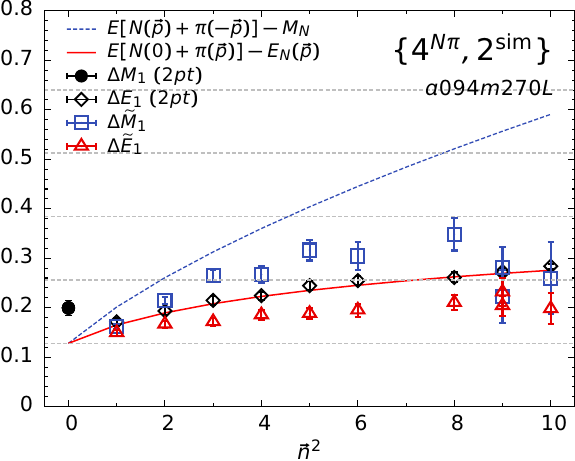} 
}
{
    \includegraphics[width=0.24\linewidth]{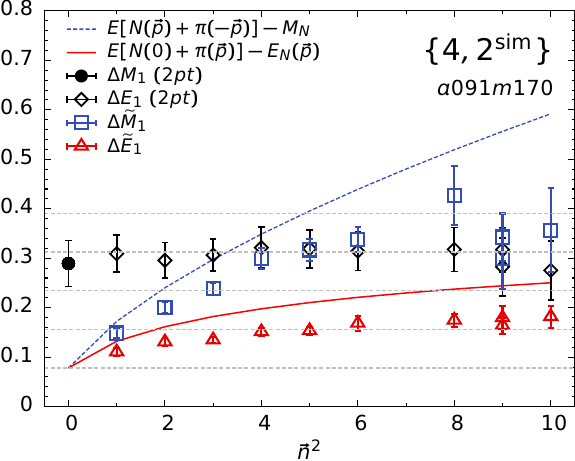}  \hspace{0.2in} 
    \includegraphics[width=0.24\linewidth]{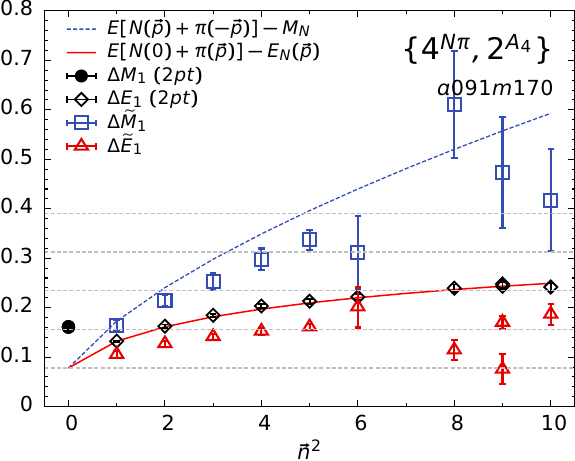}  \hspace{0.2in} 
    \includegraphics[width=0.24\linewidth]{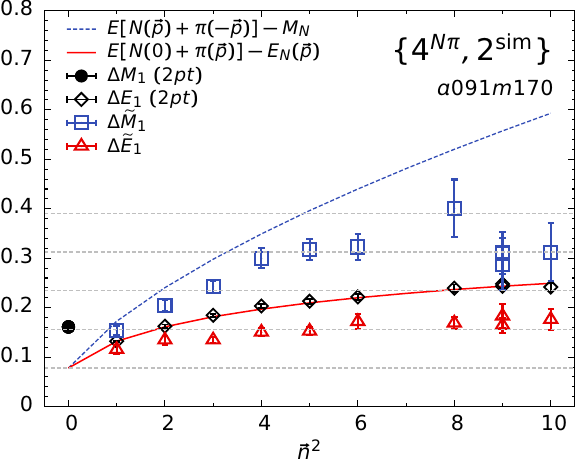} 
}
{
    \includegraphics[width=0.24\linewidth]{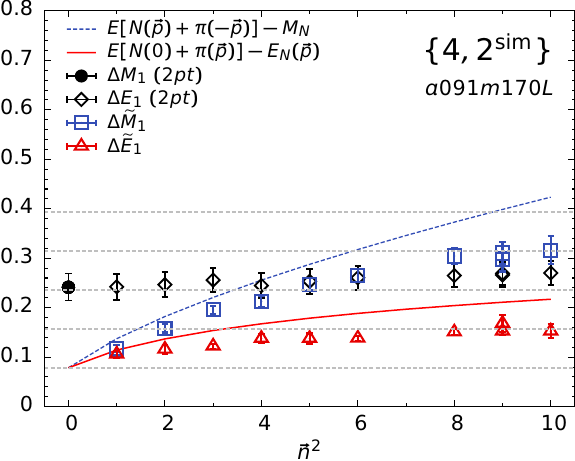}  \hspace{0.2in} 
    \includegraphics[width=0.24\linewidth]{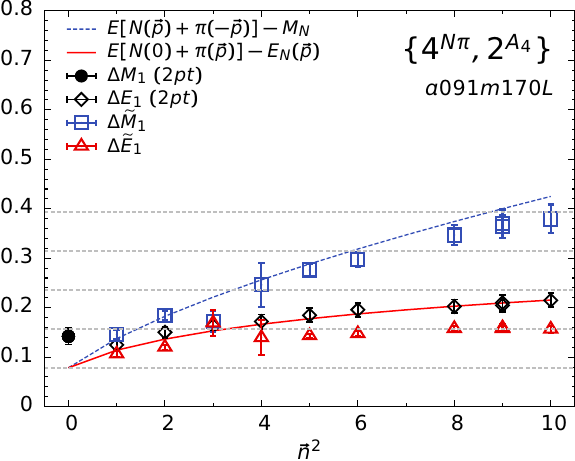}  \hspace{0.2in} 
    \includegraphics[width=0.24\linewidth]{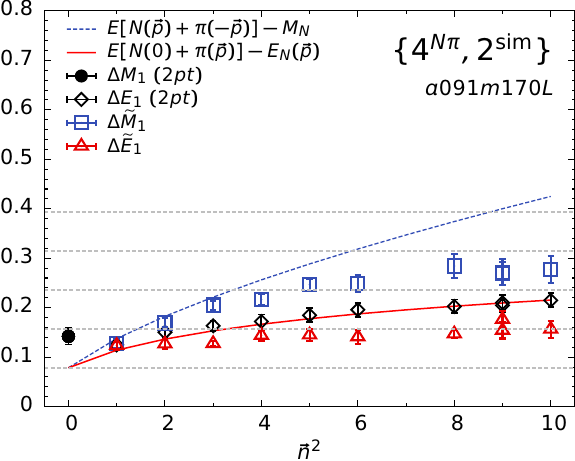} 
}
{
    \includegraphics[width=0.24\linewidth]{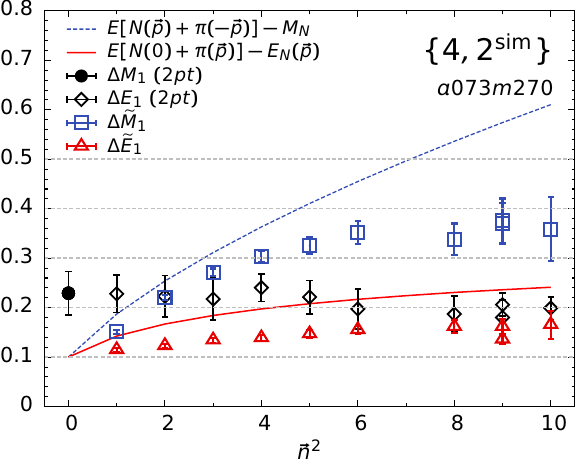}  \hspace{0.2in} 
    \includegraphics[width=0.24\linewidth]{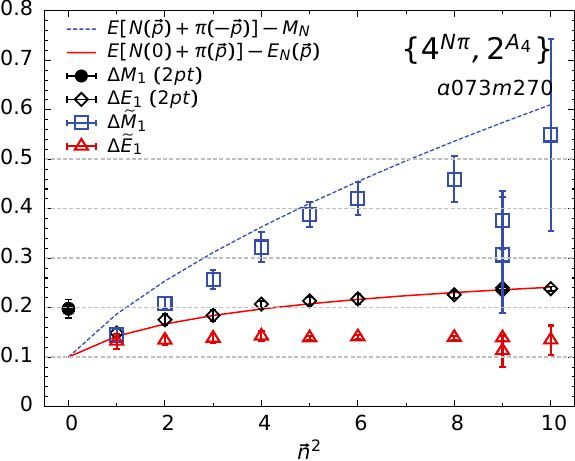}  \hspace{0.2in} 
    \includegraphics[width=0.24\linewidth]{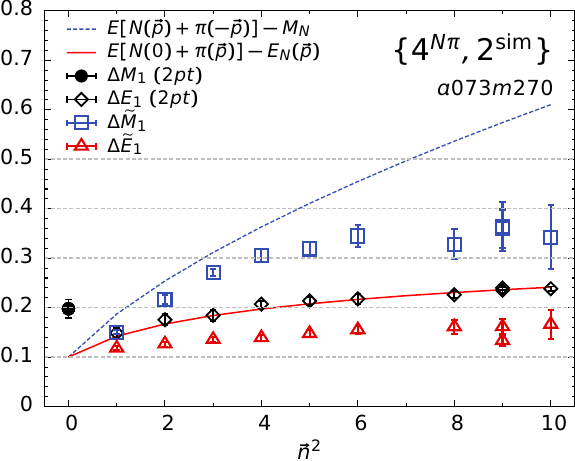} 
}
{
    \includegraphics[width=0.24\linewidth]{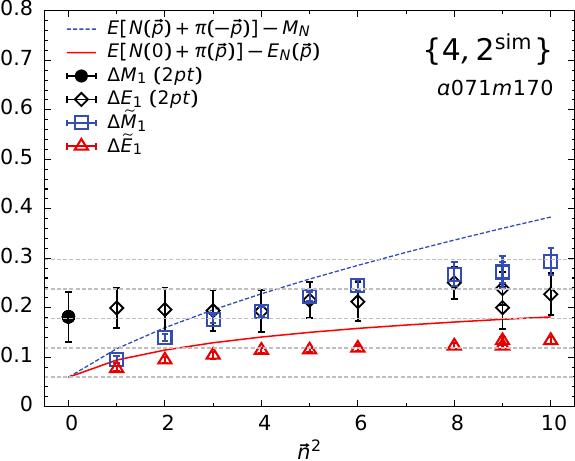}  \hspace{0.2in} 
    \includegraphics[width=0.24\linewidth]{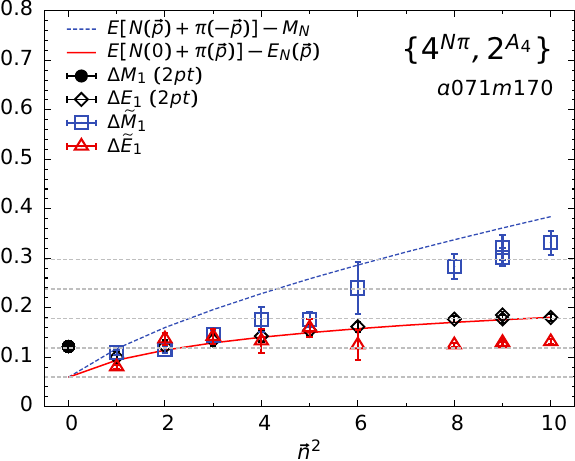}  \hspace{0.2in} 
    \includegraphics[width=0.24\linewidth]{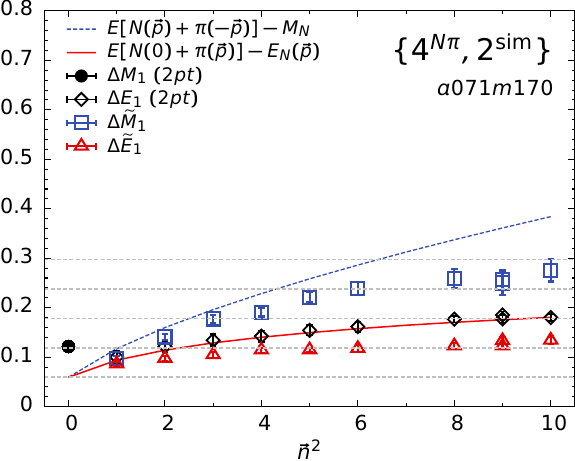} 
}
\vspace{-0.1in}
\caption{Mass gaps in the axial channel from various fits plotted
  versus the momentum transfer in units of ${\bm n}^2$ for six
  ensembles.  The $a \Delta M_1$ (black filled circles) and $a \Delta
  E_1$ (black diamonds) are from fits to the two-point function using
  strategy $\{4\}$ (left panel), and $\{4^{N\pi}\}$ that uses a prior
  with a narrow width for the energy of a noninteracting $N(\bm
  0) \pi({\bm q})$ state (middle and right panels).  The output of the $\{2^{\rm
  sim}\}$ (or $\{2^{A_4}\}$) fits are $a \Delta {\widetilde M}_1$
  (blue squares) and $a \Delta {\widetilde E}_1$ (red triangles).  The
  dotted blue line is calculated assuming $a \Delta M_1$ is given by a
  noninteracting $N(\bm q) \pi(-{\bm q})$ state, while the red dotted
  line shows the $a \Delta E_1$ for a noninteracting $N(\bm
  0) \pi({\bm q})$ state.\looseness-1
 \label{fig:AFF-deltaM}}
\end{figure*}

The data for the energy gaps, $a \Delta {\widetilde M}_1$ and $a
\Delta {\widetilde E}_1$, obtained using three strategies $\{4,2^{\rm
  sim}\}$, $\{4^{N\pi},2^{A_4}\}$, and $\{4^{N\pi},2^{\rm sim}\}$ are presented in
Fig.~\ref{fig:AFF-deltaM} and compared against the values obtained
assuming that the excited states on the two sides of the operator are
$N(\bm q) \pi(-{\bm q})$ (blue dotted lines) and $N(\bm 0) \pi({\bm
  q})$ (red dotted lines), respectively.  The data exhibit the
following features: 
\begin{itemize}
\item
The energy gaps given by the fits to the three-point functions, $a
\Delta {\widetilde M}_1$ (blue squares) and $a \Delta {\widetilde
  E}_1$ (red triangles), differ significantly, depending on the momentum
transfer $\bm q $, and the difference increases with $\bm q$. \looseness-1
\item
The rough agreement between the blue dotted line and blue squares and
the red dotted line and red triangles improves as $M_\pi$ decreases
and indicates that $a \Delta {\widetilde M}_1$ and $a \Delta
{\widetilde E}_1$ correspond to $N(\bm q) \pi(-{\bm q})$ and $N(\bm 0)
\pi({\bm q})$ excited states, respectively. The agreement was found to
be even better for the physical mass ensemble investigated in
Ref.~\cite{Jang:2019vkm} using the clover-on-HISQ formulation.
\item
The values of $a \Delta { M}_1$ (black filled circle) and $a \Delta
{E}_1$ (black diamonds) obtained from $\{4\}$ and $\{4^{N\pi}\}$ fits
(left versus the right two panels) to the two-point function have a smaller 
difference.
\item
The agreement between the $a \Delta { E}_1$ (black diamonds) from the
$\{4^{N\pi}\}$ fits to the two-point function and the dotted red line
showing the energy of the noninteracting $N(\bm 0) \pi(\bm q)$ state
is by construction since the latter is used as a prior for $a { E}_1$ 
in the $\{4^{N\pi}\}$ fit.
\end{itemize}
The identification of $N(\bm 0) \pi({\bm q})$ and $N(-\bm
q) \pi({\bm q})$ as the leading excited states on the two sides of
the operator insertion is consistent with the predictions of chiral
perturbation theory~\cite{Bar:2018xyi,Bar:2019gfx,Hansen:2016qoz}.

An important consequence of the energy gaps, $a \Delta {\widetilde
  M}_1$ and $a \Delta {\widetilde E}_1$, being different and
corresponding to different momentum dependent excited states, $N(\bm
q) \pi(-{\bm q})$ versus $N(\bm 0) \pi({\bm q})$, is that their mass
gaps cannot be determined straightforwardly from fits to the two-point
function. For example, for our calculations, to get the mass gaps for
the ten $N(\bm q) \pi(-{\bm q})$ states from the $\bm p = 0$
correlator is unrealistic, even with a variational ansatz.  As shown
by the onset of the plateau in the effective mass plots in
Fig.~\ref{fig:2ptCOMP}, the ground state dominates at $\tau \gtrsim
1$~fm, i.e., the plateau starts at $9 < \tau_{\rm start} < 14$ in the
ensembles we have analyzed. Thus, the number of earlier time slices
sensitive to, and available for determining excited-state parameters
are 6--11, which restricts the analysis to a maximum of four states, including radial excitations.
Second, at these short times, the contributions of the full set of
excited states are still significant and even the first excited state
parameters, $M_1$ and $E_1$, extracted from the fit are typically
larger and $\tau_{\rm min}$ dependent. Third, these four-state fits
(as well as the three-state fits) have exposed flat directions in the fit
parameters leading to a large space of values with roughly similar
$\chi^2/$dof as illustrated in Fig.~\ref{fig:2ptCOMP}. In short, 
fits to the
data show many equally good solutions and the output values are
heavily influenced by the priors used to stabilize the fits.

To resolve a light excited state such as $N(\bm 0) \pi(\bm q)$, which
has a mass of about 1200~MeV as $\bm q \to 0$, from the ground state from fits to the
two-point function requires very high precision data at large enough
$\tau$ by which the higher states have died out sufficiently. In our
setup, this occurs for $\tau \gtrsim 1$~fm. Isolating two (actually a whole tower as
$\bm q \to 0$) states from the ``plateau'' region at $\tau \gtrsim
1$~fm will be challenging.  In short, our work suggests that determining the
masses and amplitudes of all the needed low-lying excited states from
fits to two-point functions constructed using a single nucleon or
multihadron interpolating operator is unlikely in the foreseeable
future.

One can improve the situation by working on anisotropic lattices
(setting the spacing in the time direction much finer than in the
three spatial directions to have more points to fit within the same physical time interval)
and/or by using a variational approach with many nucleon interpolating
operators, including relevant multihadron operators with the same
quantum numbers. The two methods have been implemented together
successfully in detailed calculations of the meson and baryon
excited-state spectra~\cite{Edwards:2011jj}. For matrix elements,
however, only exploratory calculations of nucleon charges using the
variational method have been
performed~\cite{Yoon:2016dij,Dragos:2016rtx}.  Each of these
approaches, unfortunately, requires additional/new simulations that are beyond
the scope of the current work.

We are, therefore, faced with the following possibilities to
systematically include all the relevant excited states to get percent
level precision:
\begin{itemize}
\item{[A]} Take only the ground state parameters from fits to the
  two-point function and leave all the excited state parameters,
  $ \Delta {\widetilde M}_i$ and $ \Delta {\widetilde E}_i$, to be
  determined from the three-point functions. This is the basis of our
  strategies $\{4^{N\pi},2^{A_4}\}$ and $\{4^{N\pi},2^{\rm sim}\}$,
  however so far we have been able to include a single excited
  state. To include the next, second, excited state with the current
  data, one could hardwire the $ \Delta {\widetilde M}_1$ and $ \Delta
  {\widetilde E}_1$, determined from a two-state fit, in a three-state
  fit with only $ \Delta {\widetilde M}_2$ and $ \Delta {\widetilde
  E}_2$ free.  Our attempts at this failed---the $\chi^2$ does
  not decrease by two units for each additional parameter as required
  to satisfy the Akaike information criteria, and the parameter values
  have over 100\% errors.  We are also not able to estimate how
  precise the data need to be for this approach to work given the
  large flat regions in the $\chi^2$ landscape, evident already by the
  range of $ \Delta {\widetilde M}_1$ and $ \Delta {\widetilde
  E}_1$ values, and the large number of possible states that could
  contribute.
\item{[B]} Assume, based on chiral perturbation theory, that $N(\bm q)
  \pi(-{\bm q})$ and $N(\bm 0) \pi({\bm q})$ are the relevant first
  excited states and hardwire their noninteracting energies for $
  \Delta { M}_1$ and $ \Delta { E}_1$ in fits to the three-point
  function. For the second and higher excited states, one can again
  resort to $\chi$PT or take the estimate of the next lowest energy level
  from fits to the two-point function. This approach has recently been
  used in Ref.~\cite{Bali:2019yiy}. In our case, the
  $\{4^{N\pi},3^{\ast}\}$ strategy is a step in this direction; 
  however, since $\{4^{N\pi},2^{A_4}\}$ and $\{4^{N\pi},2^{\rm sim}\}$
  do a better job of satisfying PCAC, one could add a third state with
  fixed $ \Delta { M}_2$ and $ \Delta { E}_2$ to the latter when
  making the fits. Our attempts at adding a third state to the
  $\{4^{N\pi},2^{\rm sim}\}$ fit led to both an overparameterization
  and essentially undetermined values for all the extra parameters.
\item{[C]} Determine the spectrum of [multihadron] excited states in
  a finite box from a variational calculation of  two-point functions
  with a large enough basis of operators and use them as priors in
  fits to the three-point functions. Our contention, based on the
  current analyses, is that, for the first excited state, the energy
  gaps will be close to those given by $\{4^{N\pi},2^{A_4}\}$ or the
  $\{4^{N\pi},2^{\rm sim}\}$ strategies and the fits to the
  three-point functions with current statistics will not be sensitive
  to the higher states.
\end{itemize}

In short, determining the spectrum of multiparticle excited states
that contribute significantly is essential for obtaining ground state
matrix elements in the axial channel. The $A_4$ correlator allows us
to nonperturbatively identify $N \pi$ as giving the leading
contribution, consistent with $\chi$PT analysis, however, more work is
needed to determine the second relevant [multiparticle] excited state, which
may be necessary to reach percent level precision.  In
Sec.~\ref{sec:VFF}, we show that similar issues need to be addressed
in the vector channel also, but the electric and magnetic form factors
are less sensitive to the values of the excited-state energies.

\section{Comparison of the axial form factors extracted using 4 strategies}
\label{sec:compAFF}

\begin{figure*}[tbp] 
\subfigure
{
    \includegraphics[width=0.235\linewidth]{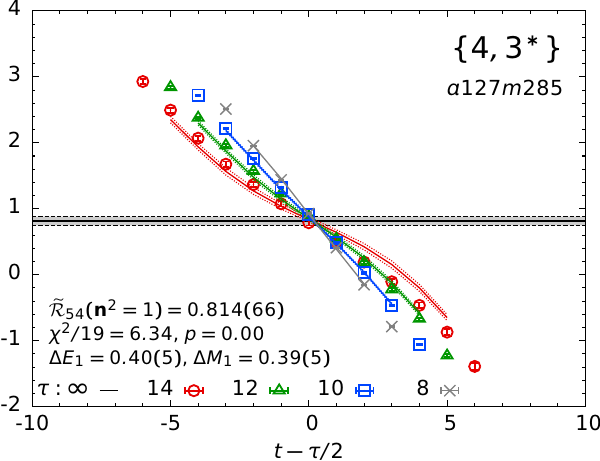}  
    \includegraphics[width=0.235\linewidth]{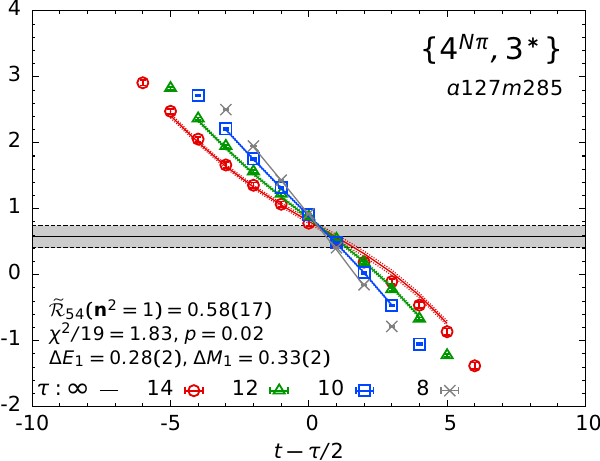} 
    \includegraphics[width=0.235\linewidth]{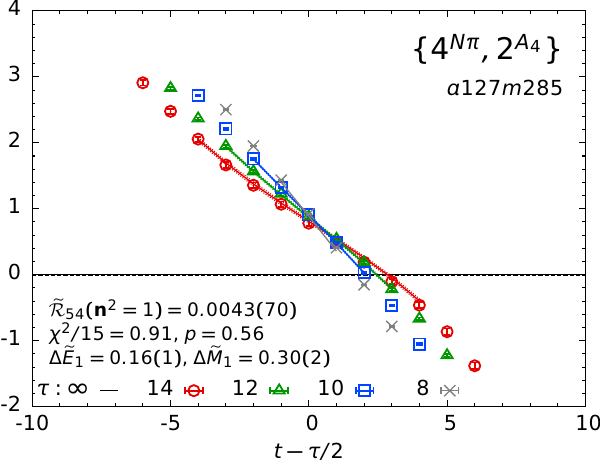} 
    \includegraphics[width=0.235\linewidth]{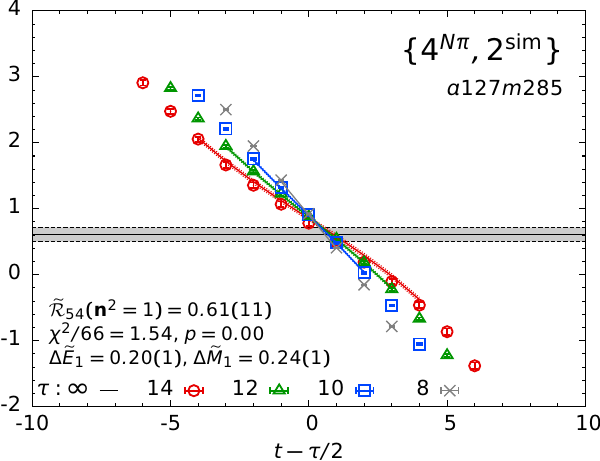} 
}
{
}
{
    \includegraphics[width=0.235\linewidth]{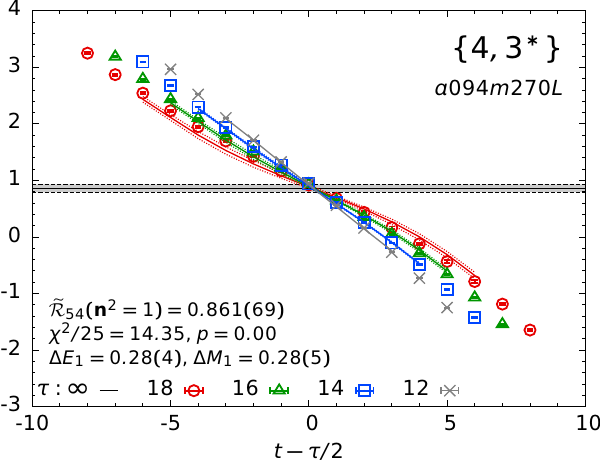}  
    \includegraphics[width=0.235\linewidth]{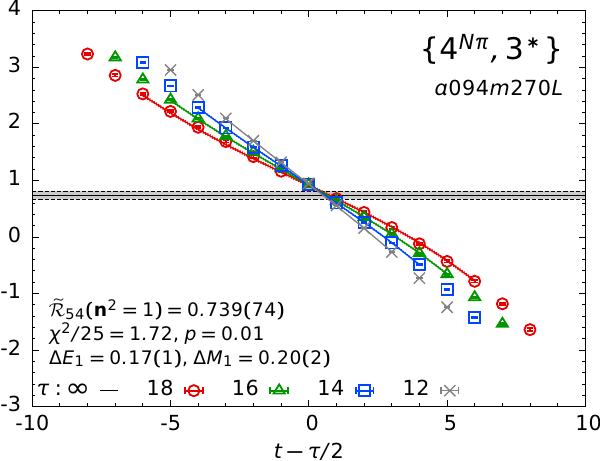} 
    \includegraphics[width=0.235\linewidth]{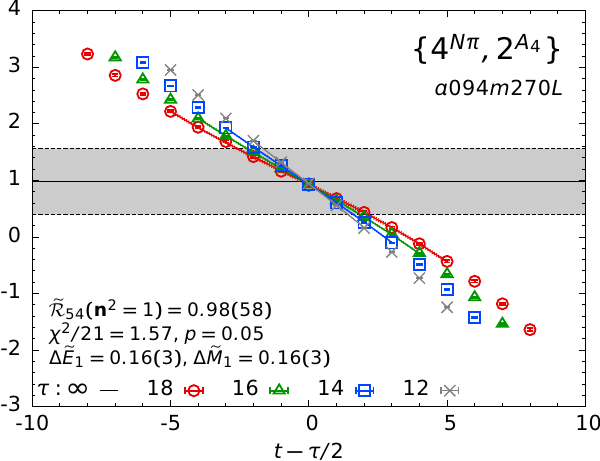} 
    \includegraphics[width=0.235\linewidth]{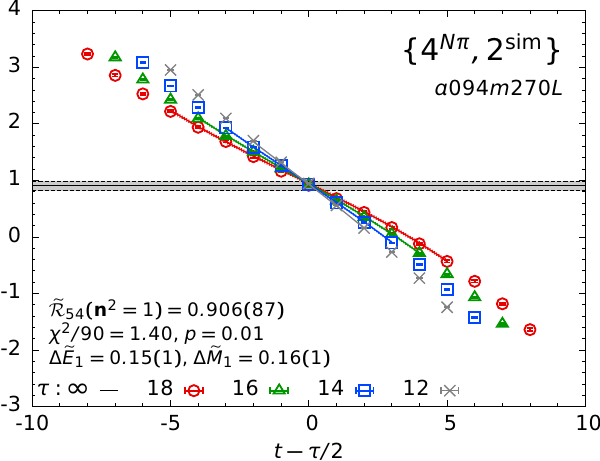} 
}
{
    \includegraphics[width=0.235\linewidth]{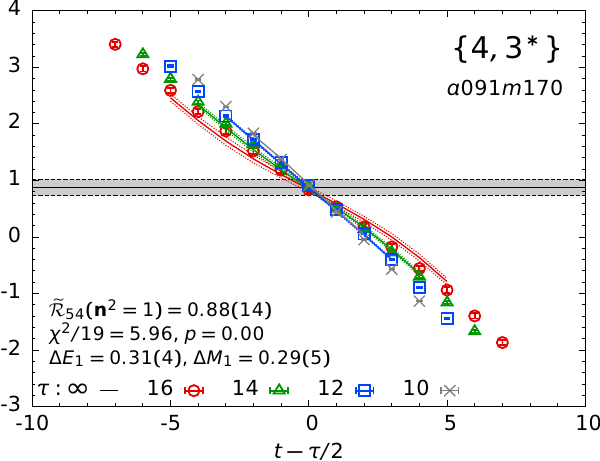}  
    \includegraphics[width=0.235\linewidth]{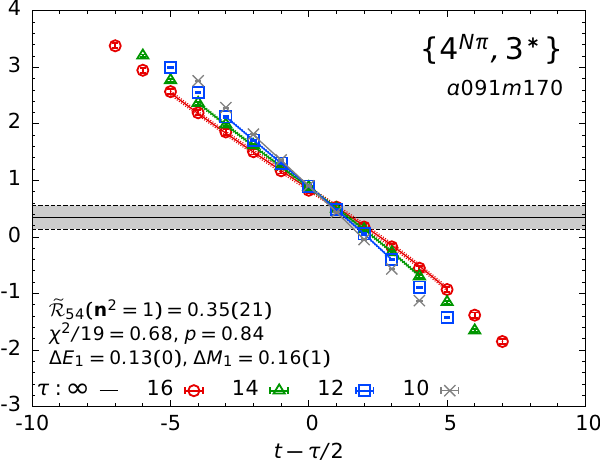} 
    \includegraphics[width=0.235\linewidth]{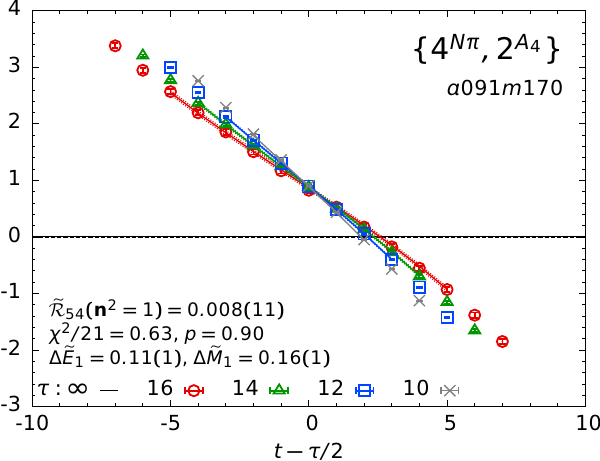} 
    \includegraphics[width=0.235\linewidth]{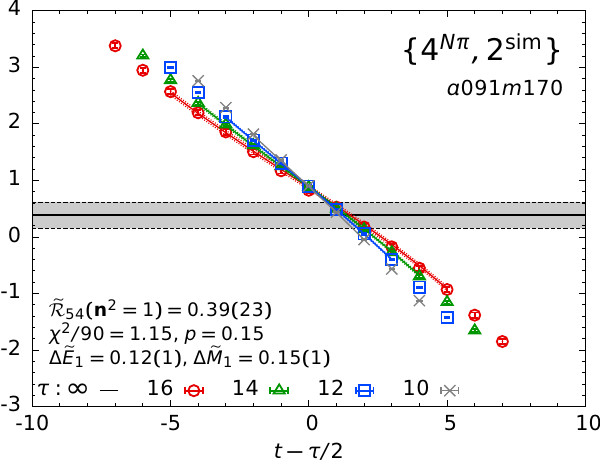} 
}
{
    \includegraphics[width=0.235\linewidth]{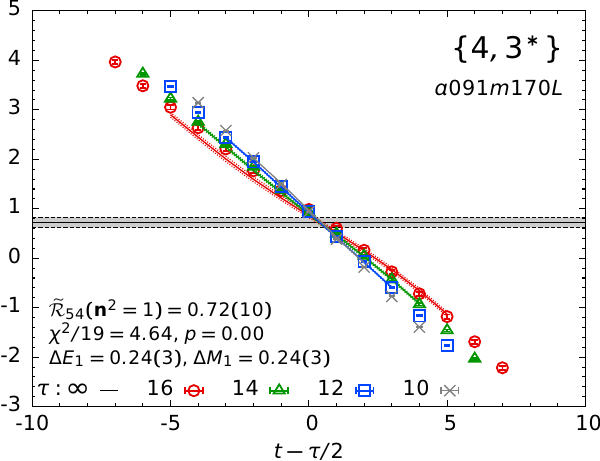} 
    \includegraphics[width=0.235\linewidth]{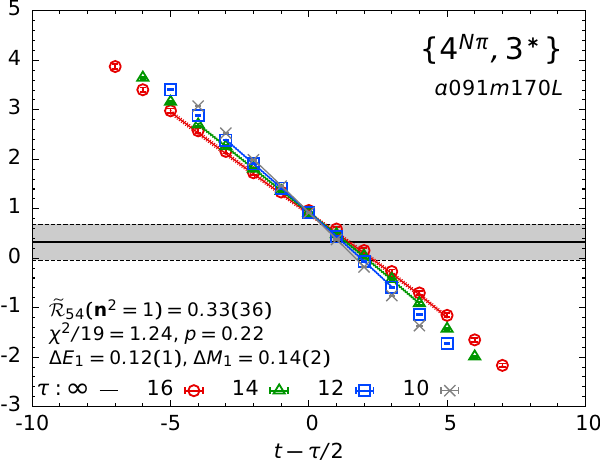} 
    \includegraphics[width=0.235\linewidth]{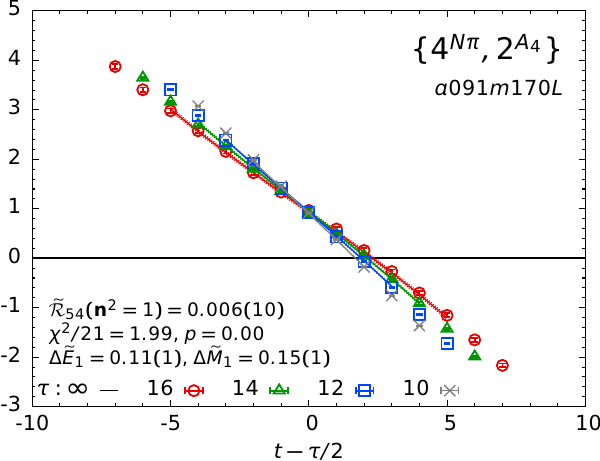} 
    \includegraphics[width=0.235\linewidth]{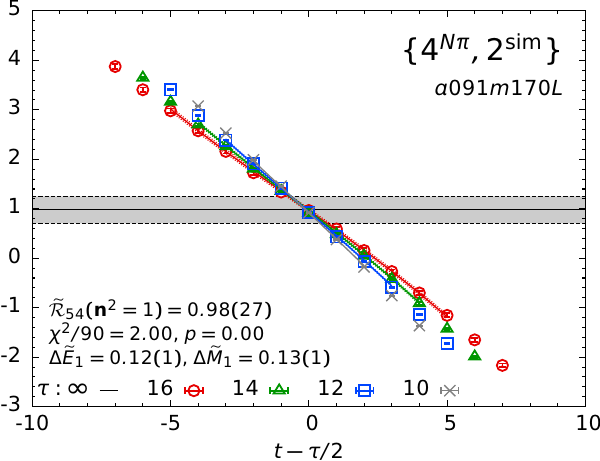} 
}
{
    \includegraphics[width=0.235\linewidth]{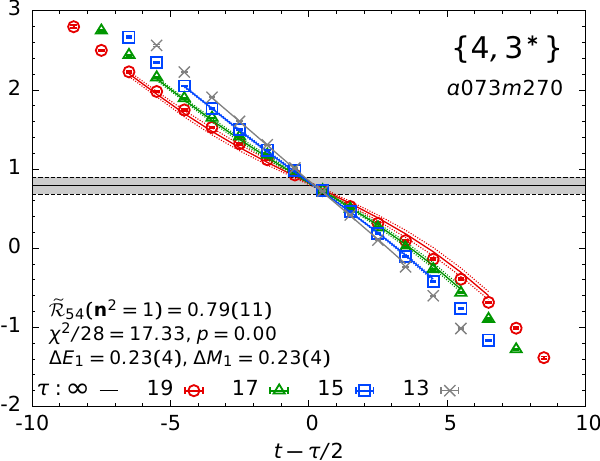}  
    \includegraphics[width=0.235\linewidth]{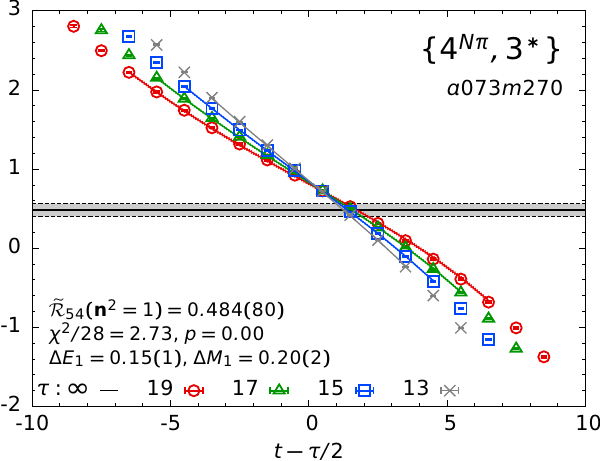} 
    \includegraphics[width=0.235\linewidth]{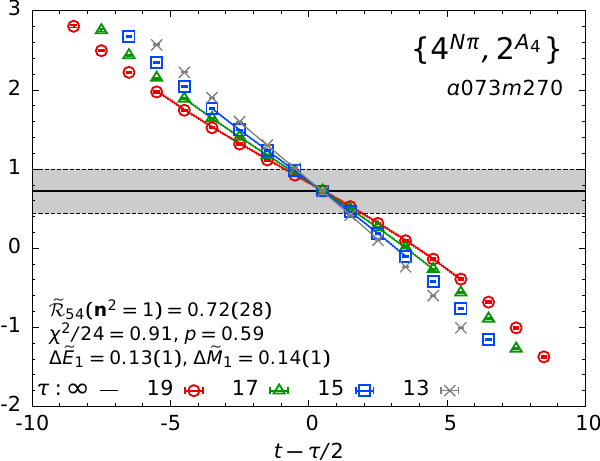} 
    \includegraphics[width=0.235\linewidth]{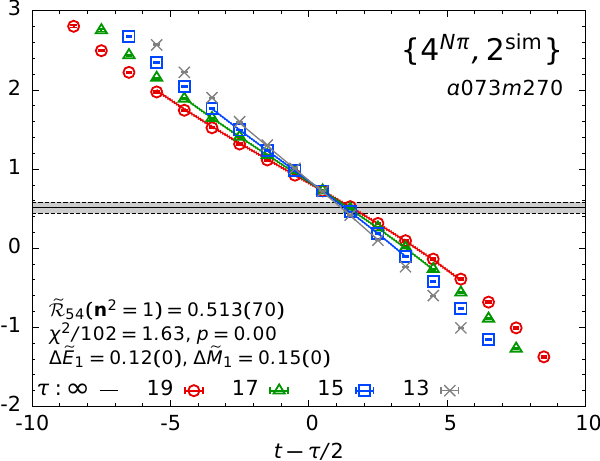} 
}
{
    \includegraphics[width=0.235\linewidth]{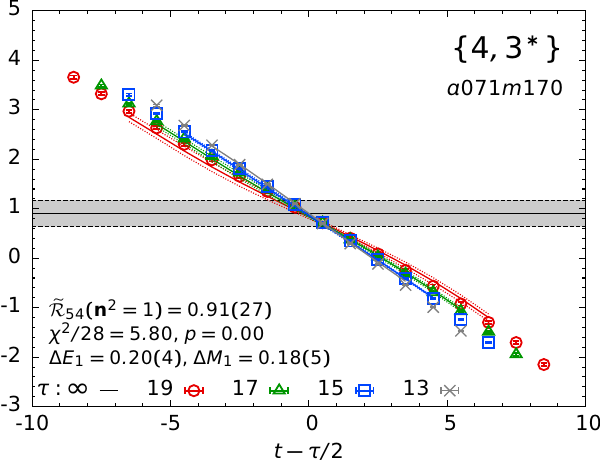}  
    \includegraphics[width=0.235\linewidth]{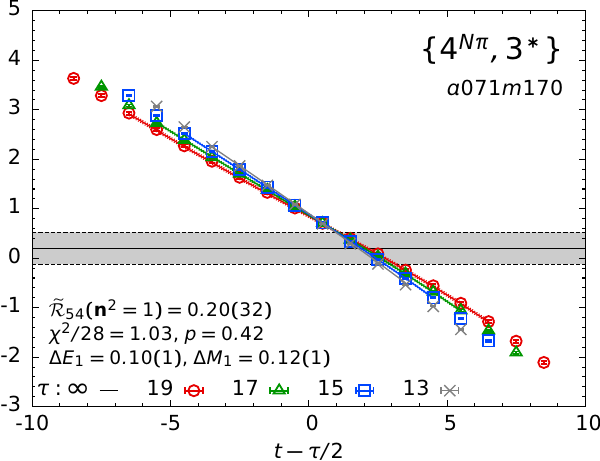} 
    \includegraphics[width=0.235\linewidth]{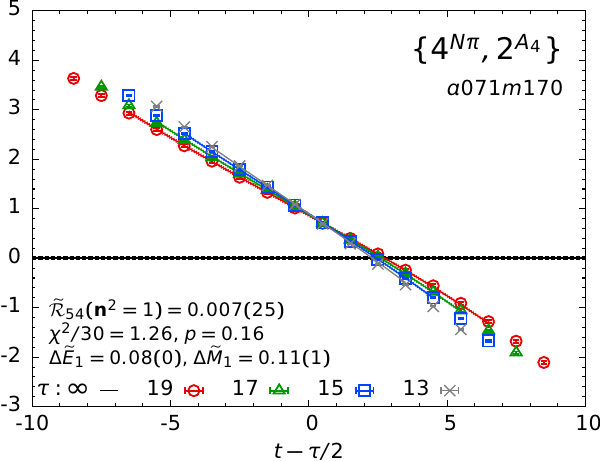} 
    \includegraphics[width=0.235\linewidth]{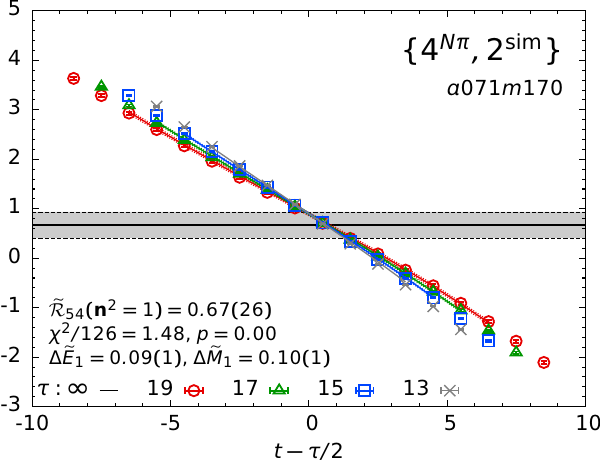} 
}
\caption{The ratio $R_{54}$, defined in
  Eq.~\protect\eqref{eq:r2ff-GPGA4}, is plotted versus the shifted
  operator insertion time $t -\tau/2$ for ${\bm n} = (0,0,1)$.
  Results of the fits with $\{4^{},3^{\ast}\}$ (left column),
  $\{4^{N\pi},3^{\ast}\}$ ( second column), $\{4^{N\pi},2^{A_4}\}$
  (third), and $\{4^{N\pi},2^{\rm sim}\}$ (right) strategies are shown
  by lines connecting the data points. The $\tau \to \infty$ value is
  shown by the gray band.  The $y$-axis interval is the same for a given
  row to facilitate comparison of the result and the error.  The
  legends give the analysis strategy, the ensemble ID, the ground
  state value (the gray band), the $\chi^2/$dof and the $p$-value of
  the fit, and the mass gaps, $\Delta M_1$ and $\Delta E_1$ (or
  $\Delta {\widetilde M}_1$ and $\Delta {\widetilde E}_1$ for $\{2^{A_4}\}$ 
or $\{2^{\rm sim}\}$ fits), of the first excited state on the two sides of the
  operator. For each $\tau$, only the data points connected by lines with the same color as
  the symbols are included in the  simultaneous fits.   \label{fig:affA4COMP}}
\end{figure*}

\begin{figure*}[tbp] 
\subfigure
{
    \includegraphics[width=0.235\linewidth]{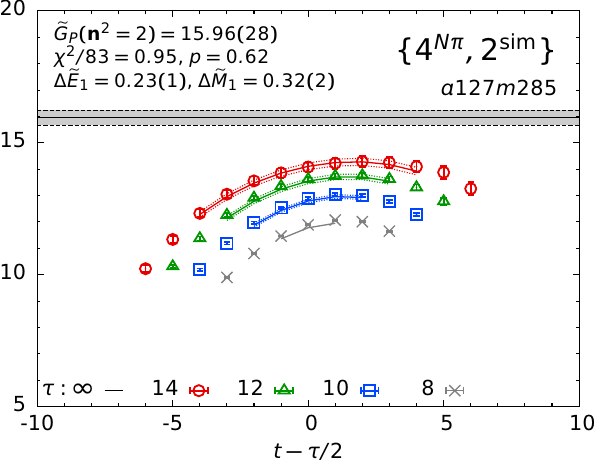} 
    \includegraphics[width=0.235\linewidth]{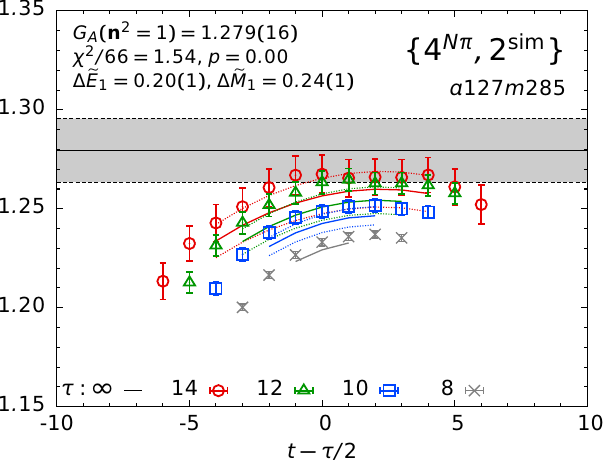}  
    \includegraphics[width=0.235\linewidth]{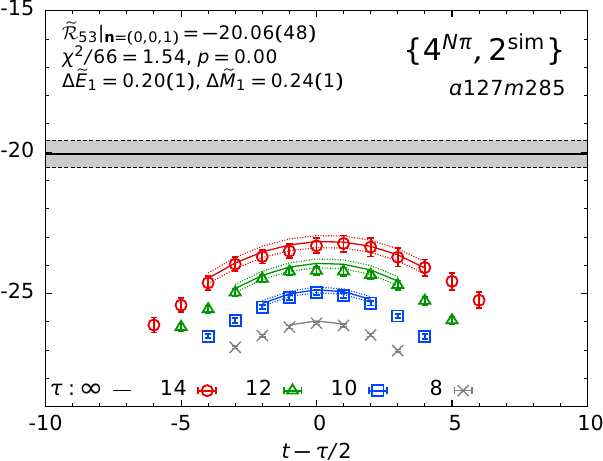} 
    \includegraphics[width=0.235\linewidth]{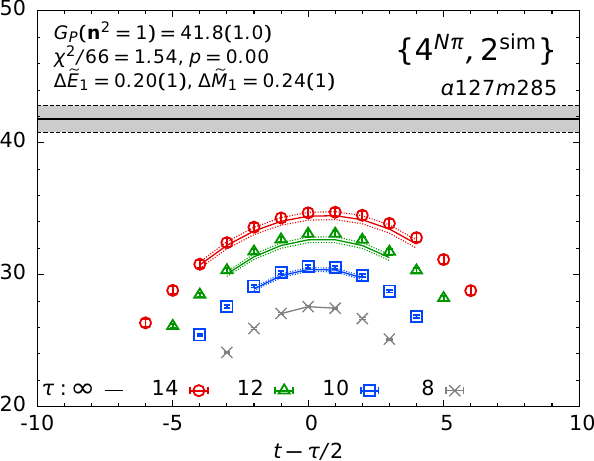} 
}
{
}
{
    \includegraphics[width=0.235\linewidth]{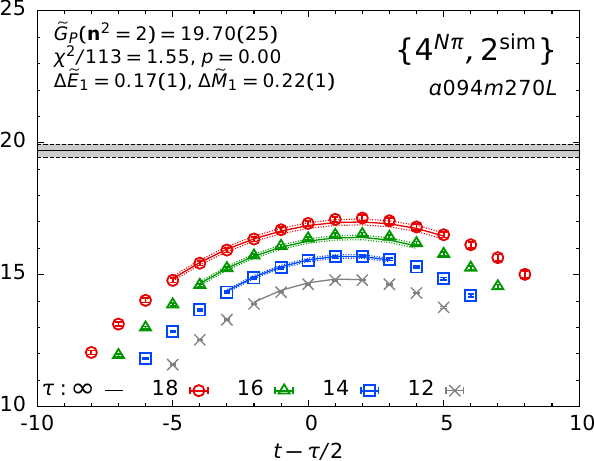} 
    \includegraphics[width=0.235\linewidth]{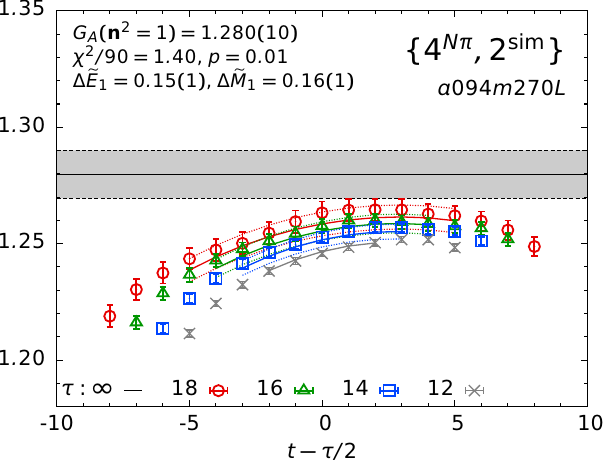}  
    \includegraphics[width=0.235\linewidth]{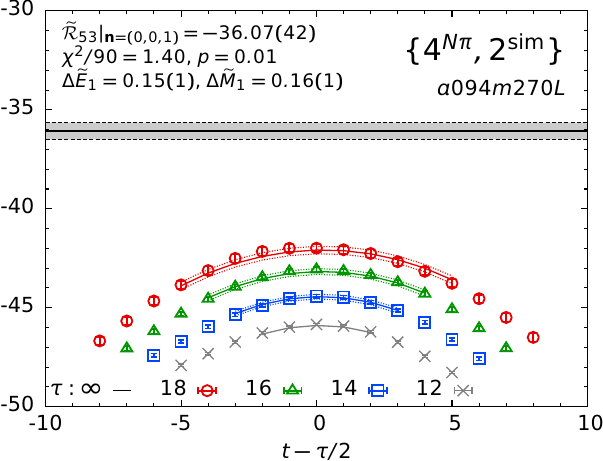} 
    \includegraphics[width=0.235\linewidth]{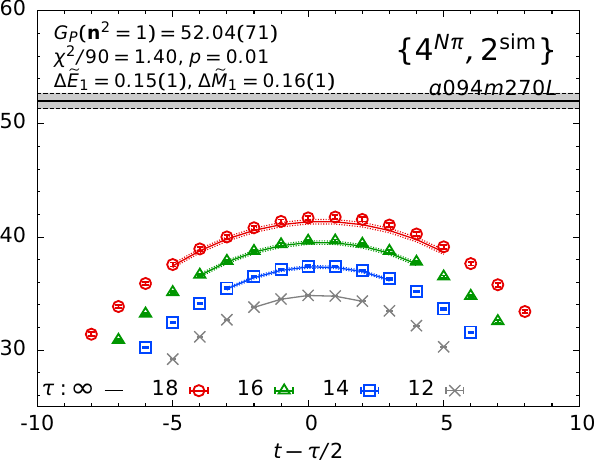} 
}
{
    \includegraphics[width=0.235\linewidth]{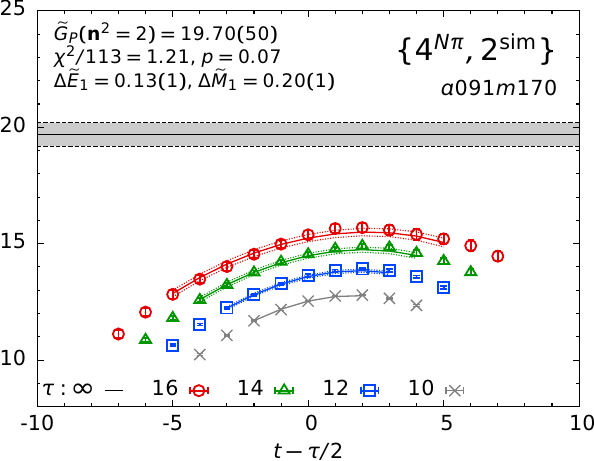} 
    \includegraphics[width=0.235\linewidth]{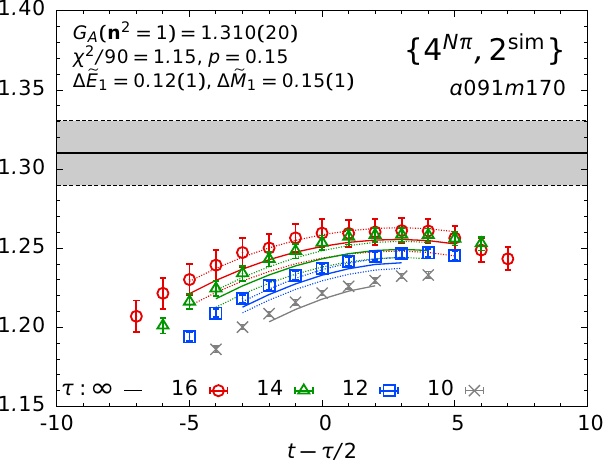}  
    \includegraphics[width=0.235\linewidth]{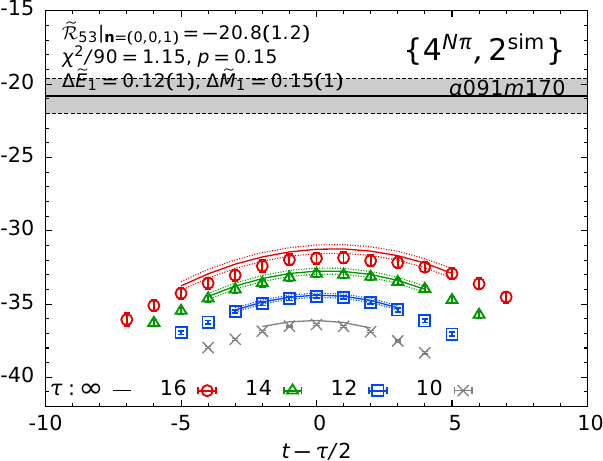} 
    \includegraphics[width=0.235\linewidth]{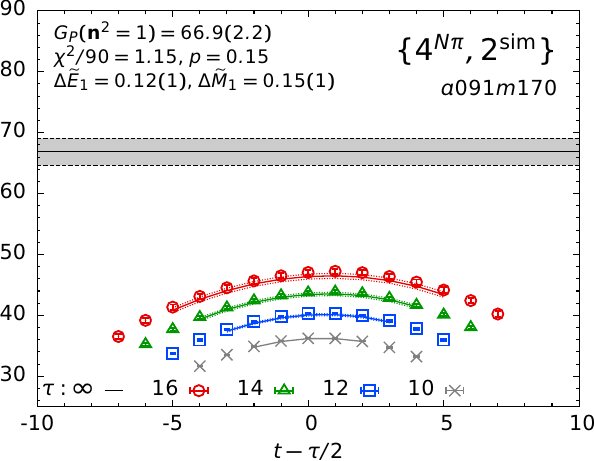} 
}
{
    \includegraphics[width=0.235\linewidth]{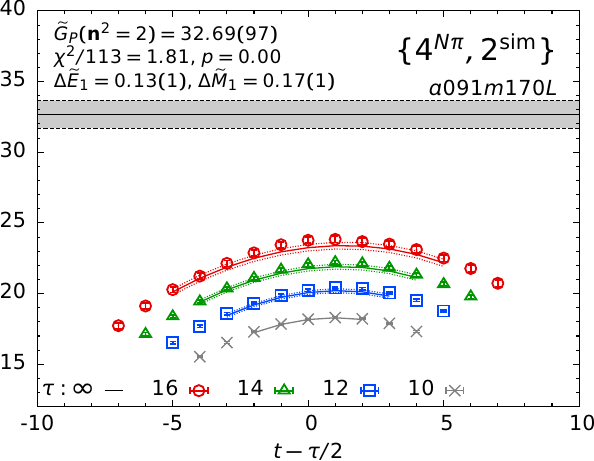} 
    \includegraphics[width=0.235\linewidth]{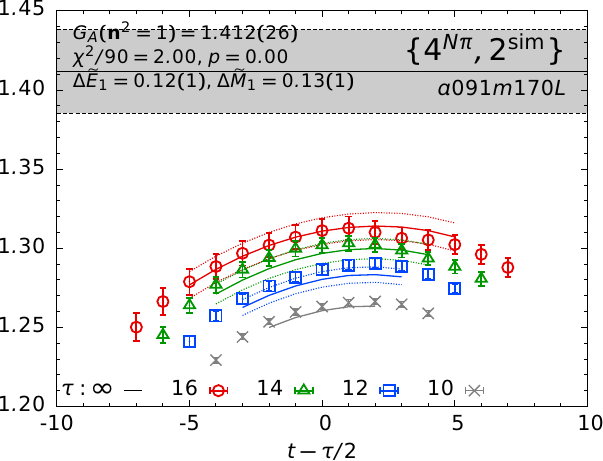}  
    \includegraphics[width=0.235\linewidth]{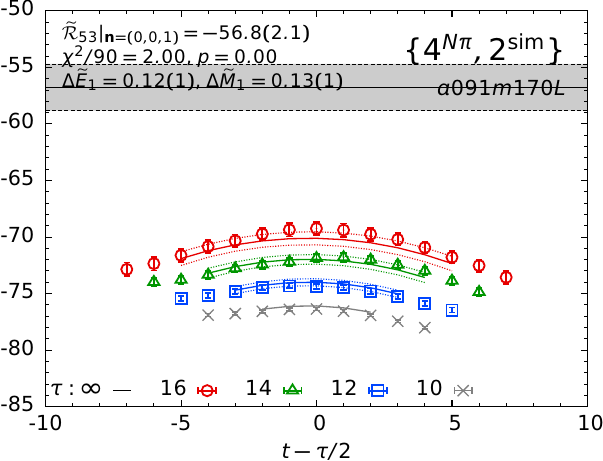} 
    \includegraphics[width=0.235\linewidth]{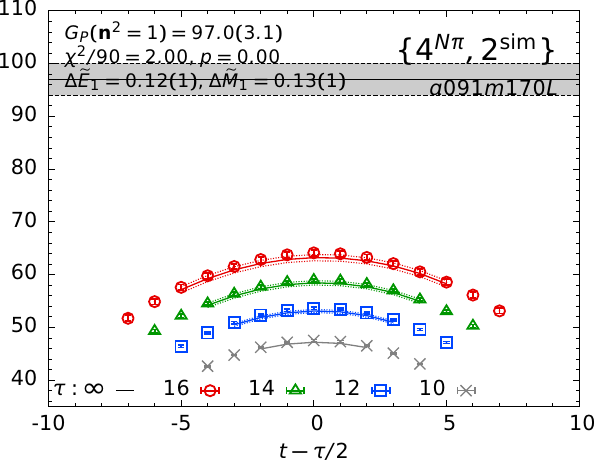} 
}
{
    \includegraphics[width=0.235\linewidth]{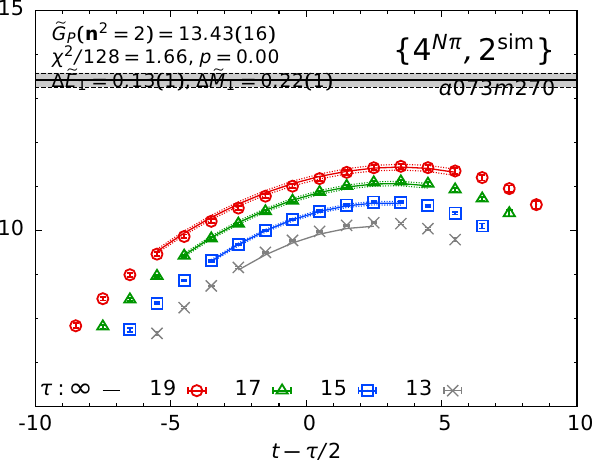} 
    \includegraphics[width=0.235\linewidth]{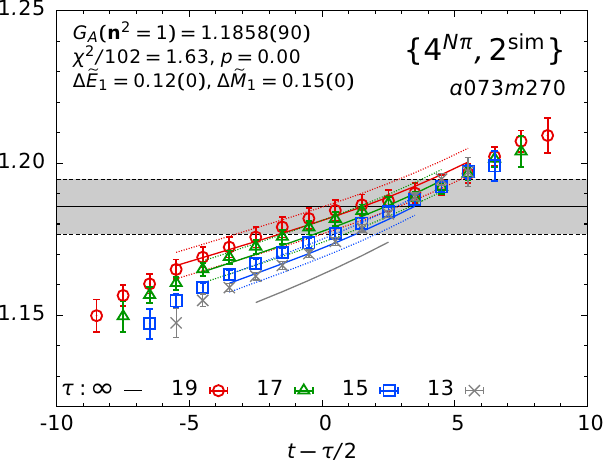}  
    \includegraphics[width=0.235\linewidth]{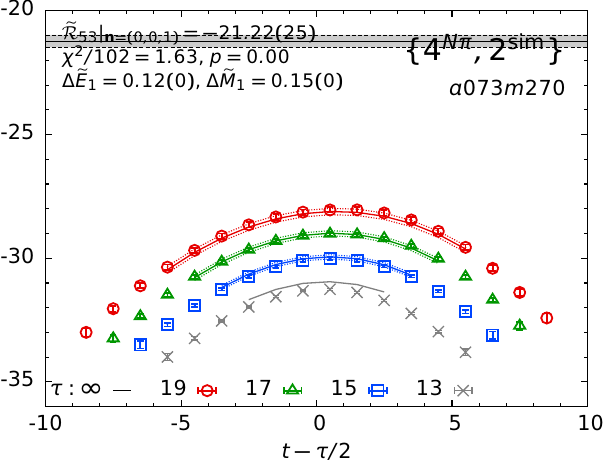} 
    \includegraphics[width=0.235\linewidth]{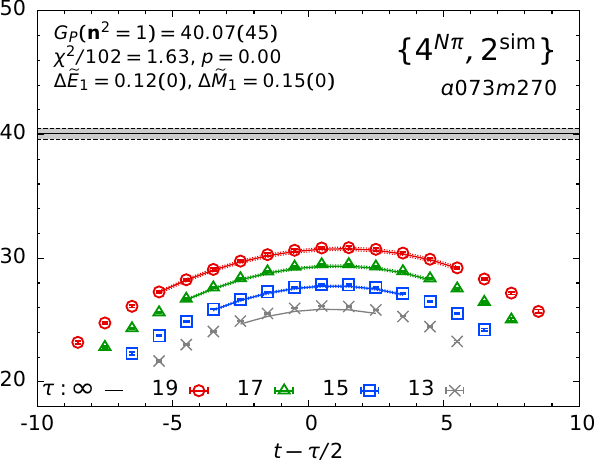} 
}
{
    \includegraphics[width=0.235\linewidth]{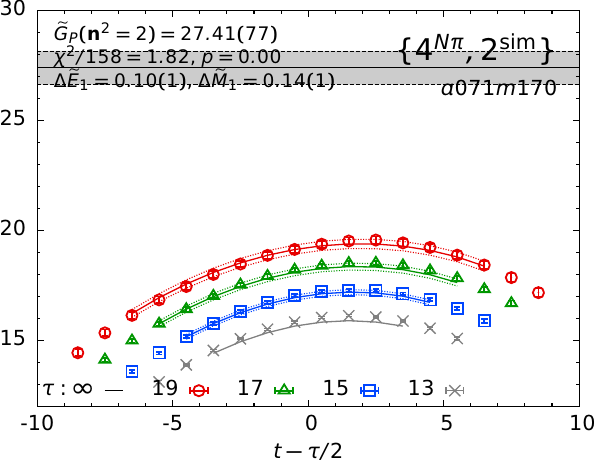} 
    \includegraphics[width=0.235\linewidth]{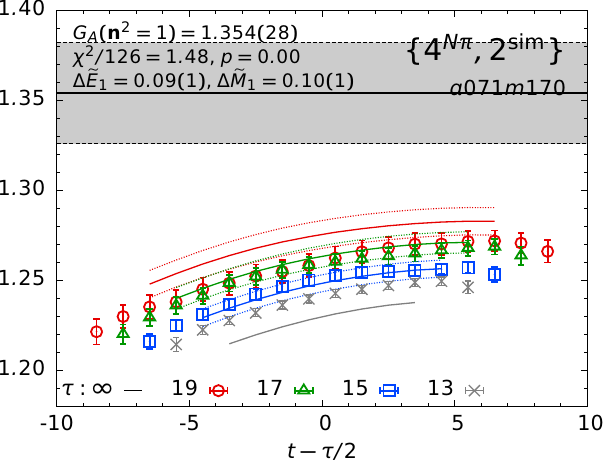}  
    \includegraphics[width=0.235\linewidth]{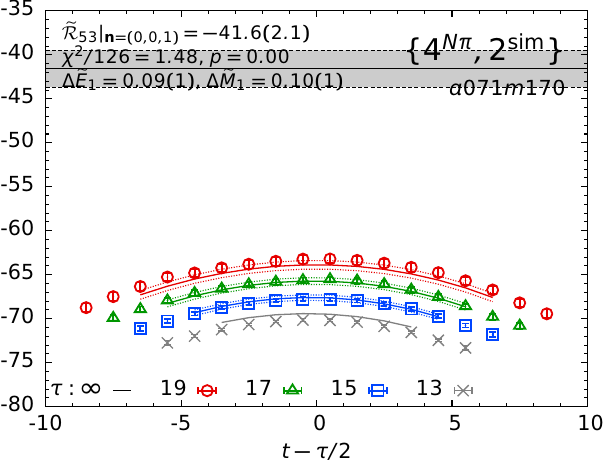} 
    \includegraphics[width=0.235\linewidth]{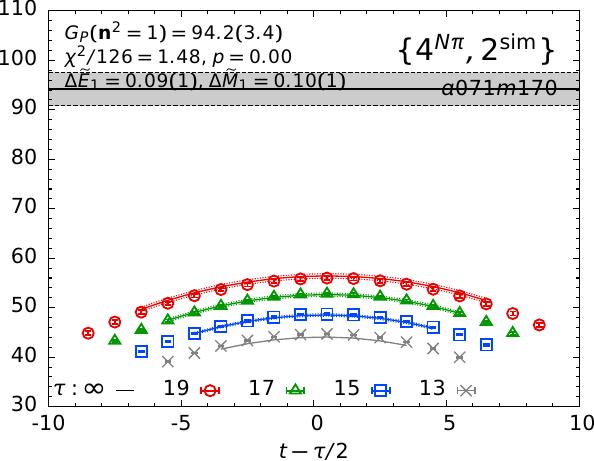} 
}
\caption{Matrix elements of the axial and pseudoscalar currents that
  give (i) ${\widetilde G}_P$ (from $R_{51}$ with ${\bm n}^2=2$ defined in
  Eq.~\protect\eqref{eq:r2ff-GPGA1}] in column one, (ii) $G_A$ (from
  $R_{53}$ with ${\bm n}^2 = 1$ and $q_z=0$ defined in
  Eq.~\protect\eqref{eq:r2ff-GPGA3}] in the second column, (iii) the
  combination $\frac{{\widetilde G}_P}{2M_N} - \frac{(M+E)}{q_3^2} G_A
  $ [from $R_{53}$ with $q_3 = (0,0,1)2\pi/La$] in the third column,
  and (iv) $G_P$ (from $R_5$ defined in
  Eq.~\protect\eqref{eq:r2ff-GP}) in the right column.  All data are
  with the $\{4^{N\pi},2^{\rm sim}\}$ strategy and plotted versus the
  shifted operator insertion time $t -\tau/2$.  The rest is the same 
  as in Fig.~\protect\ref{fig:affA4COMP}. 
\label{fig:aff4STcomp}}
\end{figure*}

\begin{figure*}[tbp] 
\subfigure
{
    \includegraphics[width=0.23\linewidth]{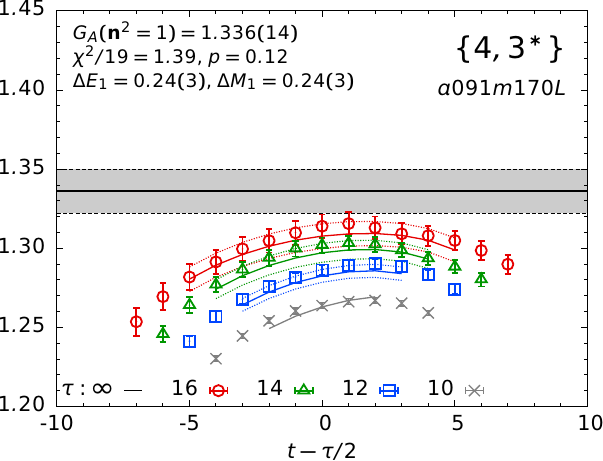}  
    \includegraphics[width=0.23\linewidth]{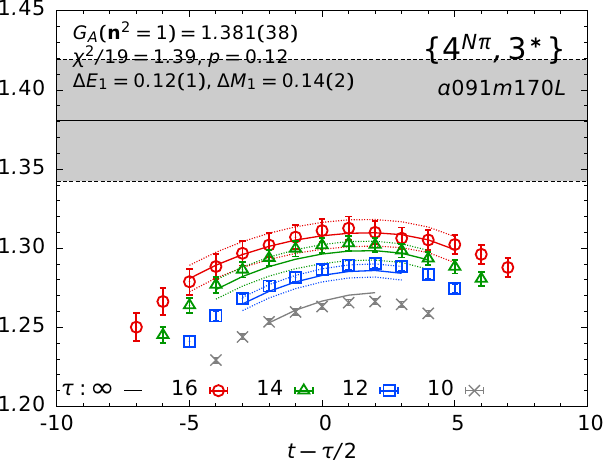}  
    \includegraphics[width=0.23\linewidth]{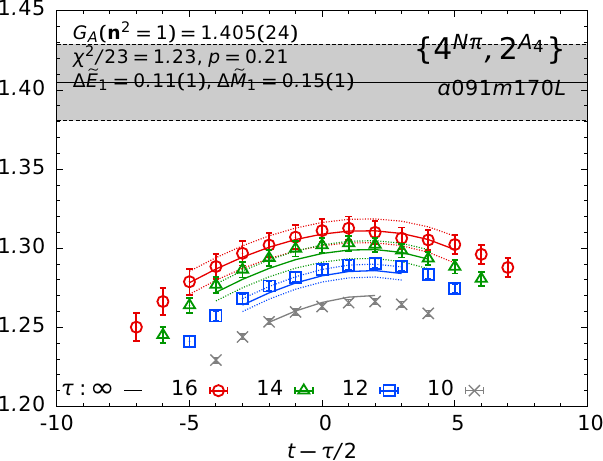}  
    \includegraphics[width=0.23\linewidth]{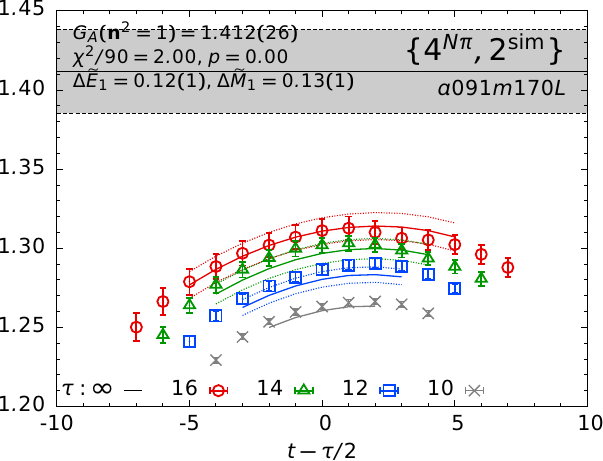}  
}
{
    \includegraphics[width=0.23\linewidth]{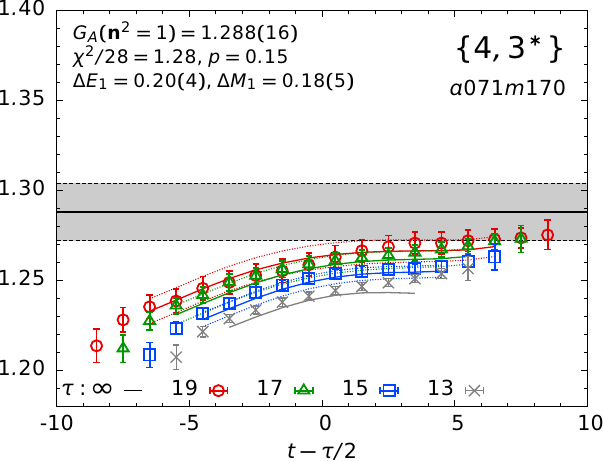}  
    \includegraphics[width=0.23\linewidth]{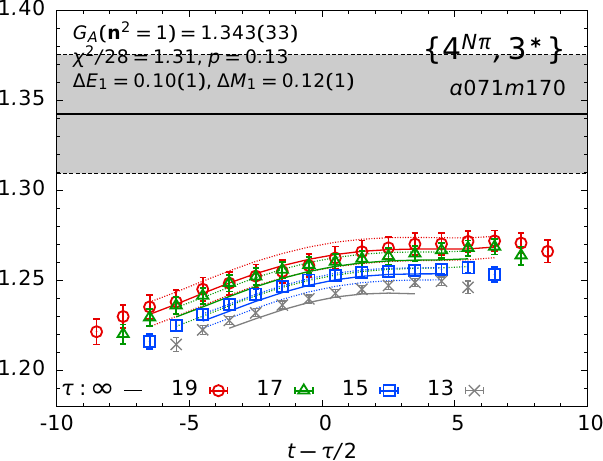}  
    \includegraphics[width=0.23\linewidth]{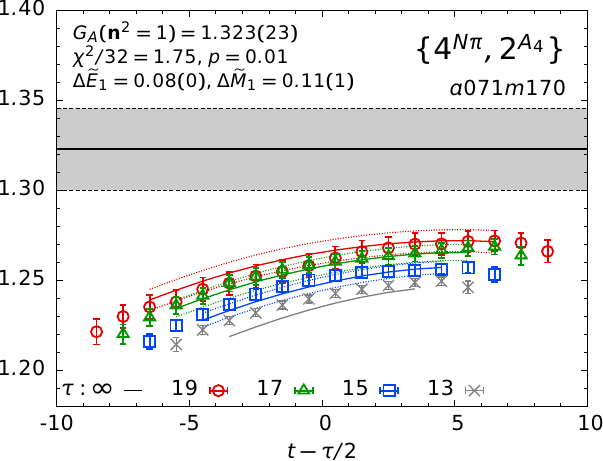}  
    \includegraphics[width=0.23\linewidth]{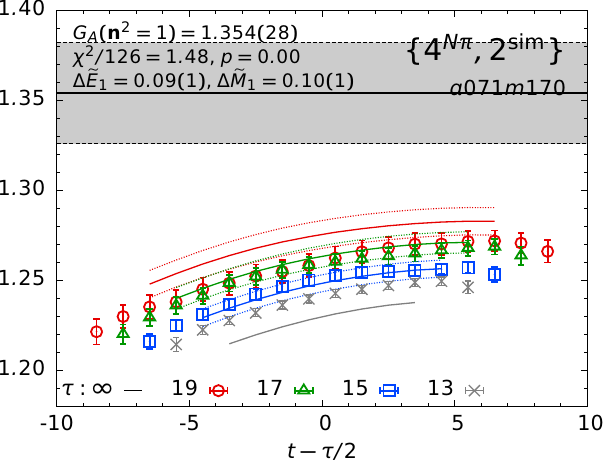}  
}
{
    \includegraphics[width=0.23\linewidth]{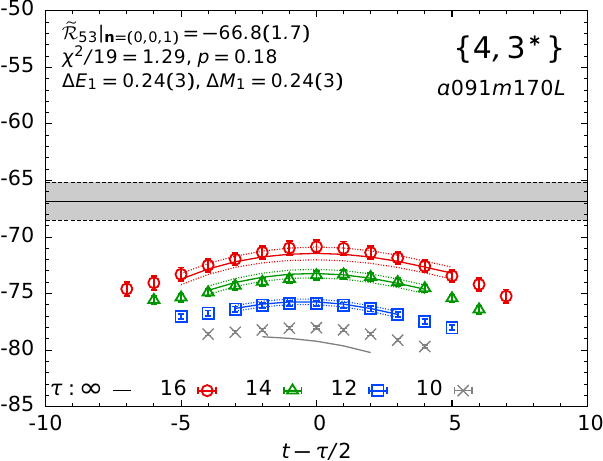} 
    \includegraphics[width=0.23\linewidth]{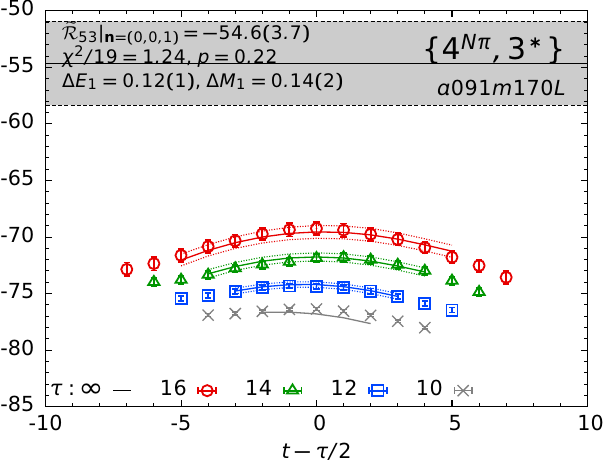} 
    \includegraphics[width=0.23\linewidth]{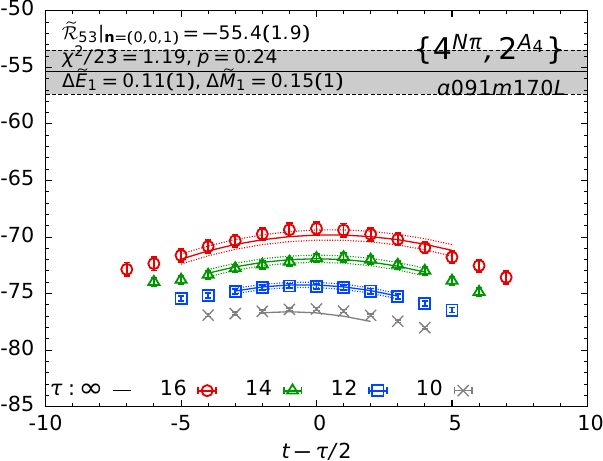} 
    \includegraphics[width=0.23\linewidth]{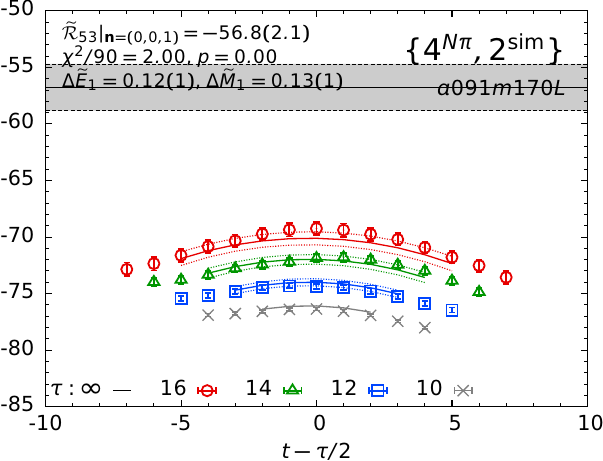} 
}
{
    \includegraphics[width=0.23\linewidth]{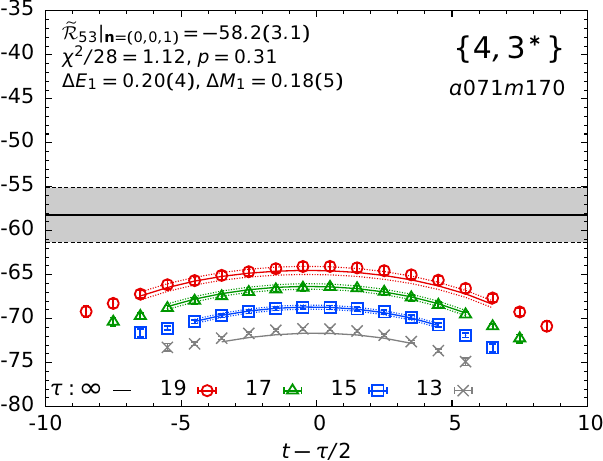} 
    \includegraphics[width=0.23\linewidth]{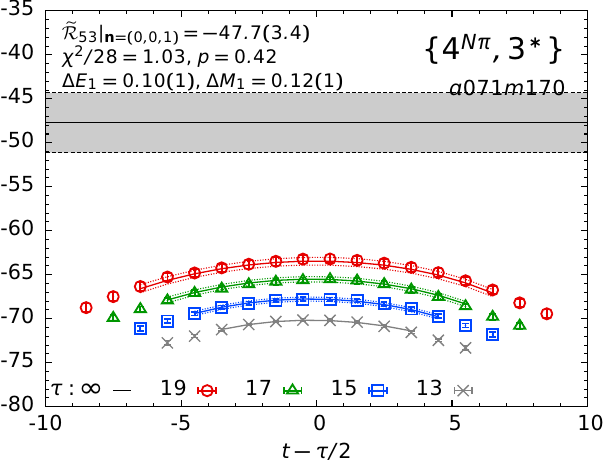} 
    \includegraphics[width=0.23\linewidth]{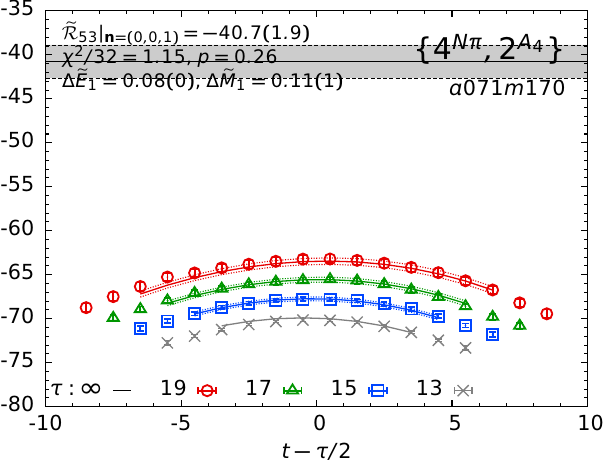} 
    \includegraphics[width=0.23\linewidth]{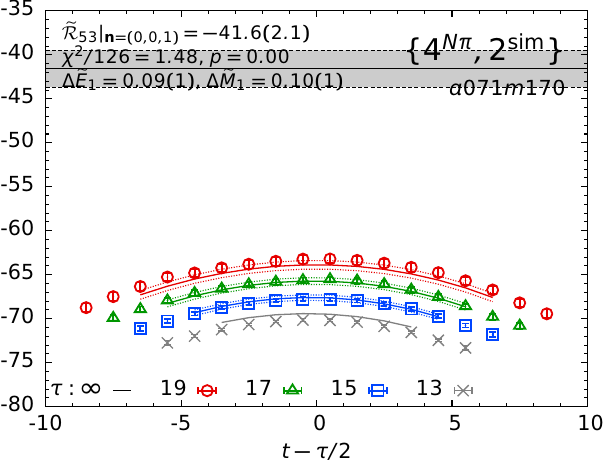} 
}
{
    \includegraphics[width=0.23\linewidth]{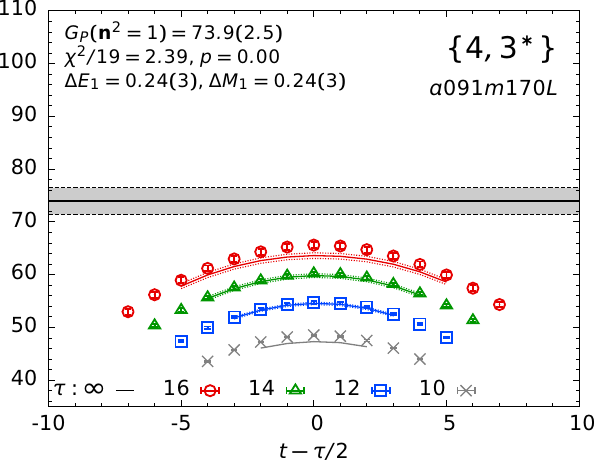} 
    \includegraphics[width=0.23\linewidth]{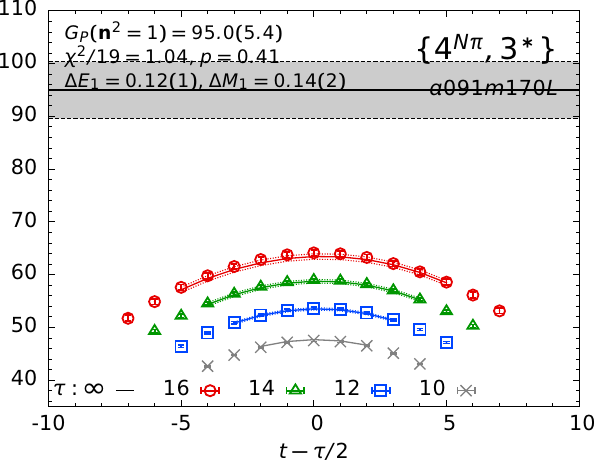} 
    \includegraphics[width=0.23\linewidth]{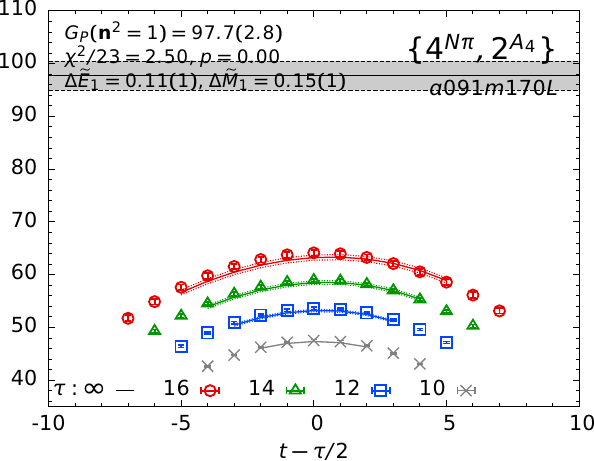} 
    \includegraphics[width=0.23\linewidth]{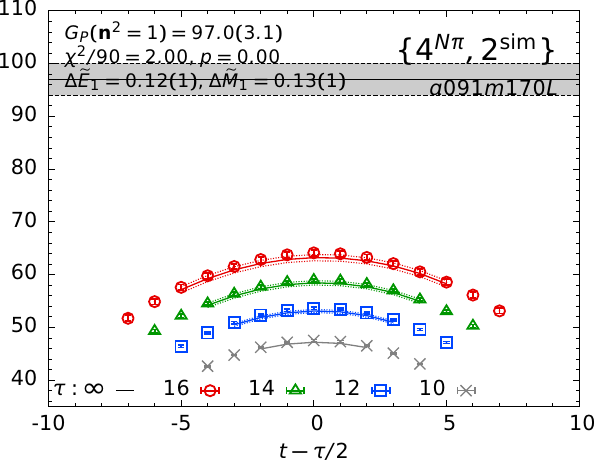} 
}
{
    \includegraphics[width=0.23\linewidth]{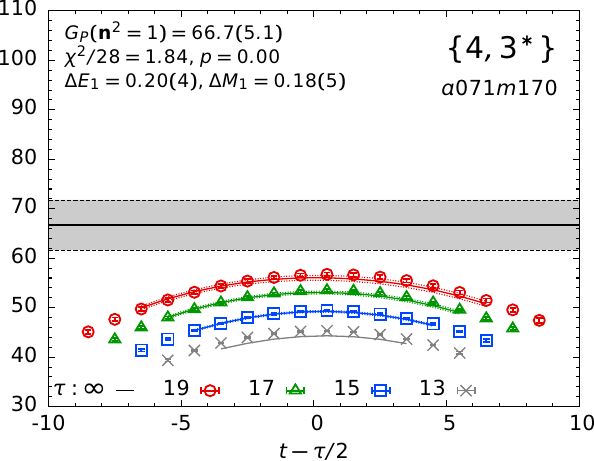} 
    \includegraphics[width=0.23\linewidth]{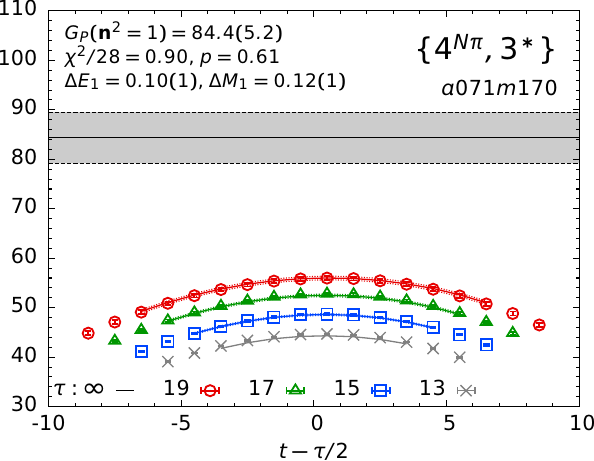} 
    \includegraphics[width=0.23\linewidth]{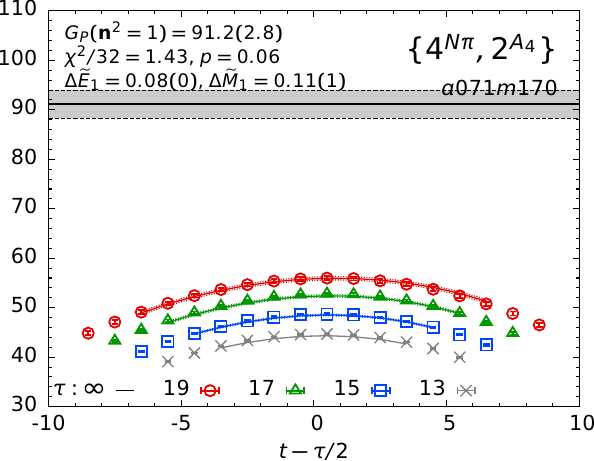} 
    \includegraphics[width=0.23\linewidth]{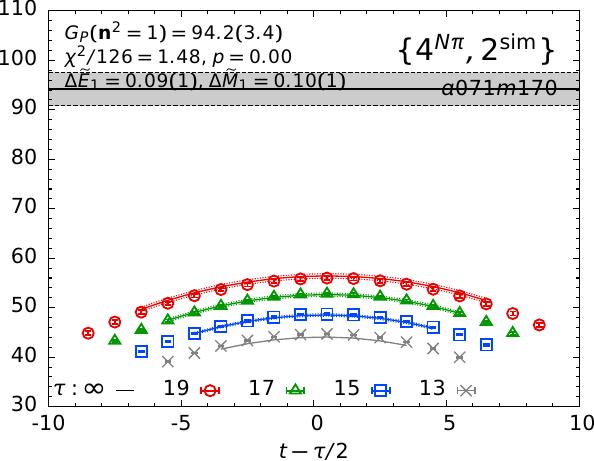} 
}
\caption{Matrix elements at momentum transfer ${\bm n}^2 = 1$ that
  give $G_A$ [from $R_{53}$ with $q_z=0$ defined in
  Eq.~\protect\eqref{eq:r2ff-GPGA3}] in rows one and two, the
  combination $\frac{{\widetilde G}_P}{2M_N} - \frac{(M+E)}{q_3^2} G_A
  $ [from $R_{53}$ with $q_3 = (0,0,1)2\pi/La$] in rows three and
  four, and $G_P$ [from $R_5$ defined in
  Eq.~\protect\eqref{eq:r2ff-GP}] in rows five and six. Data from the
  $a091m170L$ (rows one, three and five) and $a071m170$ (rows two,
  four and six) ensembles are plotted versus the shifted operator
  insertion time $t -\tau/2$. The four panels in each row show the
  data and fits from the four strategies, $\{4^{},3^{\ast}\}$ (left),
  $\{4^{N\pi},3^{\ast}\}$ (second), $\{4^{N\pi},2^{A_4}\}$ (third),
  and $\{4^{N\pi},2^{\rm sim}\}$ (right).  The $y$-axis interval is
  chosen to be the same for each row to facilitate comparison of
  the result and the error.  The rest is the same
  as in Fig.~\protect\ref{fig:affA4COMP}.
  \label{fig:aff4Npisimcomp}}
\end{figure*}

\begin{figure*}[tb] 
\subfigure
{
    \includegraphics[width=0.44\linewidth]{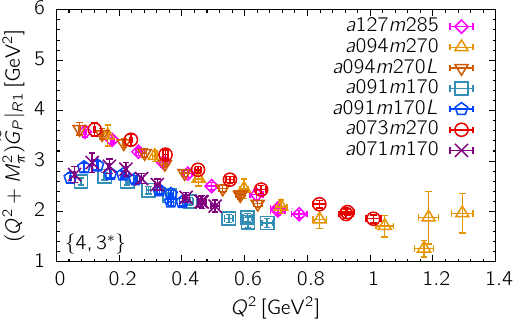}  
    \includegraphics[width=0.44\linewidth]{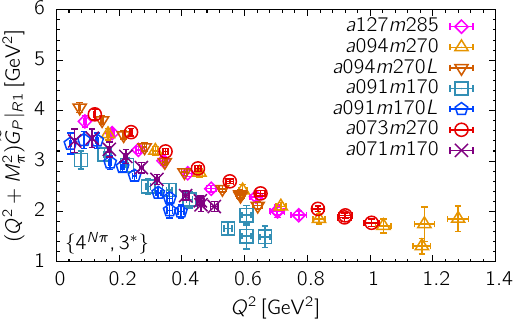}  
}
\subfigure
{
    \includegraphics[width=0.44\linewidth]{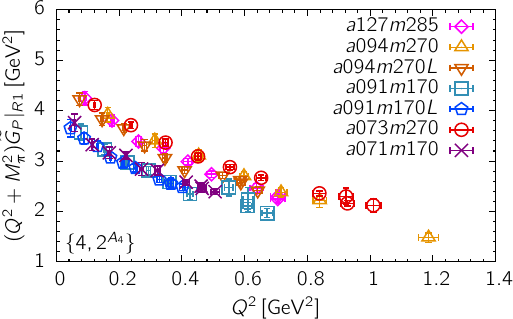}  
    \includegraphics[width=0.44\linewidth]{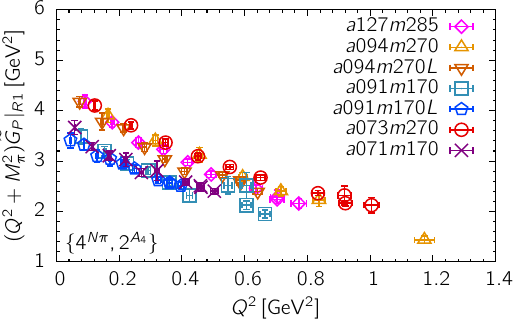}  
}
\subfigure
{
    \includegraphics[width=0.44\linewidth]{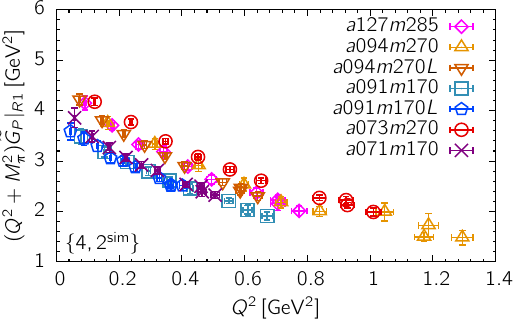}  
    \includegraphics[width=0.44\linewidth]{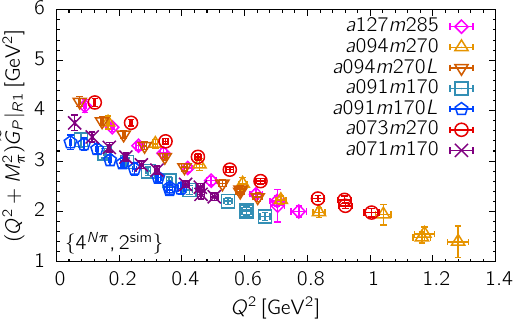}  
}
\caption{The data for $(Q^2 + M_N^2) {\widetilde G}_P(Q^2)$ from the
  seven ensembles are plotted versus $Q^2$. According to the pion-pole
  dominance hypothesis, Eq.~\protect\eqref{eq:PPD}, the result should
  be a smooth monotonic function that is proportional to $G_A(Q^2)$.
  The data from the $\{4,3^\ast\}$ and $\{4^{N\pi},3^\ast\}$
  strategies on the $M_\pi = 170$~MeV ensembles (top two panels) show
  deviations from this expectation at small $Q^2$. Also, the ``lines''
  of data from a given ensemble move up slightly as $a\to 0$ and down as
  $M_\pi \to 135$~MeV. The labels specify the analysis strategy and
  the ensemble ID.
\label{fig:PPD}}
\end{figure*}

This appendix contains the  data for the axial form factors
obtained from four strategies used to remove ESC: $\{4^{},3^{\ast}\}$,
$\{4^{N\pi},3^{\ast}\}$, $\{4^{N\pi},2^{A_4}\}$, and
$\{4^{N\pi},2^{\rm sim}\}$.  The renormalized axial form factors
$Z_AG_A$ and $Z_A \widetilde G_P$ and the unrenormalized $G_P$ are
given in Tables~\ref{tab:GA-renormalized},~\ref{tab:GPt-renormalized}
and~\ref{tab:GP-unrenormalized}, respectively. Data for the left-hand
side of Eq.~\protect\eqref{eq:testPCAC}, which by the PCAC relation
should equal unity, are presented in
Table~\ref{tab:testPCAC}. Figure~\ref{fig:affA4COMP} shows the data
for $R_{54}$, defined in Eq.~\protect\eqref{eq:r2ff-GPGA4}, for six
ensembles and compares the fits with the four strategies.  A
comparison of three matrix elements that give $\widetilde G_P$, $G_A$,
and $G_P$ obtained using the four strategies is shown in
Fig.~\ref{fig:aff4STcomp} for the $a091m170L$ and $a071m170$
ensembles.  Each row in Fig.~\ref{fig:aff4Npisimcomp} compares the
results of the fits to data obtained using the four strategies to remove
ESC. The six rows show data for the $a091m170L$ (rows one, three, and
five) and $a071m170$ (rows two, four and six) ensembles and the three
ratios: $R_{53}$ defined in Eqs~\protect\eqref{eq:r2ff-GPGA3} for two
different ${\bm n}^2 = 1$ momentum channels, and for $R_5$ defined in
Eq.~\protect\eqref{eq:r2ff-GP}.  Figure~\ref{fig:PPD} shows that the
data for $(Q^2 + M_N^2) {\widetilde G}_P(Q^2)$ are almost linear and
monotonic versus $Q^2$ on all seven ensembles except at small $Q^2$
for the $\{4,3^\ast\}$, and to a lesser extent for
$\{4^{N\pi},3^\ast\}$, strategy on the $M_\pi = 170$~MeV ensembles
(data in the upper two panels).

  \begin{table*}[t]  
  \begin{ruledtabular}
    \begin{tabular}{c |rrrr | rrrr }
$\bm {n}$ & $\{4,3^*\}$ & $\{4^{N\pi},3^*\}$ & $\{4^{N\pi},2^{A_4}\}$ & $\{4^{N\pi},2^\text{sim}\}$ & $\{4,3^*\}$ & $\{4^{N\pi},3^*\}$ & $\{4^{N\pi},2^{A_4}\}$ & $\{4^{N\pi},2^\text{sim}\}$\\ \hline 
\hline & \multicolumn{4}{c|}{$a127m285$} & \multicolumn{4}{c}{$a094m270$} \\ \hline
$(1,0,0)$ & 1.128(18) & 1.136(20) & 1.152(20) & 1.128(22)\,[1.54] & 1.009(20) & 1.008(20) & 1.014(20) & 1.007(21)\,[1.33]\\
$(1,1,0)$ & 1.021(17) & 1.023(18) & 1.031(17) & 1.011(18)\,[0.95] & 0.864(23) & 0.866(16) & 0.884(18) & 0.878(17)\,[1.23]\\
$(1,1,1)$ & 0.921(15) & 0.918(16) & 0.923(16) & 0.915(17)\,[0.82] & 0.743(24) & 0.747(15) & 0.763(26) & 0.752(21)\,[1.19]\\
$(2,0,0)$ & 0.853(16) & 0.848(16) & 0.858(18) & 0.856(19)\,[1.59] & 0.656(40) & 0.674(23) & 0.714(20) & 0.701(23)\,[1.31]\\
$(2,1,0)$ & 0.785(14) & 0.779(14) & 0.800(15) & 0.786(16)\,[1.22] & 0.589(26) & 0.601(14) & 0.619(28) & 0.615(16)\,[0.98]\\
$(2,1,1)$ & 0.720(15) & 0.712(14) & 0.747(16) & 0.716(16)\,[1.25] & 0.537(26) & 0.549(16) & 0.588(29) & 0.566(15)\,[1.12]\\
$(2,2,0)$ & 0.639(15) & 0.635(14) & 0.641(18) & 0.627(18)\,[1.15] & 0.482(32) & 0.490(24) &           & 0.519(27)\,[1.36]\\
$(2,2,1)$ & 0.592(16) & 0.585(13) & 0.608(23) & 0.587(21)\,[1.18] & 0.424(27) & 0.436(23) &           & 0.460(13)\,[1.26]\\
$(3,0,0)$ & 0.614(22) & 0.608(25) & 0.627(31) & 0.618(59)\,[1.38] & 0.542(84) & 0.521(50) &           & 0.448(19)\,[1.12]\\
$(3,1,0)$ & 0.570(16) & 0.563(15) & 0.585(17)  & 0.560(29)\,[1.39] & 0.489(53) & 0.485(33) &           & 0.430(36)\,[1.19]\\
\hline & \multicolumn{4}{c|}{$a094m270L$} & \multicolumn{4}{c}{$a091m170$} \\ \hline
$(1,0,0)$ & 1.124(19) & 1.134(21) & 1.134(20) & 1.134(20)\,[1.40] & 1.122(19) & 1.153(30) & 1.167(23) & 1.156(25)\,[1.15]\\
$(1,1,0)$ & 1.030(17) & 1.031(18) & 1.027(22) & 1.030(18)\,[1.55] & 1.018(17) & 1.020(29) & 1.028(20) & 1.020(21)\,[1.21]\\
$(1,1,1)$ & 0.951(16) & 0.945(17) & 0.963(17) & 0.952(17)\,[1.57] & 0.937(16) & 0.932(31) & 0.948(20) & 0.937(21)\,[1.14]\\
$(2,0,0)$ & 0.889(16) & 0.876(17) & 0.886(16) & 0.887(16)\,[1.58] & 0.873(17) & 0.849(32) & 0.894(21) & 0.893(22)\,[1.28]\\
$(2,1,0)$ & 0.828(16) & 0.815(15) & 0.827(14) & 0.834(15)\,[1.48] & 0.813(16) & 0.789(26) & 0.828(18) & 0.830(19)\,[1.92]\\
$(2,1,1)$ & 0.776(15) & 0.761(15) & 0.771(14) & 0.773(15)\,[1.50] & 0.755(16) & 0.728(29) & 0.764(22) & 0.765(17)\,[1.63]\\
$(2,2,0)$ & 0.695(15) & 0.680(15) & 0.715(15) & 0.699(14)\,[1.24] & 0.660(18) & 0.595(34) & 0.741(34) & 0.707(23)\,[2.15]\\
$(2,2,1)$ & 0.659(14) & 0.647(14) & 0.675(15) & 0.652(14)\,[1.14] & 0.632(17) & 0.600(36) & 0.678(29) & 0.636(18)\,[1.51]\\
$(3,0,0)$ & 0.662(15) & 0.637(16) & 0.687(20) & 0.652(19)\,[1.10] & 0.603(28) & 0.513(49) & 0.753(78) & 0.627(28)\,[1.13]\\
$(3,1,0)$ & 0.627(15) & 0.601(15) & 0.623(16) & 0.618(18)\,[1.44] & 0.607(21) & 0.552(37) & 0.638(22) & 0.616(22)\,[1.23]\\
\hline & \multicolumn{4}{c|}{$a091m170L$} & \multicolumn{4}{c}{$a073m270$} \\ \hline
$(1,0,0)$ & 1.169(22) & 1.208(39) & 1.229(29) & 1.236(30)\,[2.00] & 1.067(14) & 1.072(15) & 1.061(15) & 1.066(15)\,[1.63]\\
$(1,1,0)$ & 1.101(20) & 1.119(29) & 1.137(28) & 1.132(29)\,[1.81] & 0.945(13) & 0.942(13) & 0.941(13) & 0.946(14)\,[1.66]\\
$(1,1,1)$ & 1.048(19) & 1.059(27) & 1.054(23) & 1.073(29)\,[2.18] & 0.841(13) & 0.834(12) & 0.847(12) & 0.850(12)\,[1.29]\\
$(2,0,0)$ & 0.972(19) & 0.945(27) & 1.013(31) & 0.997(26)\,[1.40] & 0.760(14) & 0.750(13) & 0.781(12) & 0.774(12)\,[1.18]\\
$(2,1,0)$ & 0.930(18) & 0.900(25) & 0.971(29) & 0.956(20)\,[2.12] & 0.699(13) & 0.691(11) & 0.725(12) & 0.712(11)\,[2.25]\\
$(2,1,1)$ & 0.889(18) & 0.851(25) & 0.933(29) & 0.905(25)\,[2.68] & 0.637(15) & 0.637(10) & 0.674(12) & 0.663(10)\,[1.73]\\
$(2,2,0)$ & 0.806(18) & 0.755(28) & 0.855(29) & 0.849(24)\,[2.42] & 0.554(15) & 0.559(11) & 0.592(12) & 0.577(11)\,[1.62]\\
$(2,2,1)$ & 0.772(18) & 0.719(29) & 0.833(29) & 0.787(21)\,[2.36] & 0.518(16) & 0.529(10) & 0.544(13) & 0.546(11)\,[1.35]\\
$(3,0,0)$ & 0.766(20) & 0.700(34) & 0.842(31) & 0.790(21)\,[1.99] & 0.520(17) & 0.521(15) & 0.540(24) & 0.547(16)\,[1.28]\\
$(3,1,0)$ & 0.735(19) & 0.666(31) & 0.815(29) & 0.773(23)\,[1.98] & 0.483(15) & 0.487(13) & 0.529(24) & 0.508(12)\,[1.57]\\
\hline & \multicolumn{4}{c|}{$a071m170$} \\ \hline
$(1,0,0)$ & 1.154(18) & 1.203(31) & 1.186(23) & 1.214(27)\,[1.48]\\
$(1,1,0)$ & 1.078(14) & 1.099(22) & 1.076(16) & 1.103(22)\,[1.82]\\
$(1,1,1)$ & 1.001(14) & 0.997(19) & 1.002(15) & 1.018(21)\,[1.43]\\
$(2,0,0)$ & 0.941(16) & 0.930(22) & 0.954(19) & 0.957(20)\,[1.47]\\
$(2,1,0)$ & 0.897(16) & 0.878(18) & 0.896(14) & 0.912(19)\,[1.92]\\
$(2,1,1)$ & 0.837(19) & 0.812(19) & 0.876(33) & 0.871(18)\,[1.69]\\
$(2,2,0)$ & 0.777(19) & 0.737(22) & 0.813(18) & 0.799(17)\,[1.73]\\
$(2,2,1)$ & 0.731(20) & 0.703(21) & 0.787(17) & 0.768(18)\,[1.63]\\
$(3,0,0)$ & 0.697(28) & 0.658(28) & 0.784(21) & 0.739(26)\,[1.67]\\
$(3,1,0)$ & 0.686(24) & 0.651(22) & 0.763(18) & 0.722(21)\,[1.97]\\
\end{tabular}
    \end{ruledtabular} \caption{Data for the renormalized axial form
      factor $Z_A G_A (Q^2)$ obtained using four strategies
      $\{4,3^*\}$, $\{4^{N\pi},3^*\}$, $\{4^{N\pi},2^{A_4}\}$,
      $\{4^{N\pi},2^\text{sim}\}$ for controlling excited-state
      contamination. The values of $Q^2$, given in
      Table~\protect\ref{tab:Q2}, for a given of value of ${\bm n}$ are different for all seven
      ensembles, so only the data with the four strategies on each
      ensemble should be compared. No reasonable fits could be made
      for the four largest $Q^2$ points for the $a094m270$ ensemble
      with the $\{4^{N\pi},2^{A_4}\}$ strategy. The $\chi^2/$dof is
      shown only for the $\{2^\text{sim}\}$ fit. In other cases, the
      result is obtained using a two-step process---first fits are
      made to remove ESC and then the overdetermined set of equations
      is solved to get the form factors. The data are arranged by
      ensemble to facilitate comparison between the four
      strategies for each $Q^2$. } 
\label{tab:GA-renormalized} 
\end{table*}
    
  \begin{table*}  
  \begin{ruledtabular}
    \begin{tabular}{c |rrrr | rrrr }
  
$\bm {n}$ & $\{4,3^*\}$ & $\{4^{N\pi},3^*\}$ & $\{4^{N\pi},2^{A_4}\}$ & $\{4^{N\pi},2^\text{sim}\}$ & $\{4,3^*\}$ & $\{4^{N\pi},3^*\}$ & $\{4^{N\pi},2^{A_4}\}$ & $\{4^{N\pi},2^\text{sim}\}$\\ \hline 
\hline & \multicolumn{4}{c|}{$a127m285$} & \multicolumn{4}{c}{$a094m270$} \\ \hline
$(1,0,0)$ & 20.82(58) & 21.99(67) & 24.29(77) & 23.78(76)\,[1.54] & 14.85(88) & 14.91(48) & 16.46(55) & 15.89(54)\,[1.33]\\
$(1,1,0)$ & 13.22(32) & 13.73(30) & 14.56(38) & 14.17(33)\,[0.95] & 8.03(27) & 8.26(22) & 8.79(29) & 8.61(23)\,[1.23]\\
$(1,1,1)$ & 9.29(23) & 9.40(20) & 9.82(29) & 9.64(24)\,[0.82] & 5.03(27) & 5.23(17) & 5.92(27) & 5.55(22)\,[1.19]\\
$(2,0,0)$ & 7.04(21) & 7.12(19) & 7.65(20) & 7.47(24)\,[1.59] & 3.66(29) & 3.63(18) & 4.03(14) & 3.83(19)\,[1.31]\\
$(2,1,0)$ & 5.50(14) & 5.52(12) & 5.95(12) & 5.73(15)\,[1.22] & 2.65(17) & 2.66(10) & 3.05(12) & 2.82(11)\,[0.98]\\
$(2,1,1)$ & 4.35(13) & 4.27(11) & 4.76(11) & 4.54(14)\,[1.25] & 2.02(20) & 2.03(11) & 2.45(14) & 2.18(11)\,[1.12]\\
$(2,2,0)$ & 3.24(13) & 3.19(10) & 3.43(10) & 3.27(13)\,[1.15] & 1.53(19) & 1.54(13) &          & 1.74(18)\,[1.36]\\
$(2,2,1)$ & 2.62(10) & 2.56(10) & 2.86(10) & 2.82(13)\,[1.18] & 1.01(13) & 1.06(12) &          & 1.20(09)\,[1.26]\\
$(3,0,0)$ & 2.57(13) & 2.55(16) & 2.91(17) & 2.69(40)\,[1.38] & 1.49(41) & 1.40(27) &          & 1.25(11)\,[1.12]\\
$(3,1,0)$ & 2.28(09) & 2.26(10) & 2.53(12) & 2.34(13)\,[1.39] & 1.44(29) & 1.37(19) &          & 1.03(23)\,[1.19]\\
\hline & \multicolumn{4}{c|}{$a094m270L$} & \multicolumn{4}{c}{$a091m170$} \\ \hline
$(1,0,0)$ & 24.84(84) & 27.72(71) & 28.51(68) & 28.53(68)\,[1.40] & 24.27(67) & 28.2(1.7) & 32.6(1.4) & 32.0(1.3)\,[1.15]\\
$(1,1,0)$ & 16.23(50) & 17.49(37) & 17.35(78) & 17.39(36)\,[1.55] & 14.79(42) & 17.17(69) & 17.53(50) & 17.31(51)\,[1.21]\\
$(1,1,1)$ & 11.70(31) & 12.26(27) & 12.73(27) & 12.30(26)\,[1.57] & 10.20(27) & 11.53(55) & 11.65(29) & 11.75(31)\,[1.14]\\
$(2,0,0)$ & 8.94(22) & 9.26(23) & 9.27(25) & 9.34(21)\,[1.58] & 7.52(23) & 7.78(47) & 8.74(23) & 8.71(26)\,[1.28]\\
$(2,1,0)$ & 7.04(17) & 7.14(15) & 7.27(16) & 7.38(16)\,[1.48] & 5.94(17) & 6.25(33) & 6.63(15) & 6.75(18)\,[1.92]\\
$(2,1,1)$ & 5.74(15) & 5.79(14) & 5.82(13) & 5.99(15)\,[1.50] & 4.84(16) & 4.94(36) & 5.12(30) & 5.37(20)\,[1.63]\\
$(2,2,0)$ & 4.08(11) & 4.07(11) & 4.53(09) & 4.25(11)\,[1.24] & 3.22(14) & 2.90(23) & 4.38(30) & 3.84(12)\,[2.15]\\
$(2,2,1)$ & 3.55(11) & 3.51(11) & 3.90(09) & 3.70(11)\,[1.14] & 2.95(12) & 3.04(31) & 3.35(14) & 3.16(13)\,[1.51]\\
$(3,0,0)$ & 3.54(12) & 3.57(13) & 3.99(11) & 3.64(12)\,[1.10] & 2.79(16) & 2.38(40) & 3.97(40) & 3.20(15)\,[1.13]\\
$(3,1,0)$ & 3.02(09) & 2.97(10) & 3.36(09) & 3.20(12)\,[1.44] & 2.53(13) & 2.16(30) & 2.81(12) & 2.74(12)\,[1.23]\\
\hline & \multicolumn{4}{c|}{$a091m170L$} & \multicolumn{4}{c}{$a073m270$} \\ \hline
$(1,0,0)$ & 36.3(1.3) & 45.3(2.8) & 46.2(2.0) & 45.7(2.0)\,[2.00] & 18.48(71) & 19.98(56) & 20.88(44) & 21.18(39)\,[1.63]\\
$(1,1,0)$ & 24.55(77) & 29.2(1.4) & 28.41(91) & 28.5(1.0)\,[1.81] & 10.98(36) & 11.46(23) & 11.89(22) & 12.05(21)\,[1.66]\\
$(1,1,1)$ & 18.27(55) & 21.45(96) & 19.54(74) & 20.40(62)\,[2.18] & 7.41(21) & 7.56(14) & 7.98(16) & 8.04(14)\,[1.29]\\
$(2,0,0)$ & 13.64(42) & 14.85(71) & 15.20(72) & 15.07(52)\,[1.40] & 5.38(12) & 5.44(11) & 5.89(13) & 5.87(11)\,[1.18]\\
$(2,1,0)$ & 11.33(29) & 12.04(43) & 12.32(32) & 12.34(39)\,[2.12] & 4.20(10) & 4.15(08) & 4.60(08) & 4.52(09)\,[2.25]\\
$(2,1,1)$ & 9.41(25) & 9.74(38) & 10.25(27) & 10.23(43)\,[2.68] & 3.35(11) & 3.26(07) & 3.69(07) & 3.59(07)\,[1.73]\\
$(2,2,0)$ & 6.78(19) & 6.70(30) & 7.41(19) & 7.55(23)\,[2.42] & 2.35(08) & 2.26(06) & 2.60(05) & 2.49(06)\,[1.62]\\
$(2,2,1)$ & 5.95(18) & 5.80(31) & 6.56(18) & 6.35(25)\,[2.36] & 1.99(07) & 1.93(06) & 2.18(05) & 2.12(06)\,[1.35]\\
$(3,0,0)$ & 5.58(22) & 5.19(38) & 6.59(24) & 6.22(26)\,[1.99] & 1.96(09) & 1.90(08) & 2.34(19) & 2.25(08)\,[1.28]\\
$(3,1,0)$ & 5.08(17) & 4.70(30) & 5.89(18) & 5.77(27)\,[1.98] & 1.71(07) & 1.65(06) & 1.97(14) & 1.84(08)\,[1.57]\\
\hline & \multicolumn{4}{c|}{$a071m170$} \\ \hline
$(1,0,0)$ & 31.8(1.8) & 39.4(2.7) & 42.5(1.6) & 43.5(1.8)\,[1.48]\\
$(1,1,0)$ & 20.9(1.4) & 24.3(1.3) & 23.12(57) & 24.46(72)\,[1.82]\\
$(1,1,1)$ & 14.73(73) & 16.46(77) & 15.82(45) & 16.66(46)\,[1.43]\\
$(2,0,0)$ & 11.37(55) & 12.41(54) & 12.16(54) & 12.36(31)\,[1.47]\\
$(2,1,0)$ & 8.86(33) & 9.61(35) & 9.29(27) & 9.79(25)\,[1.92]\\
$(2,1,1)$ & 7.26(31) & 7.57(29) & 8.08(56) & 8.10(23)\,[1.69]\\
$(2,2,0)$ & 5.17(21) & 5.27(21) & 5.88(14) & 5.77(14)\,[1.73]\\
$(2,2,1)$ & 4.50(23) & 4.55(20) & 5.16(12) & 5.10(16)\,[1.63]\\
$(3,0,0)$ & 4.44(25) & 4.36(27) & 5.07(14) & 4.79(19)\,[1.67]\\
$(3,1,0)$ & 3.95(23) & 3.95(19) & 4.53(10) & 4.31(14)\,[1.97]\\
\end{tabular}
    \end{ruledtabular}
      \caption{Data for the renormalized induced pseudoscalar form factor, 
       $Z_A \widetilde G_P(Q^2)$, obtained using the four strategies $\{4,3^*\}$,
        $\{4^{N\pi},3^*\}$, $\{4^{N\pi},2^{A_4}\}$,
        $\{4^{N\pi},2^\text{sim}\}$ for controlling excited-state
        contamination. The rest is the same as in Table~\protect\ref{tab:GA-renormalized}.
}
    \label{tab:GPt-renormalized}
    \end{table*}
    
    
  \begin{table*}  
    \begin{ruledtabular}
      \setlength{\tabcolsep}{0.5pt}
    \begin{tabular}{c |rrrr | rrrr }
  
$\bm {n}$ & $\{4,3^*\}$ & $\{4^{N\pi},3^*\}$ & $\{4^{N\pi},2^{A_4}\}$ & $\{4^{N\pi},2^\text{sim}\}$ & $\{4,3^*\}$ & $\{4^{N\pi},3^*\}$ & $\{4^{N\pi},2^{A_4}\}$ & $\{4^{N\pi},2^\text{sim}\}$\\ \hline 
\hline & \multicolumn{4}{c|}{$a127m285$} & \multicolumn{4}{c}{$a094m270$} \\ \hline
$(1,0,0)$ & 36.0(9)\,[3.89] & 38.6(8)\,[1.89] & 42.0(1.0)\,[4.74] & 41.8(1.0)\,[1.54] & 28.5(2.1)\,[1.11] & 28.5(7)\,[1.09] & 31.0(9)\,[2.48] & 30.3(8)\,[1.33]\\
$(1,1,0)$ & 23.4(5)\,[2.23] & 24.2(3)\,[1.07] & 25.6(5)\,[2.06] & 25.1(4)\,[0.95] & 15.6(5)\,[1.07] & 16.0(3)\,[0.80] & 17.2(5)\,[0.95] & 16.7(4)\,[1.23]\\
$(1,1,1)$ & 17.1(4)\,[0.98] & 17.3(3)\,[0.87] & 18.1(4)\,[0.55] & 17.8(3)\,[0.82] & 10.3(3)\,[1.00] & 10.6(2)\,[0.95] & 11.6(4)\,[1.14] & 11.1(4)\,[1.19]\\
$(2,0,0)$ & 13.0(3)\,[1.25] & 13.1(2)\,[0.98] & 14.2(3)\,[0.62] & 13.9(4)\,[1.59] & 7.2(4)\,[1.10] & 7.2(3)\,[1.16] & 8.0(2)\,[1.38] & 7.6(3)\,[1.31]\\
$(2,1,0)$ & 10.3(2)\,[1.35] & 10.3(1)\,[1.22] & 11.1(2)\,[1.19] & 10.8(2)\,[1.22] & 6.1(3)\,[0.99] & 6.0(2)\,[1.03] & 7.0(3)\,[0.91] & 6.2(3)\,[0.98]\\
$(2,1,1)$ & 8.6(2)\,[1.32] & 8.5(1)\,[1.31] & 9.3(2)\,[1.46] & 8.9(2)\,[1.25] & 4.6(4)\,[1.93] & 4.6(2)\,[1.94] & 5.2(2)\,[1.85] & 4.8(2)\,[1.12]\\
$(2,2,0)$ & 6.1(1)\,[1.08] & 6.0(1)\,[1.05] & 6.6(2)\,[1.14] & 6.3(2)\,[1.15] & 3.4(5)\,[1.08] & 3.4(3)\,[1.09] &         & 4.0(4)\,[1.36]\\
$(2,2,1)$ & 5.5(2)\,[0.98] & 5.4(1)\,[0.96] & 5.9(2)\,[1.05] & 5.8(2)\,[1.18] & 2.7(5)\,[1.03] & 2.7(3)\,[1.02] &         & 2.8(2)\,[1.26]\\
$(3,0,0)$ & 5.4(2)\,[0.68] & 5.3(2)\,[0.65] & 6.1(3)\,[0.57] & 5.7(6)\,[1.38] & 2.0(8)\,[0.69] & 2.3(5)\,[0.68] &         & 2.6(4)\,[1.12]\\
$(3,1,0)$ & 4.8(2)\,[1.02] & 4.7(2)\,[1.02] & 5.5(2)\,[1.22] & 5.0(3)\,[1.39] & 2.4(3)\,[1.19] & 2.4(5)\,[1.19]  &         & 2.7(7)\,[1.19]\\
\hline & \multicolumn{4}{c|}{$a094m270L$} & \multicolumn{4}{c}{$a091m170$} \\ \hline
$(1,0,0)$ & 44.5(1.5)\,[3.97] & 49.9(8)\,[1.62] & 52.1(1.1)\,[1.55] & 52.0(8)\,[1.40] & 50.8(1.5)\,[2.43] & 65.7(1.4)\,[0.92] & 67.9(2.6)\,[1.11] & 66.9(2.4)\,[1.15]\\
$(1,1,0)$ & 29.9(8)\,[2.82] & 32.3(4)\,[1.36] & 33.2(2.0)\,[3.33] & 32.5(4)\,[1.55] & 31.4(8)\,[2.84] & 36.5(9)\,[1.46] & 37.7(9)\,[1.12] & 37.1(10)\,[1.21]\\
$(1,1,1)$ & 21.9(5)\,[2.01] & 23.0(3)\,[1.28] & 24.0(3)\,[2.78] & 23.3(3)\,[1.57] & 22.0(5)\,[2.13] & 24.6(7)\,[0.87] & 25.7(5)\,[0.77] & 26.2(7)\,[1.14]\\
$(2,0,0)$ & 17.0(3)\,[1.32] & 17.7(2)\,[0.99] & 17.9(4)\,[1.05] & 18.0(3)\,[1.58] & 16.9(4)\,[1.05] & 18.9(8)\,[0.61] & 19.0(5)\,[0.72] & 19.1(5)\,[1.28]\\
$(2,1,0)$ & 13.9(2)\,[1.53] & 14.1(2)\,[0.82] & 14.4(2)\,[1.82] & 14.6(2)\,[1.48] & 13.3(3)\,[1.83] & 13.6(5)\,[1.21] & 15.3(3)\,[1.43] & 15.5(4)\,[1.92]\\
$(2,1,1)$ & 11.6(2)\,[1.47] & 11.7(2)\,[0.94] & 11.9(1)\,[2.56] & 12.1(2)\,[1.50] & 11.1(3)\,[1.35] & 10.9(8)\,[1.14] & 12.0(6)\,[2.12] & 12.3(4)\,[1.63]\\
$(2,2,0)$ & 8.6(1)\,[0.59] & 8.5(1)\,[0.44] & 9.3(1)\,[0.84] & 8.9(1)\,[1.24] & 8.0(3)\,[1.94] & 7.6(6)\,[1.78] & 10.2(6)\,[2.00] & 9.1(3)\,[2.15]\\
$(2,2,1)$ & 7.5(1)\,[1.26] & 7.5(1)\,[1.07] & 8.1(1)\,[0.96] & 7.9(1)\,[1.14] & 6.7(2)\,[1.32] & 5.7(6)\,[1.13] & 7.8(3)\,[1.85] & 7.4(3)\,[1.51]\\
$(3,0,0)$ & 7.5(2)\,[0.76] & 7.6(2)\,[0.72] & 8.4(2)\,[0.85] & 7.7(2)\,[1.10] & 7.4(3)\,[1.07] & 7.4(8)\,[1.09] & 10.3(1.6)\,[1.43] & 7.8(4)\,[1.13]\\
$(3,1,0)$ & 6.5(1)\,[1.25] & 6.4(3)\,[0.97] & 7.3(2)\,[1.10] & 7.0(2)\,[1.44] & 6.3(3)\,[1.32] & 6.5(7)\,[1.29] & 6.8(3)\,[1.08] & 6.9(3)\,[1.23]\\
\hline & \multicolumn{4}{c|}{$a091m170L$} & \multicolumn{4}{c}{$a073m270$} \\ \hline
$(1,0,0)$ & 73.9(2.3)\,[2.39] & 95.0(5.1)\,[1.04] & 97.7(3.2)\,[2.50] & 97.0(3.4)\,[2.00] & 34.2(1.4)\,[3.18] & 37.2(9)\,[1.33] & 39.1(6)\,[2.03] & 40.1(5)\,[1.63]\\
$(1,1,0)$ & 50.3(1.5)\,[2.88] & 61.4(3.1)\,[1.54] & 60.6(1.4)\,[1.95] & 60.4(1.8)\,[1.81] & 20.8(7)\,[3.20] & 21.8(3)\,[1.58] & 22.6(3)\,[1.09] & 23.0(2)\,[1.66]\\
$(1,1,1)$ & 37.5(1.0)\,[3.48] & 44.8(1.9)\,[2.12] & 42.1(1.1)\,[5.28] & 44.0(1.1)\,[2.18] & 14.7(4)\,[2.04] & 15.0(2)\,[1.32] & 15.7(2)\,[0.89] & 15.7(2)\,[1.29]\\
$(2,0,0)$ & 30.1(8)\,[1.52] & 34.2(1.4)\,[1.15] & 33.9(1.5)\,[1.02] & 34.1(1.1)\,[1.40] & 10.9(2)\,[2.27] & 11.1(1)\,[1.72] & 11.7(2)\,[1.22] & 11.8(2)\,[1.18]\\
$(2,1,0)$ & 24.7(6)\,[2.07] & 27.5(10)\,[1.45] & 27.5(6)\,[1.57] & 27.2(8)\,[2.12] & 8.5(1)\,[0.99] & 8.5(1)\,[0.91] & 9.3(1)\,[1.68] & 9.2(1)\,[2.25]\\
$(2,1,1)$ & 20.7(5)\,[1.78] & 22.7(7)\,[1.51] & 23.0(6)\,[1.91] & 23.3(9)\,[2.68] & 7.0(1)\,[1.19] & 6.9(1)\,[1.25] & 7.7(1)\,[1.84] & 7.5(1)\,[1.73]\\
$(2,2,0)$ & 15.3(3)\,[2.41] & 16.0(5)\,[1.99] & 17.0(4)\,[1.98] & 17.2(6)\,[2.42] & 5.0(1)\,[1.03] & 5.0(1)\,[0.94] & 5.6(1)\,[1.97] & 5.4(1)\,[1.62]\\
$(2,2,1)$ & 13.6(3)\,[1.85] & 14.3(4)\,[1.71] & 15.0(4)\,[1.54] & 15.0(6)\,[2.36] & 4.4(2)\,[0.75] & 4.2(1)\,[0.77] & 4.9(1)\,[1.36] & 4.7(1)\,[1.35]\\
$(3,0,0)$ & 13.8(3)\,[2.10] & 14.3(6)\,[1.91] & 15.5(5)\,[2.04] & 14.8(5)\,[1.99] & 4.3(2)\,[0.94] & 4.2(2)\,[0.95] & 5.0(3)\,[1.26] & 4.9(2)\,[1.28]\\
$(3,1,0)$ & 11.9(3)\,[1.74] & 12.1(5)\,[1.58] & 13.5(4)\,[2.00] & 13.4(6)\,[1.98] & 3.8(2)\,[0.78] & 3.8(2)\,[0.79] & 4.5(3)\,[1.14] & 4.2(2)\,[1.57]\\
\hline & \multicolumn{4}{c|}{$a071m170$} \\ \hline
$(1,0,0)$ & 66.7(4.7)\,[1.84] & 84.4(5.2)\,[0.90] & 91.2(3.0)\,[1.43] & 94.2(3.5)\,[1.48]\\
$(1,1,0)$ & 42.7(2.8)\,[1.98] & 50.3(2.5)\,[1.28] & 49.7(10)\,[6.23] & 52.0(1.4)\,[1.82]\\
$(1,1,1)$ & 31.1(1.7)\,[1.68] & 35.2(1.5)\,[1.31] & 34.4(7)\,[3.93] & 36.0(10)\,[1.43]\\
$(2,0,0)$ & 24.3(1.2)\,[1.60] & 26.6(10)\,[1.47] & 26.4(10)\,[1.74] & 27.6(7)\,[1.47]\\
$(2,1,0)$ & 19.2(6)\,[1.97] & 20.7(6)\,[1.64] & 20.7(4)\,[6.03] & 22.4(6)\,[1.92]\\
$(2,1,1)$ & 15.9(6)\,[1.24] & 16.7(5)\,[1.08] & 17.9(1.2)\,[1.58] & 18.2(5)\,[1.69]\\
$(2,2,0)$ & 11.7(5)\,[1.64] & 12.2(4)\,[1.24] & 13.6(3)\,[1.29] & 13.5(4)\,[1.73]\\
$(2,2,1)$ & 10.4(4)\,[0.68] & 10.6(4)\,[0.57] & 11.9(3)\,[0.89] & 12.1(3)\,[1.63]\\
$(3,0,0)$ & 10.5(5)\,[1.24] & 10.3(5)\,[1.17] & 11.5(3)\,[1.11] & 11.5(4)\,[1.67]\\
$(3,1,0)$ & 9.2(4)\,[1.19] & 9.3(5)\,[1.16] & 10.5(3)\,[1.31] & 10.4(4)\,[1.97]\\
\end{tabular}
    \end{ruledtabular}
      \caption{Data for the unrenormalized pseudoscalar form factor $
        G_P(Q^2)$ obtained using four strategies $\{4,3^*\}$,
        $\{4^{N\pi},3^*\}$, $\{4^{N\pi},2^{A_4}\}$,
        $\{4^{N\pi},2^\text{sim}\}$ for controlling excited-state
        contamination. The numbers within the square brackets are the $\chi^2/$dof of
        the fit.  }
    \label{tab:GP-unrenormalized}
 \end{table*}
    
 \begin{table*}   
  \begin{ruledtabular}
    \begin{tabular}{c |rrrr | rrrr }
$\vec{n}$ & $\{4,3^*\}$ & $\{4^{N\pi},3^*\}$ & $\{4^{N\pi},2^{A_4}\}$ & $\{4^{N\pi},2^\text{sim}\}$ & $\{4,3^*\}$ & $\{4^{N\pi},3^*\}$ & $\{4^{N\pi},2^{A_4}\}$ & $\{4^{N\pi},2^\text{sim}\}$\\ \hline 
\hline & \multicolumn{4}{c|}{$a127m285$} & \multicolumn{4}{c}{$a094m270$} \\ \hline
$(1,0,0)$ & 0.876(21) & 0.931(16) & 1.007(20) & 1.015(20)\,[1.54] & 0.930(68) & 0.931(21) & 1.016(25) & 0.992(21)\,[1.33]\\
$(1,1,0)$ & 0.926(19) & 0.965(11) & 1.014(17) & 1.008(13)\,[0.95] & 0.951(41) & 0.972(16) & 1.016(24) & 1.000(15)\,[1.23]\\
$(1,1,1)$ & 0.959(23) & 0.980(13) & 1.019(19) & 1.009(13)\,[0.82] & 0.946(35) & 0.973(19) & 1.074(31) & 1.025(25)\,[1.19]\\
$(2,0,0)$ & 0.964(28) & 0.985(15) & 1.049(25) & 1.027(21)\,[1.59] & 0.982(39) & 0.937(26) & 0.982(18) & 0.950(22)\,[1.31]\\
$(2,1,0)$ & 0.970(26) & 0.984(13) & 1.033(14) & 1.012(15)\,[1.22] & 0.948(41) & 0.923(21) & 1.027(53) & 0.954(26)\,[0.98]\\
$(2,1,1)$ & 0.966(31) & 0.963(15) & 1.021(14) & 1.017(19)\,[1.25] & 0.909(54) & 0.888(30) & 0.993(26) & 0.921(33)\,[1.12]\\
$(2,2,0)$ & 1.004(45) & 0.998(18) & 1.063(26) & 1.036(28)\,[1.15] & 0.936(76) & 0.915(50) &           & 0.981(69)\,[1.36]\\
$(2,2,1)$ & 0.967(42) & 0.963(21) & 1.029(25) & 1.050(45)\,[1.18] & 0.794(80) & 0.796(63) &           & 0.846(47)\,[1.26]\\
$(3,0,0)$ & 0.916(30) & 0.919(31) & 1.020(42) & 0.957(64)\,[1.38] & 0.89(17)  & 0.86(12)  &           & 0.905(65)\,[1.12]\\
$(3,1,0)$ & 0.947(33) & 0.955(26) & 1.034(36)  & 0.998(73)\,[1.39] & 1.04(12)  & 0.984(88) &           & 0.86(13)\,[1.19]\\
\hline & \multicolumn{4}{c|}{$a094m270L$} & \multicolumn{4}{c}{$a091m170$} \\ \hline
$(1,0,0)$ & 0.858(29) & 0.957(17) & 0.991(13) & 0.991(10)\,[1.40] & 0.723(19) & 0.856(36) & 0.945(28) & 0.938(25)\,[1.15]\\
$(1,1,0)$ & 0.914(29) & 0.992(14) & 0.998(34) & 0.989(08)\,[1.55] & 0.824(25) & 0.968(29) & 0.983(21) & 0.978(20)\,[1.21]\\
$(1,1,1)$ & 0.939(26) & 0.996(13) & 1.017(10) & 0.994(08)\,[1.57] & 0.859(22) & 0.988(36) & 0.986(19) & 1.007(19)\,[1.14]\\
$(2,0,0)$ & 0.947(24) & 1.001(13) & 0.993(15) & 0.999(10)\,[1.58] & 0.862(24) & 0.937(37) & 0.990(18) & 0.988(19)\,[1.28]\\
$(2,1,0)$ & 0.949(25) & 0.984(11) & 0.988(11) & 0.994(08)\,[1.48] & 0.881(25) & 0.967(39) & 0.982(15) & 0.996(18)\,[1.92]\\
$(2,1,1)$ & 0.952(25) & 0.983(13) & 0.977(12) & 1.001(10)\,[1.50] & 0.905(27) & 0.965(50) & 0.959(35) & 1.001(26)\,[1.63]\\
$(2,2,0)$ & 0.946(27) & 0.969(14) & 1.022(09) & 0.985(12)\,[1.24] & 0.877(35) & 0.888(56) & 1.071(51) & 0.983(24)\,[2.15]\\
$(2,2,1)$ & 0.951(32) & 0.963(17) & 1.020(12) & 1.006(18)\,[1.14] & 0.921(39) & 1.001(74) & 0.986(25) & 0.992(36)\,[1.51]\\
$(3,0,0)$ & 0.940(32) & 0.992(19) & 1.025(16) & 0.989(23)\,[1.10] & 0.923(51) & 0.94(12) & 1.058(08) & 1.020(41)\,[1.13]\\
$(3,1,0)$ & 0.920(26) & 0.948(18) & 1.036(22) & 0.998(27)\,[1.44] & 0.908(46) & 0.867(88) & 0.963(27) & 0.972(36)\,[1.23]\\
\hline & \multicolumn{4}{c|}{$a091m170L$} & \multicolumn{4}{c}{$a073m270$} \\ \hline
$(1,0,0)$ & 0.710(22) & 0.897(44) & 0.903(22) & 0.889(22)\,[2.00] & 0.855(33) & 0.926(21) & 0.980(10) & 0.993(06)\,[1.63]\\
$(1,1,0)$ & 0.811(25) & 0.986(48) & 0.947(18) & 0.952(22)\,[1.81] & 0.909(34) & 0.955(16) & 0.993(09) & 1.001(06)\,[1.66]\\
$(1,1,1)$ & 0.861(25) & 1.034(42) & 0.952(21) & 0.977(17)\,[2.18] & 0.936(33) & 0.965(14) & 1.002(10) & 1.004(06)\,[1.29]\\
$(2,0,0)$ & 0.884(27) & 1.027(51) & 0.975(29) & 0.985(20)\,[1.40] & 0.935(19) & 0.960(10) & 0.997(12) & 1.004(08)\,[1.18]\\
$(2,1,0)$ & 0.917(26) & 1.041(44) & 0.985(15) & 0.999(21)\,[2.12] & 0.944(25) & 0.946(10) & 0.998(08) & 0.998(09)\,[2.25]\\
$(2,1,1)$ & 0.926(25) & 1.034(40) & 0.988(17) & 1.019(26)\,[2.68] & 0.959(38) & 0.933(14) & 0.997(09) & 0.987(10)\,[1.73]\\
$(2,2,0)$ & 0.934(25) & 1.019(41) & 0.991(17) & 1.016(18)\,[2.42] & 0.969(41) & 0.925(14) & 1.004(09) & 0.984(15)\,[1.62]\\
$(2,2,1)$ & 0.946(27) & 1.022(45) & 0.992(18) & 1.019(30)\,[2.36] & 0.962(40) & 0.912(17) & 1.001(12) & 0.974(18)\,[1.35]\\
$(3,0,0)$ & 0.900(26) & 0.950(45) & 0.989(22) & 0.994(28)\,[1.99] & 0.939(32) & 0.910(25) & 1.077(76) & 1.025(19)\,[1.28]\\
$(3,1,0)$ & 0.928(25) & 0.978(44) & 0.994(20) & 1.027(31)\,[1.98] & 0.963(34) & 0.919(22) & 1.010(44) & 0.980(36)\,[1.57]\\
\hline & \multicolumn{4}{c|}{$a071m170$} \\ \hline
$(1,0,0)$ & 0.723(46) & 0.885(45) & 0.968(23) & 0.972(25)\,[1.48]\\
$(1,1,0)$ & 0.833(59) & 0.971(44) & 0.950(14) & 0.979(17)\,[1.82]\\
$(1,1,1)$ & 0.878(54) & 1.004(43) & 0.964(17) & 0.998(15)\,[1.43]\\
$(2,0,0)$ & 0.915(57) & 1.029(39) & 0.985(28) & 1.001(14)\,[1.47]\\
$(2,1,0)$ & 0.898(41) & 1.012(33) & 0.961(16) & 0.998(13)\,[1.92]\\
$(2,1,1)$ & 0.920(45) & 1.006(33) & 0.995(34) & 1.004(15)\,[1.69]\\
$(2,2,0)$ & 0.892(30) & 0.976(29) & 0.987(15) & 0.987(13)\,[1.73]\\
$(2,2,1)$ & 0.914(34) & 0.976(31) & 0.990(15) & 1.004(15)\,[1.63]\\
$(3,0,0)$ & 0.956(61) & 1.007(47) & 0.981(16) & 0.984(18)\,[1.67]\\
$(3,1,0)$ & 0.934(48) & 1.000(36) & 0.978(16) & 0.986(19)\,[1.97]\\
\end{tabular}
  \caption{Check of the PCAC relation between the axial and
    pseudoscalar form factors given in Eq.~\protect\eqref{eq:testPCAC}
    for four strategies used to remove ESC. Since PCAC is an operator
    relation, deviations from unity should  only be due to
    discretization errors.}  
\label{tab:testPCAC} 
\end{ruledtabular} 
\end{table*}

\onecolumngrid\hrule width0pt\twocolumngrid\cleardoublepage

\section{Comparison of electric and magnetic form factors extracted using 4 strategies}
\label{sec:TcompVFF}

\begin{figure*}[tbhp] 
\subfigure
{
    \includegraphics[width=0.24\linewidth]{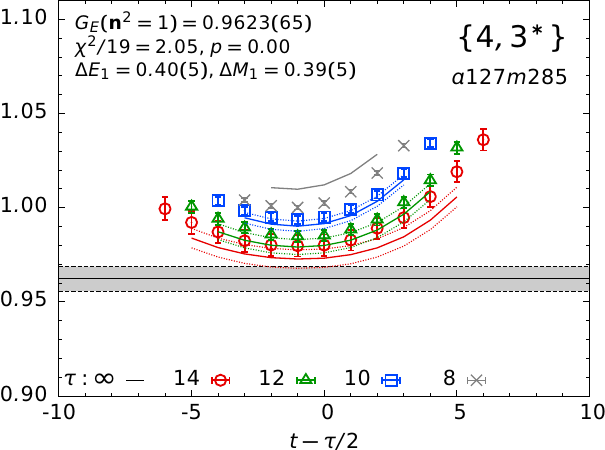}  
    \includegraphics[width=0.24\linewidth]{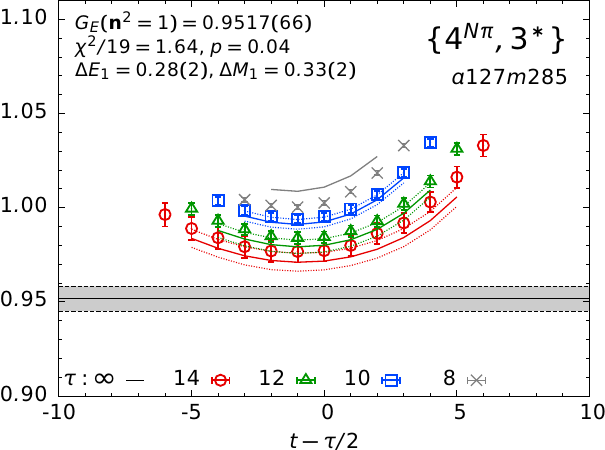} 
    \includegraphics[width=0.24\linewidth]{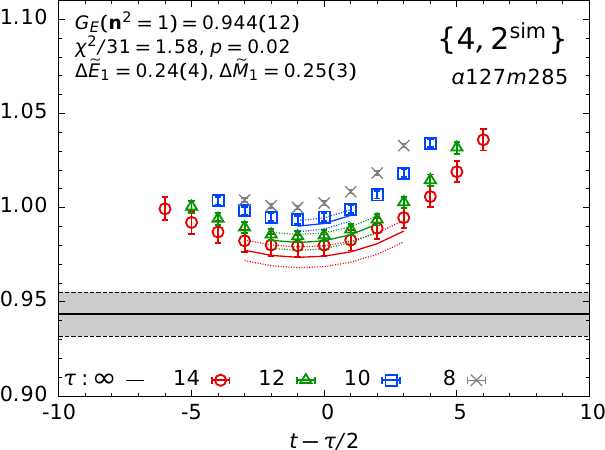} 
    \includegraphics[width=0.24\linewidth]{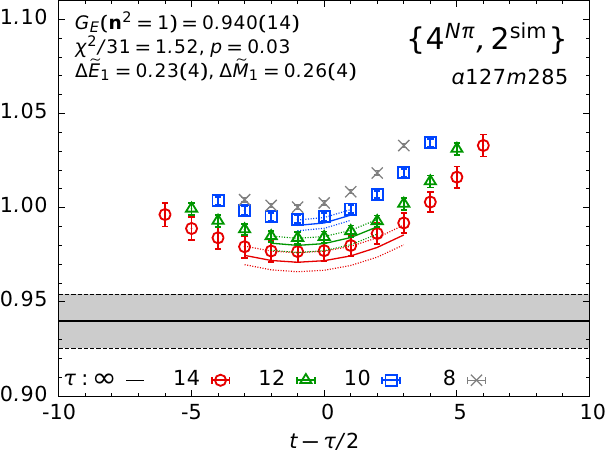} 
}
{
    \includegraphics[width=0.24\linewidth]{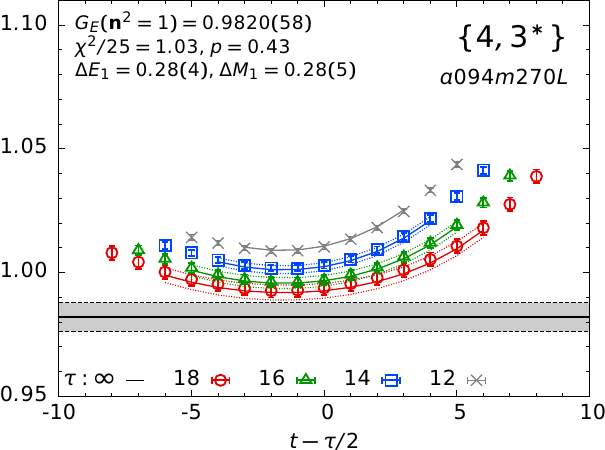}  
    \includegraphics[width=0.24\linewidth]{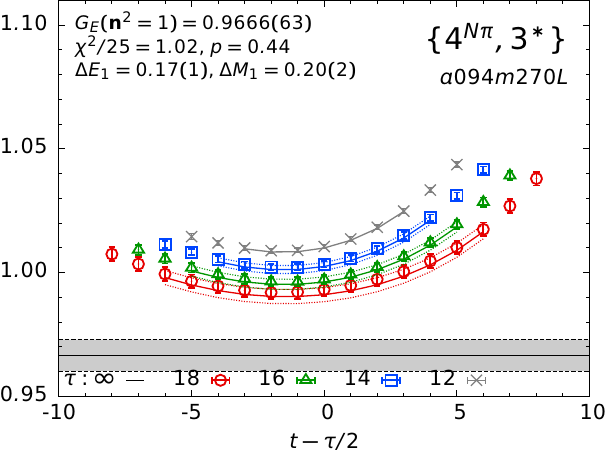} 
    \includegraphics[width=0.24\linewidth]{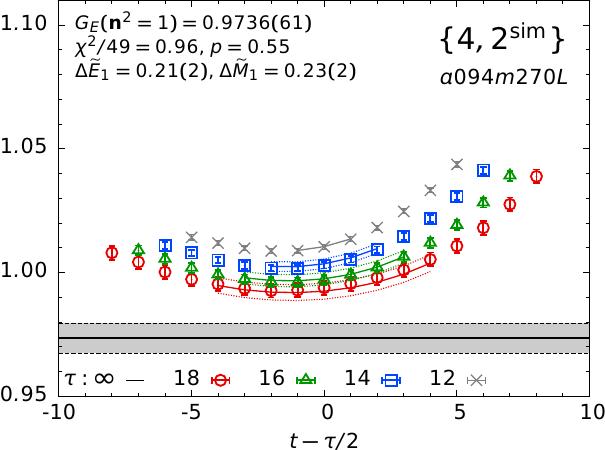} 
    \includegraphics[width=0.24\linewidth]{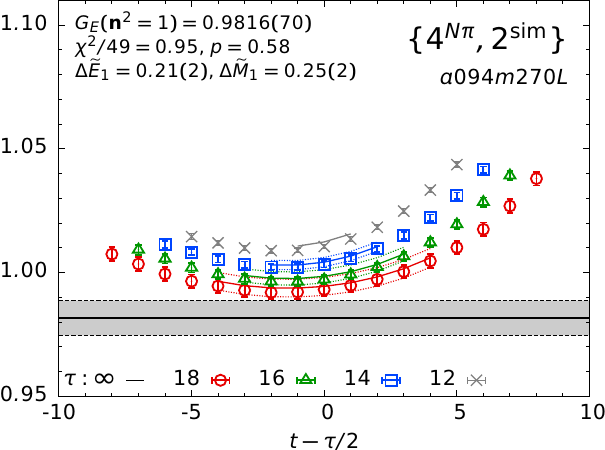} 
}
{
}
{
    \includegraphics[width=0.24\linewidth]{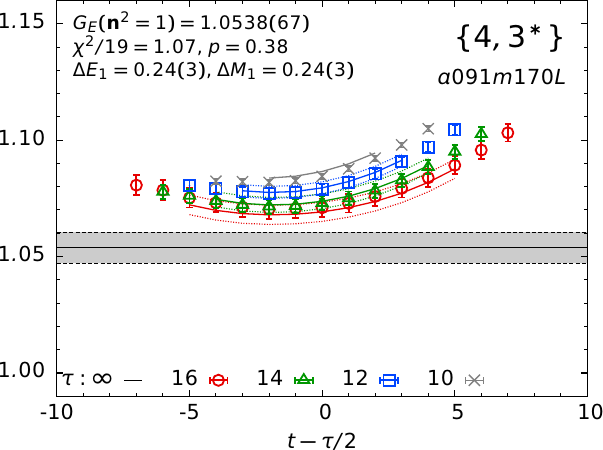} 
    \includegraphics[width=0.24\linewidth]{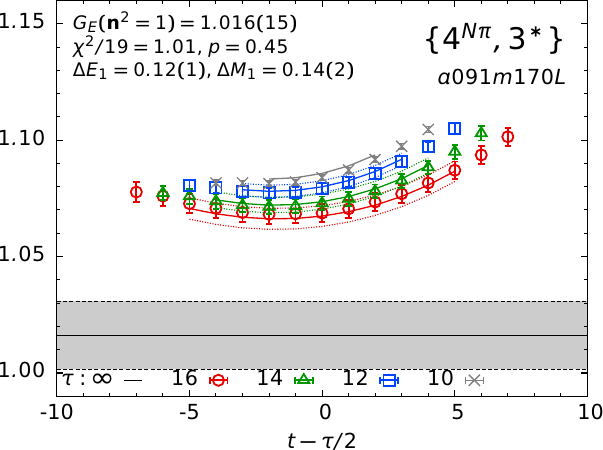} 
    \includegraphics[width=0.24\linewidth]{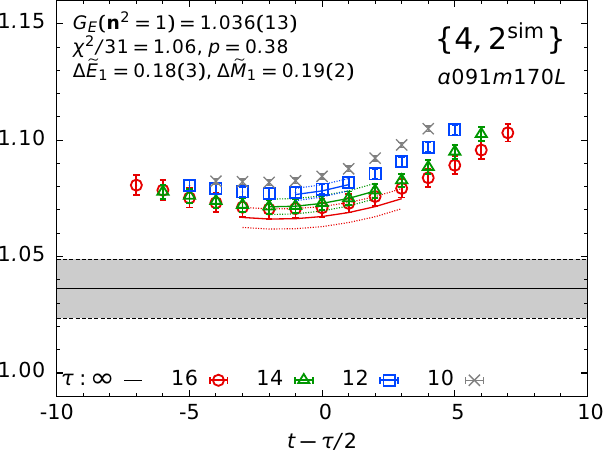} 
    \includegraphics[width=0.24\linewidth]{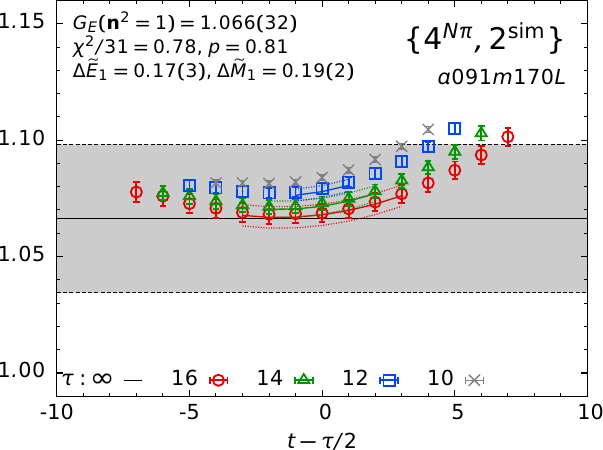} 
}
{
    \includegraphics[width=0.24\linewidth]{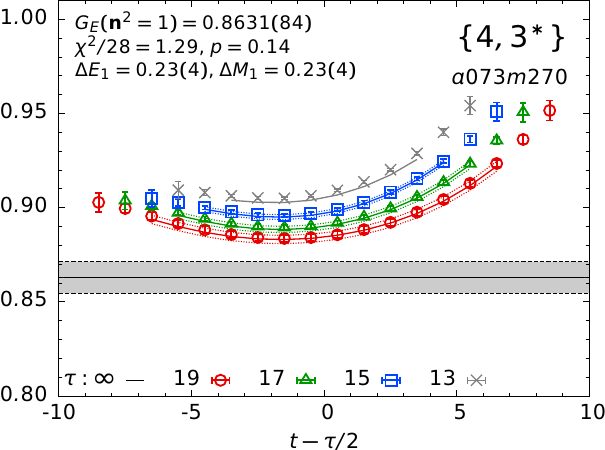}  
    \includegraphics[width=0.24\linewidth]{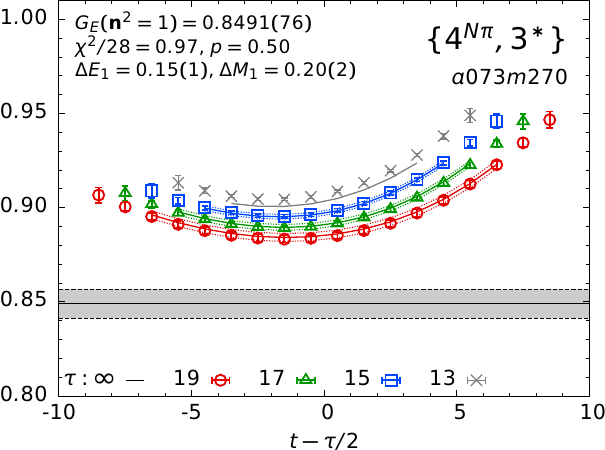} 
    \includegraphics[width=0.24\linewidth]{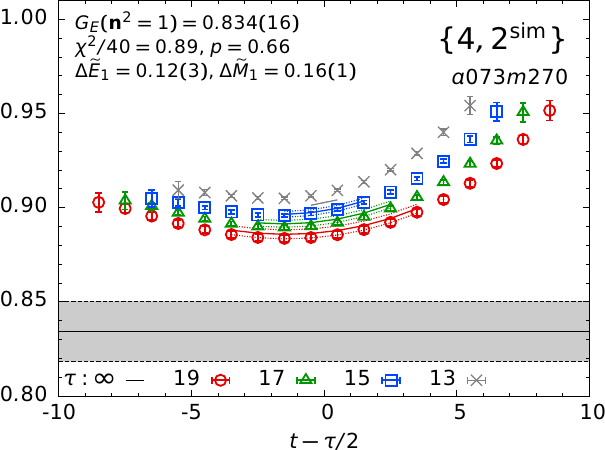} 
    \includegraphics[width=0.24\linewidth]{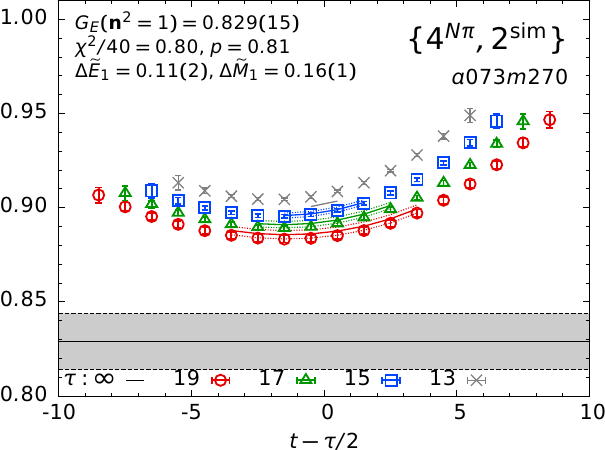} 
}
{
    \includegraphics[width=0.24\linewidth]{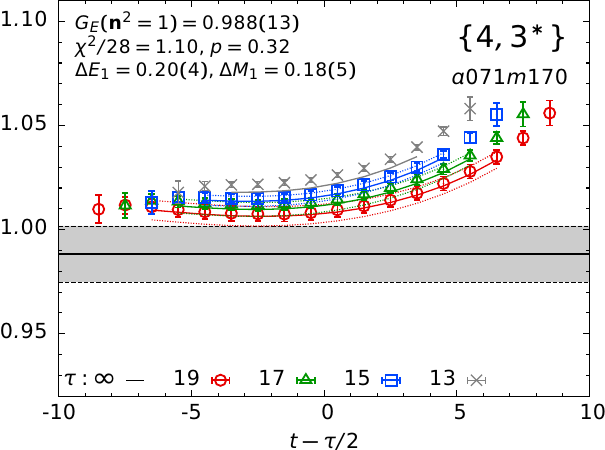}  
    \includegraphics[width=0.24\linewidth]{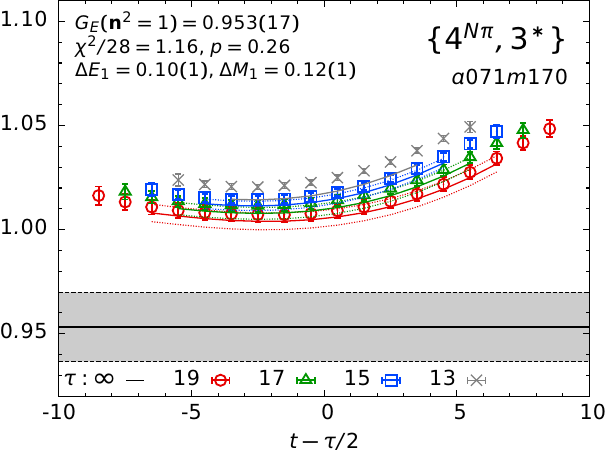} 
    \includegraphics[width=0.24\linewidth]{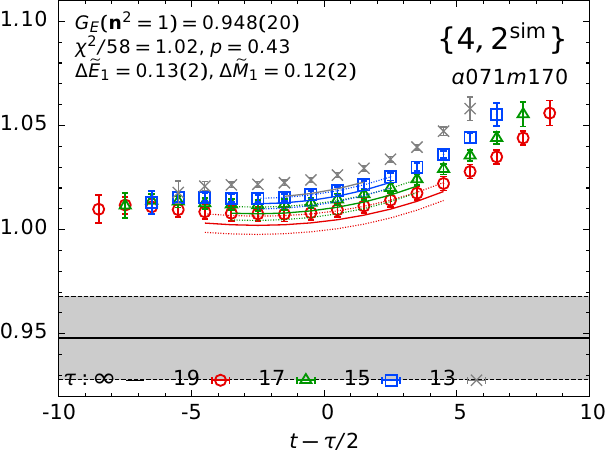} 
    \includegraphics[width=0.24\linewidth]{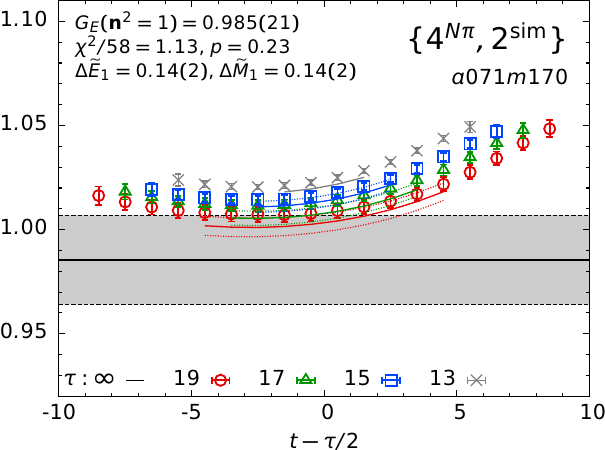} 
}
\caption{$G_E({\bm n} = (1,0,0))$ obtained from the ratio $R_{4}$
  defined in Eq.~\protect\eqref{eq:GE4} 
  for five  ensembles plotted versus the shifted operator insertion point
  $t -\tau/2$.  The panels in the left column show the fits with $\{4,3^\ast\}$, the
  second with $\{4^{N\pi},3^\ast\}$, the third with $\{4,2^{\rm
    sim}\}$ and the right column with $\{4^{N\pi},2^{\rm sim}\}$ strategies.  The
  interval along the $y$-axis is the same for a given row to facilitate comparison
  of the result and the error.  The rest is the same
  as in Fig.~\protect\ref{fig:affA4COMP}.
  \label{fig:GECOMP}}
\end{figure*}

\begin{figure*}[tbp] 
\subfigure
{
    \includegraphics[width=0.24\linewidth]{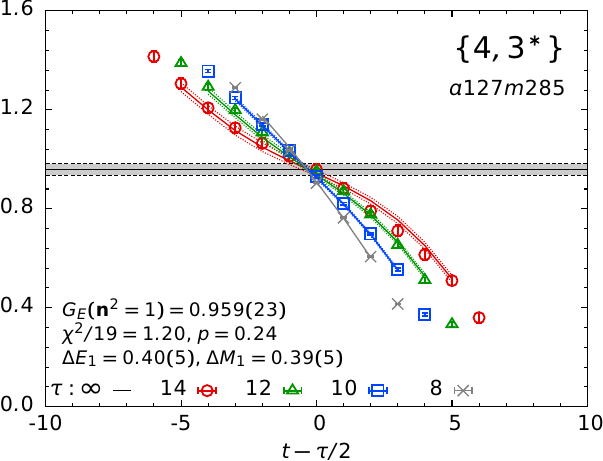}  
    \includegraphics[width=0.24\linewidth]{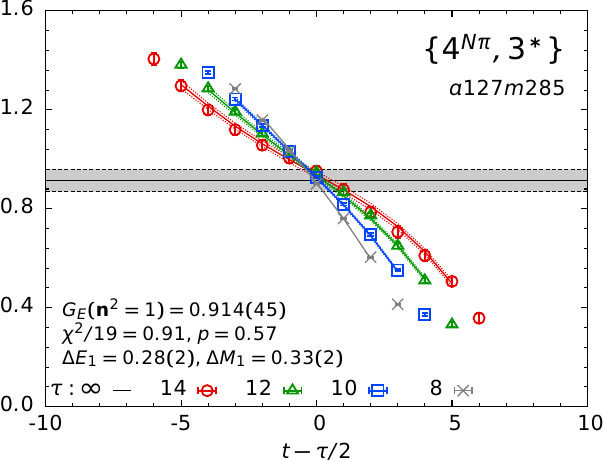} 
    \includegraphics[width=0.24\linewidth]{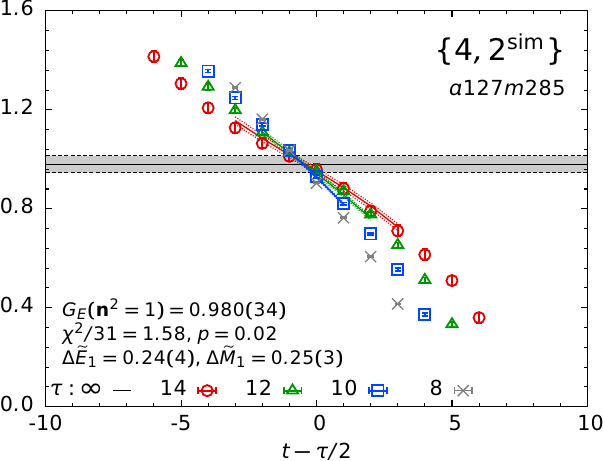} 
    \includegraphics[width=0.24\linewidth]{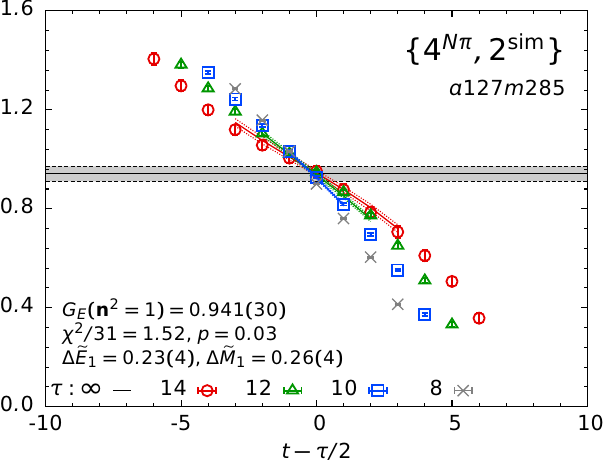} 
}
{
    \includegraphics[width=0.24\linewidth]{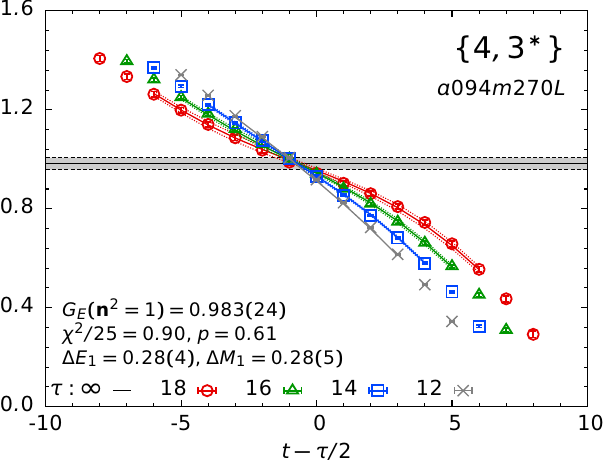}  
    \includegraphics[width=0.24\linewidth]{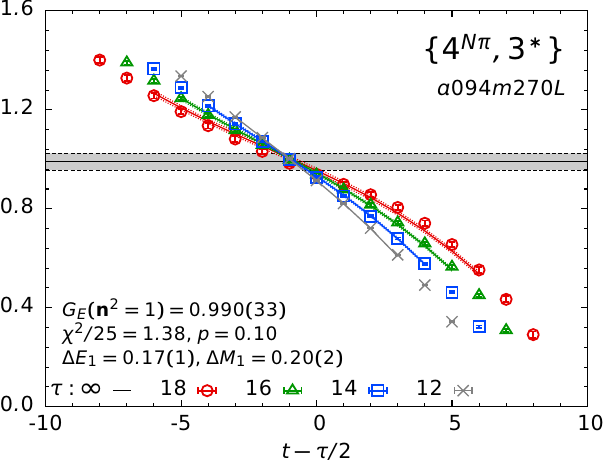} 
    \includegraphics[width=0.24\linewidth]{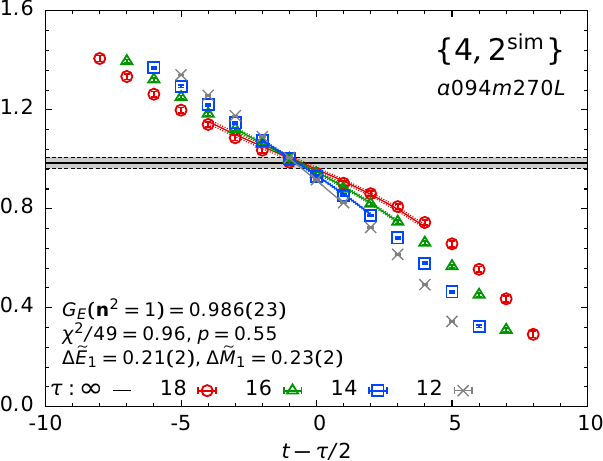} 
    \includegraphics[width=0.24\linewidth]{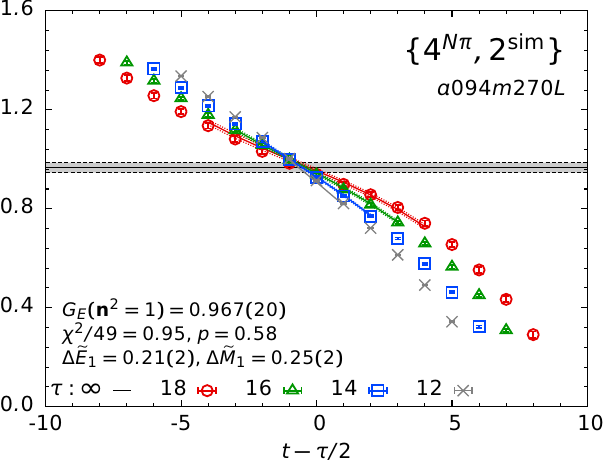} 
}
{
}
{
    \includegraphics[width=0.24\linewidth]{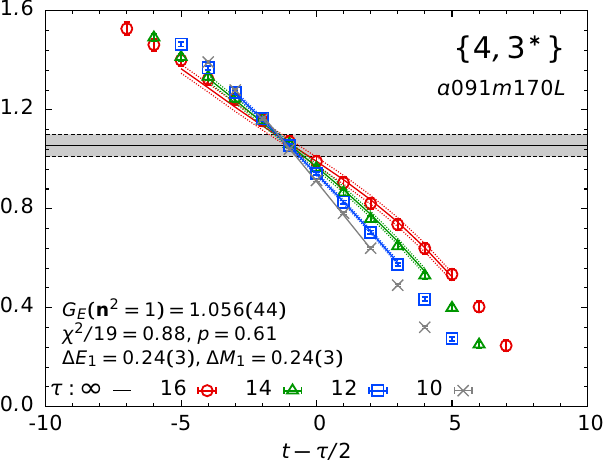} 
    \includegraphics[width=0.24\linewidth]{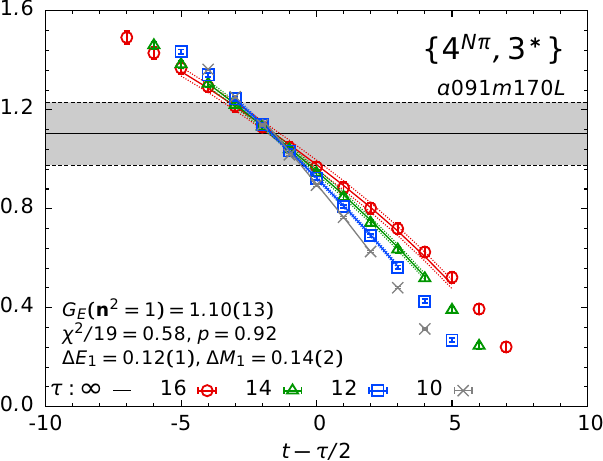} 
    \includegraphics[width=0.24\linewidth]{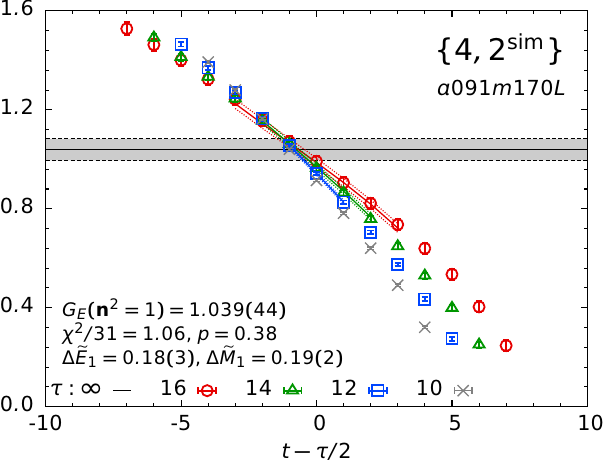} 
    \includegraphics[width=0.24\linewidth]{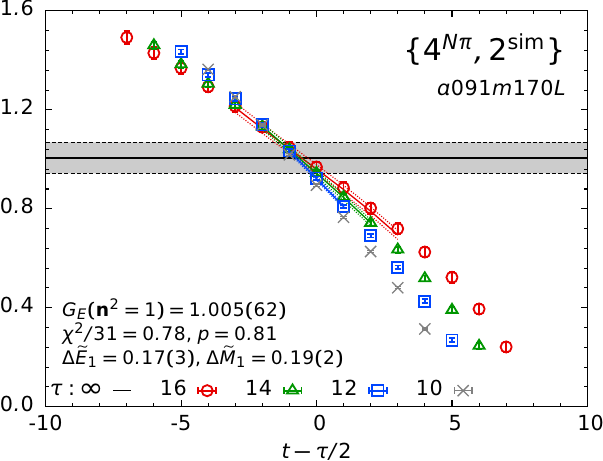} 
}
{
    \includegraphics[width=0.24\linewidth]{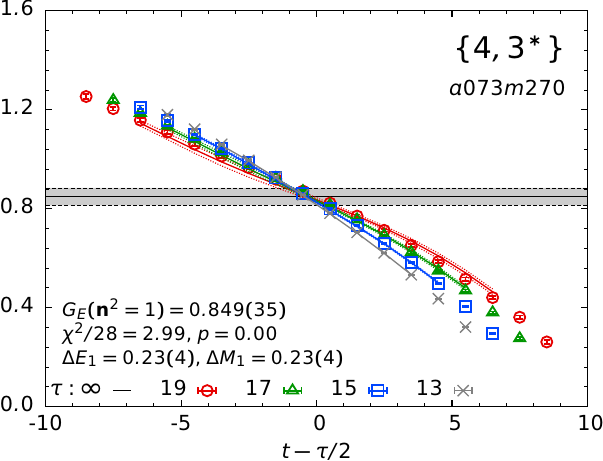}  
    \includegraphics[width=0.24\linewidth]{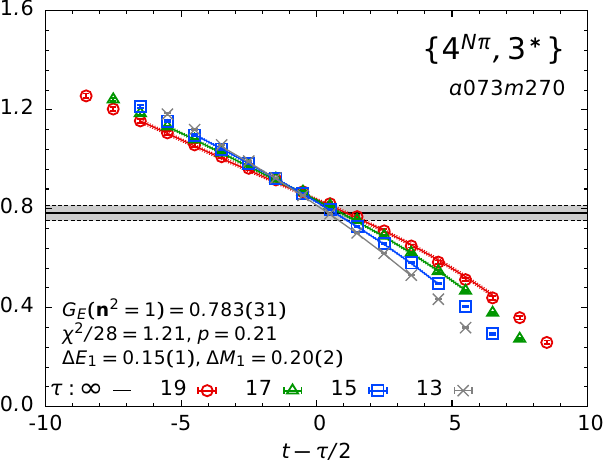} 
    \includegraphics[width=0.24\linewidth]{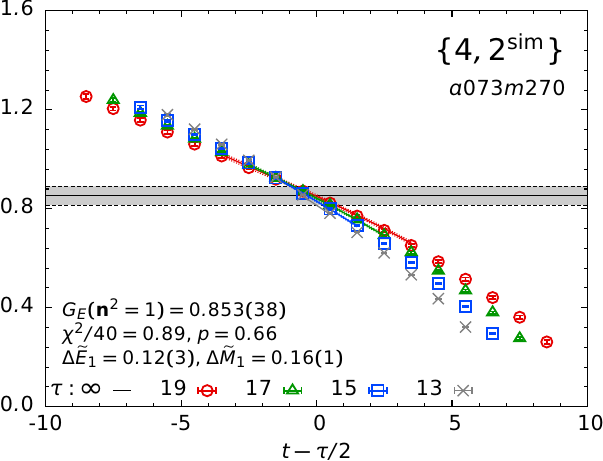} 
    \includegraphics[width=0.24\linewidth]{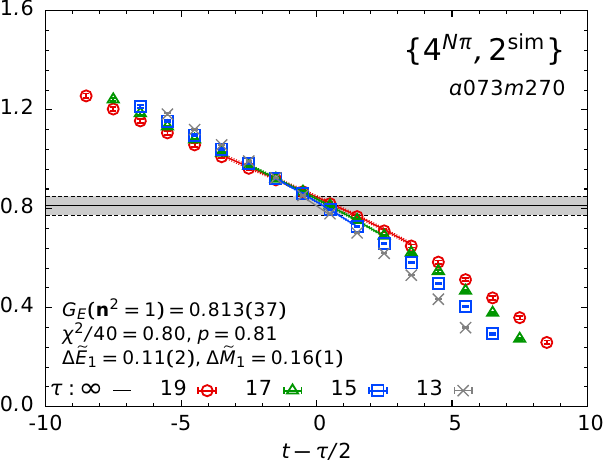} 
}
{
    \includegraphics[width=0.24\linewidth]{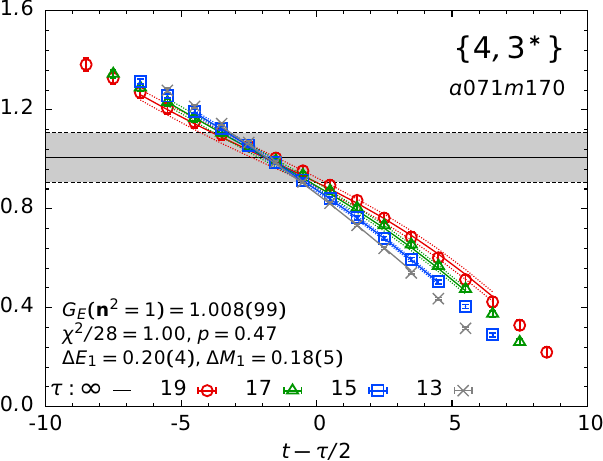}  
    \includegraphics[width=0.24\linewidth]{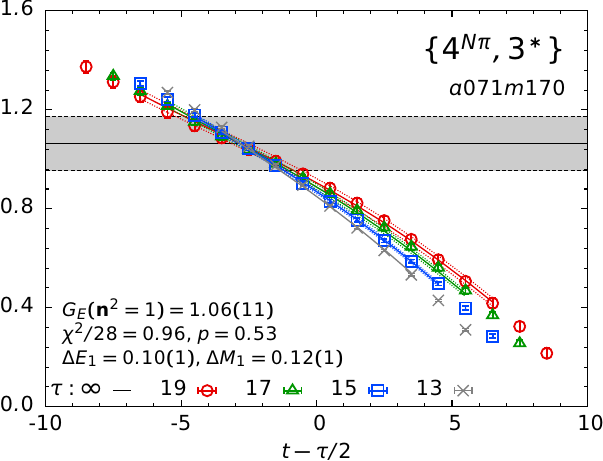} 
    \includegraphics[width=0.24\linewidth]{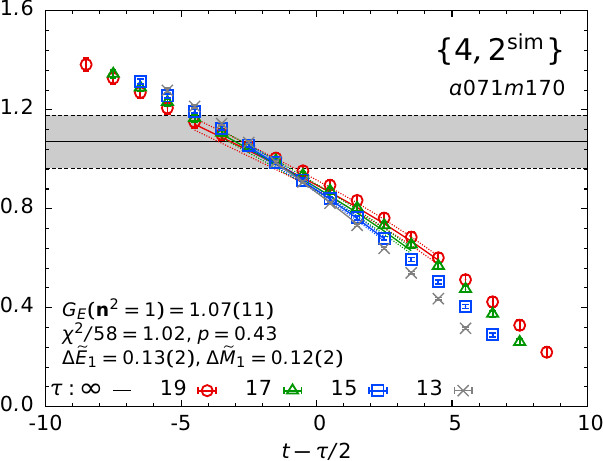} 
    \includegraphics[width=0.24\linewidth]{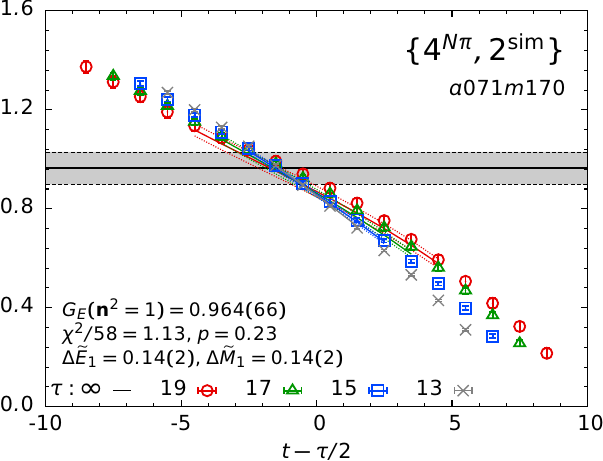} 
}
\caption{$G_E({\bm n} = (1,0,0))$, obtained from $\Im V_i$ (see
  Eq.~\protect\eqref{eq:GE1}) plotted versus the
  shifted operator insertion point $t -\tau/2$.  The rest is the same
  as in Fig.~\protect\ref{fig:affA4COMP}.
  \label{fig:ViCOMP}}
\end{figure*}

\begin{figure*}[tbp] 
\subfigure
{
    \includegraphics[width=0.24\linewidth]{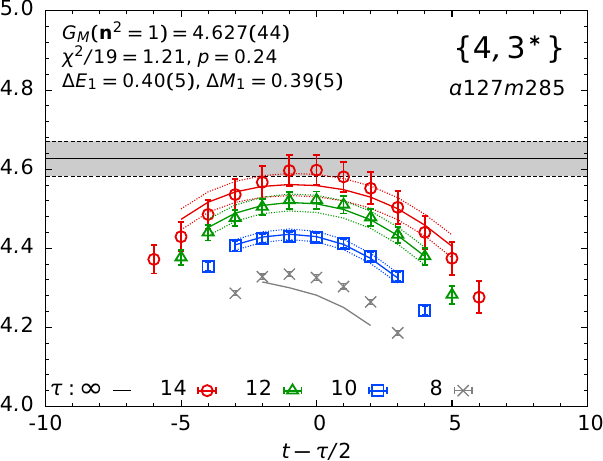}  
    \includegraphics[width=0.24\linewidth]{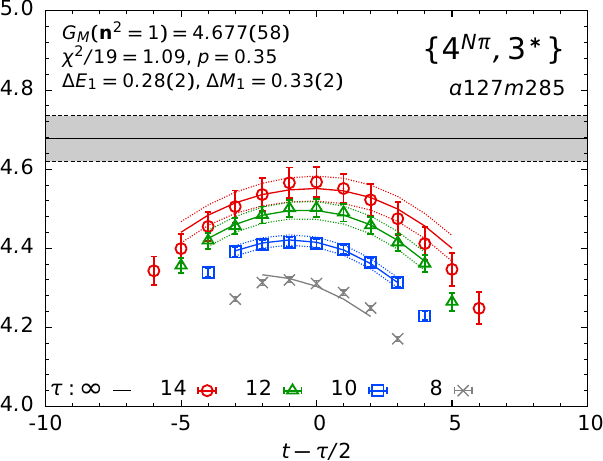}  
    \includegraphics[width=0.24\linewidth]{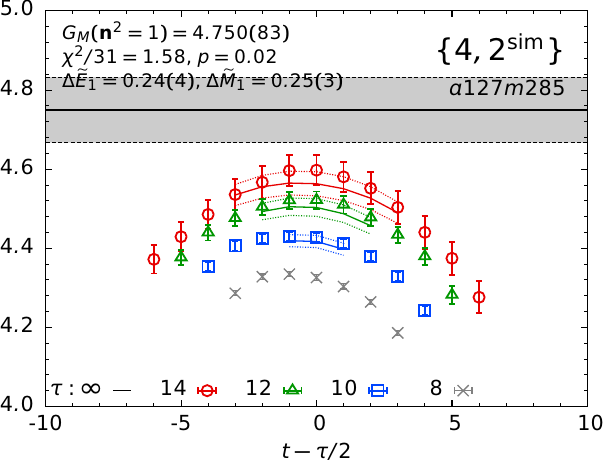} 
    \includegraphics[width=0.24\linewidth]{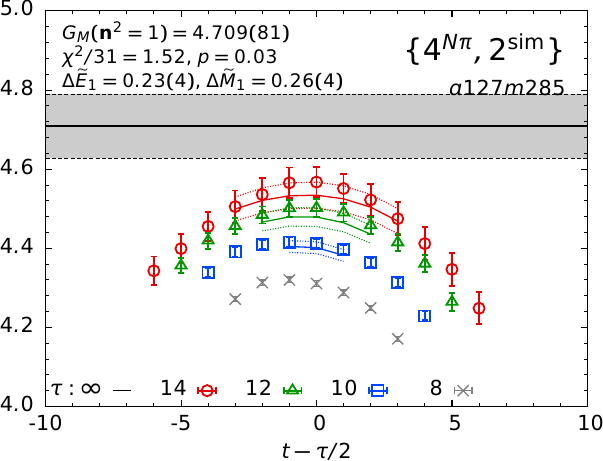} 
}
{
    \includegraphics[width=0.24\linewidth]{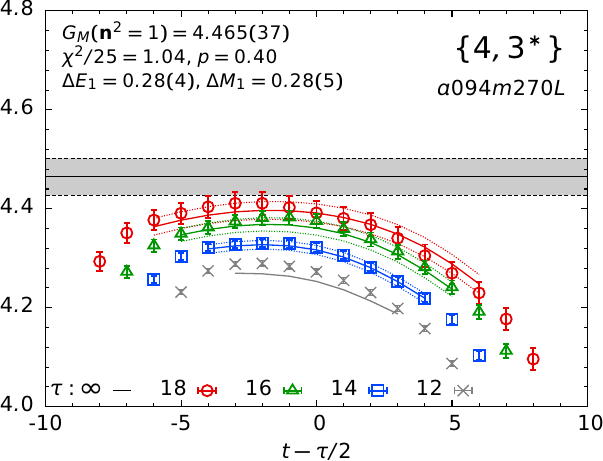}  
    \includegraphics[width=0.24\linewidth]{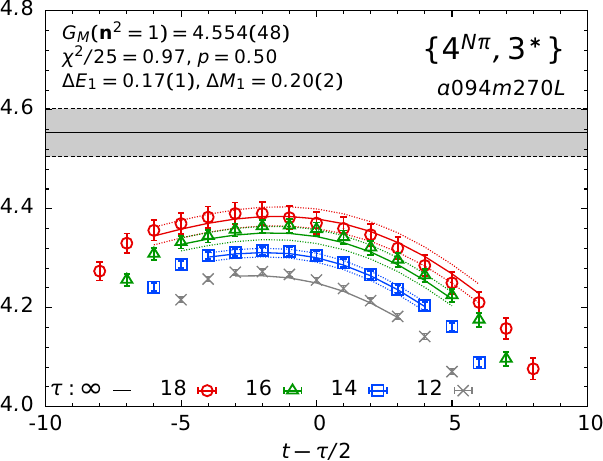} 
    \includegraphics[width=0.24\linewidth]{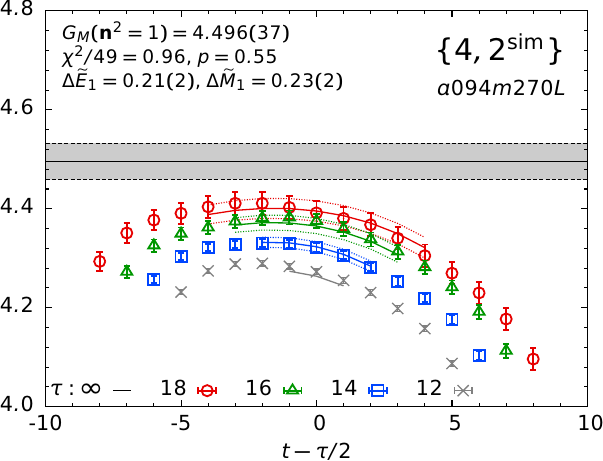} 
    \includegraphics[width=0.24\linewidth]{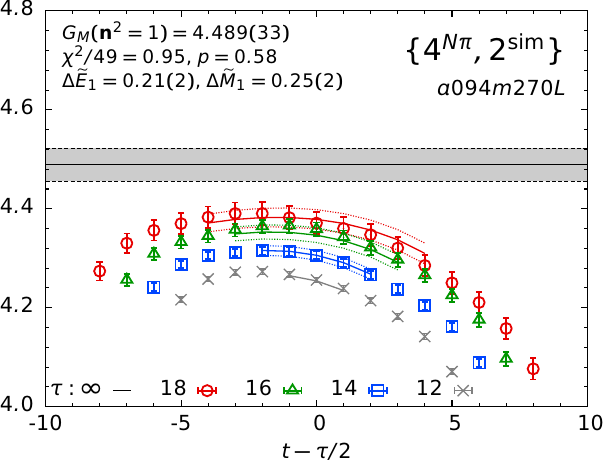} 
}
{
}
{
    \includegraphics[width=0.24\linewidth]{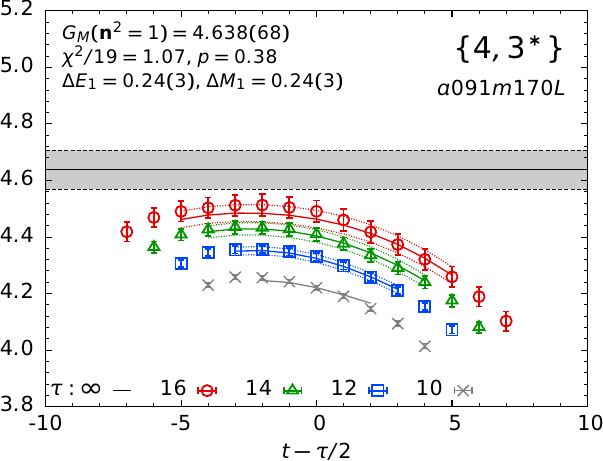} 
    \includegraphics[width=0.24\linewidth]{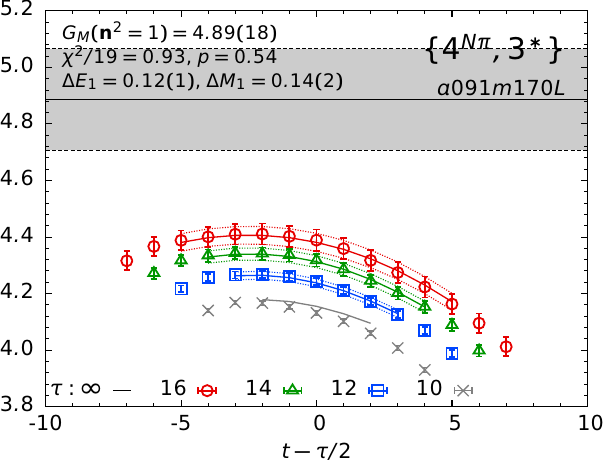} 
    \includegraphics[width=0.24\linewidth]{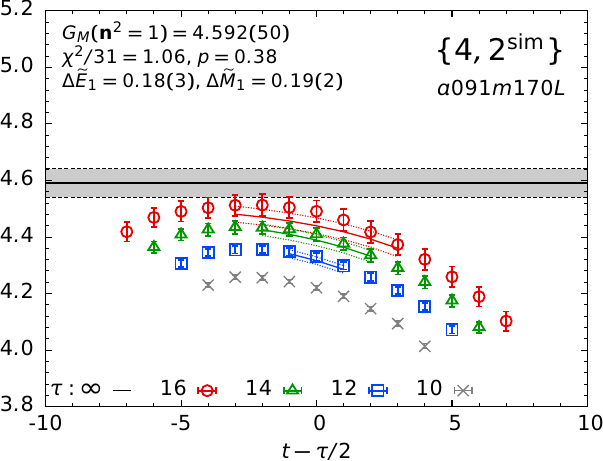} 
    \includegraphics[width=0.24\linewidth]{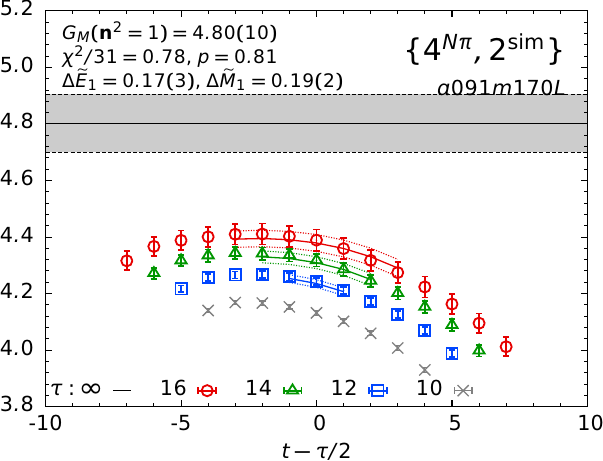} 
}
{
    \includegraphics[width=0.24\linewidth]{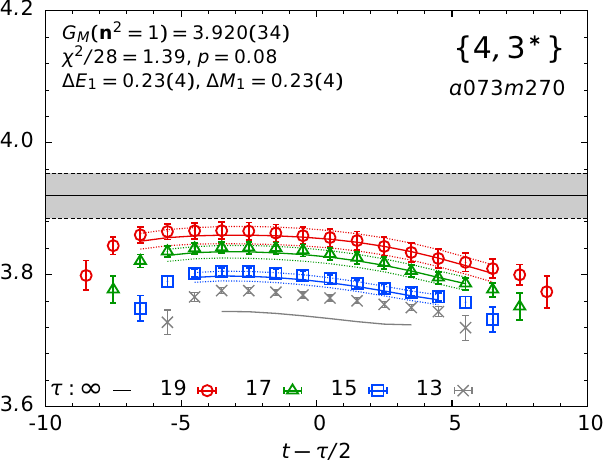}  
    \includegraphics[width=0.24\linewidth]{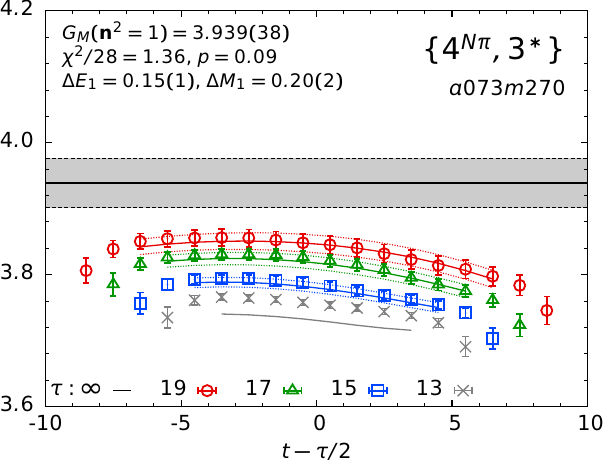} 
    \includegraphics[width=0.24\linewidth]{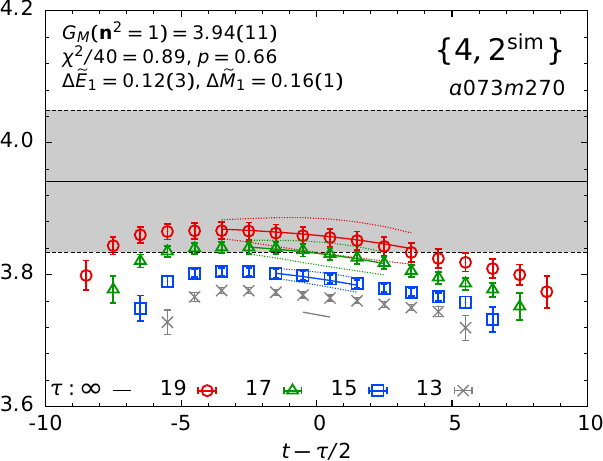} 
    \includegraphics[width=0.24\linewidth]{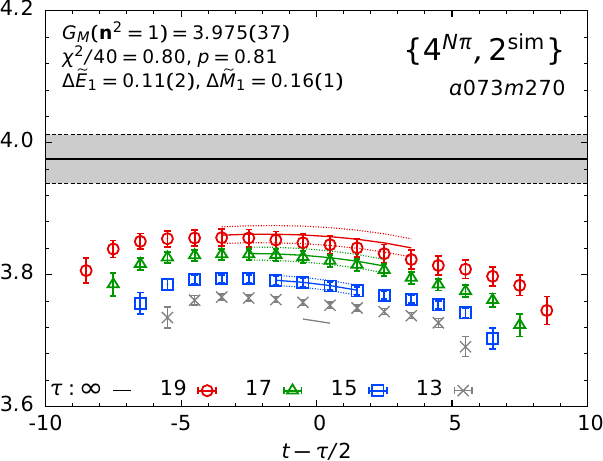} 
}
{
    \includegraphics[width=0.24\linewidth]{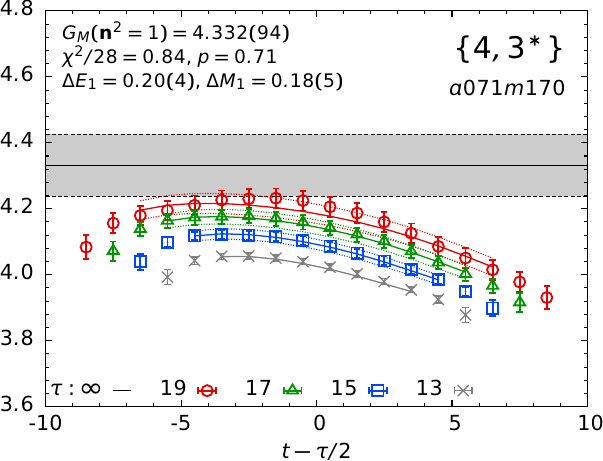}  
    \includegraphics[width=0.24\linewidth]{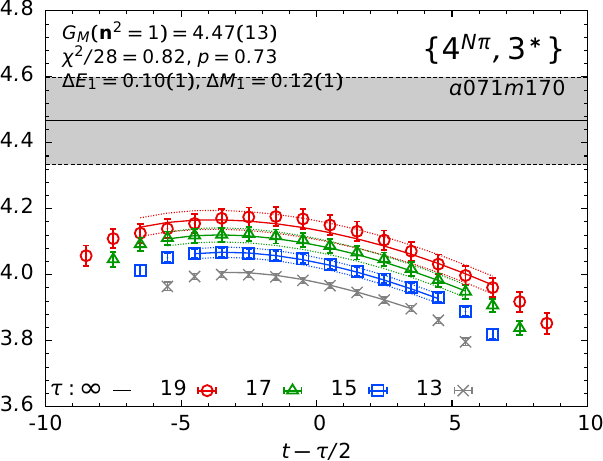} 
    \includegraphics[width=0.24\linewidth]{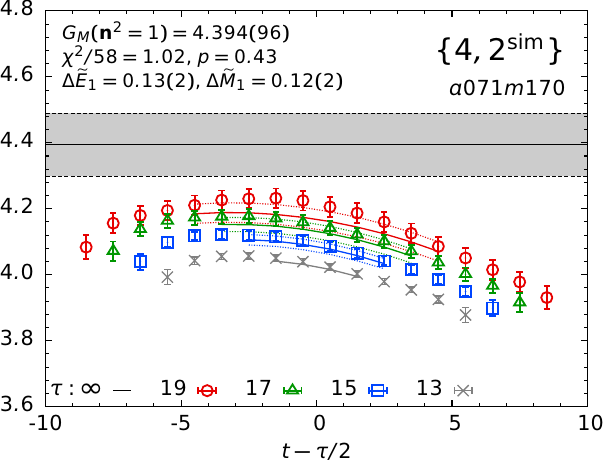} 
    \includegraphics[width=0.24\linewidth]{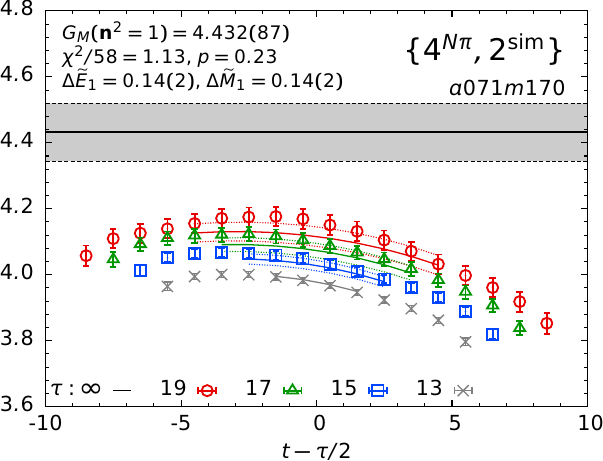} 
}
\caption{$G_M({\bm n} = (1,0,0))$, obtained from $\Re V_i$ [see
  Eq.~\protect\eqref{eq:GM1}] plotted versus the
  shifted operator insertion point $t -\tau/2$.  The rest is the same
  as in Fig.~\protect\ref{fig:affA4COMP}.
  \label{fig:GMCOMP}}
\end{figure*}

\begin{figure*}[tbp] 
\subfigure
{
    \includegraphics[width=0.43\linewidth]{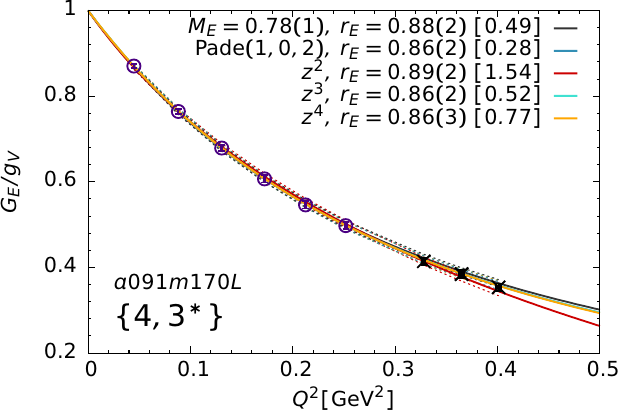}  
    \includegraphics[width=0.43\linewidth]{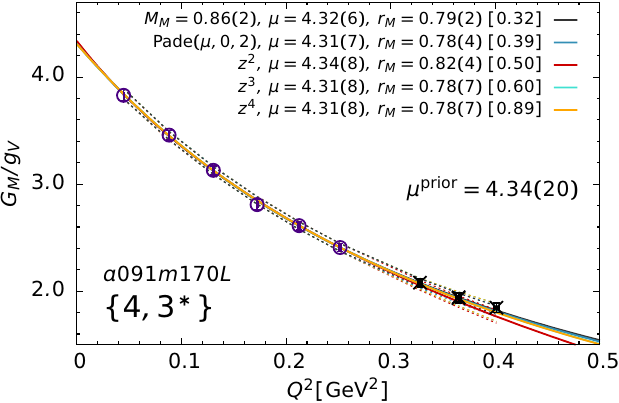}  
}
{
    \includegraphics[width=0.43\linewidth]{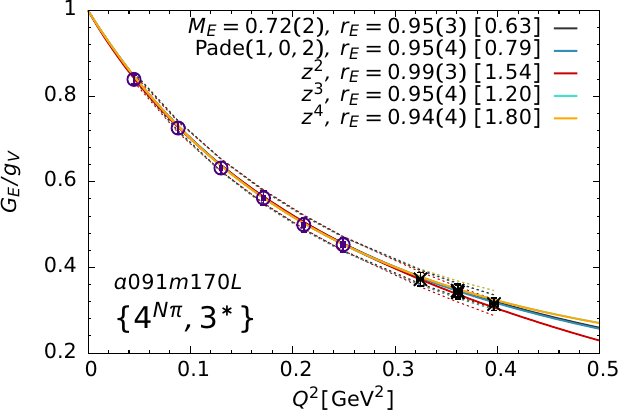}  
    \includegraphics[width=0.43\linewidth]{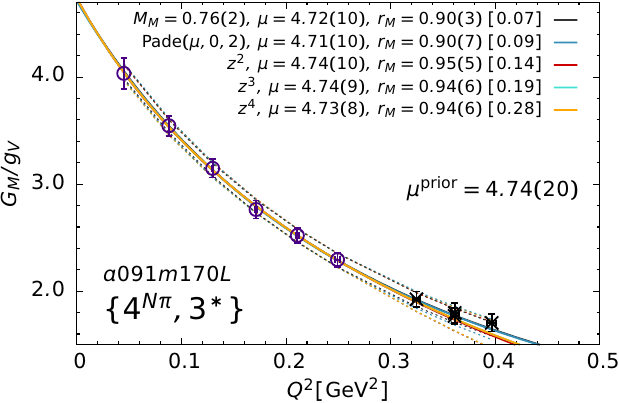}  
}
{
    \includegraphics[width=0.43\linewidth]{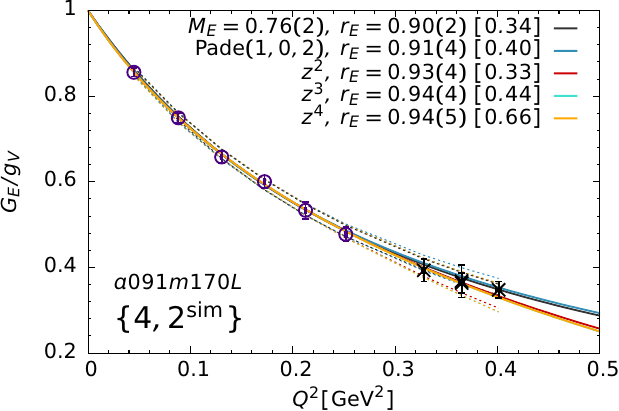}  
    \includegraphics[width=0.43\linewidth]{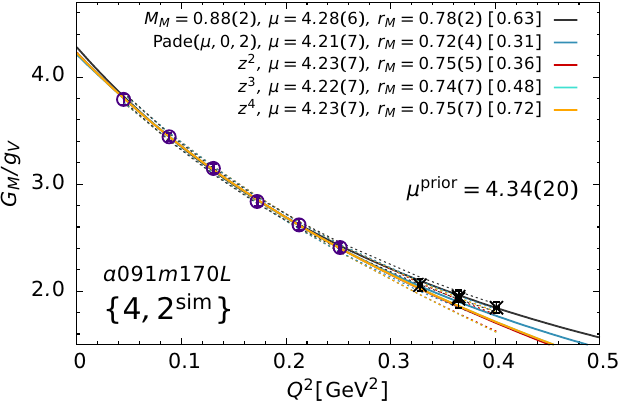}  
}
{
    \includegraphics[width=0.43\linewidth]{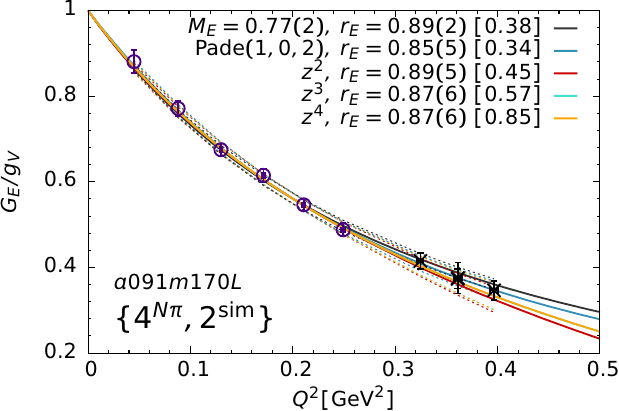}  
    \includegraphics[width=0.43\linewidth]{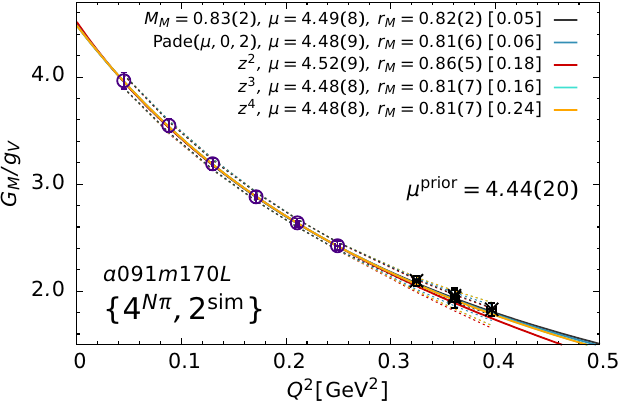}  
}
\caption{Data for the renormalized electric (left) and magnetic (right) form
  factors from the $a091m170L$ ensemble. All fits are made to the
  lowest six $Q^2$ points (open circles) and the remaining four points 
  not fit are shown by the symbol cross. Error bands are shown only
  over the range of the data for clarity. The prior and its width,
  $\mu^{\rm prior}$, used in the fits to $G_M$ is given in each panel
  and explained in the text. The top line of the labels gives the
  results of the dipole fit ($M_E$, $\langle r_E \rangle$) or ($M_M$,
  $\mu$ and $ \langle r_M \rangle$). Lines 2--5 give $\langle
  r_E \rangle$ or ($\mu$ and $ \langle r_M \rangle$) from the $P_2$
  Pad\'e and the $z^{\{2,3,4\}}$ fits. In
  each case, the $\chi^2$/dof of the fits are given within the square
  brackets.  The four rows show data from the four strategies
  $\{4^{},3^{\ast}\}$, $\{4^{N\pi},3^{\ast}\}$, $\{4^{},2^{\rm sim}\}$
  and $\{4^{N\pi},2^{\rm sim}\}$ defined in the
  text.\looseness-1 \label{fig:VFF-Q2D7}}
\end{figure*}

\begin{figure*}[tbp] 
\subfigure
{
    \includegraphics[width=0.43\linewidth]{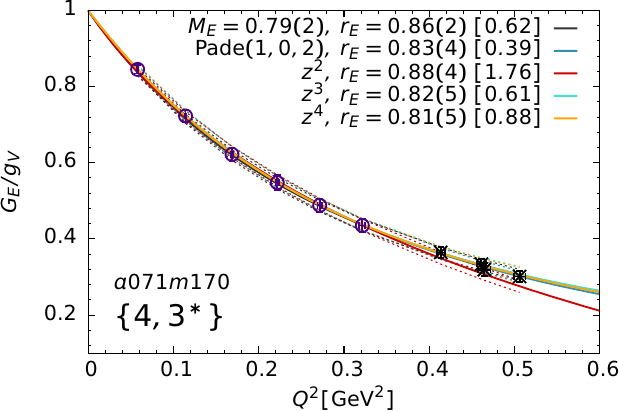}  
    \includegraphics[width=0.43\linewidth]{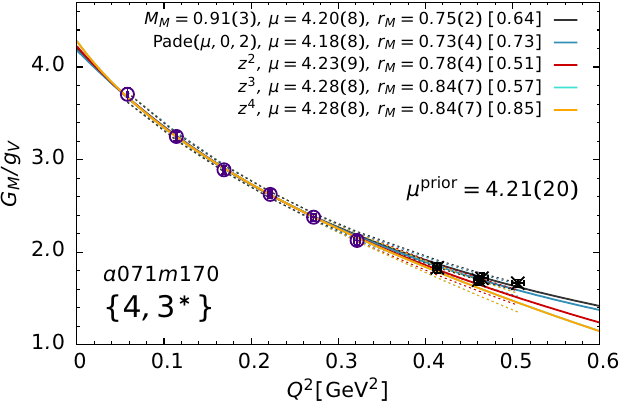}  
}
{
    \includegraphics[width=0.43\linewidth]{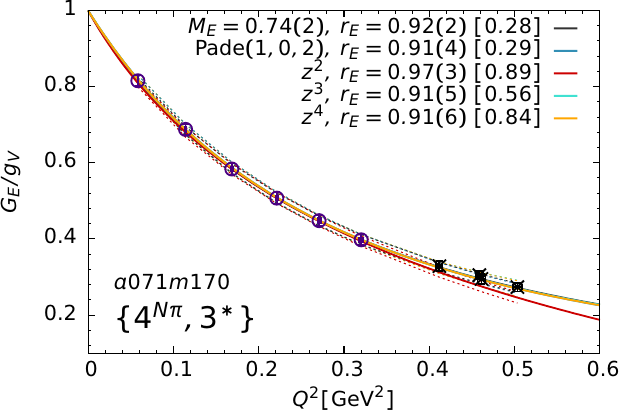}  
    \includegraphics[width=0.43\linewidth]{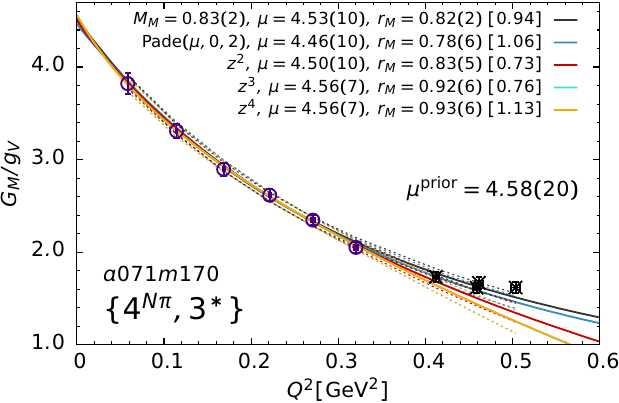}  
}
{
    \includegraphics[width=0.43\linewidth]{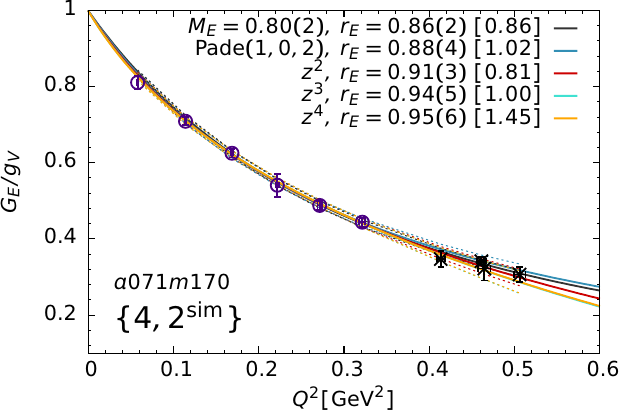}  
    \includegraphics[width=0.43\linewidth]{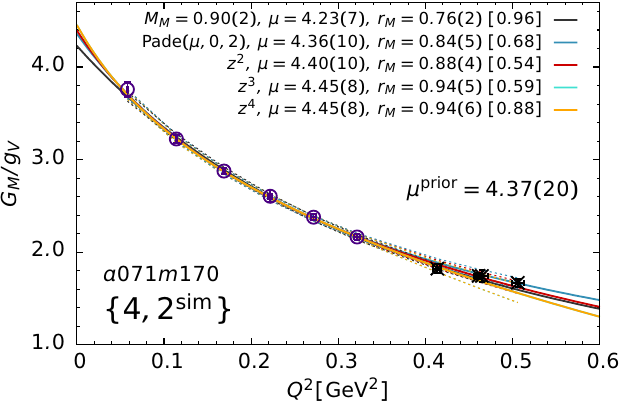}  
}
{
    \includegraphics[width=0.43\linewidth]{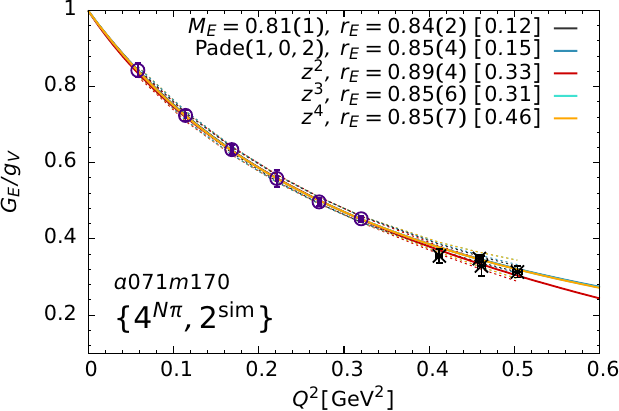}  
    \includegraphics[width=0.43\linewidth]{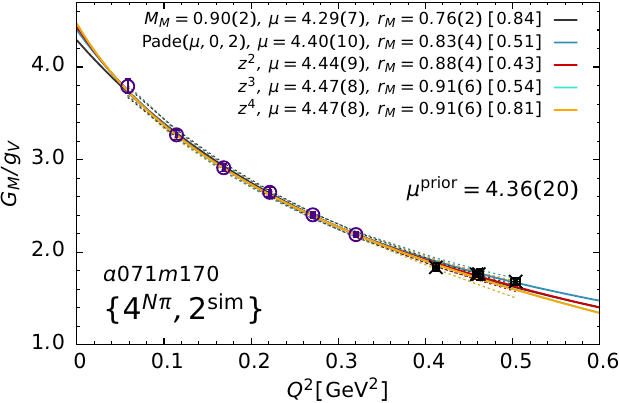}  
}
\caption{The data for the renormalized electric (left) and magnetic (right) form
  factors from the $a071m170$ ensemble fit using the dipole, $P_2$ Pad\'e 
  and $z^{2,3,4}$ ansatz. The rest is the same as in
  Fig.~\protect\ref{fig:VFF-Q2D7}.
  \label{fig:VFF-Q2E7}}
\end{figure*}

In this appendix, we show in Figs.~\ref{fig:GECOMP},~\ref{fig:ViCOMP}
and~\ref{fig:GMCOMP} the ratios defined in
Eqs.~\eqref{eq:GM1}--\eqref{eq:GE4} that give $G_E^{V_4}$,
$G_E^{V_i}$, and $G_M^{V_i}$. The four panels in each 
row show the results for the ground state matrix element obtained using 
the four ESC strategies, $\{4,3^*\}$,
$\{4^{N\pi},3^*\}$, $\{4,2^{\rm sim}\}$,
$\{4^{N\pi},2^\text{sim}\}$. The renormalized electric and
magnetic form factors are given in
Tables~\ref{tab:FF-GE4strategies},~\ref{tab:FF-GEi4strategies},
and~\ref{tab:FF-GM4strategies}. Each panel in Figs.~\ref{fig:VFF-Q2D7}
and~\ref{fig:VFF-Q2E7} shows the dipole, Pad\'e  and $z$-expansion fits to these data and gives the values of $\rEsq,\ \rMsq, \ \mu$ obtained. 
Data from the four strategies are shown in the four rows in each 
figure, and for the $a091m170L$ and $a071m170$ ensembles 
in the two figures.

  \begin{table*}   
  \vspace{-15mm}
  \begin{ruledtabular}
    \begin{tabular}{c|lllllll}
$\bm {n}$ & $a127m285$ & $a094m270$ & $a094m270L$ & $a091m170$ & $a091m170L$ & $a073m270$ & $a071m170$\\ \hline\hline 
& \multicolumn{7}{c}{Strategy $\{4,3^\ast\}$} \\ \hline
$(1,0,0)$ & 0.764(05)[2.05] & 0.642(20)[1.18] & 0.817(04)[1.03] & 0.791(06)[1.06] & 0.870(05)[1.07] & 0.737(07)[1.29] & 0.845(11)[1.10]\\
$(1,1,0)$ & 0.609(06)[1.21] & 0.461(21)[1.56] & 0.680(06)[0.90] & 0.643(08)[1.04] & 0.764(06)[0.90] & 0.569(11)[0.99] & 0.723(15)[1.17]\\
$(1,1,1)$ & 0.495(08)[1.47] & 0.356(16)[0.95] & 0.576(08)[1.32] & 0.541(09)[0.63] & 0.679(07)[0.79] & 0.454(12)[1.35] & 0.622(18)[1.25]\\
$(2,0,0)$ & 0.417(08)[1.04] & 0.245(21)[0.91] & 0.499(08)[1.03] & 0.464(10)[0.95] & 0.607(08)[1.71] & 0.376(09)[1.16] & 0.548(20)[0.91]\\
$(2,1,0)$ & 0.356(08)[1.01] & 0.214(14)[1.02] & 0.436(08)[1.44] & 0.401(09)[1.44] & 0.546(09)[1.24] & 0.312(10)[1.42] & 0.488(17)[0.88]\\
$(2,1,1)$ & 0.303(10)[1.33] & 0.184(11)[1.13] & 0.385(08)[1.69] & 0.355(09)[1.09] & 0.498(09)[0.81] & 0.262(11)[1.56] & 0.435(18)[0.95]\\
$(2,2,0)$ & 0.234(09)[0.77] & 0.137(16)[1.02] & 0.310(08)[1.35] & 0.276(11)[1.52] & 0.414(09)[1.54] & 0.196(11)[1.21] & 0.364(14)[0.89]\\
$(2,2,1)$ & 0.208(10)[1.04] & 0.128(14)[1.12] & 0.278(08)[1.62] & 0.251(09)[0.47] & 0.385(09)[0.91] & 0.176(09)[2.24] & 0.333(13)[0.62]\\
$(3,0,0)$ & 0.212(10)[1.23] & 0.094(43)[1.33] & 0.284(08)[1.31] & 0.236(16)[1.11] & 0.384(09)[1.26] & 0.180(09)[0.97] & 0.319(18)[0.52]\\
$(3,1,0)$ & 0.188(09)[0.78] & 0.125(24)[1.41] & 0.261(07)[1.31] & 0.222(11)[1.42] & 0.353(09)[1.35] & 0.156(08)[1.07] & 0.301(14)[0.48]\\
\hline & \multicolumn{7}{c}{Strategy $\{4^{N\pi},3^\ast\}$} \\ \hline
$(1,0,0)$ & 0.755(05)[1.64] & 0.643(09)[1.15] & 0.804(04)[1.02] & 0.750(10)[0.94] & 0.839(11)[1.01] & 0.725(06)[0.97] & 0.815(14)[1.16]\\
$(1,1,0)$ & 0.599(05)[1.04] & 0.465(07)[1.24] & 0.664(05)[0.89] & 0.593(09)[1.01] & 0.726(13)[0.87] & 0.557(06)[1.02] & 0.688(14)[1.19]\\
$(1,1,1)$ & 0.487(05)[1.49] & 0.357(07)[0.89] & 0.559(05)[1.27] & 0.492(11)[0.85] & 0.632(15)[0.76] & 0.442(07)[1.59] & 0.584(15)[1.46]\\
$(2,0,0)$ & 0.409(05)[1.01] & 0.268(10)[0.96] & 0.481(05)[0.98] & 0.416(12)[0.88] & 0.562(16)[1.92] & 0.364(06)[1.11] & 0.507(16)[1.05]\\
$(2,1,0)$ & 0.350(05)[1.03] & 0.224(06)[1.04] & 0.422(05)[1.19] & 0.353(10)[1.18] & 0.499(17)[1.43] & 0.306(05)[1.58] & 0.448(14)[0.96]\\
$(2,1,1)$ & 0.300(05)[1.38] & 0.192(07)[1.14] & 0.371(05)[1.53] & 0.317(12)[0.79] & 0.453(17)[0.93] & 0.263(05)[2.01] & 0.398(14)[1.12]\\
$(2,2,0)$ & 0.233(05)[0.76] & 0.141(12)[1.01] & 0.302(05)[1.14] & 0.230(16)[1.09] & 0.372(16)[1.73] & 0.201(05)[1.47] & 0.328(13)[0.88]\\
$(2,2,1)$ & 0.207(05)[1.03] & 0.131(10)[1.12] & 0.271(05)[1.49] & 0.228(14)[0.55] & 0.346(15)[1.05] & 0.184(04)[2.52] & 0.305(11)[0.67]\\
$(3,0,0)$ & 0.207(08)[1.20] & 0.122(22)[1.37] & 0.269(07)[1.24] & 0.213(24)[1.03] & 0.342(16)[1.33] & 0.184(07)[0.97] & 0.295(14)[0.55]\\
$(3,1,0)$ & 0.185(07)[0.76] & 0.122(27)[1.42] & 0.248(06)[1.09] & 0.212(17)[1.75] & 0.314(15)[1.48] & 0.160(06)[1.18] & 0.273(12)[0.48]\\
\hline & \multicolumn{7}{c}{Strategy $\{4,2^\text{sim}\}$} \\ \hline
$(1,0,0)$ & 0.749(09)[1.58] & 0.603(41)[1.07] & 0.810(05)[0.96] & 0.778(11)[1.48] & 0.856(10)[1.06] & 0.713(14)[0.89] & 0.811(16)[1.02]\\
$(1,1,0)$ & 0.589(14)[1.19] & 0.424(52)[0.92] & 0.671(07)[0.83] & 0.625(19)[0.55] & 0.749(12)[0.98] & 0.549(11)[1.17] & 0.709(10)[1.57]\\
$(1,1,1)$ & 0.481(17)[1.44] & 0.362(22)[1.02] & 0.562(11)[1.08] & 0.539(15)[0.85] & 0.658(14)[1.09] & 0.450(11)[1.04] & 0.625(11)[1.35]\\
$(2,0,0)$ & 0.416(14)[1.19] & 0.16(15)[1.31] & 0.492(09)[1.11] & 0.433(27)[0.82] & 0.600(14)[0.93] & 0.362(18)[1.35] & 0.541(30)[0.91]\\
$(2,1,0)$ & 0.350(14)[1.03] & 0.218(33)[0.92] & 0.407(16)[1.06] & 0.347(43)[0.77] & 0.533(20)[1.27] & 0.311(11)[0.97] & 0.488(11)[1.31]\\
$(2,1,1)$ & 0.294(17)[1.36] & 0.104(44)[1.41] & 0.349(24)[1.30] & 0.322(32)[0.70] & 0.477(17)[1.30] & 0.254(21)[0.87] & 0.444(12)[1.20]\\
$(2,2,0)$ & 0.236(19)[0.90] & 0.158(03)[1.28] & 0.281(24)[1.13] & 0.232(83)[1.24] & 0.393(26)[1.29] & 0.180(25)[0.76] & 0.347(22)[1.10]\\
$(2,2,1)$ & 0.172(61)[1.91] & 0.115(24)[1.04] & 0.179(60)[1.06] & 0.219(47)[0.89] & 0.363(23)[1.41] & 0.182(17)[0.79] & 0.341(11)[1.02]\\
$(3,0,0)$ & 0.08(26)[1.21] & 0.062(52)[1.60] & 0.242(29)[1.25] & 0.224(66)[1.33] & 0.368(38)[1.26] & 0.177(25)[0.95] & 0.322(31)[0.66]\\
$(3,1,0)$ & 0.17(15)[0.91] & 0.121(04)[1.74] & 0.14(16)[1.01] & 0.214(39)[1.04] & 0.348(19)[1.29] & 0.167(16)[1.13] & 0.306(20)[0.70]\\
\hline & \multicolumn{7}{c}{Strategy $\{4^{N\pi},2^\text{sim}\}$} \\ \hline
$(1,0,0)$ & 0.746(11)[1.52] & 0.593(43)[1.09] & 0.816(06)[0.95] & 0.790(15)[1.59] & 0.881(27)[0.78] & 0.708(13)[0.80] & 0.843(18)[1.13]\\
$(1,1,0)$ & 0.590(14)[1.21] & 0.427(41)[0.89] & 0.676(07)[0.88] & 0.641(15)[0.60] & 0.771(17)[0.81] & 0.551(11)[1.19] & 0.724(14)[1.62]\\
$(1,1,1)$ & 0.478(19)[1.53] & 0.357(21)[1.01] & 0.567(15)[1.18] & 0.548(11)[0.86] & 0.675(14)[1.11] & 0.450(10)[1.08] & 0.635(12)[1.16]\\
$(2,0,0)$ & 0.416(14)[1.19] & 0.216(90)[1.36] & 0.494(10)[1.16] & 0.441(36)[0.94] & 0.615(16)[0.92] & 0.364(18)[1.37] & 0.559(22)[0.89]\\
$(2,1,0)$ & 0.349(15)[1.06] & 0.220(26)[0.93] & 0.408(18)[1.14] & 0.358(48)[0.80] & 0.546(14)[1.27] & 0.312(11)[0.96] & 0.497(10)[1.29]\\
$(2,1,1)$ & 0.295(15)[1.44] & 0.112(44)[1.38] & 0.351(28)[1.41] & 0.336(41)[0.76] & 0.488(16)[1.35] & 0.259(19)[0.86] & 0.453(08)[1.16]\\
$(2,2,0)$ & 0.233(16)[1.00] & 0.154(04)[1.50] & 0.247(31)[1.07] & 0.17(16)[0.89] & 0.415(19)[1.36] & 0.185(23)[0.76] & 0.355(18)[1.11]\\
$(2,2,1)$ & 0.169(63)[1.95] & 0.135(08)[1.06] & 0.177(51)[1.33] & 0.219(48)[0.96] & 0.373(23)[1.44] & 0.186(14)[0.78] & 0.348(09)[1.01]\\
$(3,0,0)$ & 0.127(70)[0.96] & 0.076(76)[1.60] & 0.246(56)[1.17] & 0.224(64)[1.34] & 0.375(36)[1.28] & 0.181(22)[0.96] & 0.329(27)[0.66]\\
$(3,1,0)$ & 0.04(28)[1.20] & 0.132(72)[1.66] & 0.211(30)[1.31] & 0.246(13)[1.10] & 0.347(22)[0.99] & 0.166(17)[1.13] & 0.313(15)[0.71]\\
\end{tabular} \vspace{5mm}
  \end{ruledtabular} \caption{Data for renormalized $G_E^{\Re
  V_4}(Q^2)/g_V$ from the seven ensembles and with the four strategies
  for controlling ESC. The $\chi^2/$dof of the fits are given within
  square parentheses, and are the same for the three quantities
  $G_E^{\Re V_4}(Q^2)$, $G_E^{\Im V_i}(Q^2)$ and $G_M^{\Re V_i}(Q^2)$ in 
  the simultaneous $\{2^{\rm sim}\}$ fits.  Only data with the four
  strategies for a given ensemble and ${\bm n}$ can be compared.} 
\label{tab:FF-GE4strategies} 
\end{table*}
  

  \begin{table*}  
  \vspace{-15mm}
  \begin{ruledtabular}
    \begin{tabular}{c|lllllll}
$\bm {n}$ & $a127m285$ & $a094m270$ & $a094m270L$ & $a091m170$ & $a091m170L$ & $a073m270$ & $a071m170$\\ \hline\hline 
& \multicolumn{7}{c}{Strategy $\{4,3^\ast\}$} \\ \hline
$(1,0,0)$ & 0.761(17)[1.20] & 0.701(89)[0.96] & 0.817(19)[0.90] & 0.816(46)[2.36] & 0.872(35)[0.88] & 0.726(30)[2.99] & 0.862(81)[1.00]\\
$(1,1,0)$ & 0.620(15)[1.23] & 0.534(52)[1.00] & 0.690(17)[0.91] & 0.679(30)[1.47] & 0.788(28)[0.58] & 0.583(25)[2.32] & 0.749(59)[1.59]\\
$(1,1,1)$ & 0.505(16)[1.24] & 0.394(25)[1.20] & 0.593(18)[1.51] & 0.591(26)[1.37] & 0.728(27)[0.85] & 0.483(23)[1.86] & 0.661(57)[1.04]\\
$(2,0,0)$ & 0.436(20)[0.85] & 0.296(38)[1.08] & 0.531(17)[1.13] & 0.518(26)[0.73] & 0.635(29)[0.81] & 0.412(16)[1.23] & 0.611(56)[1.00]\\
$(2,1,0)$ & 0.374(16)[1.48] & 0.254(25)[0.92] & 0.466(14)[1.19] & 0.461(22)[1.27] & 0.598(24)[0.81] & 0.339(16)[1.15] & 0.528(34)[1.11]\\
$(2,1,1)$ & 0.318(17)[1.90] & 0.185(25)[1.48] & 0.419(15)[1.34] & 0.418(20)[0.81] & 0.564(22)[1.20] & 0.295(18)[1.35] & 0.484(34)[0.97]\\
$(2,2,0)$ & 0.248(20)[1.54] & 0.163(27)[1.25] & 0.345(13)[0.98] & 0.335(21)[0.96] & 0.473(25)[1.07] & 0.216(16)[1.11] & 0.427(27)[1.16]\\
$(2,2,1)$ & 0.218(18)[1.90] & 0.095(36)[1.04] & 0.311(15)[0.99] & 0.309(19)[0.77] & 0.439(20)[0.69] & 0.201(15)[1.42] & 0.376(24)[0.78]\\
$(3,0,0)$ & 0.221(17)[1.43] & 0.082(66)[1.20] & 0.322(16)[1.31] & 0.262(34)[0.92] & 0.435(28)[0.87] & 0.212(16)[1.04] & 0.378(38)[0.60]\\
$(3,1,0)$ & 0.212(21)[0.61] & 0.211(93)[1.56] & 0.302(13)[1.00] & 0.280(30)[1.31] & 0.397(29)[1.23] & 0.162(18)[0.89] & 0.346(28)[0.58]\\
\hline & \multicolumn{7}{c}{Strategy $\{4^{N\pi},3^\ast\}$} \\ \hline
$(1,0,0)$ & 0.725(34)[0.91] & 0.635(33)[1.00] & 0.823(27)[1.38] & 0.94(12)[1.74] & 0.91(10)[0.58] & 0.669(26)[1.21] & 0.910(93)[0.96]\\
$(1,1,0)$ & 0.609(15)[0.91] & 0.498(20)[1.01] & 0.694(20)[0.69] & 0.709(63)[1.09] & 0.860(69)[0.57] & 0.567(15)[1.31] & 0.793(56)[1.67]\\
$(1,1,1)$ & 0.505(13)[1.09] & 0.375(19)[1.20] & 0.601(17)[0.97] & 0.624(55)[1.36] & 0.802(63)[0.74] & 0.477(12)[1.52] & 0.707(58)[1.26]\\
$(2,0,0)$ & 0.430(15)[0.81] & 0.317(23)[1.08] & 0.531(17)[1.16] & 0.443(51)[0.66] & 0.682(62)[0.77] & 0.405(12)[1.01] & 0.659(51)[0.98]\\
$(2,1,0)$ & 0.371(12)[1.49] & 0.259(16)[0.94] & 0.471(13)[0.90] & 0.490(44)[0.66] & 0.639(51)[0.80] & 0.343(09)[1.04] & 0.528(32)[1.24]\\
$(2,1,1)$ & 0.320(11)[1.90] & 0.198(18)[1.49] & 0.427(13)[1.06] & 0.453(45)[0.57] & 0.604(44)[1.22] & 0.310(09)[1.43] & 0.496(35)[1.09]\\
$(2,2,0)$ & 0.250(12)[1.53] & 0.167(21)[1.24] & 0.353(11)[0.93] & 0.351(42)[0.68] & 0.492(48)[1.14] & 0.233(08)[1.15] & 0.423(29)[1.15]\\
$(2,2,1)$ & 0.219(13)[1.89] & 0.113(26)[1.05] & 0.321(13)[0.88] & 0.317(52)[0.79] & 0.449(38)[0.75] & 0.217(10)[1.45] & 0.374(31)[0.80]\\
$(3,0,0)$ & 0.217(19)[1.42] & 0.095(97)[1.23] & 0.319(17)[1.25] & 0.198(23)[0.87] & 0.432(55)[0.81] & 0.220(16)[1.05] & 0.402(38)[0.59]\\
$(3,1,0)$ & 0.206(23)[0.59] & 0.16(12)[1.57] & 0.306(18)[0.88] & 0.296(72)[1.37] & 0.378(54)[1.21] & 0.176(17)[0.93] & 0.330(42)[0.61]\\
\hline & \multicolumn{7}{c}{Strategy $\{4,2^\text{sim}\}$} \\ \hline
$(1,0,0)$ & 0.778(27)[1.58] & 0.585(78)[1.07] & 0.820(19)[0.96] & 0.815(42)[1.48] & 0.858(36)[1.06] & 0.729(32)[0.89] & 0.916(89)[1.02]\\
$(1,1,0)$ & 0.615(19)[1.19] & 0.448(48)[0.92] & 0.680(18)[0.83] & 0.631(36)[0.55] & 0.767(28)[0.98] & 0.560(19)[1.17] & 0.755(54)[1.57]\\
$(1,1,1)$ & 0.488(24)[1.44] & 0.379(35)[1.02] & 0.569(20)[1.08] & 0.550(25)[0.85] & 0.690(25)[1.09] & 0.458(17)[1.04] & 0.653(41)[1.35]\\
$(2,0,0)$ & 0.436(17)[1.19] & 0.15(16)[1.31] & 0.504(15)[1.11] & 0.453(35)[0.82] & 0.610(24)[0.93] & 0.376(20)[1.35] & 0.555(61)[0.91]\\
$(2,1,0)$ & 0.359(18)[1.03] & 0.225(36)[0.92] & 0.419(21)[1.06] & 0.388(48)[0.77] & 0.562(23)[1.27] & 0.326(14)[0.97] & 0.522(29)[1.31]\\
$(2,1,1)$ & 0.290(26)[1.36] & 0.003(04)[1.41] & 0.361(28)[1.30] & 0.346(39)[0.70] & 0.525(22)[1.30] & 0.263(24)[0.87] & 0.476(26)[1.20]\\
$(2,2,0)$ & 0.241(25)[0.90] & 0.170(05)[1.28] & 0.292(27)[1.13] & 0.239(86)[1.24] & 0.424(27)[1.29] & 0.197(27)[0.76] & 0.398(34)[1.10]\\
$(2,2,1)$ & 0.152(94)[1.91] & 0.002(04)[1.04] & 0.202(64)[1.06] & 0.270(39)[0.89] & 0.406(22)[1.41] & 0.194(18)[0.79] & 0.360(20)[1.02]\\
$(3,0,0)$ & 0.09(26)[1.21] & 0.002(05)[1.60] & 0.258(34)[1.25] & 0.282(30)[1.33] & 0.406(38)[1.26] & 0.193(26)[0.95] & 0.338(39)[0.66]\\
$(3,1,0)$ & 0.196(84)[0.91] & 0.117(09)[1.74] & 0.19(13)[1.01] & 0.251(45)[1.04] & 0.376(25)[1.29] & 0.167(20)[1.13] & 0.350(23)[0.70]\\
\hline & \multicolumn{7}{c}{Strategy $\{4^{N\pi},2^\text{sim}\}$} \\ \hline
$(1,0,0)$ & 0.747(25)[1.52] & 0.527(62)[1.09] & 0.804(17)[0.95] & 0.818(55)[1.59] & 0.830(49)[0.78] & 0.694(33)[0.80] & 0.825(56)[1.13]\\
$(1,1,0)$ & 0.608(17)[1.21] & 0.434(41)[0.89] & 0.680(17)[0.88] & 0.650(34)[0.60] & 0.805(45)[0.81] & 0.556(17)[1.19] & 0.759(37)[1.62]\\
$(1,1,1)$ & 0.486(21)[1.53] & 0.365(26)[1.01] & 0.576(24)[1.18] & 0.566(22)[0.86] & 0.732(39)[1.11] & 0.456(15)[1.08] & 0.676(29)[1.16]\\
$(2,0,0)$ & 0.434(14)[1.19] & 0.21(10)[1.36] & 0.506(16)[1.16] & 0.470(49)[0.94] & 0.649(33)[0.92] & 0.378(20)[1.37] & 0.582(28)[0.89]\\
$(2,1,0)$ & 0.359(16)[1.06] & 0.228(30)[0.93] & 0.424(23)[1.14] & 0.411(48)[0.80] & 0.600(26)[1.27] & 0.328(13)[0.96] & 0.536(18)[1.29]\\
$(2,1,1)$ & 0.298(20)[1.44] & 0.003(04)[1.38] & 0.367(33)[1.41] & 0.369(40)[0.76] & 0.560(27)[1.35] & 0.270(22)[0.86] & 0.490(17)[1.16]\\
$(2,2,0)$ & 0.242(20)[1.00] & 0.166(03)[1.50] & 0.262(33)[1.07] & 0.18(17)[0.89] & 0.460(25)[1.36] & 0.207(23)[0.76] & 0.412(25)[1.11]\\
$(2,2,1)$ & 0.161(95)[1.95] & 0.04(30)[1.06] & 0.177(56)[1.33] & 0.287(34)[0.96] & 0.431(25)[1.44] & 0.201(13)[0.78] & 0.370(14)[1.01]\\
$(3,0,0)$ & 0.127(98)[0.96] & 0.003(09)[1.60] & 0.272(53)[1.17] & 0.291(27)[1.34] & 0.412(41)[1.28] & 0.199(23)[0.96] & 0.351(29)[0.66]\\
$(3,1,0)$ & 0.15(17)[1.20] & 0.109(56)[1.66] & 0.251(40)[1.31] & 0.282(17)[1.10] & 0.401(31)[0.99] & 0.168(19)[1.13] & 0.358(18)[0.71]\\
\end{tabular} \vspace{5mm}
  \end{ruledtabular}
  \caption{Data for the renormalized $G_E^{\Im V_i}(Q^2)/g_V$ from the seven ensembles and with the 
    four strategies for controlling ESC. The rest is the same as in
    Table~\protect\ref{tab:FF-GE4strategies}. }
   \label{tab:FF-GEi4strategies}
  \end{table*}

  \begin{table*}  
  \begin{ruledtabular}
    \begin{tabular}{c|lllllll}
$\bm {n}$ & $a127m285$ & $a094m270$ & $a094m270L$ & $a091m170$ & $a091m170L$ & $a073m270$ & $a071m170$\\ \hline\hline 
& \multicolumn{7}{c}{Strategy $\{4,3^\ast\}$} \\ \hline
$(1,0,0)$ & 3.671(33)[1.21] & 3.072(58)[0.71] & 3.713(27)[1.04] & 3.498(42)[1.27] & 3.830(54)[1.07] & 3.349(26)[1.39] & 3.705(68)[0.84]\\
$(1,1,0)$ & 3.071(26)[1.21] & 2.320(71)[0.42] & 3.193(22)[0.80] & 2.962(30)[0.75] & 3.458(43)[1.71] & 2.724(16)[0.96] & 3.250(45)[1.12]\\
$(1,1,1)$ & 2.620(28)[1.18] & 1.834(68)[0.62] & 2.790(23)[0.91] & 2.566(68)[0.93] & 3.129(40)[1.53] & 2.269(22)[0.88] & 2.889(44)[0.97]\\
$(2,0,0)$ & 2.231(20)[0.97] & 1.593(69)[1.35] & 2.471(24)[1.07] & 2.302(37)[0.66] & 2.809(49)[1.01] & 1.963(22)[1.21] & 2.624(45)[0.98]\\
$(2,1,0)$ & 1.967(25)[0.67] & 1.333(48)[0.84] & 2.203(27)[0.81] & 2.032(29)[0.75] & 2.612(37)[1.53] & 1.691(24)[0.91] & 2.377(37)[0.99]\\
$(2,1,1)$ & 1.756(30)[1.00] & 1.146(51)[1.01] & 1.969(31)[1.65] & 1.808(34)[1.03] & 2.408(37)[1.16] & 1.471(33)[1.15] & 2.127(50)[0.94]\\
$(2,2,0)$ & 1.419(35)[0.85] & 0.88(15)[0.85] & 1.656(29)[1.12] & 1.527(34)[0.73] & 2.080(38)[1.27] & 1.202(29)[0.90] & 1.830(54)[1.18]\\
$(2,2,1)$ & 1.312(33)[1.83] & 0.817(67)[0.75] & 1.500(35)[1.69] & 1.364(41)[1.20] & 1.944(40)[1.31] & 1.069(36)[0.65] & 1.696(48)[1.32]\\
$(3,0,0)$ & 1.280(39)[1.21] & 0.85(20)[1.46] & 1.556(28)[1.01] & 1.442(51)[1.27] & 1.935(48)[1.34] & 1.101(31)[1.05] & 1.718(57)[0.77]\\
$(3,1,0)$ & 1.198(35)[0.97] & 0.82(14)[1.80] & 1.430(29)[1.09] & 1.283(50)[0.40] & 1.845(45)[1.51] & 1.019(29)[1.14] & 1.664(40)[0.87]\\
\hline & \multicolumn{7}{c}{Strategy $\{4^{N\pi},3^\ast\}$} \\ \hline
$(1,0,0)$ & 3.711(52)[1.09] & 3.054(47)[0.72] & 3.787(43)[0.97] & 3.69(15)[1.26] & 4.04(14)[0.93] & 3.365(34)[1.36] & 3.82(12)[0.82]\\
$(1,1,0)$ & 3.082(33)[1.17] & 2.337(40)[0.44] & 3.214(36)[0.79] & 2.960(86)[0.69] & 3.546(95)[1.54] & 2.723(21)[0.95] & 3.308(77)[1.09]\\
$(1,1,1)$ & 2.614(31)[1.22] & 1.842(38)[0.62] & 2.785(31)[0.89] & 2.535(83)[0.84] & 3.148(86)[1.45] & 2.254(19)[0.88] & 2.896(70)[0.95]\\
$(2,0,0)$ & 2.215(29)[0.98] & 1.638(51)[1.37] & 2.447(31)[1.01] & 2.260(92)[0.55] & 2.763(85)[0.98] & 1.949(21)[1.22] & 2.615(66)[0.95]\\
$(2,1,0)$ & 1.949(24)[0.72] & 1.356(25)[0.85] & 2.162(25)[0.71] & 1.983(65)[0.77] & 2.521(68)[1.39] & 1.679(17)[0.92] & 2.346(52)[0.99]\\
$(2,1,1)$ & 1.738(24)[1.03] & 1.172(31)[1.01] & 1.917(25)[1.61] & 1.714(73)[1.05] & 2.294(66)[1.04] & 1.471(17)[1.25] & 2.051(54)[0.96]\\
$(2,2,0)$ & 1.402(25)[0.90] & 0.895(50)[0.83] & 1.617(24)[1.11] & 1.439(81)[0.69] & 1.924(72)[1.11] & 1.211(18)[0.85] & 1.733(56)[1.20]\\
$(2,2,1)$ & 1.295(27)[1.81] & 0.839(43)[0.75] & 1.447(24)[1.63] & 1.187(78)[1.08] & 1.769(74)[1.11] & 1.088(19)[0.69] & 1.616(54)[1.33]\\
$(3,0,0)$ & 1.255(39)[1.22] & 0.879(99)[1.45] & 1.516(35)[1.01] & 1.44(15)[1.29] & 1.797(90)[1.24] & 1.109(28)[1.06] & 1.670(65)[0.73]\\
$(3,1,0)$ & 1.175(32)[1.02] & 0.839(80)[1.81] & 1.369(29)[1.12] & 1.182(93)[0.42] & 1.704(81)[1.30] & 1.028(25)[1.13] & 1.618(56)[0.87]\\
\hline & \multicolumn{7}{c}{Strategy $\{4,2^\text{sim}\}$} \\ \hline
$(1,0,0)$ & 3.769(68)[1.58] & 3.088(64)[1.07] & 3.739(34)[0.96] & 3.563(53)[1.48] & 3.792(47)[1.06] & 3.368(92)[0.89] & 3.759(80)[1.02]\\
$(1,1,0)$ & 3.095(41)[1.19] & 2.327(76)[0.92] & 3.209(28)[0.83] & 2.968(57)[0.55] & 3.442(36)[0.98] & 2.709(23)[1.17] & 3.222(43)[1.57]\\
$(1,1,1)$ & 2.608(32)[1.44] & 1.865(56)[1.02] & 2.799(20)[1.08] & 2.562(31)[0.85] & 3.147(44)[1.09] & 2.260(24)[1.04] & 2.876(37)[1.35]\\
$(2,0,0)$ & 2.223(28)[1.19] & 1.36(29)[1.31] & 2.476(24)[1.11] & 2.314(30)[0.82] & 2.838(39)[0.93] & 1.929(33)[1.35] & 2.603(62)[0.91]\\
$(2,1,0)$ & 1.975(31)[1.03] & 1.345(64)[0.92] & 2.179(35)[1.06] & 1.951(76)[0.77] & 2.619(50)[1.27] & 1.687(25)[0.97] & 2.377(32)[1.31]\\
$(2,1,1)$ & 1.747(44)[1.36] & 0.90(21)[1.41] & 1.934(49)[1.30] & 1.772(69)[0.70] & 2.408(39)[1.30] & 1.470(53)[0.87] & 2.164(32)[1.20]\\
$(2,2,0)$ & 1.445(62)[0.90] & 0.965(21)[1.28] & 1.606(57)[1.13] & 1.571(25)[1.24] & 2.057(54)[1.29] & 1.170(62)[0.76] & 1.821(47)[1.10]\\
$(2,2,1)$ & 1.23(18)[1.91] & 0.73(15)[1.04] & 1.30(14)[1.06] & 1.31(14)[0.89] & 1.948(51)[1.41] & 1.114(47)[0.79] & 1.744(32)[1.02]\\
$(3,0,0)$ & 1.14(34)[1.21] & 1.66(67)[1.60] & 1.486(64)[1.25] & 1.43(12)[1.33] & 1.929(84)[1.26] & 1.104(57)[0.95] & 1.743(65)[0.66]\\
$(3,1,0)$ & 1.15(54)[0.91] & 0.825(14)[1.74] & 1.14(39)[1.01] & 1.298(75)[1.04] & 1.845(52)[1.29] & 1.062(35)[1.13] & 1.668(41)[0.70]\\
\hline & \multicolumn{7}{c}{Strategy $\{4^{N\pi},2^\text{sim}\}$} \\ \hline
$(1,0,0)$ & 3.737(67)[1.52] & 3.035(88)[1.09] & 3.733(29)[0.95] & 3.609(56)[1.59] & 3.966(78)[0.78] & 3.396(35)[0.80] & 3.791(77)[1.13]\\
$(1,1,0)$ & 3.086(34)[1.21] & 2.312(73)[0.89] & 3.207(26)[0.88] & 3.000(35)[0.60] & 3.548(65)[0.81] & 2.709(23)[1.19] & 3.267(50)[1.62]\\
$(1,1,1)$ & 2.612(35)[1.53] & 1.854(58)[1.01] & 2.796(25)[1.18] & 2.584(32)[0.86] & 3.190(54)[1.11] & 2.261(24)[1.08] & 2.912(41)[1.16]\\
$(2,0,0)$ & 2.216(28)[1.19] & 1.49(16)[1.36] & 2.476(24)[1.16] & 2.314(52)[0.94] & 2.882(55)[0.92] & 1.932(34)[1.37] & 2.646(53)[0.89]\\
$(2,1,0)$ & 1.967(32)[1.06] & 1.357(49)[0.93] & 2.174(35)[1.14] & 1.965(82)[0.80] & 2.639(41)[1.27] & 1.690(25)[0.96] & 2.404(32)[1.29]\\
$(2,1,1)$ & 1.741(41)[1.44] & 0.98(22)[1.38] & 1.930(54)[1.41] & 1.796(80)[0.76] & 2.425(41)[1.35] & 1.482(46)[0.86] & 2.189(31)[1.16]\\
$(2,2,0)$ & 1.426(57)[1.00] & 0.971(20)[1.50] & 1.540(68)[1.07] & 1.39(27)[0.89] & 2.097(48)[1.36] & 1.187(52)[0.76] & 1.837(45)[1.11]\\
$(2,2,1)$ & 1.20(17)[1.95] & 0.843(50)[1.06] & 1.22(15)[1.33] & 1.29(15)[0.96] & 1.957(56)[1.44] & 1.126(35)[0.78] & 1.761(31)[1.01]\\
$(3,0,0)$ & 1.15(16)[0.96] & 2.3(1.4)[1.60] & 1.48(11)[1.17] & 1.43(13)[1.34] & 1.937(95)[1.28] & 1.115(46)[0.96] & 1.762(59)[0.66]\\
$(3,1,0)$ & 0.7(1.1)[1.20] & 0.86(35)[1.66] & 1.355(63)[1.31] & 1.354(32)[1.10] & 1.830(64)[0.99] & 1.065(36)[1.13] & 1.683(38)[0.71]\\
\end{tabular} 
  \end{ruledtabular}
  \caption{Data for the renormalized $G_M^{\Re V_i}(Q^2)/g_V$ from the seven ensembles and with the 
    four strategies for controlling ESC. The rest is the same as in
    Table~\protect\ref{tab:FF-GE4strategies}. }
   \label{tab:FF-GM4strategies}
  \end{table*}
  
%


\onecolumngrid\hrule width0pt\twocolumngrid\cleardoublepage

\section{Chiral-Continuum-Finite-Volume Fits}
\label{sec:appendixCCFV}

\begin{figure*}[h] 
\subfigure
{
    \includegraphics[width=0.32\linewidth]{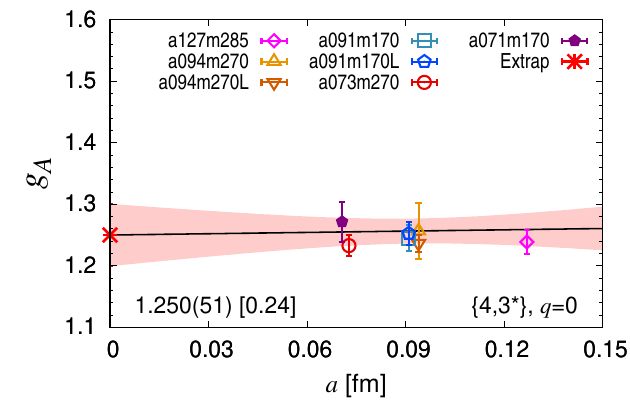}  
    \includegraphics[width=0.32\linewidth]{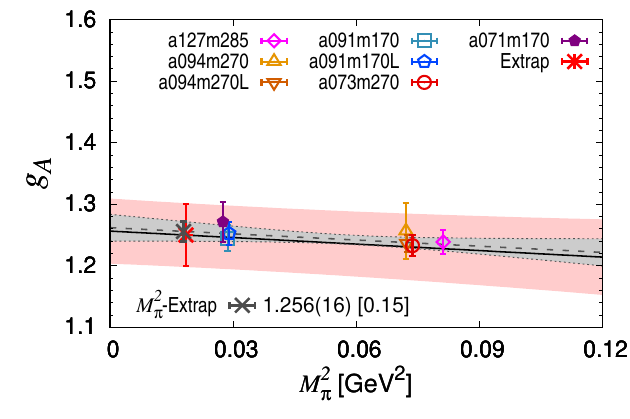}  
    \includegraphics[width=0.32\linewidth]{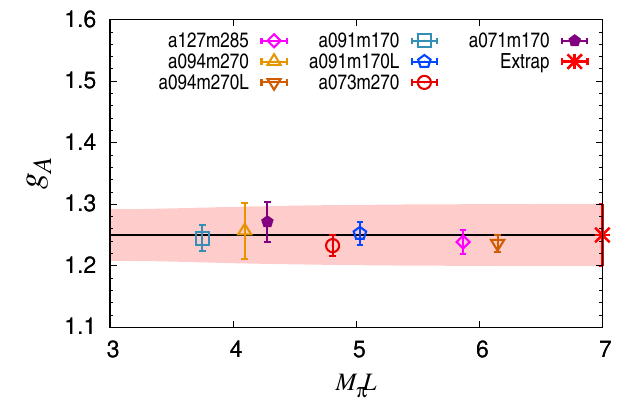}  
}
{
    \includegraphics[width=0.32\linewidth]{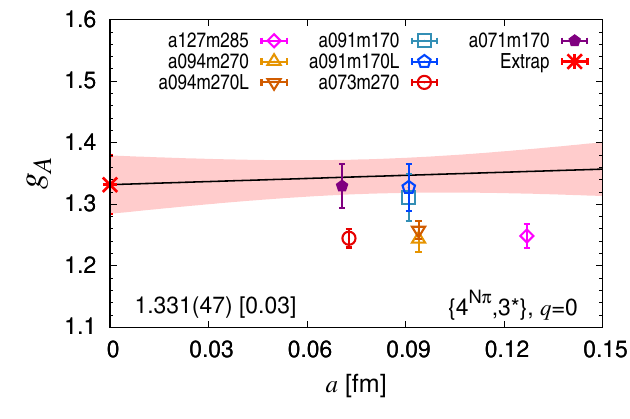} 
    \includegraphics[width=0.32\linewidth]{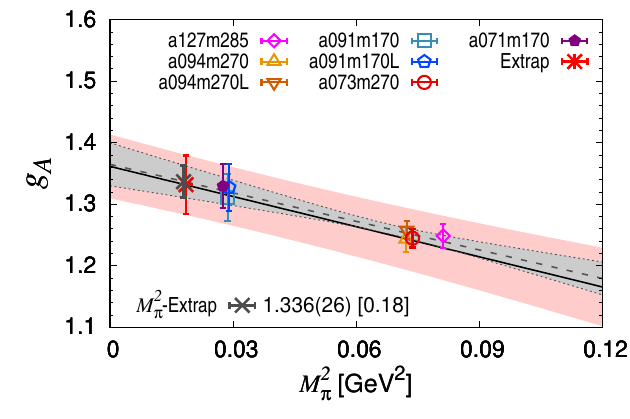} 
    \includegraphics[width=0.32\linewidth]{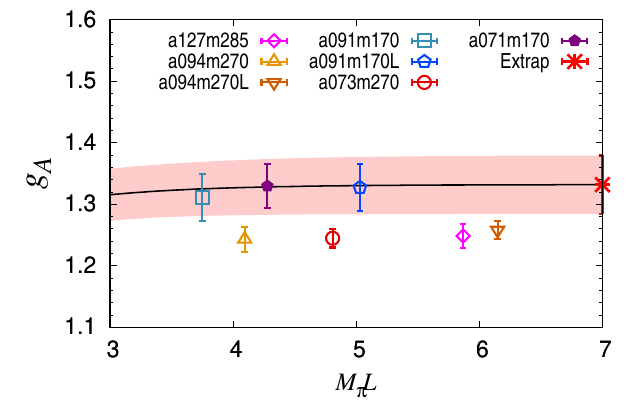} 
}
{
    \includegraphics[width=0.32\linewidth]{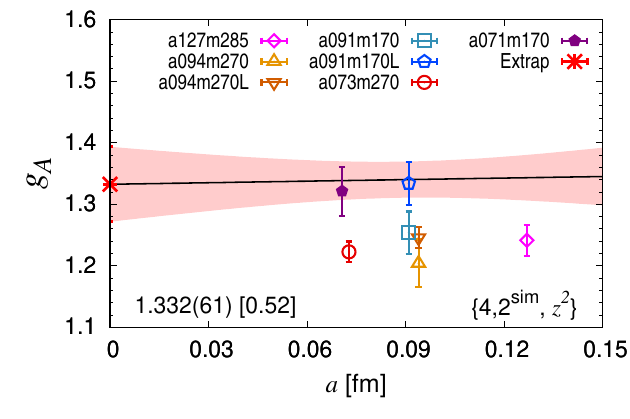} 
    \includegraphics[width=0.32\linewidth]{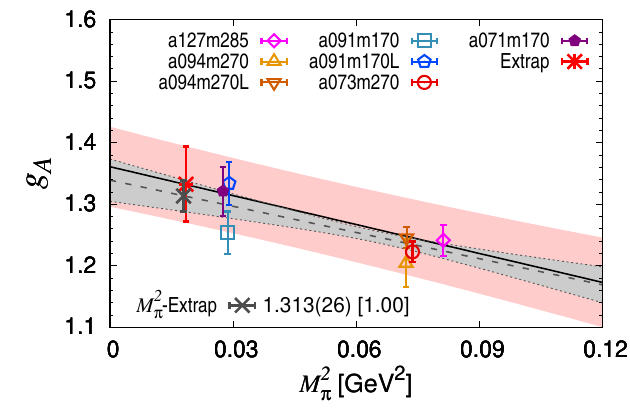} 
    \includegraphics[width=0.32\linewidth]{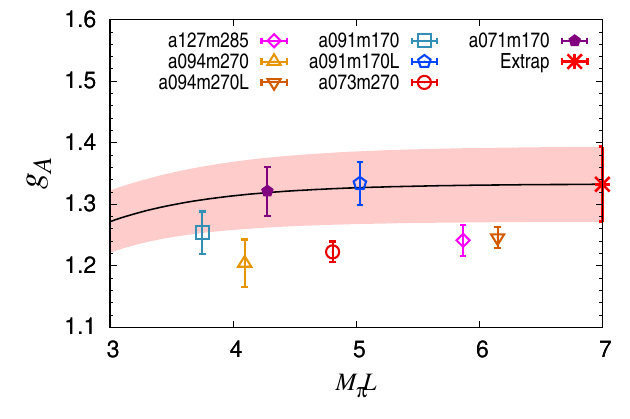} 
}
{
    \includegraphics[width=0.32\linewidth]{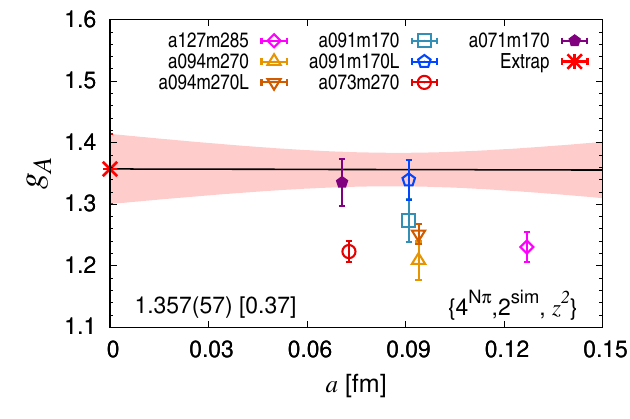} 
    \includegraphics[width=0.32\linewidth]{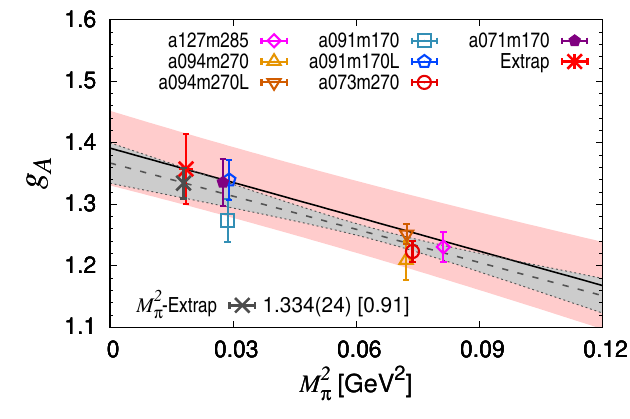} 
    \includegraphics[width=0.32\linewidth]{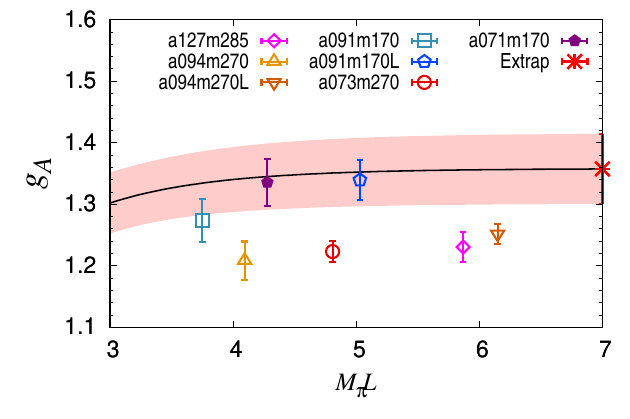} 
}
{
    \includegraphics[width=0.32\linewidth]{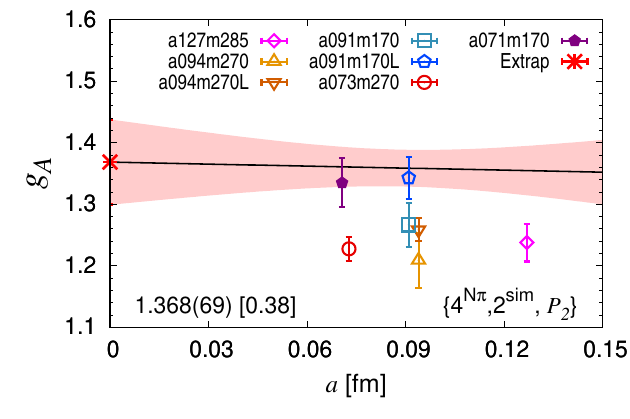} 
    \includegraphics[width=0.32\linewidth]{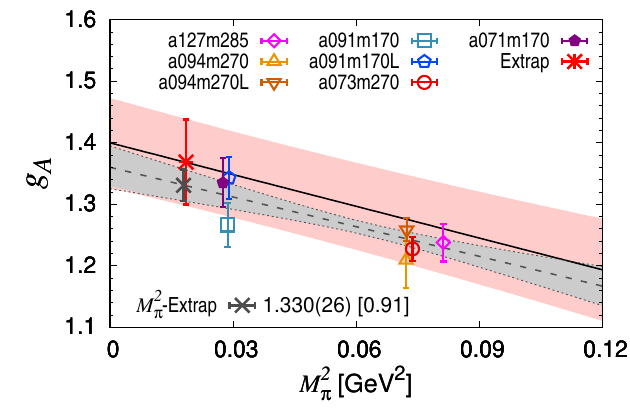} 
    \includegraphics[width=0.32\linewidth]{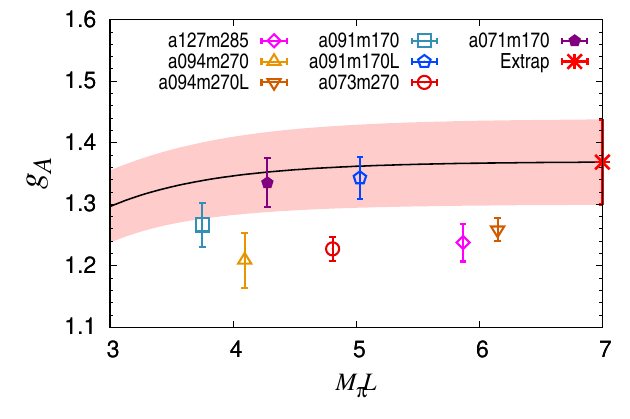} 
}
\caption{The CCFV extrapolation of the renormalized (${\rm Z}_2$ method)
  isovector axial charge $g_{A}^{u-d}$ for five strategies:
  $\{4,3^\ast\}$ (top row), $\{4^{N\pi},3^\ast\}$ (second row),
  $\{4,2^{\rm sim},z^2\}$ (third row), $\{4^{N\pi},2^{\rm sim},z^2\}$
  (fourth row) and $\{4^{N\pi},2^{\rm sim},P_2\}$ (fifth row).  In each
  panel, the result of the simultaneous fit in $\{a,M_\pi,M_\pi L\}$
  is shown by the pink band, and plotted versus $a$ (left panel),
  $M_\pi^2$ (middle) and $M_\pi L$ (right) with the other two
  variables in each case set to their physical value. The result of
  the CCFV fit at the physical point is shown by the red star (label
  Extrap) and the value and $\chi^2$/dof given in the left panel.  The gray band is the
  result of a chiral fit only with the physical point marked with a
  black cross (label $M_\pi^2$-Extrap) and the value given in the
  middle panel.  \looseness-1 \label{fig:CCFVgA}}
\end{figure*}

\begin{figure*}[tbp] 
\subfigure
{
    \includegraphics[width=0.32\linewidth]{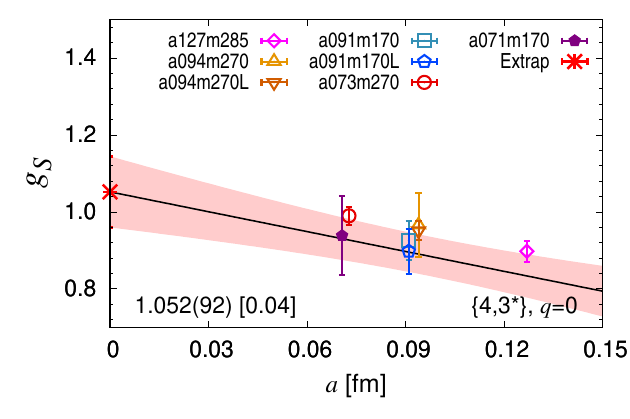}  
    \includegraphics[width=0.32\linewidth]{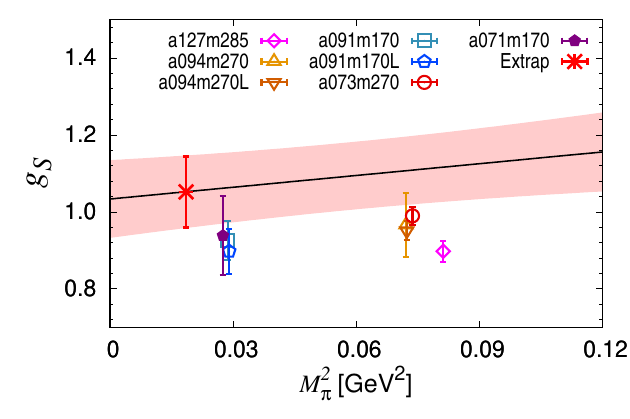}  
    \includegraphics[width=0.32\linewidth]{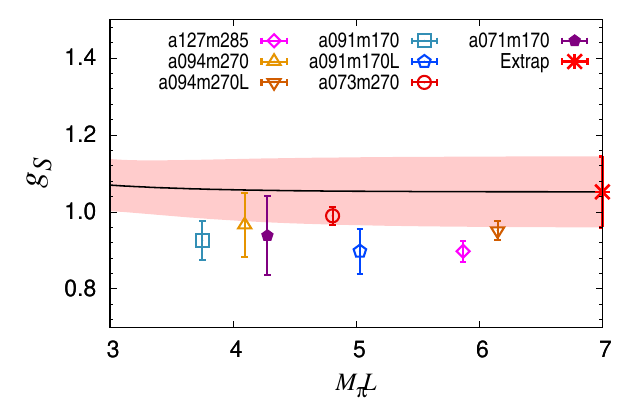}  
}
{
    \includegraphics[width=0.32\linewidth]{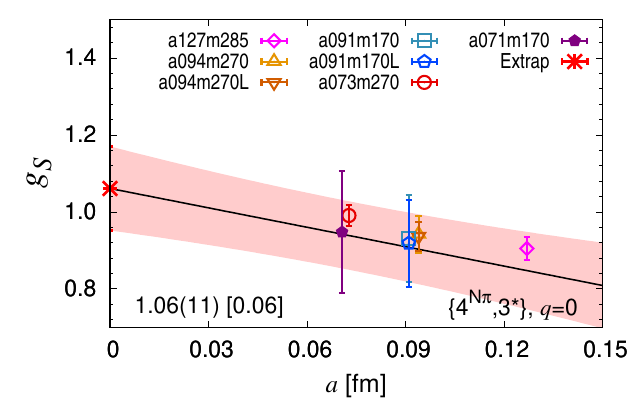} 
    \includegraphics[width=0.32\linewidth]{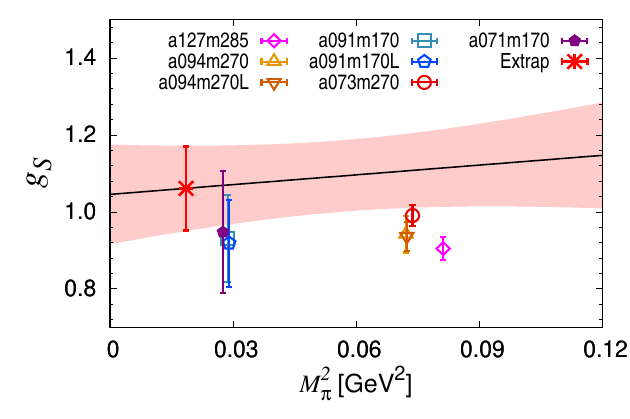} 
    \includegraphics[width=0.32\linewidth]{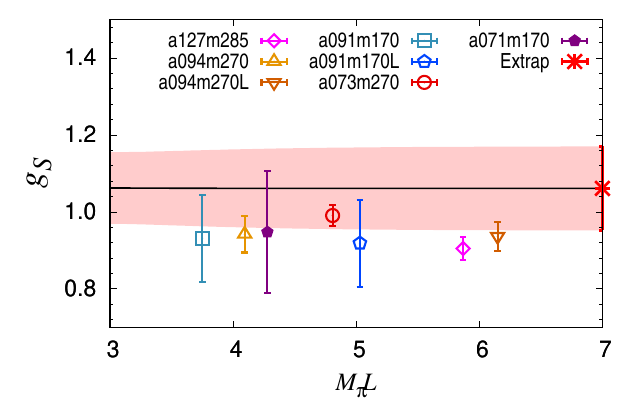} 
}
{
    \includegraphics[width=0.32\linewidth]{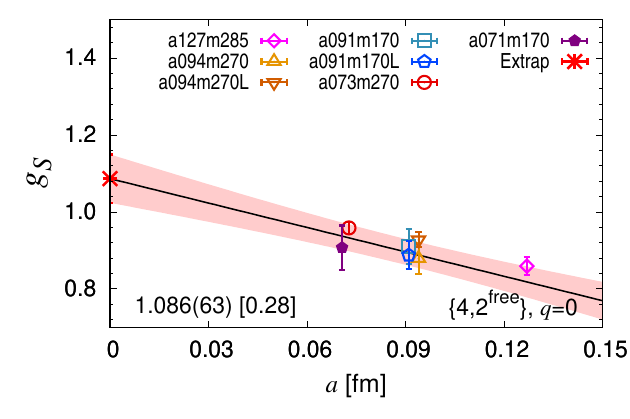} 
    \includegraphics[width=0.32\linewidth]{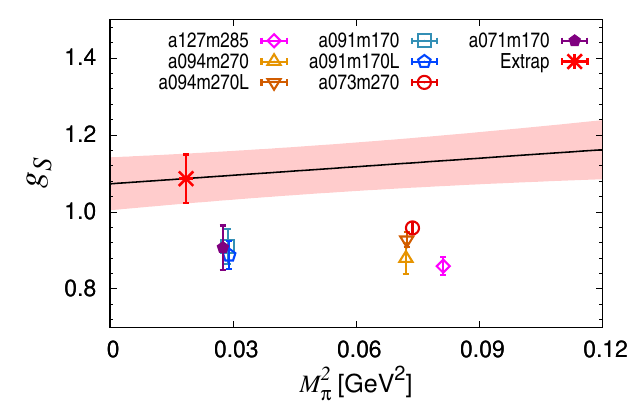} 
    \includegraphics[width=0.32\linewidth]{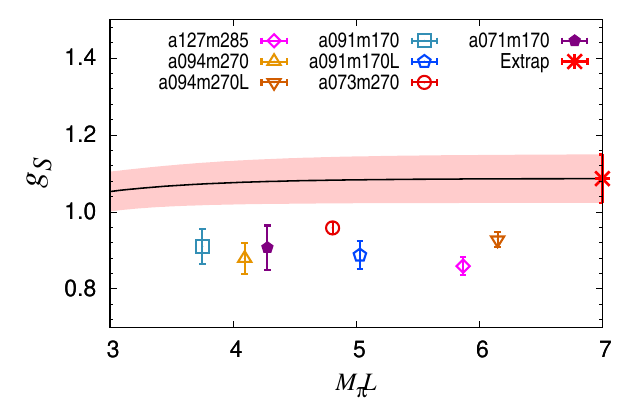} 
}
{
    \includegraphics[width=0.32\linewidth]{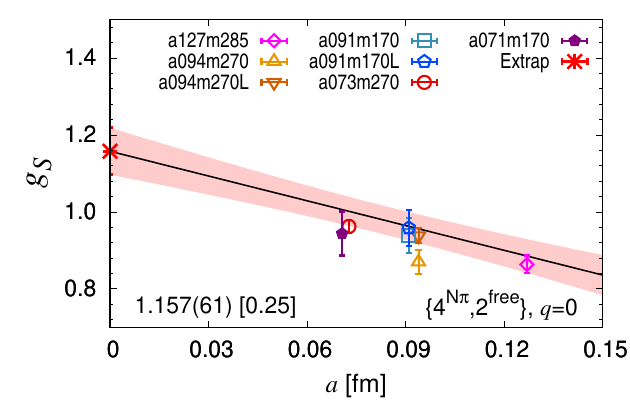} 
    \includegraphics[width=0.32\linewidth]{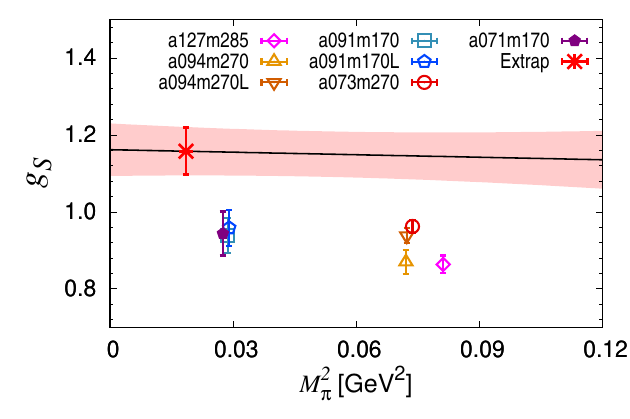} 
    \includegraphics[width=0.32\linewidth]{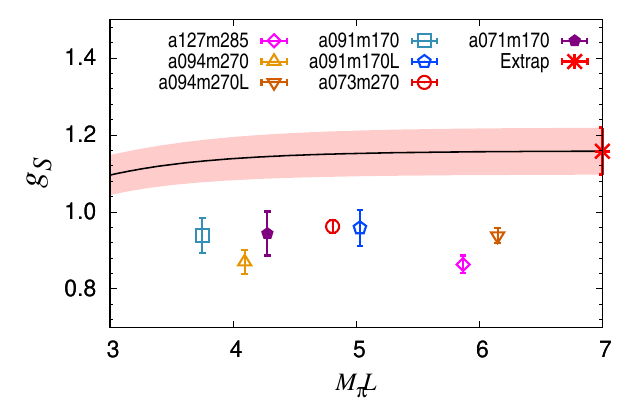} 
}
\caption{The CCFV extrapolation of the renormalized (${\rm Z}_1$ method)
  isovector scalar charge $g_{S}^{u-d}$ for the four strategies, to remove ESC:
  $\{4,3^\ast\}$ (top row), $\{4^{N\pi},3^\ast\}$ (second row),
  $\{4,2^{\rm free}\}$ (third row) and $\{4^{N\pi},2^{\rm free}\}$
  (bottom row).  The rest is the same as in
  Fig.~\protect\ref{fig:CCFVgA}.
  \label{fig:CCFVgS}}
\end{figure*}

\begin{figure*}[tbp] 
\subfigure
{
    \includegraphics[width=0.32\linewidth]{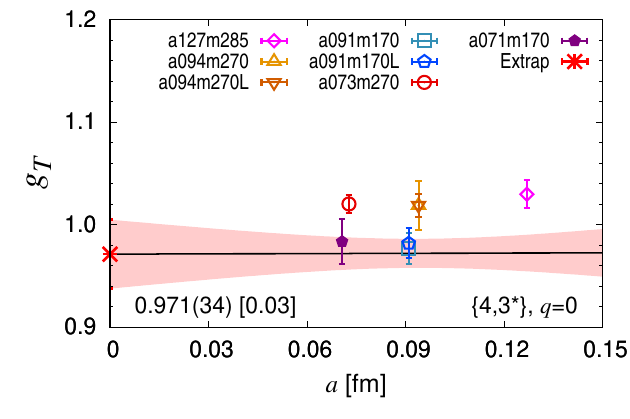}  
    \includegraphics[width=0.32\linewidth]{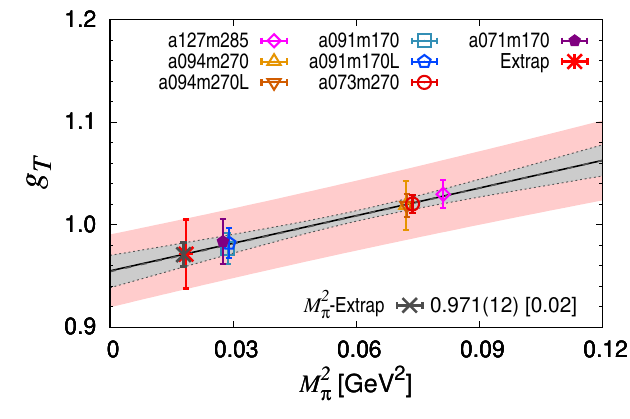}  
    \includegraphics[width=0.32\linewidth]{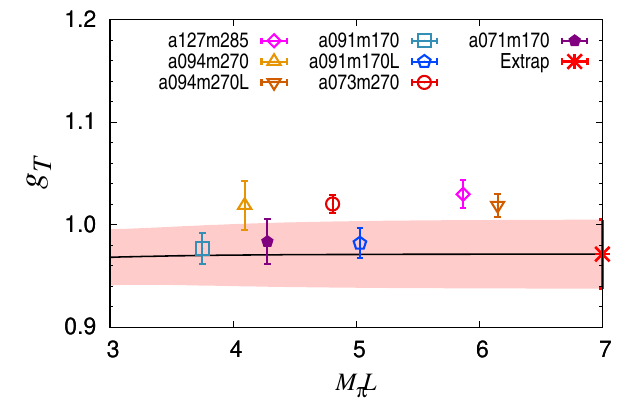}  
}
{
    \includegraphics[width=0.32\linewidth]{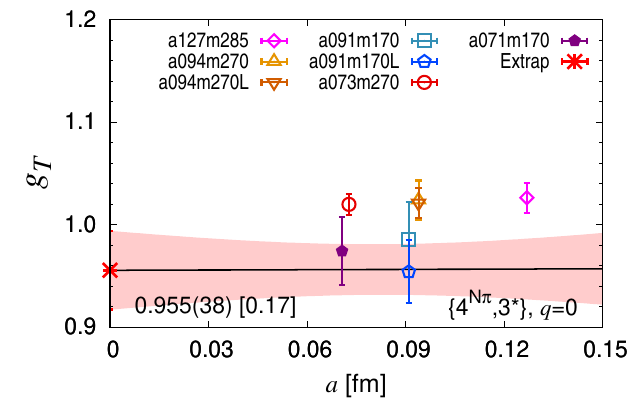} 
    \includegraphics[width=0.32\linewidth]{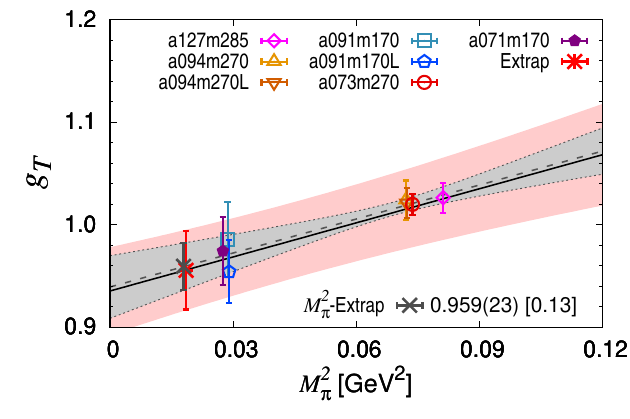} 
    \includegraphics[width=0.32\linewidth]{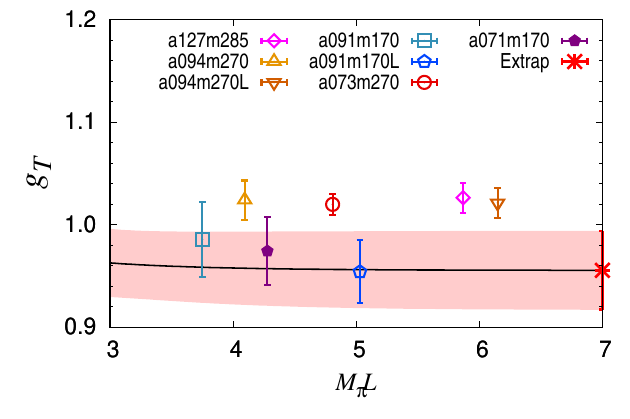} 
}
{
    \includegraphics[width=0.32\linewidth]{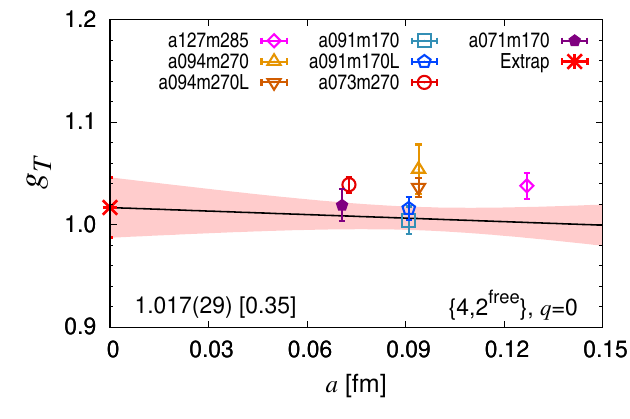} 
    \includegraphics[width=0.32\linewidth]{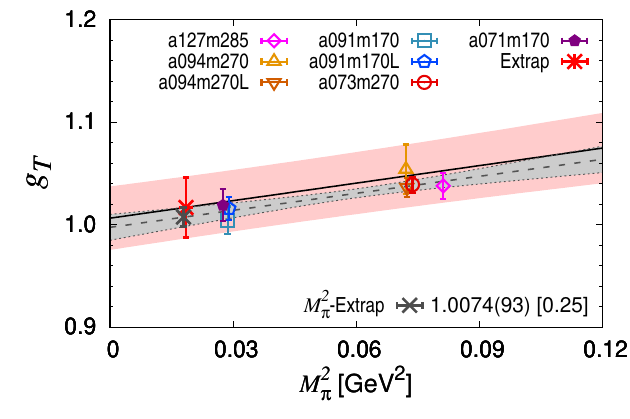} 
    \includegraphics[width=0.32\linewidth]{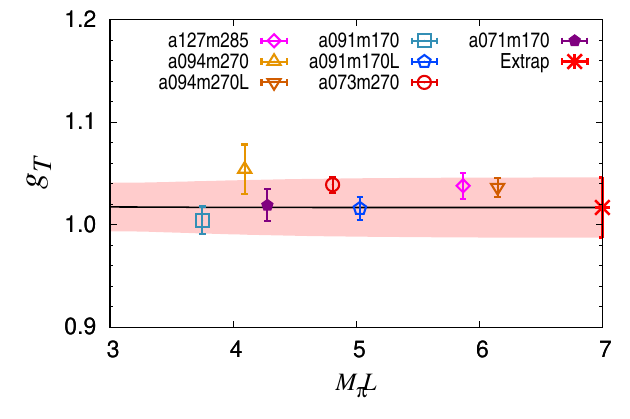} 
}
{
    \includegraphics[width=0.32\linewidth]{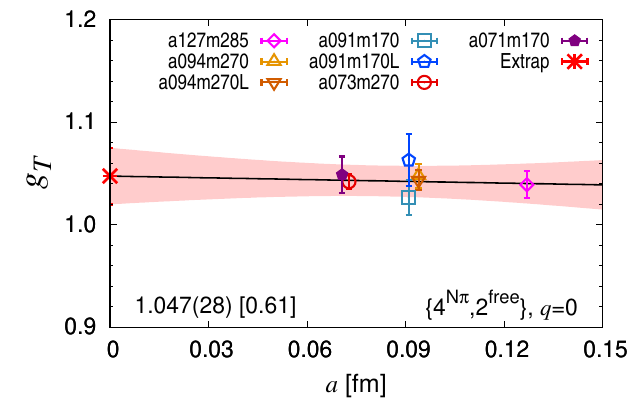} 
    \includegraphics[width=0.32\linewidth]{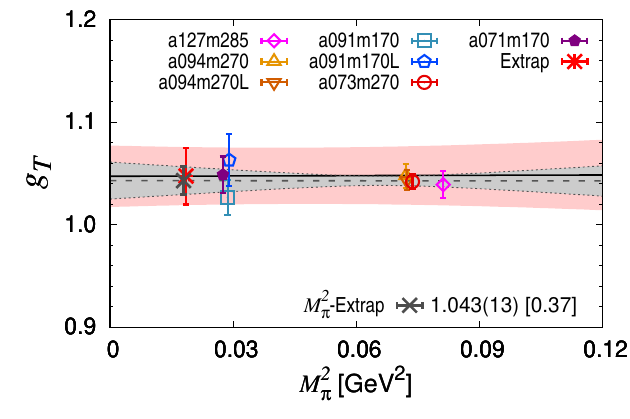} 
    \includegraphics[width=0.32\linewidth]{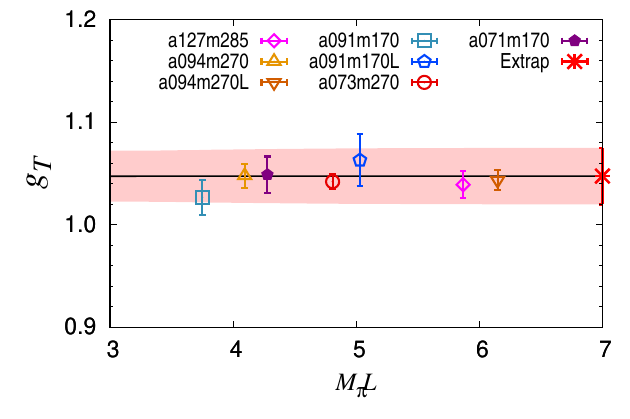} 
}
\caption{The CCFV extrapolation of the renormalized (${\rm Z}_2$
  method) isovector tensor charge $g_{T}^{u-d}$ for the four 
  strategies to remove ESC: 
  $\{4,3^\ast\}$ (top row), $\{4^{N\pi},3^\ast\}$ (second row),
  $\{4,2^{\rm free}\}$ (third row) and $\{4^{N\pi},2^{\rm free}\}$
  (bottom row).  The rest is the same as in
  Fig.~\protect\ref{fig:CCFVgA}.
  \label{fig:CCFVgT}}
\end{figure*}

\begin{figure*}[tbp] 
\subfigure
{
    \includegraphics[width=0.32\linewidth]{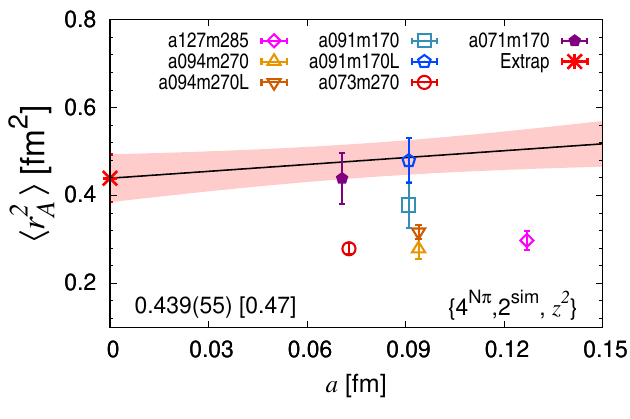}  
    \includegraphics[width=0.32\linewidth]{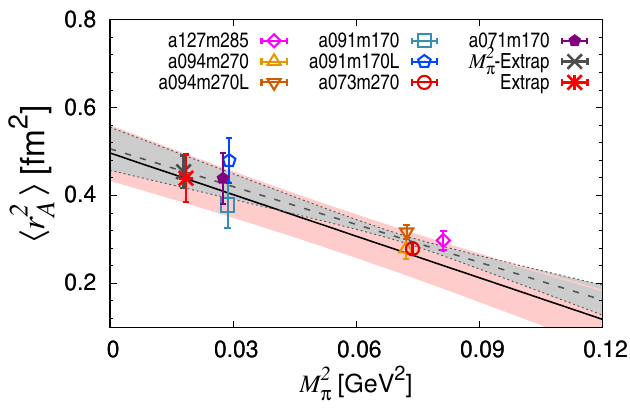}  
    \includegraphics[width=0.32\linewidth]{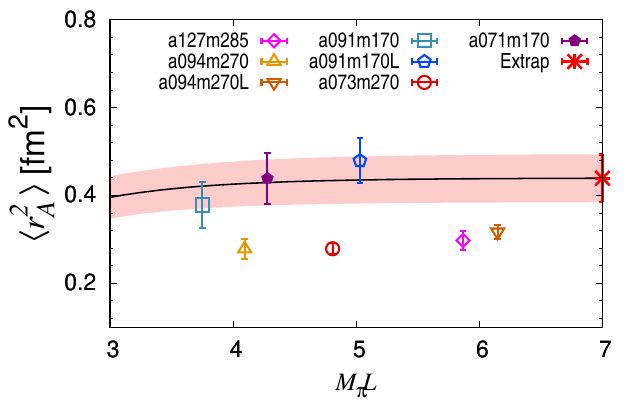}  
}
\subfigure
{
    \includegraphics[width=0.32\linewidth]{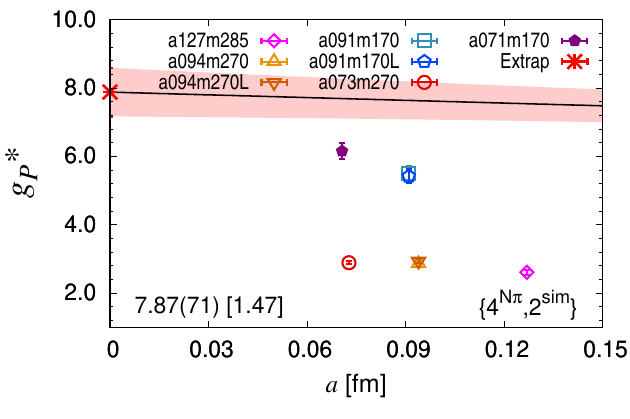}  
    \includegraphics[width=0.32\linewidth]{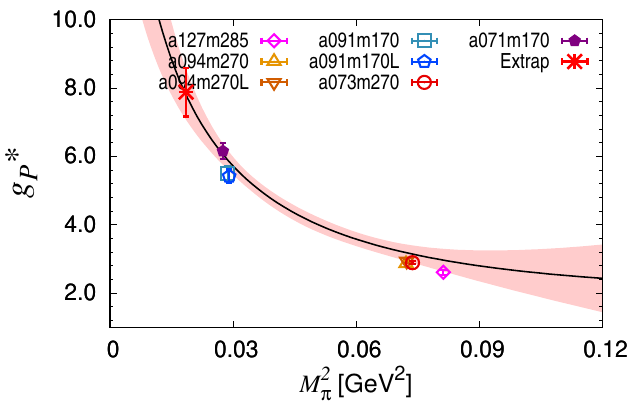}  
    \includegraphics[width=0.32\linewidth]{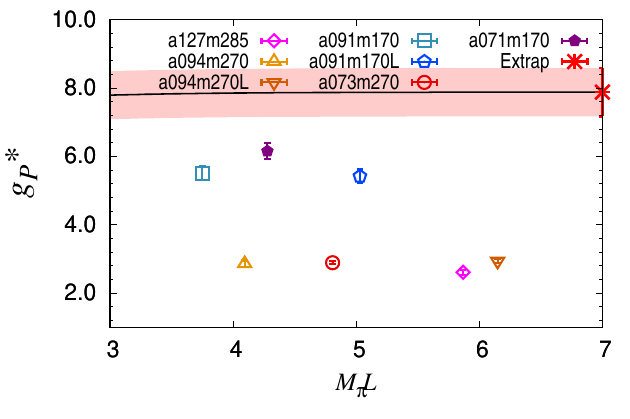}  
}
\subfigure
{
    \includegraphics[width=0.32\linewidth]{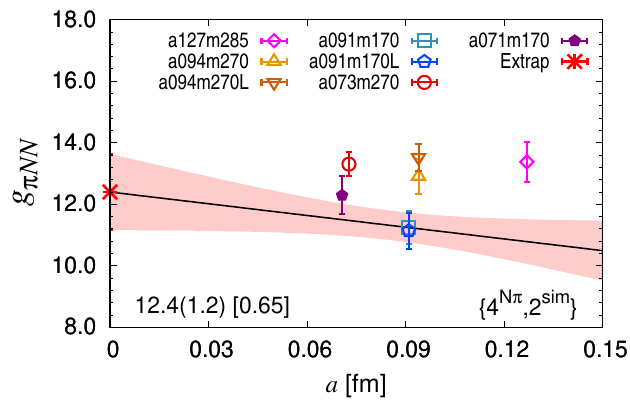}  
    \includegraphics[width=0.32\linewidth]{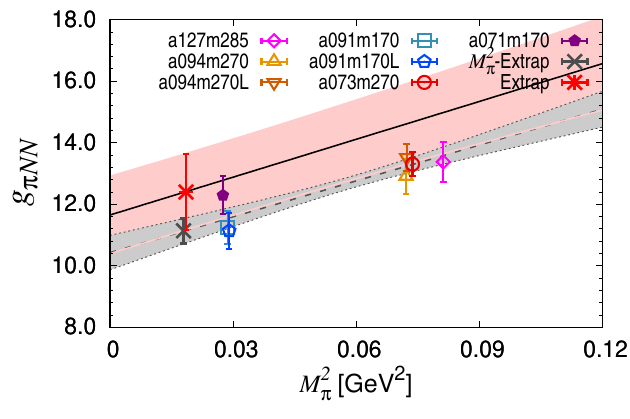}  
    \includegraphics[width=0.32\linewidth]{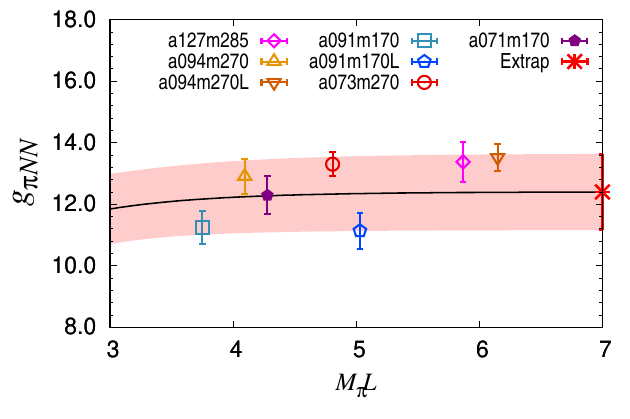}  
}
\caption{The CCFV extrapolation of the axial charge radius squared
  $\langle r_A^2 \rangle$(top row), the induced pseudoscalar charge
  $g_P^\ast|_{{\rm Z}_2}$ (middle row), and the pion-nucleon coupling
  $g_{\pi N N}|_{{\rm Z}_2}$ (bottom row). The data for $\langle r_A^2
  \rangle$ are obtained using the $z^2$-fit to parameterize the $Q^2$
  behavior. Data for $g_P^\ast|_{{\rm Z}_2}$ and $g_{\pi N N}|_{{\rm Z}_2}$ are
  obtained using the pion-pole dominance ansatz given in
  Eq.~\eqref{eq:GPt_fit} to fit ${\widetilde G}_P$. Data for all three quantities are with the
  $\{4^{N\pi},2^{\rm sim}\}$ strategy. The rest is the same as in
  Fig.~\protect\ref{fig:CCFVgA}. 
  \label{fig:CCFV-AFF}}
\end{figure*}

\begin{figure*}[tbp] 
\subfigure
{
    \includegraphics[width=0.32\linewidth]{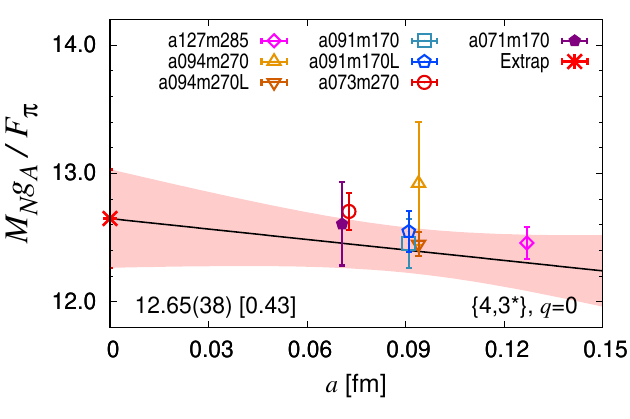}  
    \includegraphics[width=0.32\linewidth]{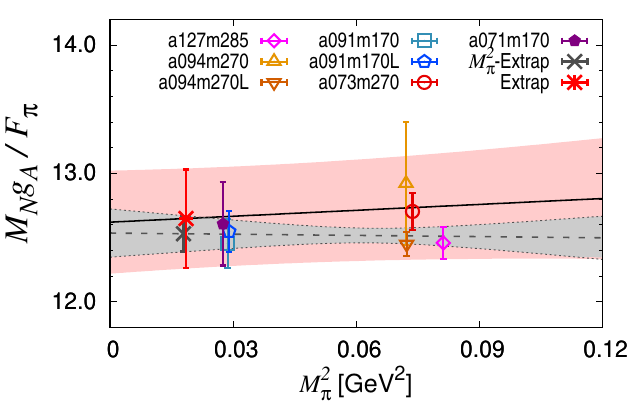}  
    \includegraphics[width=0.32\linewidth]{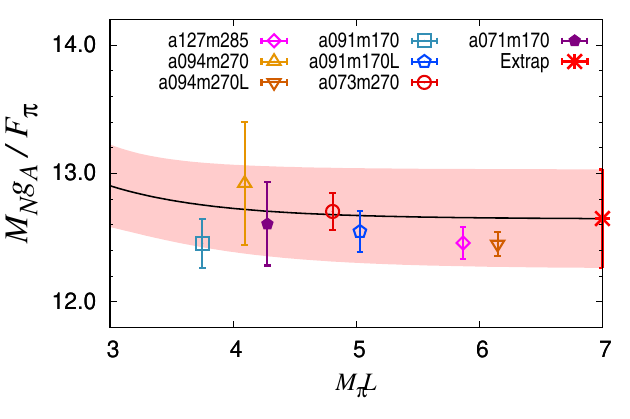}  
}
{
    \includegraphics[width=0.32\linewidth]{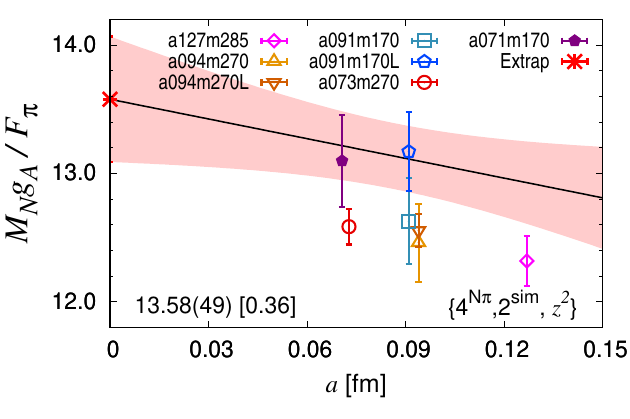}  
    \includegraphics[width=0.32\linewidth]{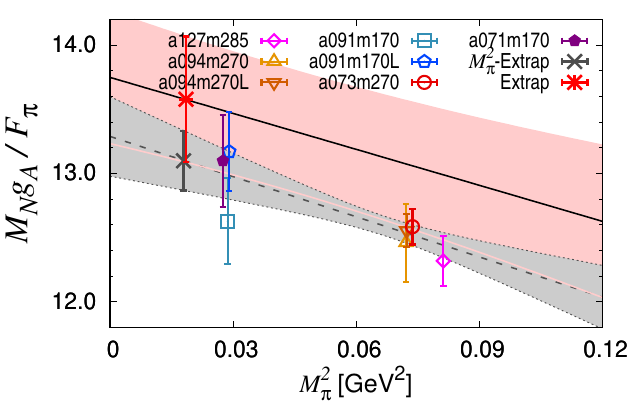}  
    \includegraphics[width=0.32\linewidth]{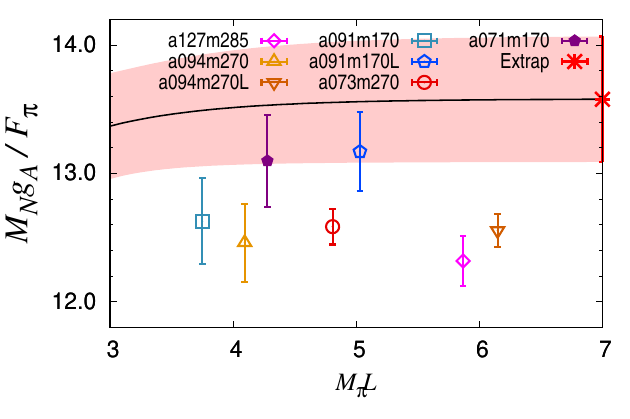}  
}
{
    \includegraphics[width=0.32\linewidth]{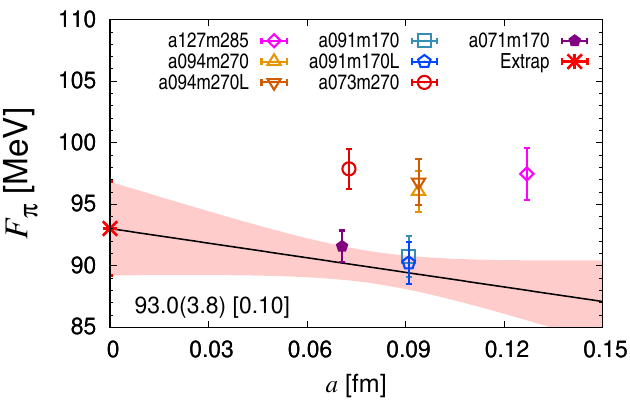}  
    \includegraphics[width=0.32\linewidth]{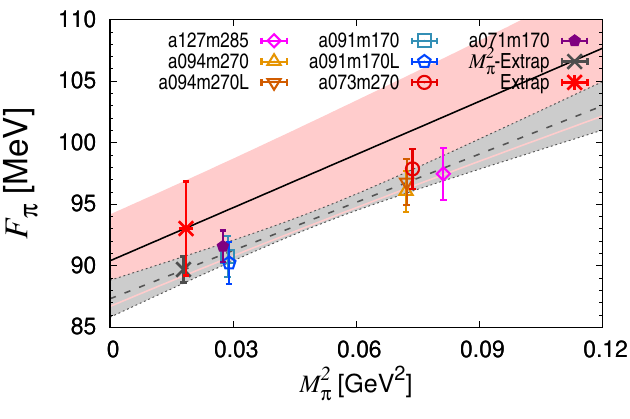}  
    \includegraphics[width=0.32\linewidth]{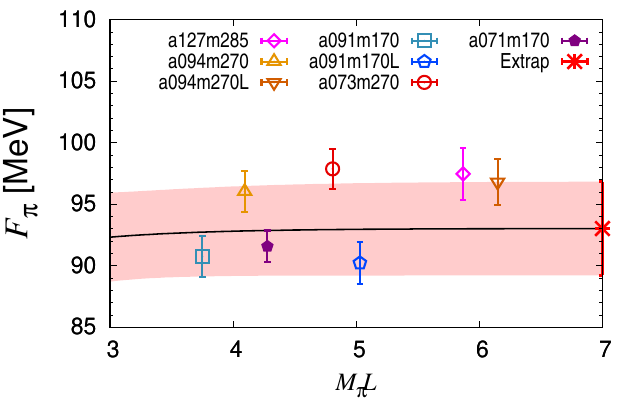}  
}
\caption{The CCFV extrapolation of the
  product $M_N g_A / F_\pi$ with $g_A$ from the $\{4,3^{\ast}\}$ (top
  row) and the $\{4^{N\pi},2^{\rm sim}\}$ (middle row) strategies. The bottom row shows 
  the fit for $F_\pi$ renormalized using the ${\rm Z}_1$  method. The rest is the same as in
  Fig.~\protect\ref{fig:CCFVgA}. 
  \label{fig:gpiNN}}
\end{figure*}

\begin{figure*}[tbp] 
\subfigure
{
    \includegraphics[width=0.28\linewidth]{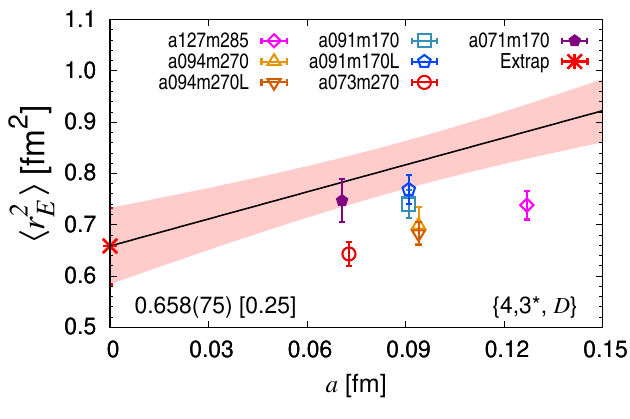}  
    \includegraphics[width=0.28\linewidth]{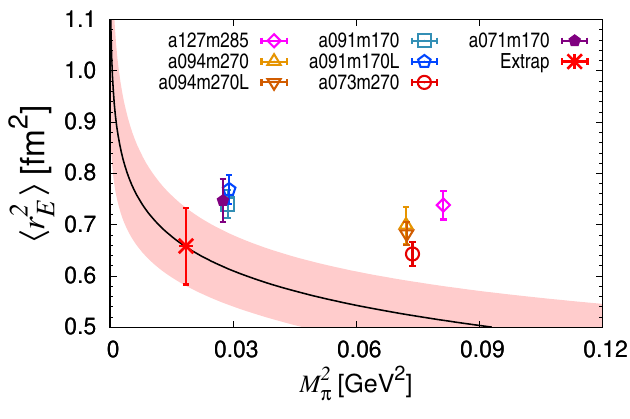}  
    \includegraphics[width=0.28\linewidth]{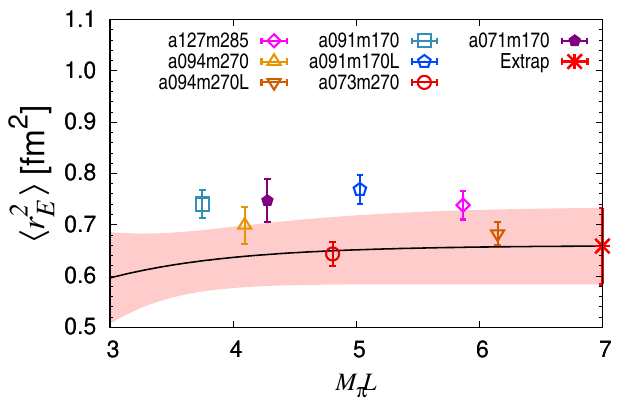}  
}
{
    \includegraphics[width=0.28\linewidth]{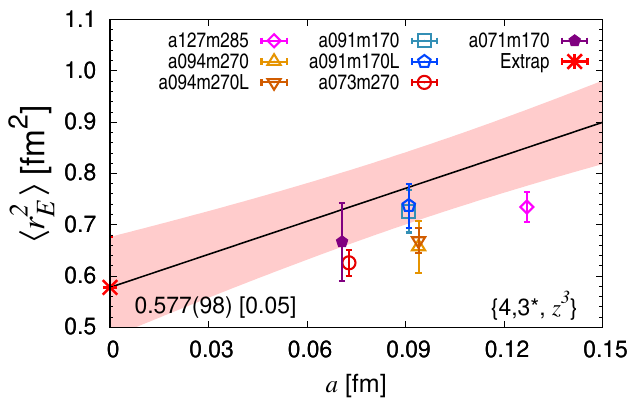}  
    \includegraphics[width=0.28\linewidth]{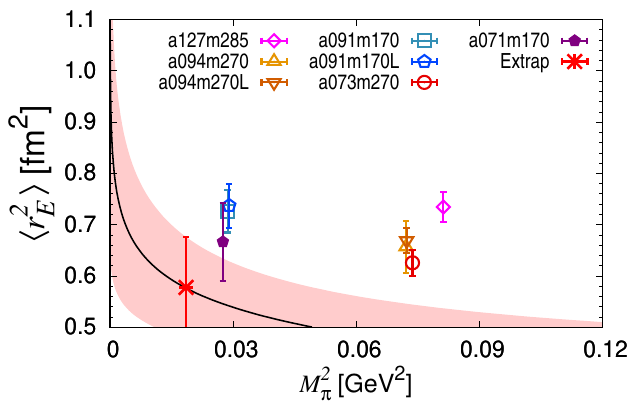}  
    \includegraphics[width=0.28\linewidth]{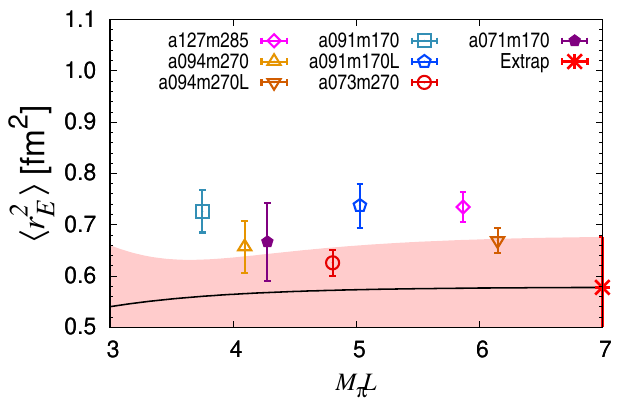}  
}
{
    \includegraphics[width=0.28\linewidth]{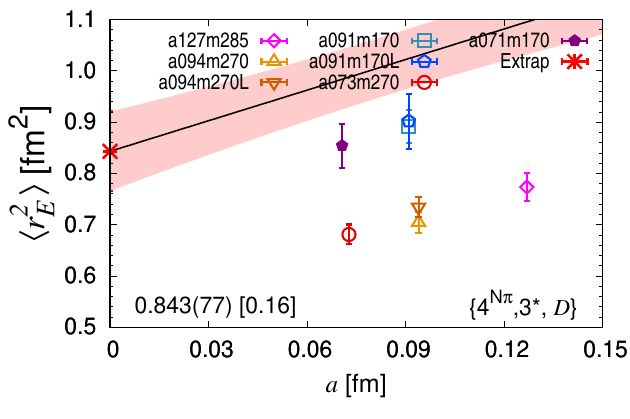}  
    \includegraphics[width=0.28\linewidth]{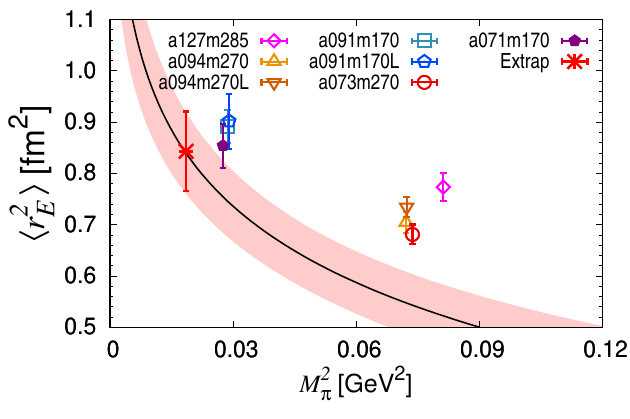}  
    \includegraphics[width=0.28\linewidth]{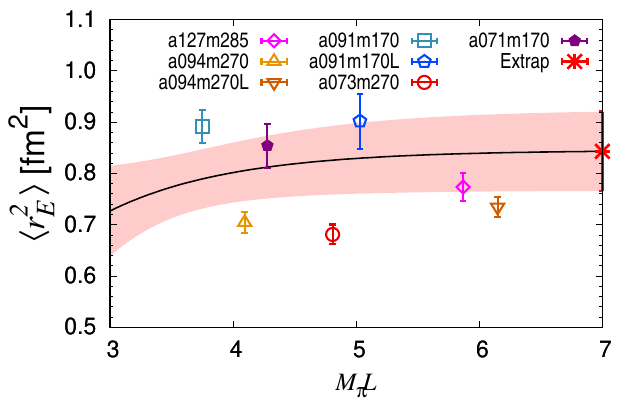}  
}
{
    \includegraphics[width=0.28\linewidth]{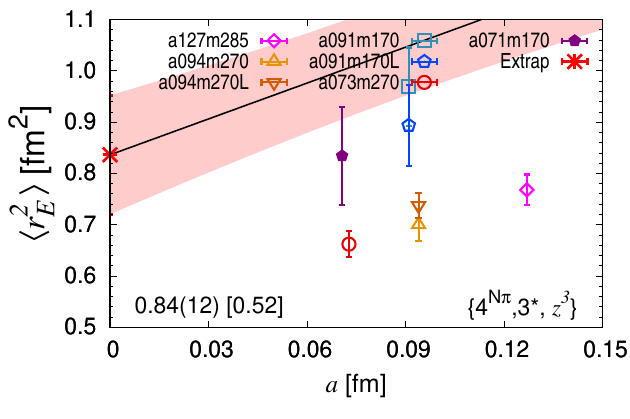}  
    \includegraphics[width=0.28\linewidth]{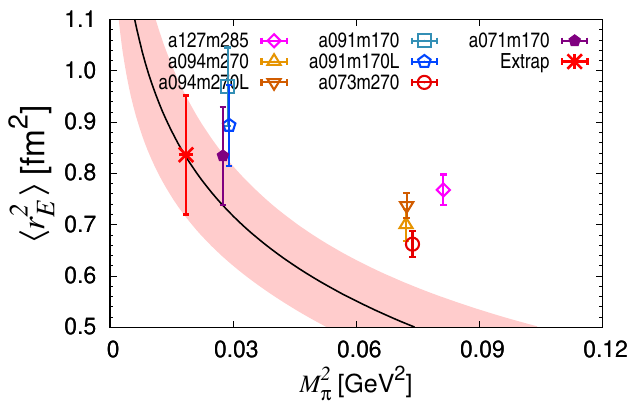}  
    \includegraphics[width=0.28\linewidth]{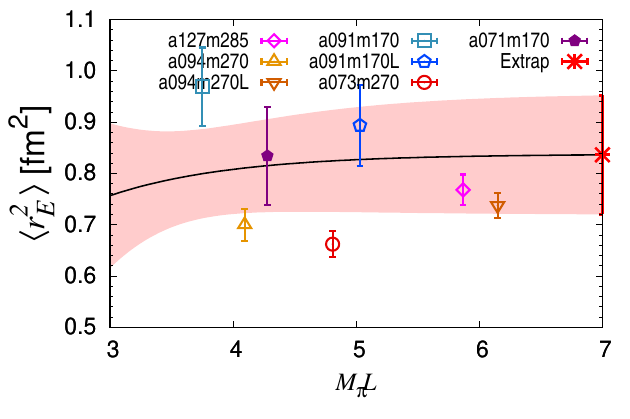}  
}
{
    \includegraphics[width=0.28\linewidth]{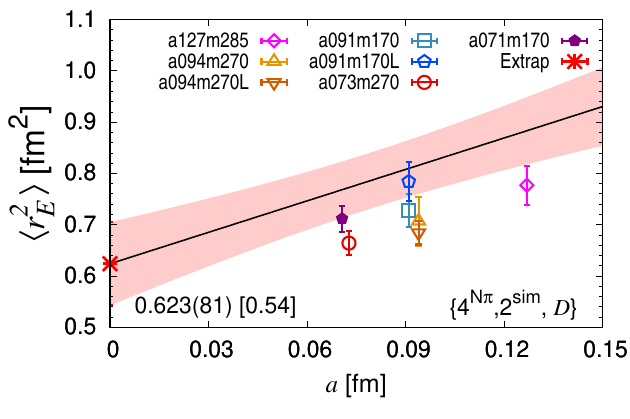}  
    \includegraphics[width=0.28\linewidth]{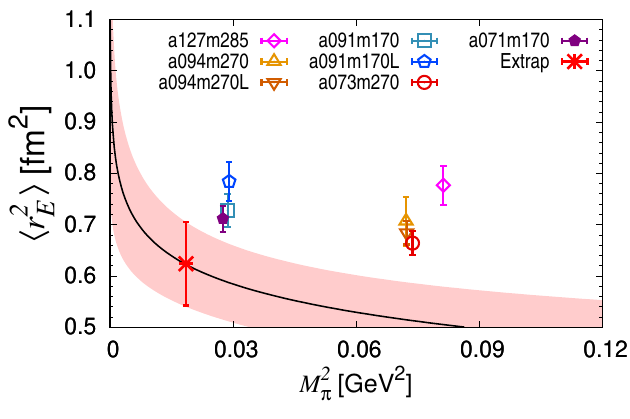}  
    \includegraphics[width=0.28\linewidth]{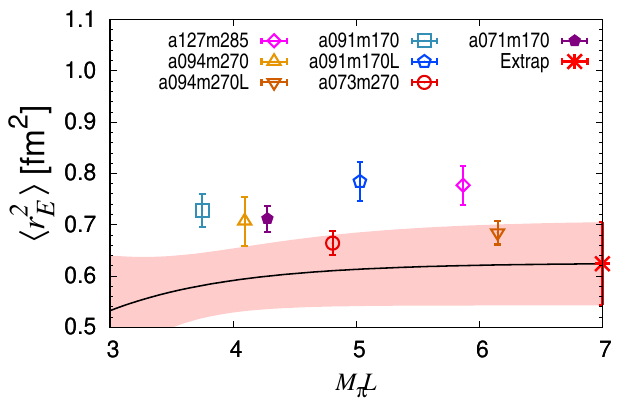}  
}
{
    \includegraphics[width=0.28\linewidth]{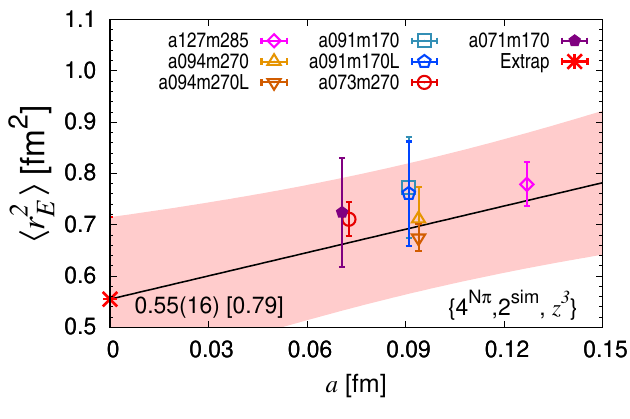}  
    \includegraphics[width=0.28\linewidth]{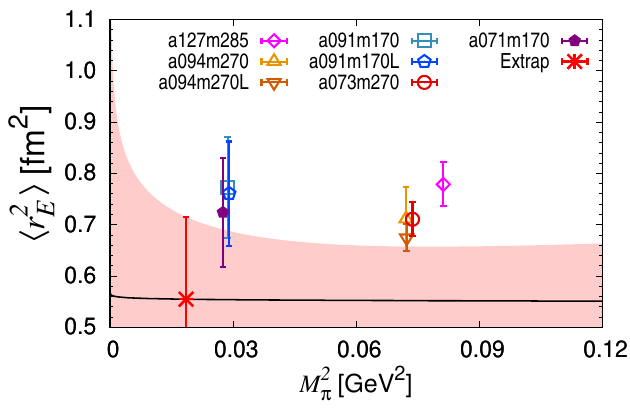}  
    \includegraphics[width=0.28\linewidth]{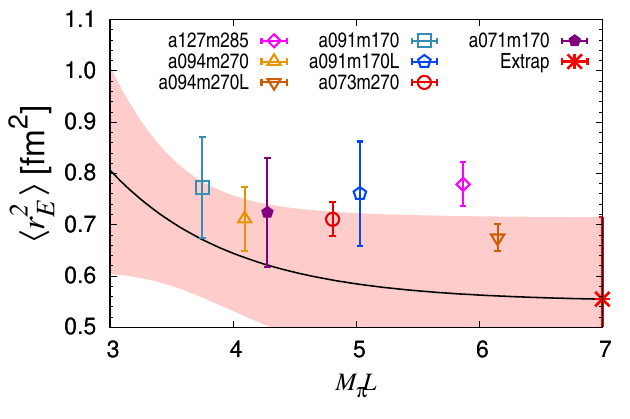}  
}
\caption{The CCFV extrapolation of the electric charge radius squared,
  $\langle r_E^2 \rangle$. Data and fits are shown for three
  strategies, $\{4^{},3^{\ast}\}$ (rows one and two),
  $\{4^{N\pi},3^{\ast}\}$ (rows three and four), and $\{4^{N\pi},2^{\rm
  sim}\}$ (rows five and six). Data from the dipole (D) fit (rows one,
  three, and five) are compared with those from $z^3$ (rows two, four, 
  and six).  Each panel shows the simultaneous (CCFV) fit in the three
  variables, $\{a,M_\pi, M_\pi L\}$, but plotted versus a single
  variable ($a$, or $M_\pi^2$, or $M_\pi L$) with the other two set to
  their physical value defined by $a=0$, $M_\pi=135$~MeV, $M_\pi
  L=\infty$.  The result and the $\chi^2$/dof of the fit are given by
  the label at the bottom left in the left panel and marked by a red
  star (``Extrap''). \label{fig:CCFV-VFF-rE}}
\end{figure*}

\begin{figure*}[tbp] 
\subfigure
{
    \includegraphics[width=0.28\linewidth]{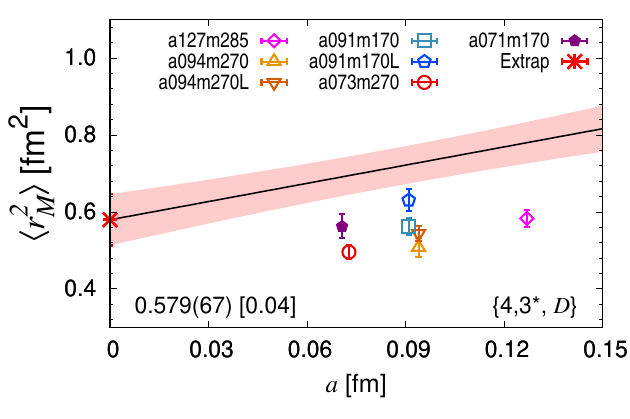}  
    \includegraphics[width=0.28\linewidth]{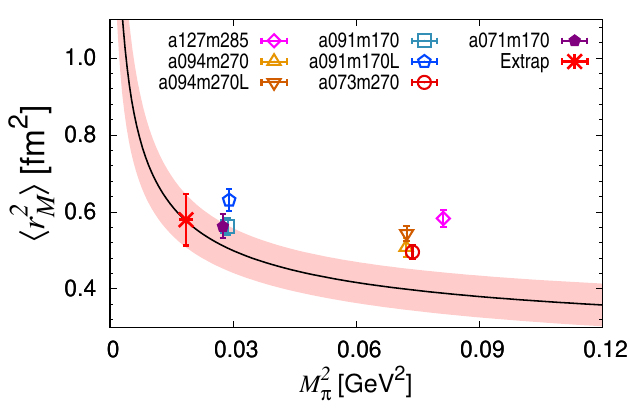}  
    \includegraphics[width=0.28\linewidth]{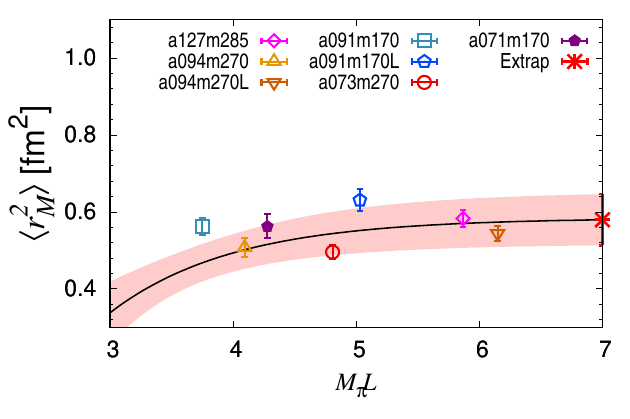}  
}
{
    \includegraphics[width=0.28\linewidth]{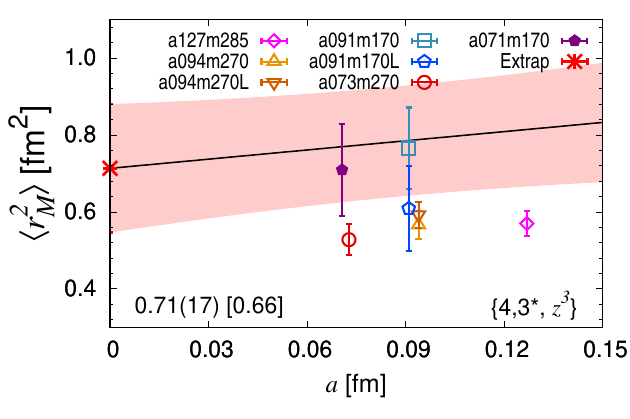}  
    \includegraphics[width=0.28\linewidth]{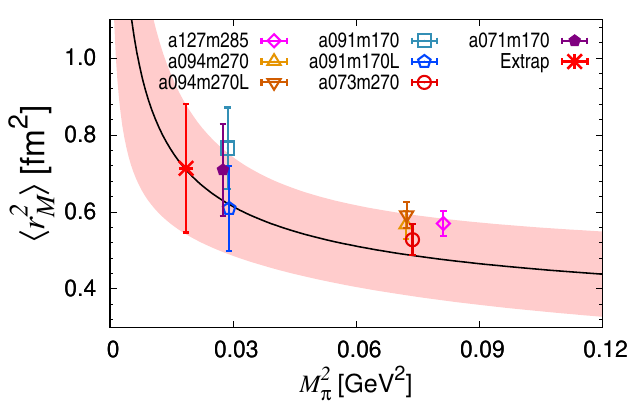}  
    \includegraphics[width=0.28\linewidth]{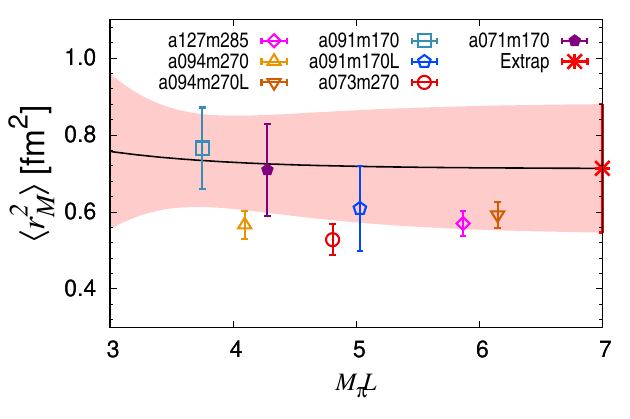}  
}
{
    \includegraphics[width=0.28\linewidth]{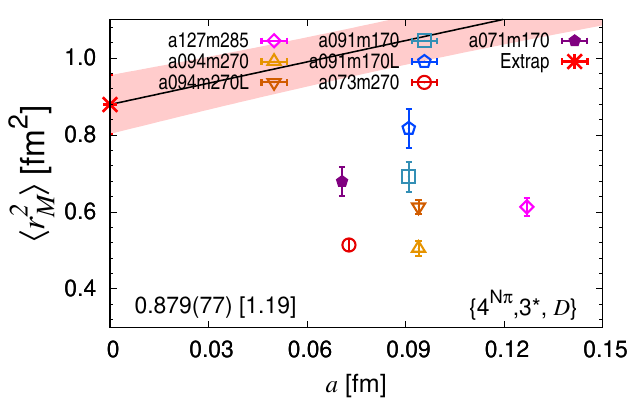}  
    \includegraphics[width=0.28\linewidth]{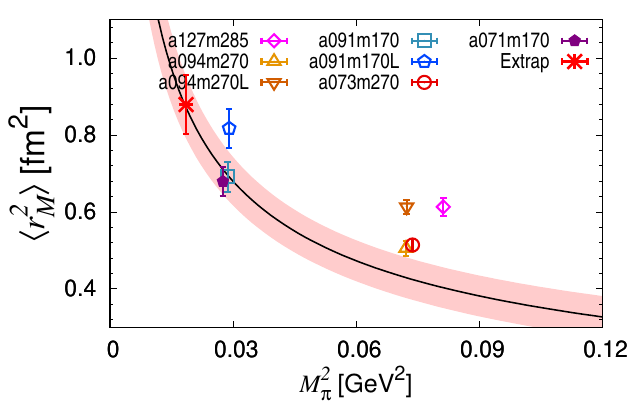}  
    \includegraphics[width=0.28\linewidth]{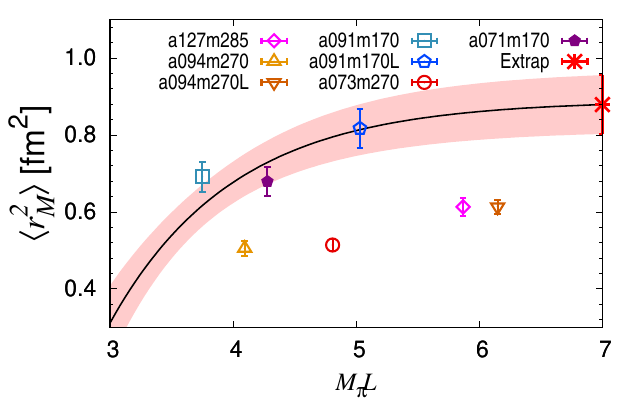}  
}
{
    \includegraphics[width=0.28\linewidth]{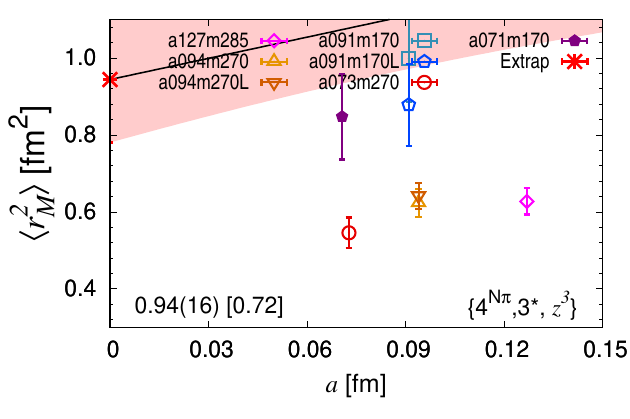}  
    \includegraphics[width=0.28\linewidth]{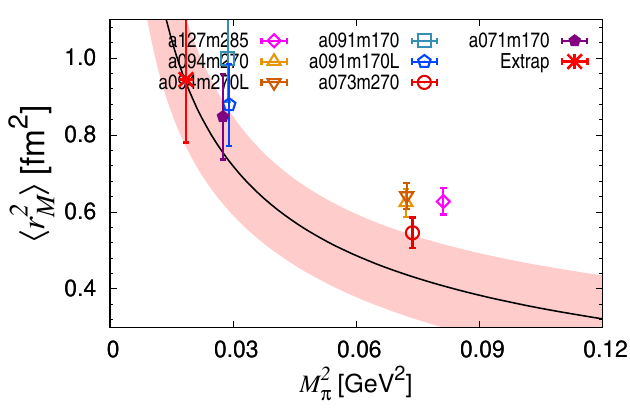}  
    \includegraphics[width=0.28\linewidth]{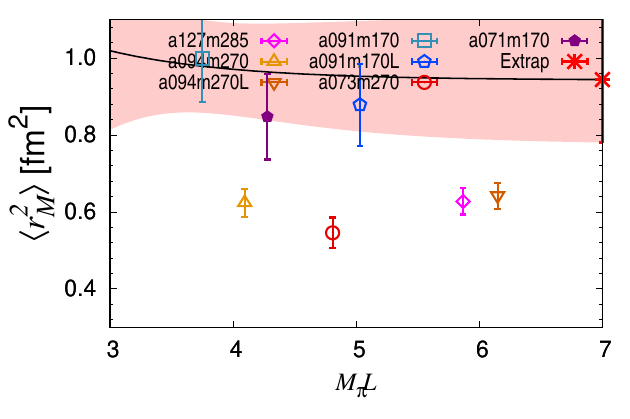}  
}
{
    \includegraphics[width=0.28\linewidth]{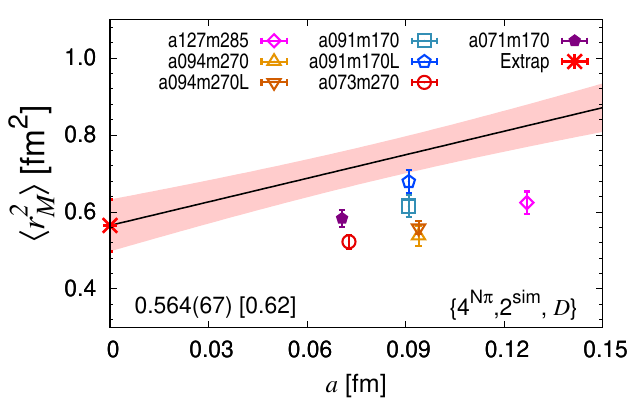}  
    \includegraphics[width=0.28\linewidth]{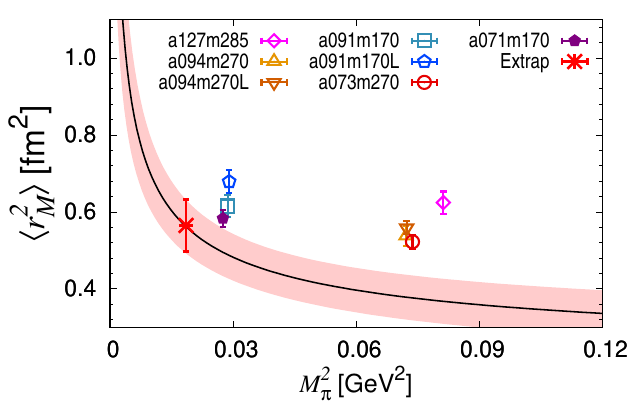}  
    \includegraphics[width=0.28\linewidth]{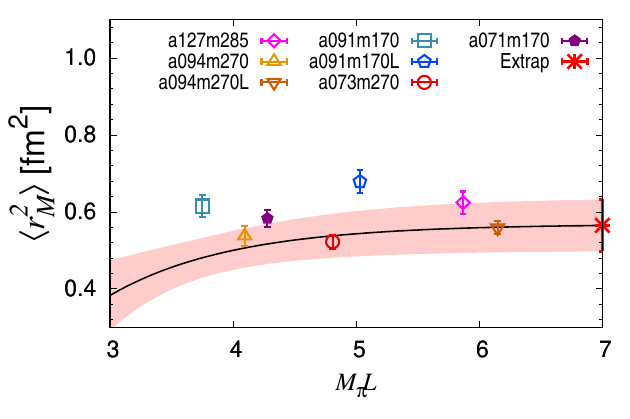}  
}
{
    \includegraphics[width=0.28\linewidth]{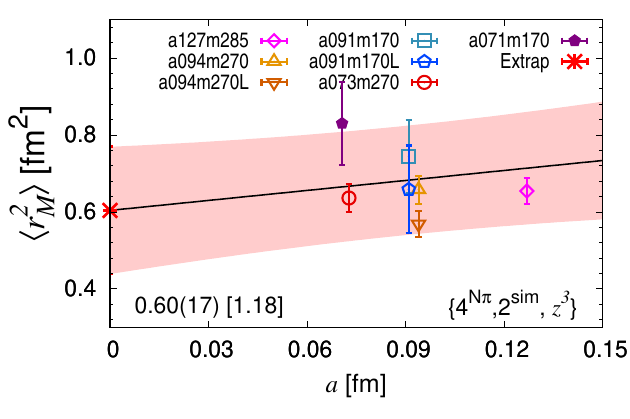}  
    \includegraphics[width=0.28\linewidth]{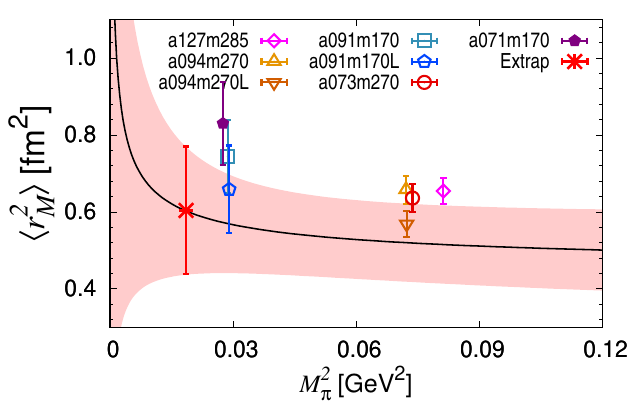}  
    \includegraphics[width=0.28\linewidth]{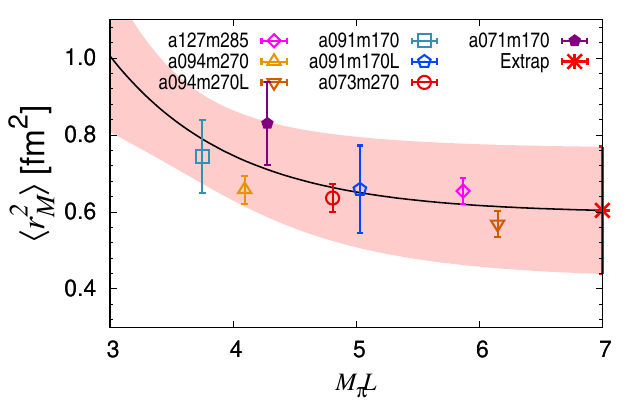}  
}
\caption{The CCFV extrapolation of the magnetic charge radius squared,
  $\langle r_M^2 \rangle$. A prior for $G_M(0)$ was used when making
  the dipole, $P_2$ and $z^3$ fits to parameterize the $Q^2$
  dependence as explained in the text. The rest is the same as in
  Fig.~\protect\ref{fig:CCFV-VFF-rE}.
  \label{fig:CCFV-VFF-rM}}
\end{figure*}

\begin{figure*}[tbp] 
\subfigure
{
    \includegraphics[width=0.28\linewidth]{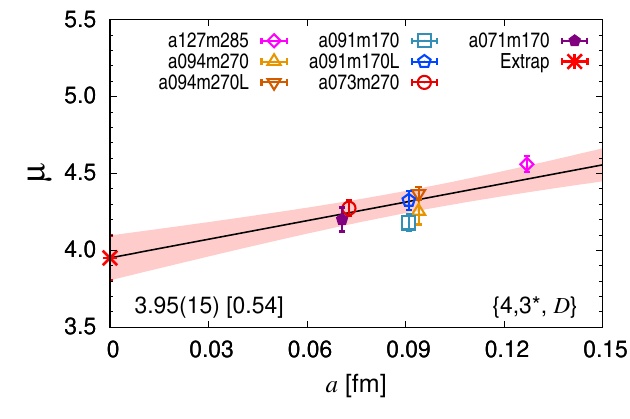}  
    \includegraphics[width=0.28\linewidth]{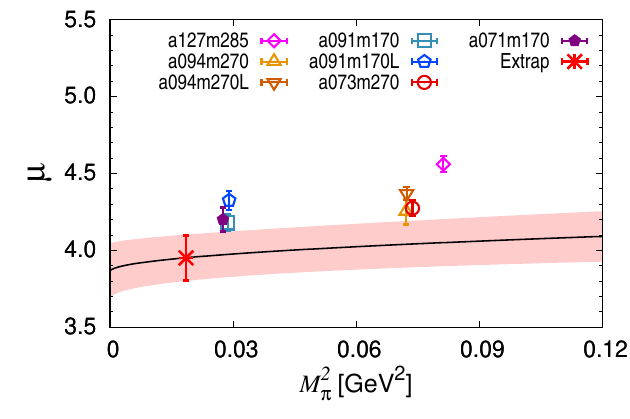}  
    \includegraphics[width=0.28\linewidth]{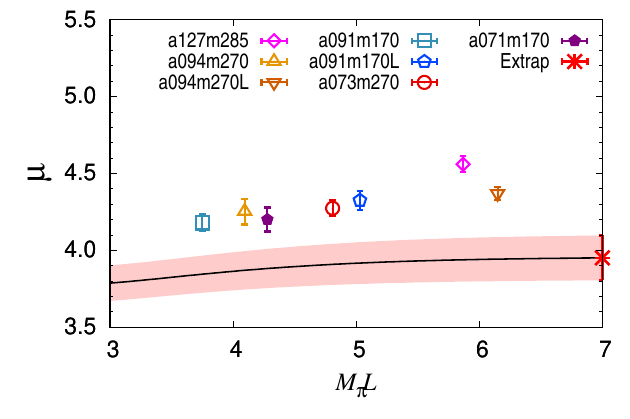}  
}
{
    \includegraphics[width=0.28\linewidth]{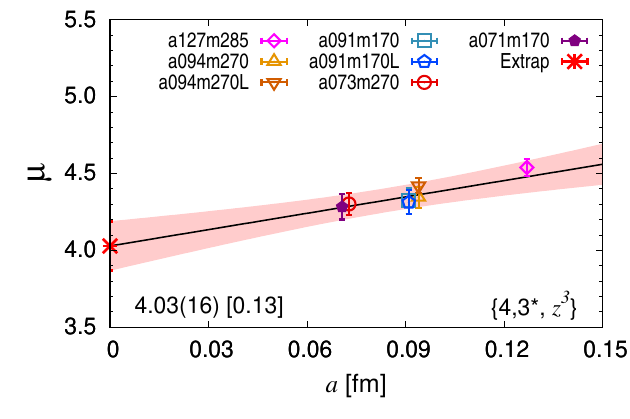}  
    \includegraphics[width=0.28\linewidth]{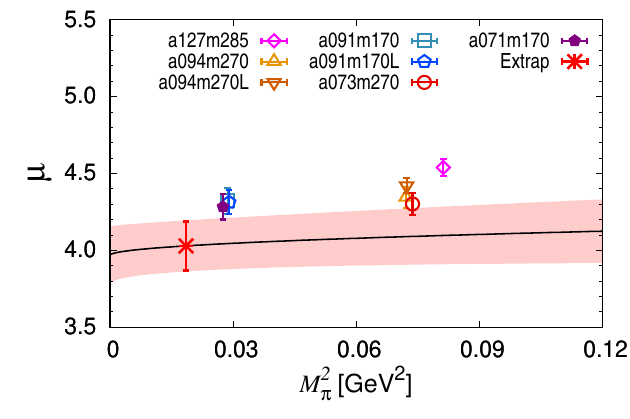}  
    \includegraphics[width=0.28\linewidth]{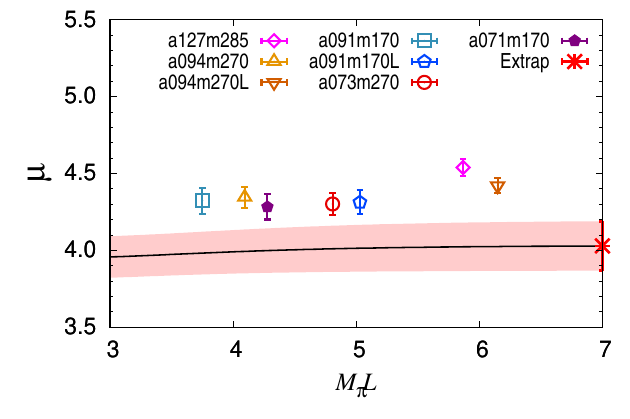}  
}
{
    \includegraphics[width=0.28\linewidth]{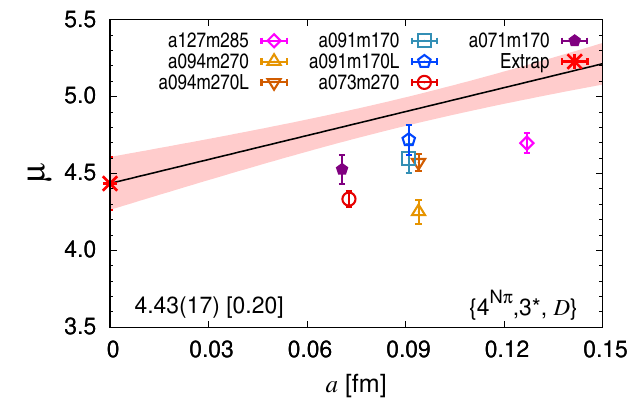}  
    \includegraphics[width=0.28\linewidth]{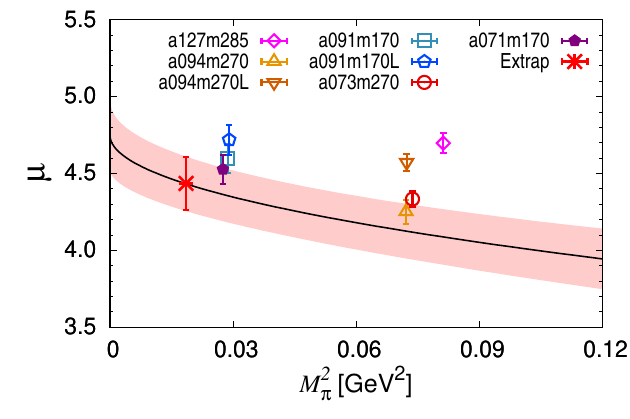}  
    \includegraphics[width=0.28\linewidth]{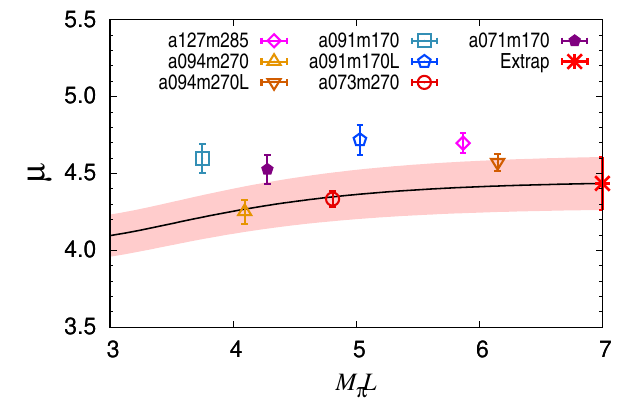}  
}
{
    \includegraphics[width=0.28\linewidth]{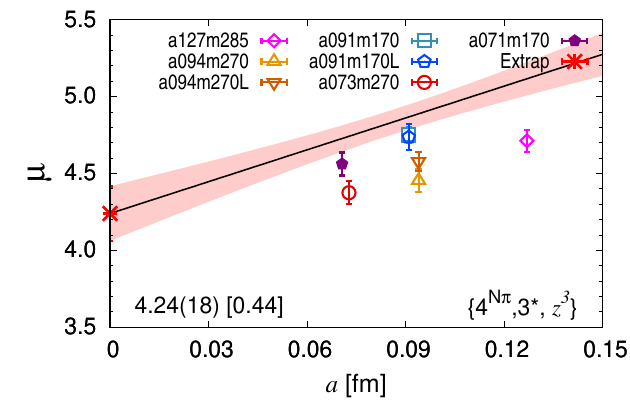}  
    \includegraphics[width=0.28\linewidth]{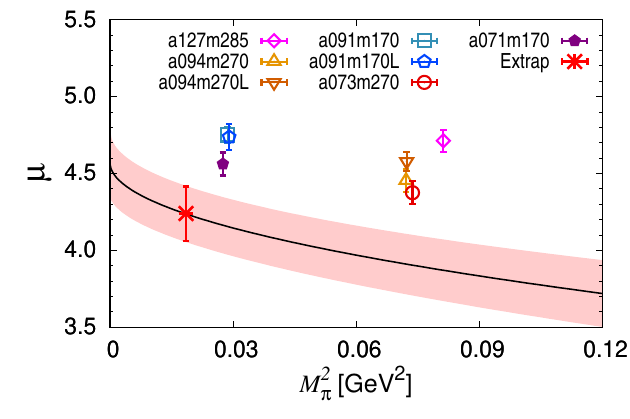}  
    \includegraphics[width=0.28\linewidth]{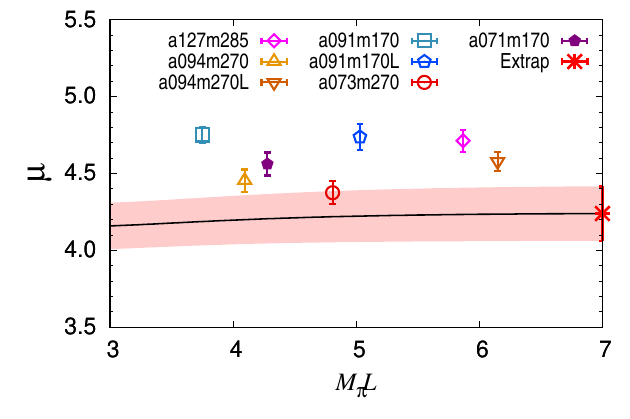}  
}
{
    \includegraphics[width=0.28\linewidth]{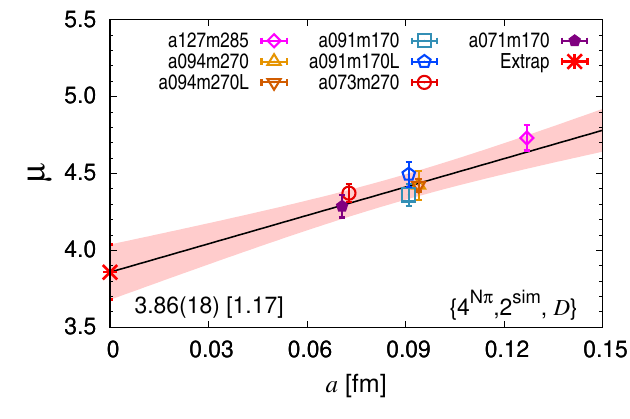}  
    \includegraphics[width=0.28\linewidth]{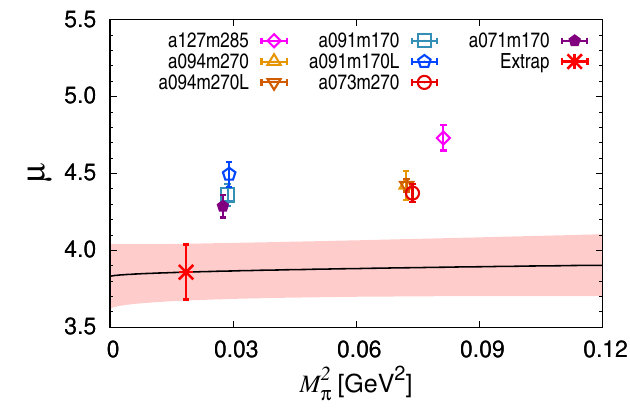}  
    \includegraphics[width=0.28\linewidth]{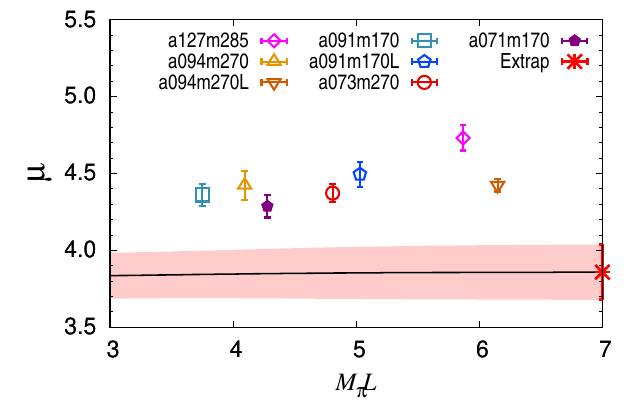}  
}
{
    \includegraphics[width=0.28\linewidth]{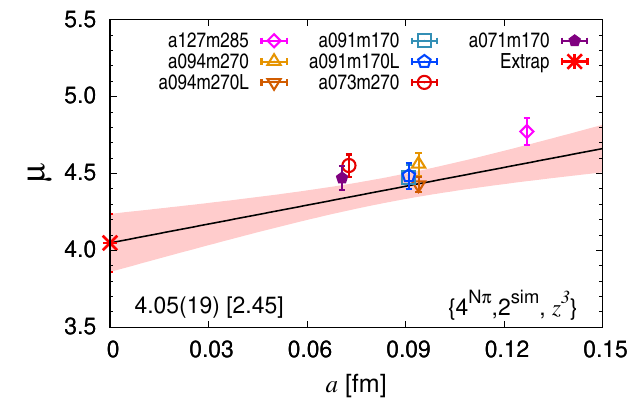}  
    \includegraphics[width=0.28\linewidth]{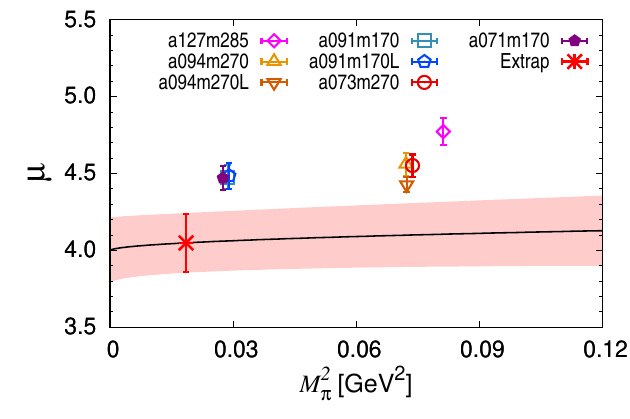}  
    \includegraphics[width=0.28\linewidth]{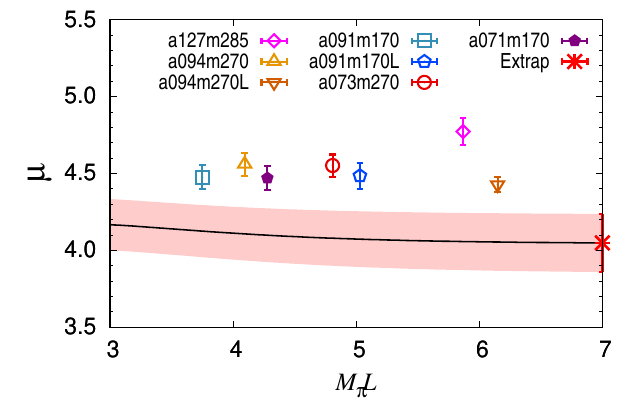}  
}
\caption{The CCFV extrapolation of the isovector magnetic dipole
  moment, $\mu^{p-n}$. A prior for $G_M(0)$ was used when making the
  dipole, $P_2$  and $z^3$ fits to parameterize the $Q^2$ dependence
  as explained in the text. The rest is the same as in
  Fig.~\protect\ref{fig:CCFV-VFF-rE}
  \label{fig:CCFV-VFF-mu}}
\end{figure*}

This appendix contains the figures showing the  CCFV fits 
made  to get the results at the physical point for various analysis 
strategies. Figures~\ref{fig:CCFVgA}--\ref{fig:CCFV-VFF-mu} show 
the data and fits for the three isovector
charges, $g_{A,S,T}^{u-d}$; the axial charge radius squared, $\langle
r_A^2 \rangle$; the induced pseudoscalar charge $g_P^\ast|_{{\rm Z}_2}$; the
pion-nucleon coupling $g_{\pi N N}|_{{\rm Z}_2}$; the product $M_N g_A /
F_\pi$; the pion decay constant, $F_\pi$; the electric and magnetic
charge radius squared, $\langle r_E^2 \rangle$ and $\langle r_M^2
\rangle$; and the magnetic moment, $\mu^{p-n}$, respectively. The extraction of 
the final results from the set of CCFV fits and the 
assessment of additional systematic 
uncertainties is presented in Sec.~\ref{sec:CCFV}.  


\onecolumngrid\hrule width0pt\twocolumngrid\cleardoublepage

\section{Variance-covariance matrices of the fits}
\label{sec:covar}

The fits versus \(Q^2/4M_N^2\) and $z$ presented in Sec.~\ref{sec:FFfits} 
and the errors on the fit parameters, calculated by propagating the errors on individual points, were done using {\tt lsqfit} \cite{lsqfit}, which calls  {\tt multifit} from the GNU scientific library~\cite{galassi2001gnu}, and {\tt gvar} \cite{gvar} routines.  In this appendix we provide the sampling variance-covariance matrices of the fits. The errors quoted in the {main text in Sec.~\ref{sec:FFfits}} are the square roots of the diagonal elements of these matrices.  

The variance-covariance matrices for the $\{4^{N\pi},2^{\rm sim},{\widehat P}_2\}$ fit to \(G_A\) given in Eq.~\eqref{eq:GAPade}  is
\begin{equation}
\begin{blockarray}{cccc}
&g_A&b_0&b_1\\
\begin{block}{l(rrr)}
g_{A\ }&1.184\times10^{-4}&1.507\times10^{-3}&-4.499\times10^{-3}\\
b_{0\ }&1.507\times10^{-3}&3.898\times10^{-2}&-1.419\times10^{-1}\\
b_{1\ }&-4.499\times10^{-3}&-1.419\times10^{-1}&6.631\times10^{-1}\\
\end{block}\\
\end{blockarray}\,,
\label{eq:GAcovPade}
\end{equation}
and for the $\{4^{N\pi},2^{\rm sim},{\widehat z}^2\}$ fit to \(G_A\) given in Eq.~\eqref{eq:GAz2} and using the notation given in 
Eq.~\eqref{eq:Zexpansion} is
\begin{equation}
\begin{blockarray}{cccc}
&a_0&a_1&a_2\\
\begin{block}{l(rrr)}
a_{0\ }&2.188\times10^{-5}&-2.238\times10^{-5}&-1.155\times10^{-4}\\
a_{1\ }&-2.238\times10^{-5}&8.549\times10^{-4}&2.769\times10^{-3}\\
a_{2\ }&-1.155\times10^{-4}&2.769\times10^{-3}&1.811\times10^{-2}\\
\end{block}\\
\end{blockarray}\,.
\label{eq:GAcovz2}
\end{equation}

\medskip

The variance-covariance matrix for the $\{4^{N\pi},3^\ast,{\widehat P}_2\}$  fit to \(G_E\) and \(G_M\) given in Eq.~\eqref{eq:GEMPade} are
\begin{equation}
\begin{blockarray}{cccc}
&g_V&b_0&b_1\\
\begin{block}{l(rrr)}
g_V &2.782\times10^{-5}&1.155\times10^{-3}&-5.206\times10^{-3}\\
b_{0\ }&1.155\times10^{-3}&8.260\times10^{-2}&-3.912\times10^{-1}\\
b_{1\ }&-5.206\times10^{-3}&-3.912\times10^{-1}&3.791\hphantom{{}\times10^{-0}}\\
\end{block}\\
\end{blockarray}\ 
\label{eq:GEcovPade}
\end{equation}
and
\begin{equation}
\begin{blockarray}{cccc}
&\mu&b_0&b_1\\
\begin{block}{l(rrr)}
\mu   &2.271\times10^{-3}&1.413\times10^{-2}&-3.931\times10^{-2}\\
b_{0\ }&1.413\times10^{-2}&1.238\times10^{-1}&-4.836\times10^{-1}\\
b_{1\ }&-3.931\times10^{-2}&-4.836\times10^{-1}&3.157\hphantom{{}\times10^{-0}}\\
\end{block}\\
\end{blockarray}\,,
\label{eq:GMcovPade}
\end{equation}
\begin{widetext}
respectively, and for the $\{4^{N\pi},3^\ast,{\widehat z}^3\}$ fit are
\begin{equation}
\begin{blockarray}{ccccc}
&a_0&a_1&a_2&a_3\\
\begin{block}{l(rrrr)}
a_{0\ \ }&6.5178\times10^{-6\quad }&1.0374\times10^{-5\quad }&5.9814\times10^{-6\quad }&1.8014\times10^{-5}\\
a_{1\ \  }&1.0374\times10^{-5\quad }&8.1815\times10^{-4\quad }&4.6176\times10^{-3\quad }&7.3053\times10^{-3}\\
a_{2\ \ }&5.9814\times10^{-6\quad }&4.6176\times10^{-3\quad }&3.4227\times10^{-2\quad }&6.2428\times10^{-2}\\
a_{3\ \ }&1.8014\times10^{-5\quad }&7.3053\times10^{-3\quad }&6.2428\times10^{-2\quad }&1.2394\times10^{-1}\\
\end{block}\\
\end{blockarray}
\label{eq:GEcovz3}
\end{equation}
and
\begin{equation}
\begin{blockarray}{ccccc}
&a_0&a_1&a_2&a_3\\
\begin{block}{l(rrrr)}
a_{0\ \ }&1.2656\times10^{-4\quad }&3.8253\times10^{-4\quad }&9.5665\times10^{-4\quad }&1.0035\times10^{-3}\\
a_{1\ \  }&3.8253\times10^{-4\quad }&1.9869\times10^{-2\quad }&1.3524\times10^{-1\quad }&2.5216\times10^{-1}\\
a_{2\ \ }&9.5665\times10^{-4\quad }&1.3524\times10^{-1\quad }&1.3230\hphantom{{}\times10^{-0\quad }}&2.7857\hphantom{{}\times10^{-0}}\\
a_{3\ \ }&1.0035\times10^{-3\quad }&2.5216\times10^{-1\quad }&2.7857\hphantom{{}\times10^{-0\quad }}&6.4023\hphantom{{}\times10^{-0}}\\
\end{block}\\
\end{blockarray}\,,
\label{eq:GMcovz3}
\end{equation}
respectively.
\end{widetext}

\bibliographystyle{apsrev4-1} 
\bibliography{ref} 

\end{document}